\documentclass[superscriptaddress,floatfix]{revtex4}
\usepackage{graphicx}
\usepackage{xspace}
\usepackage{color}
\usepackage{lineno}

\newcommand{\pT} {$p_{T}$\xspace}
\newcommand{\sqrts} {$\sqrt{s_{_{NN}}}$\xspace}

\newcommand{\raa}{R_{\rm AA}}

\newcommand{\piplus}{$\pi^{+}$}
\newcommand{\piminus}{$\pi^{-}$}
\newcommand{\pinull}{$\pi^{0}$}

\newcommand {\pp}      {\mbox{p+p}\xspace}

\begin{document}
\title{An Experimental Exploration of the QCD Phase Diagram: The Search for the Critical Point and the Onset of De-confinement}

\affiliation{Argonne National Laboratory, Argonne, Illinois 60439, USA}
\affiliation{University of Birmingham, Birmingham, United Kingdom}
\affiliation{Brookhaven National Laboratory, Upton, New York 11973, USA}
\affiliation{University of California, Berkeley, California 94720, USA}
\affiliation{University of California, Davis, California 95616, USA}
\affiliation{University of California, Los Angeles, California 90095, USA}
\affiliation{Universidade Estadual de Campinas, Sao Paulo, Brazil}
\affiliation{University of Illinois at Chicago, Chicago, Illinois 60607, USA}
\affiliation{Creighton University, Omaha, Nebraska 68178, USA}
\affiliation{Czech Technical University in Prague, FNSPE, Prague, 115 19, Czech Republic}
\affiliation{Nuclear Physics Institute AS CR, 250 68 \v{R}e\v{z}/Prague, Czech Republic}
\affiliation{University of Frankfurt, Frankfurt, Germany}
\affiliation{Institute of Physics, Bhubaneswar 751005, India}
\affiliation{Indian Institute of Technology, Mumbai, India}
\affiliation{Indiana University, Bloomington, Indiana 47408, USA}
\affiliation{Alikhanov Institute for Theoretical and Experimental Physics, Moscow, Russia}
\affiliation{University of Jammu, Jammu 180001, India}
\affiliation{Joint Institute for Nuclear Research, Dubna, 141 980, Russia}
\affiliation{Kent State University, Kent, Ohio 44242, USA}
\affiliation{University of Kentucky, Lexington, Kentucky, 40506-0055, USA}
\affiliation{Institute of Modern Physics, Lanzhou, China}
\affiliation{Lawrence Berkeley National Laboratory, Berkeley, California 94720, USA}
\affiliation{Massachusetts Institute of Technology, Cambridge, MA 02139-4307, USA}
\affiliation{Max-Planck-Institut f\"ur Physik, Munich, Germany}
\affiliation{Michigan State University, East Lansing, Michigan 48824, USA}
\affiliation{Moscow Engineering Physics Institute, Moscow Russia}
\affiliation{City College of New York, New York City, New York 10031, USA}
\affiliation{NIKHEF and Utrecht University, Amsterdam, The Netherlands}
\affiliation{Ohio State University, Columbus, Ohio 43210, USA}
\affiliation{Old Dominion University, Norfolk, VA, 23529, USA}
\affiliation{Panjab University, Chandigarh 160014, India}
\affiliation{Pennsylvania State University, University Park, Pennsylvania 16802, USA}
\affiliation{Institute of High Energy Physics, Protvino, Russia}
\affiliation{Purdue University, West Lafayette, Indiana 47907, USA}
\affiliation{Pusan National University, Pusan, Republic of Korea}
\affiliation{University of Rajasthan, Jaipur 302004, India}
\affiliation{Rice University, Houston, Texas 77251, USA}
\affiliation{Universidade de Sao Paulo, Sao Paulo, Brazil}
\affiliation{University of Science \& Technology of China, Hefei 230026, China}
\affiliation{Shandong University, Jinan, Shandong 250100, China}
\affiliation{Shanghai Institute of Applied Physics, Shanghai 201800, China}
\affiliation{SUBATECH, Nantes, France}
\affiliation{Texas A\&M University, College Station, Texas 77843, USA}
\affiliation{University of Texas, Austin, Texas 78712, USA}
\affiliation{Tsinghua University, Beijing 100084, China}
\affiliation{United States Naval Academy, Annapolis, MD 21402, USA}
\affiliation{Valparaiso University, Valparaiso, Indiana 46383, USA}
\affiliation{Variable Energy Cyclotron Centre, Kolkata 700064, India}
\affiliation{Warsaw University of Technology, Warsaw, Poland}
\affiliation{University of Washington, Seattle, Washington 98195, USA}
\affiliation{Wayne State University, Detroit, Michigan 48201, USA}
\affiliation{Institute of Particle Physics, CCNU (HZNU), Wuhan 430079, China}
\affiliation{Yale University, New Haven, Connecticut 06520, USA}
\affiliation{University of Zagreb, Zagreb, HR-10002, Croatia}

\author{M.~M.~Aggarwal}\affiliation{Panjab University, Chandigarh 160014, India}
\author{Z.~Ahammed}\affiliation{Variable Energy Cyclotron Centre, Kolkata 700064, India}
\author{A.~V.~Alakhverdyants}\affiliation{Joint Institute for Nuclear Research, Dubna, 141 980, Russia}
\author{I.~Alekseev~~}\affiliation{Alikhanov Institute for Theoretical and Experimental Physics, Moscow, Russia}
\author{B.~D.~Anderson}\affiliation{Kent State University, Kent, Ohio 44242, USA}
\author{D.~Arkhipkin}\affiliation{Brookhaven National Laboratory, Upton, New York 11973, USA}
\author{G.~S.~Averichev}\affiliation{Joint Institute for Nuclear Research, Dubna, 141 980, Russia}
\author{J.~Balewski}\affiliation{Massachusetts Institute of Technology, Cambridge, MA 02139-4307, USA}
\author{L.~S.~Barnby}\affiliation{University of Birmingham, Birmingham, United Kingdom}
\author{S.~Baumgart}\affiliation{Yale University, New Haven, Connecticut 06520, USA}
% ...

\collaboration{STAR Collaboration}\noaffiliation 
\begin{abstract}\end{abstract}\maketitle

\newpage
\newpage

\tableofcontents
\newpage

\section{Introduction}\label{introduction}

\noindent    RHIC has uncovered an exciting new state of matter, which has partonic degrees of freedom, at \sqrts $\sim 62$ to 200 GeV, the strongly coupled Quark Gluon Plasma (sQGP),. The RHIC-wide consensus on the status of this discovery as of a few years ago is documented in the various white-papers from the RHIC experiments~\cite{Adams:2005dq,Adcox:2004mh,Back:2004je,Arsene:2004fa}.  In spite of impressive progress since 2001, as noted in the white-papers several of the most important questions that motivated the construction of RHIC are not yet fully answered, and maybe it is not a coincidence that the available RHIC parameter space has not yet been fully explored.

~

\noindent The QCD phase diagram lies at the heart of what the RHIC Physics Program is all about~\cite{Stephanov:2004wx,Mohanty:2009vb}. While RHIC has been operating very successfully at or close to its maximum energy for almost a decade, it has  become clear that this collider can also be operated at lower energies down to  \sqrts = 5 GeV without extensive upgrades. From this purely empirical perspective, an exploration of the full region of \sqrts available at the RHIC facility is surely imperative.  The STAR detector, due to its large uniform acceptance and excellent particle identification capabilities, is uniquely positioned to carry out this program in depth and detail. The first exploratory beam energy scan (BES) run at RHIC took place in 2010 (Run 10), since several  STAR upgrades, most importantly a full barrel Time of Flight detector,  are now completed which add new capabilities important for the interesting physics at BES energies.   Our results at top energies suggest that a new form of matter, the sQGP, is created and that it is locally equilibrated early-on because of its observed hydrodynamic expansion patterns. Also, hadrochemical species equilibrium is observed just after hadronization. It appears that the transition to this state, at these high temperatures and low $\mu_{B}$, is a crossover~\cite{Brown:1990ev}, i.e. a smooth, continuous transition from a QGP to hadrons.  Theoretical model calculations predict that at lower temperatures and high baryon chemical potentials this cross-over will become a first order phase transition~\cite{Asakawa:1989bq,Barducci:1989eu,Barducci:1989wi,Barducci:1993bh,Berges:1998rc,Halasz:1998qr,Scavenius:2000qd,Antoniou:2002xq,Hatta:2002sj} resulting in a critical point occurring at intermediate temperatures and baryon chemical potentials~\cite{Stephanov:2004wx}.  

~

\noindent The data taken via a RHIC BES will be used to explore several of the open questions in the field heavy-ion physics. Among the questions are:

\begin{enumerate}

\item Can we see evidence of a Critical Point (CP)?
\item Can we see evidence of a  phase transition?
\item What is the evolution with \sqrts of the medium that we produce? i.e. how do the results that  indicate the presence of the sQGP turn off as  \sqrts  is reduced?
\end{enumerate}

~

\begin{figure}[htbp]
\begin{center}
\includegraphics[width=0.5\textwidth]{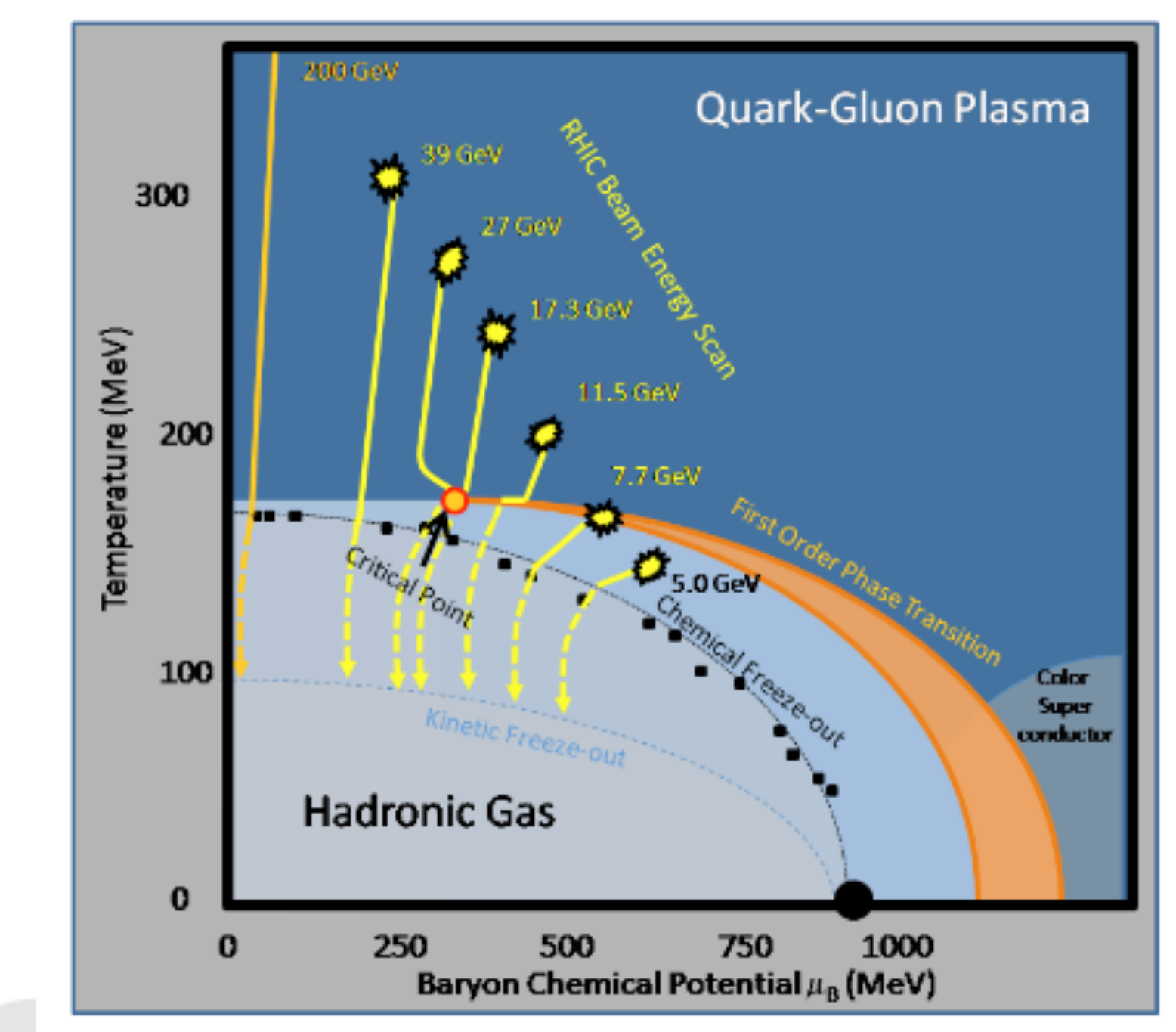}
\caption{A schematic of the  phase diagram of nuclear matter. The location of the CP is  placed  within the RHIC BES range. Lattice QCD estimates~\cite{Karsch:2003va,Fodor:2004nz,Gavai:2004sd} indicate that the CP falls within the interval 250 $<$$\mu_{B}$$<$ 450 MeV. The black closed circles are current heavy-ion experimental calculations of the chemical freeze-out temperature, T$_{{\rm ch}}$, and $\mu_{B}$ based on statistical model fits to the measured particle ratios. The yellow curves show the estimated trajectories of the  possible collision energies at RHIC. }
\label{Fig:PhaseDiagram}
\end{center}
\end{figure}

\noindent  At the forefront of this list is the search for evidence of a CP and/or its associated first order phase transition in the phase diagram of nuclear matter, Fig~\ref{Fig:PhaseDiagram}.  Of course, for any  discussion about a phase diagram to be valid, we also have to conclusively answer another of the above questions: whether collisions at RHIC form a thermodynamic state? While recent progress in lattice QCD and model calculations is indeed impressive, the location of phase boundaries between hadronic gas and the sQGP  and the exact position, in T and $\mu_{B}$, of the hypothesized critical point remain unknown ~\cite{Karsch:2003va,Fodor:2004nz,Gavai:2004sd}. It therefore falls upon the experiments  to find observational evidence of its existence.  However, the available theoretical estimates indicate that the critical point might be in the region of the phase diagram probed by heavy ion experiments (see for example ~\cite{Fodor:2009ax,deForcrand:2006pv,Gavai:2008zr,Schmidt:2008cf}), and in particular within the collision energy range probable at RHIC.  The proposed Beam Energy Scan (BES) from \sqrts=5-200 GeV at RHIC is motivated to a considerable extent by this exciting  possibility of uncovering evidence of a critical point and/or its associated first order phase transition line. \footnote{The BES had a very successful start in run 10 with data being recorded at \sqrts = 62, 39, 11.5 and 7.7 GeV.}

~

\noindent Theory has predicted several signatures of a first order phase transition and for the  CP (see for example ~\cite{Stephanov:1998dy, Gazdzicki:2003bb, Gorenstein:2003hk, Ollitrault:1992bk, Kolb:2000sd, Shuryak:2005vk, Bratkovskaya:2004nd} and references therein). Most of these result in increased fluctuations  when the freeze-out trajectory passes through the CP or strong variations as a function of \sqrts if the collision energies sampled encompass  the range in $\mu_{B}$ where the CP occurs~\cite{Stephanov:1998dy,Ejiri:2005wq}.  However, the magnitude of these oscillations and the probability of their survival through the final re-scatterings in the hadronic state have yet to be rigorously calculated. There is also the question of hard and semi-hard processes obscuring the signals of the CP. Some hydrodynamical calculations suggest that the CP acts as an attractor~\cite{Asakawa:2008ti}. This means that as long as the thermalized medium produced has initial conditions  close to the CP, as  it evolves with time its trajectory through the T, $\mu_{B}$ phase space will be focussed  towards the CP. Such a focussing effect has already been discussed in the context of liquid-gas nuclear transitions~\cite{Stephanov:1999zu}. This attraction means that the impact due to the  inability of the  theoretical calculations to pin-point  the CP's exact location is minimized, at least experimentally, as we only ``have to get close" to the correct collision energy to ``land" at the CP. A finite extension of the critical domain in the $\mu_{B}$, T diagram implies that there are unlikely to be any sharp discontinuities.  The size of the correlation lengths, and thus fluctuations, are also restricted, due to the finite system size effects, to $\lesssim $ 6 fm. A phenomenon known as ``critical slowing down" is predicted to cause the correlation lengths to be at most 2 fm~\cite{Berdnikov:1999ph, Asakawa:2005hw}. Despite these  problems in precise predictions it is imperative that RHIC lead the way in attempting to find the CP.

~

\noindent Establishing the validity of  the CP prediction, or even bounding the region where it is sited by proving the existence of both a cross-over {\it AND} a first/second order transition, would place RHIC results into text books.  Such a result would be as seminal as proving the formation of the sQGP.

~

\noindent In the following sections we describe: 

\begin{enumerate}
\item   Current proposed measurements, with estimations of the accuracy of the measurements given an assumed event count at each \sqrts. These measurements include novel extensions to those already made at the SPS. These are only possible due to the improved coverage,  particle identification (PID) and proposed improved statistics over the previous experiments.
\item The preliminary results from analysis of the data from the RHIC low energy test runs.
\item The available data to use  as the p+p baselines.
\item The STAR detector including newly  available upgrades pertinent for the energy scan.  This section includes detailed discussion of our particle identification abilities, and our particle acceptance and reconstruction  efficiencies as a function of  multiplicity and  \pT.
\item A summary including our proposed running plan
 
\end{enumerate}

\noindent We conclude with a run plan for data taking that would allow us to accomplish all our initial  goals, including those mentioned above which are not directly related to the CP search. Should  evidence of the CP, first order phase transition, or other unexpected results be observed we would then  propose further dedicated running at, and bracketing, the relevant collision energies.

\section{Proposed Physics Measurements}\label{Sec:Measurements}

\noindent The physics to be extracted from the BES have been split  into five  subsections.  First we focus on those measurements that seem most likely to provide evidence of the Critical Point. These are predominantly fluctuation measures. The second subsection covers analyses designed to instead provide evidence for a first order phase transition. The signatures of the sQGP are covered in sub-section three followed by a sub-section on particle production studies designed, in the main, to probe if the medium produced in these collisions is indeed in thermal equilibrium. In the fifth and final subsection the study of a potentially sensitive measure of local strong parity violation is presented.

~

\noindent When discussing performing an energy scan at RHIC the question often arises: why embark on such a scan when data has already been taken at the SPS? This section   explains how the scan at RHIC will improve on the data taken at CERN. Firstly such a program will enable measurements to be made from the SPS energy region on up to the top RHIC energies with the same detector. Secondly, since this is a collider experiment the same uniform acceptance will occur at each energy point. Finally, not only will the statistics be better due to the higher acceptance of the STAR detector, but also, there will be cleaner, and more
extensive PID capabilities.  This  allows us to not only repeat those measurements performed at the SPS in much finer detail but also to enhance the studies by performing numerous differential measures. It is most likely only by looking differentially  that the signals produced by passing close to the CP will be extracted.  When one is forced to integrate measures, due, for instance to small statistics or limited acceptance, much information is lost. 

~

\noindent Many of the potential signals proposed to be resulting from the CP  and/or a first order phase transition revolve around fluctuation measures. Discussed below are the measures we currently  propose.  Of course there is still the possibility of performing other studies that emerge from more details discussions and continued development of the theory.

\subsection{Locating the Critical Point}\label{SubSec:CriticalPoint}

\subsubsection{Fluctuation Measures}\label{SubSubSec:Fluctations}

\begin{figure}[htb]
	\begin{minipage}{0.46\linewidth}
		\begin{center}
			\includegraphics[width=0.8\linewidth]{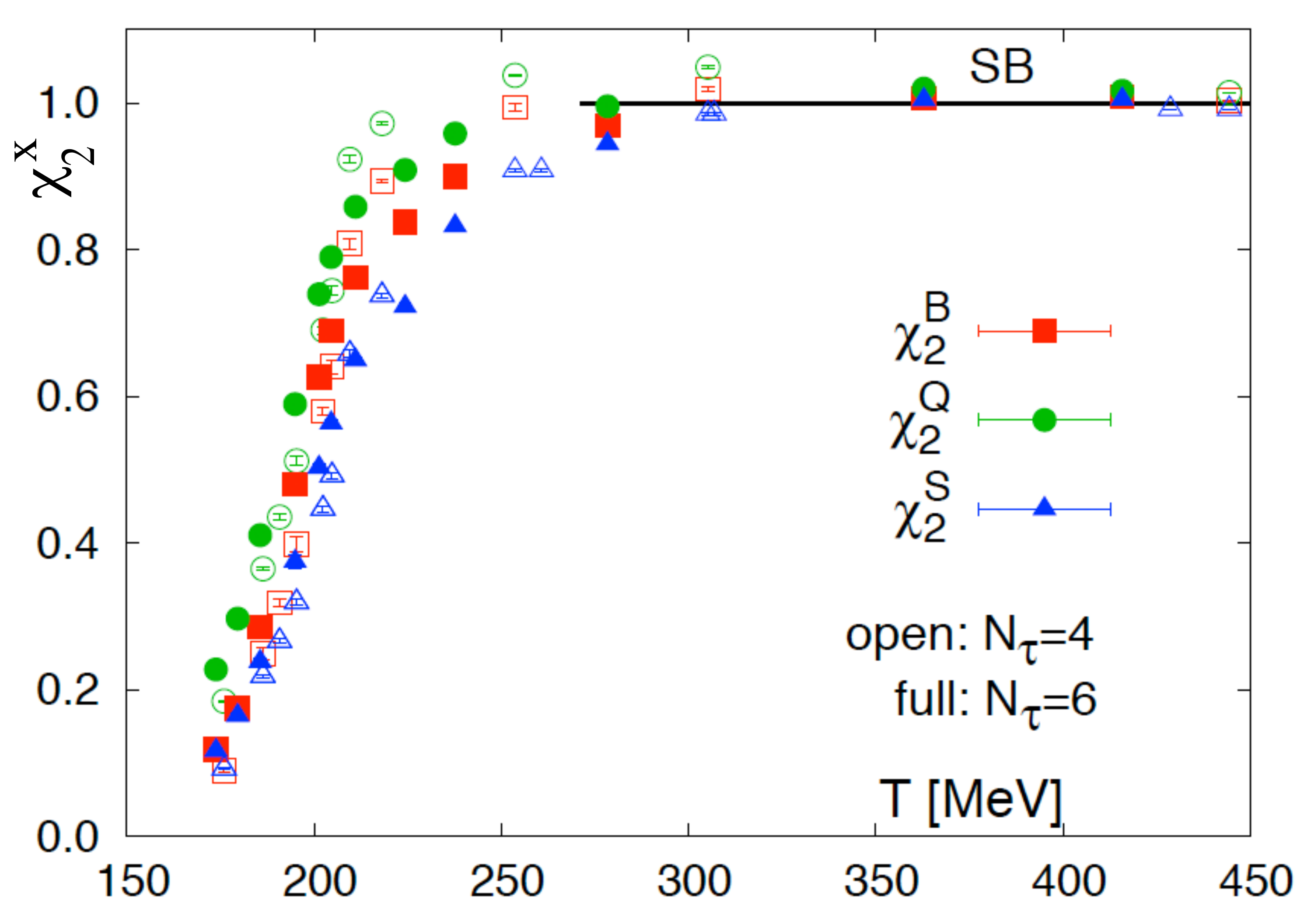}
			\caption{Quadratic fluctuations of baryon number, electric charge and strangeness. All quantities have been normalized to the corresponding free quark gas values and are for $\mu_{B}$=0 and T$_{C}$=200 MeV~\cite{Cheng:2008zh}.}
			\label{Fig:LatticeQCD1}
		\end{center}
	\end{minipage}
	\hspace{1cm}
	\begin{minipage}{0.46\linewidth}
		\begin{center}
			\includegraphics[width=0.8\linewidth]{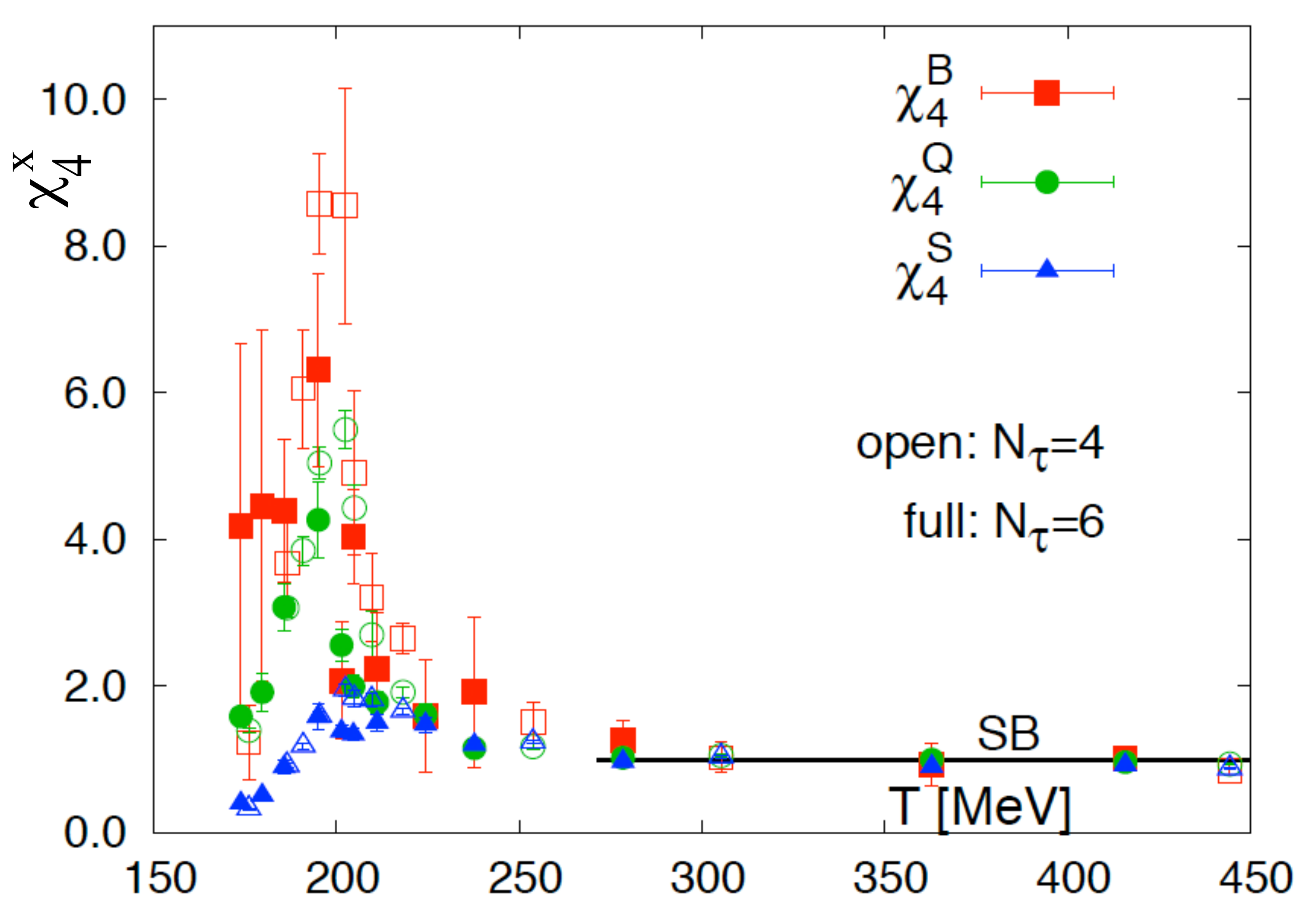}
			\caption{Quartic fluctuations of baryon number, electric charge and strangeness. All quantities have been normalized to the corresponding free quark gas values and are for $\mu_{B}$=0 and T$_{C}$=200 MeV~\cite{Cheng:2008zh}.}
			\label{Fig:LatticeQCD2}
		\end{center}
	\end{minipage}
\end{figure}

 \noindent The characteristic signature of the existence of a CP is an increase of fluctuations ~\cite{Koch:2008ia}.  For instance, Lattice  QCD calculations \cite{Cheng:2008zh} indicate large fluctuations in the derivatives of the partition functions with respect to baryon, charge,  and strangeness chemical potentials as a function of the temperature of the system.   Of particular interest are the moments of the charge/baryon number/strangeness fluctuations which are obtained theoretically from the second (quadratic) and fourth (quartic)  derivatives of the logarithm of the QCD partition function ($\chi_2$ and $\chi_4$ respectively), Fig.~\ref{Fig:LatticeQCD1} and Fig.~\ref{Fig:LatticeQCD2}.  These fluctuations can be related to event-by-event moments of various observables in heavy-ion collisions. Fig.~\ref{Fig:LatticeQCD1} and Fig.~\ref{Fig:LatticeQCD2} are calculated for $\mu_B$ = 0, such fluctuations are expected to diverge at the Critical Point.
 
 ~
 
 \noindent Measures of particular interest in STAR are fluctuations in the $\langle p_{T} \rangle$,  the  K/$\pi$, p/$\pi$, and K/p ratios,  and v$_{2}$. Also of interest are the high moments of the net-protons.

\subsubsection*{$\langle p_{T}\rangle$ Fluctuations}\label{SubSubSec:pTFluc}

\begin{figure}[htb]
\begin{minipage}{0.46\linewidth}
\begin{center}
\includegraphics[width=0.9\linewidth]{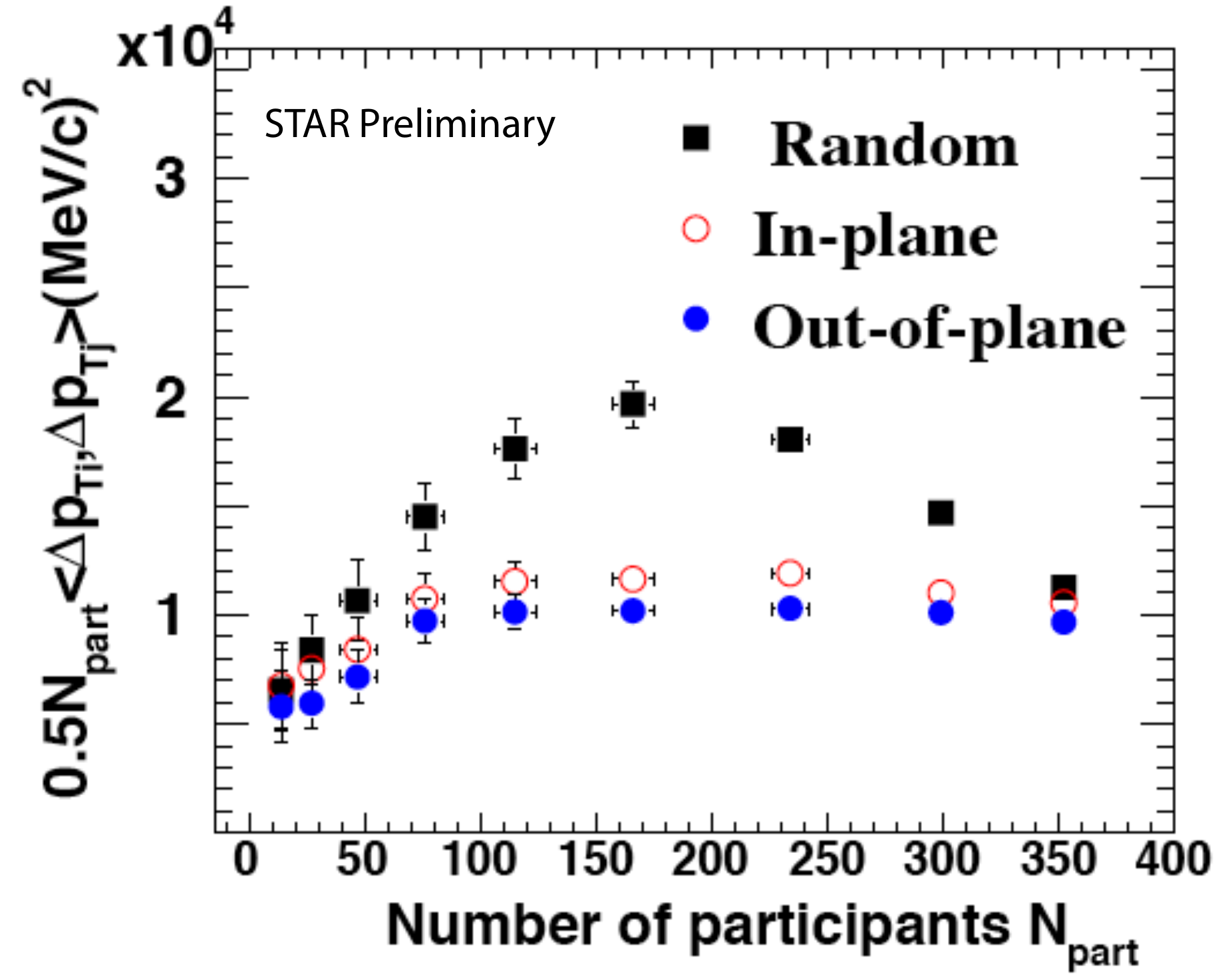}
\end{center}
\caption{The $\langle p_{T}\rangle$ fluctuations as a function of N$_{part}$ due to the presence of elliptic flow. The data are from  Au+Au collisions at \sqrts=200 GeV. From ~\cite{Voloshin:2005aa}.}
\label{Fig:v2pTFluct}
\end{minipage}
\hspace{1cm}
\begin{minipage}{0.46\linewidth}
\begin{center}
\includegraphics[width=0.8\linewidth]{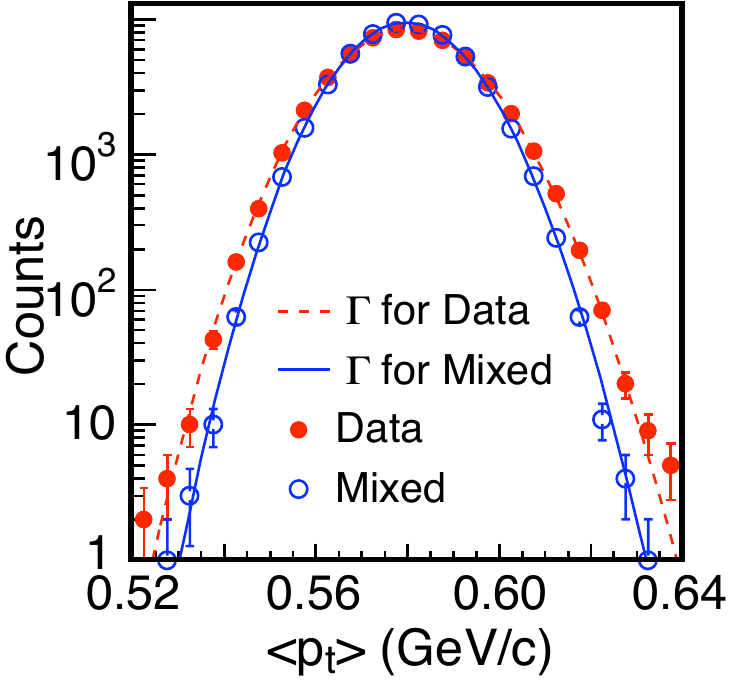}
\end{center}
\caption{The measured event-by-event $\langle p_{T}\rangle$ for data (red solid circles), and from event mixing (blue open circles) the fit to the data is shown as the red dashed curve, while that for the mixed events is the blue solid curve. The data is from \sqrts =200 GeV Au+Au collisions. From \cite{Adams:2005ka}.}
\label{Fig:MeanpTGauss}
\end{minipage}
\end{figure}

\noindent $\langle p_{T} \rangle$  fluctuations are  challenging as there are a number of effects that can swamp the signal. For instance elliptic flow can cause a non-statistical fluctuation of the $\langle p_{T}\rangle$ if the experiment does not have 2$\pi$ acceptance. Since the plane of the collision varies event to event if the acceptance is limited one is forced to measure at a random angle to the event plane each event. The  $\langle p_{T}\rangle$ in the plane is expected larger than the  $\langle p_{T}\rangle$ out of the plane, therefore while the \pT fluctuations in and out of the plane are independently small the average of a random angle creates an artificially large apparent fluctuation. This is illustrated in Fig.~\ref{Fig:v2pTFluct}, and shows that for mid-peripheral collisions fluctuations become very significant for Au-Au collisions at \sqrts=200 GeV.

\begin{figure}[htb]
\begin{minipage}{0.46\linewidth}
\begin{center}
\includegraphics[width=0.9\linewidth]{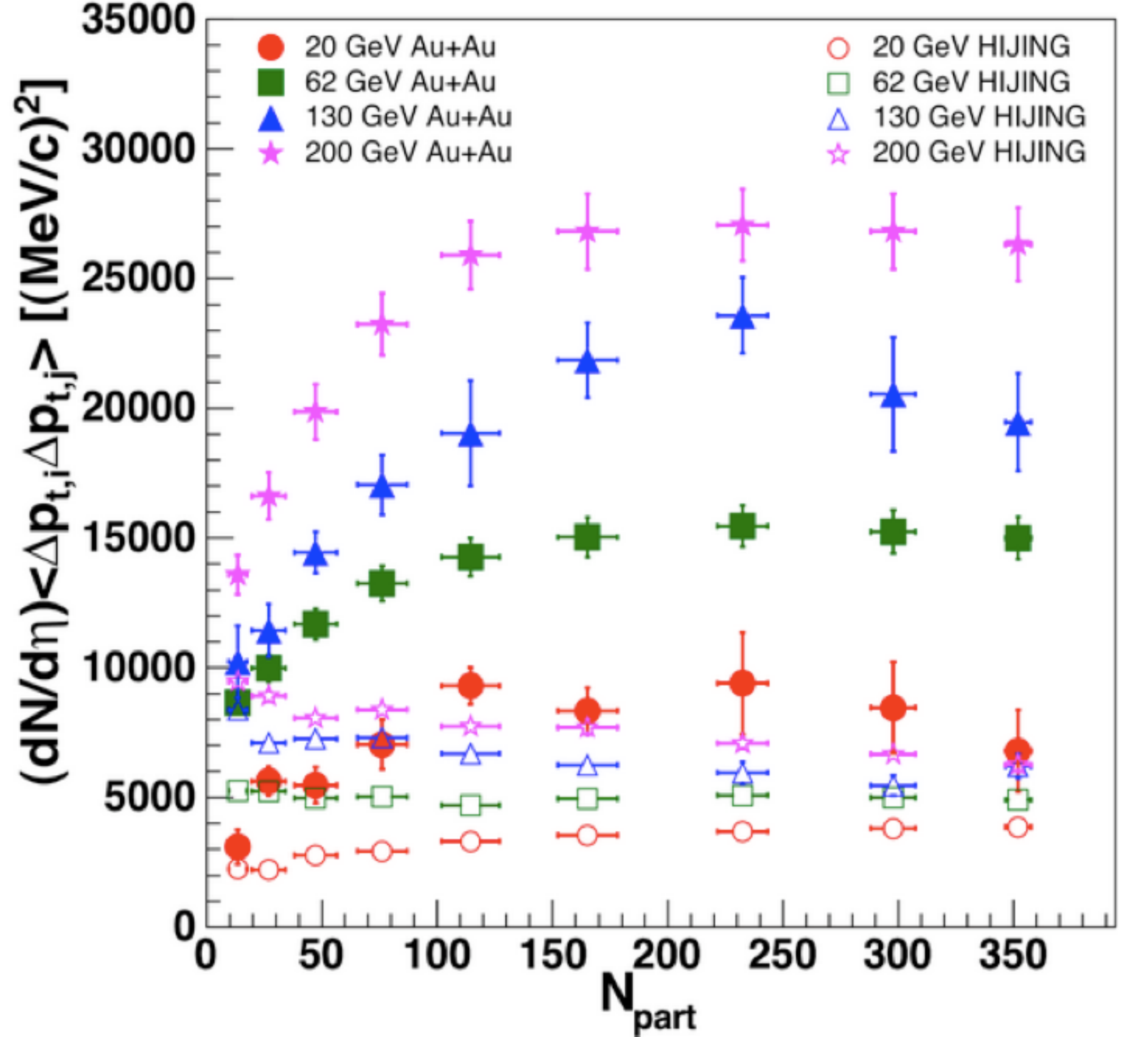}
\caption{$\langle p_{T} \rangle$  fluctuations as a function of N$_{part}$ for Au+Au collisions at various \sqrts. Also shown as the open symbols are the predictions from HIJING at the same energies~\cite{Adams:2005ka}.}
\label{Fig:ptfluctNpart}
\end{center}
\end{minipage}
\hspace{1cm}
\begin{minipage}{0.46\linewidth}
\begin{center}
\includegraphics[width=0.8\linewidth]{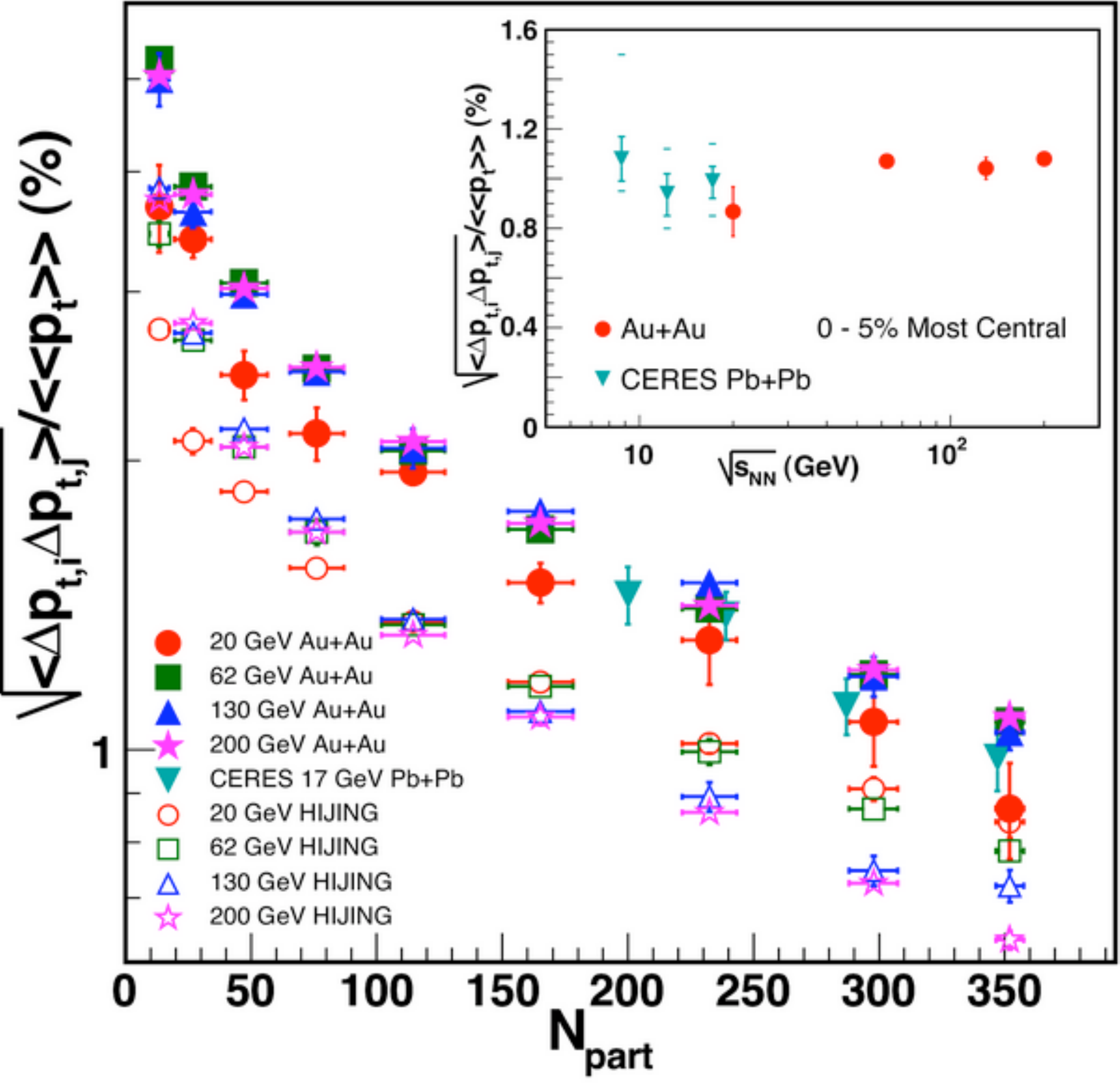}
\caption{Scaled $\langle p_{T} \rangle$ fluctuations as a function of N$_{part}$.  Also shown as the open symbols are the predictions from HIJING at the same energies. The inset shows a comparison of the STAR data to CERES as a function of \sqrts for the most central events. \cite{Adams:2005ka}.}
\label{Fig:ScaledpTFluct}
\end{center}
\end{minipage}
\end{figure}

\noindent  The $\langle p_{T} \rangle$   is measured for all events and also estimated for mixed events. The results are compared and any difference between the data and mixed events is an indication of non-statistical fluctuations, Fig.~\ref{Fig:MeanpTGauss}. Current results show significant non-statistical fluctuations, as measured via the covariance of the two particle transverse momentum correlation measure 
$\langle \Delta p_{T,i}\Delta p_{T,j}\rangle$ at all energies, Fig.~\ref{Fig:ptfluctNpart}~\cite{Adams:2005ka}. They increase with \sqrts and are larger than those predicted by HIJING (open symbols in Fig.~\ref{Fig:ptfluctNpart}).  The data of Fig.~\ref{Fig:ptfluctNpart}  show an initial rapid rise  as a function of centrality before plateauing around N$_{part} \sim$ 150, with a common  turning point  for all measured energies. Again HIJING fails to reproduce this trend being essentially  centrality independent. If one scales the fluctuations by the $\langle p_{T} \rangle$  a different picture emerges, Fig.~\ref{Fig:ScaledpTFluct}. The energy dependence is removed and the result is now inversely proportional to the centrality. HIJING now reproduces the centrality trend but still underestimates the magnitude. The CERES data confirm the energy independence of this measure as shown for the most central data in the inset of  Fig.~\ref{Fig:ScaledpTFluct}~\cite{Adams:2005ka}.

~

\begin{figure}[htb]
\begin{minipage}{0.46\linewidth}
\begin{center}
\includegraphics[width=0.85\linewidth]{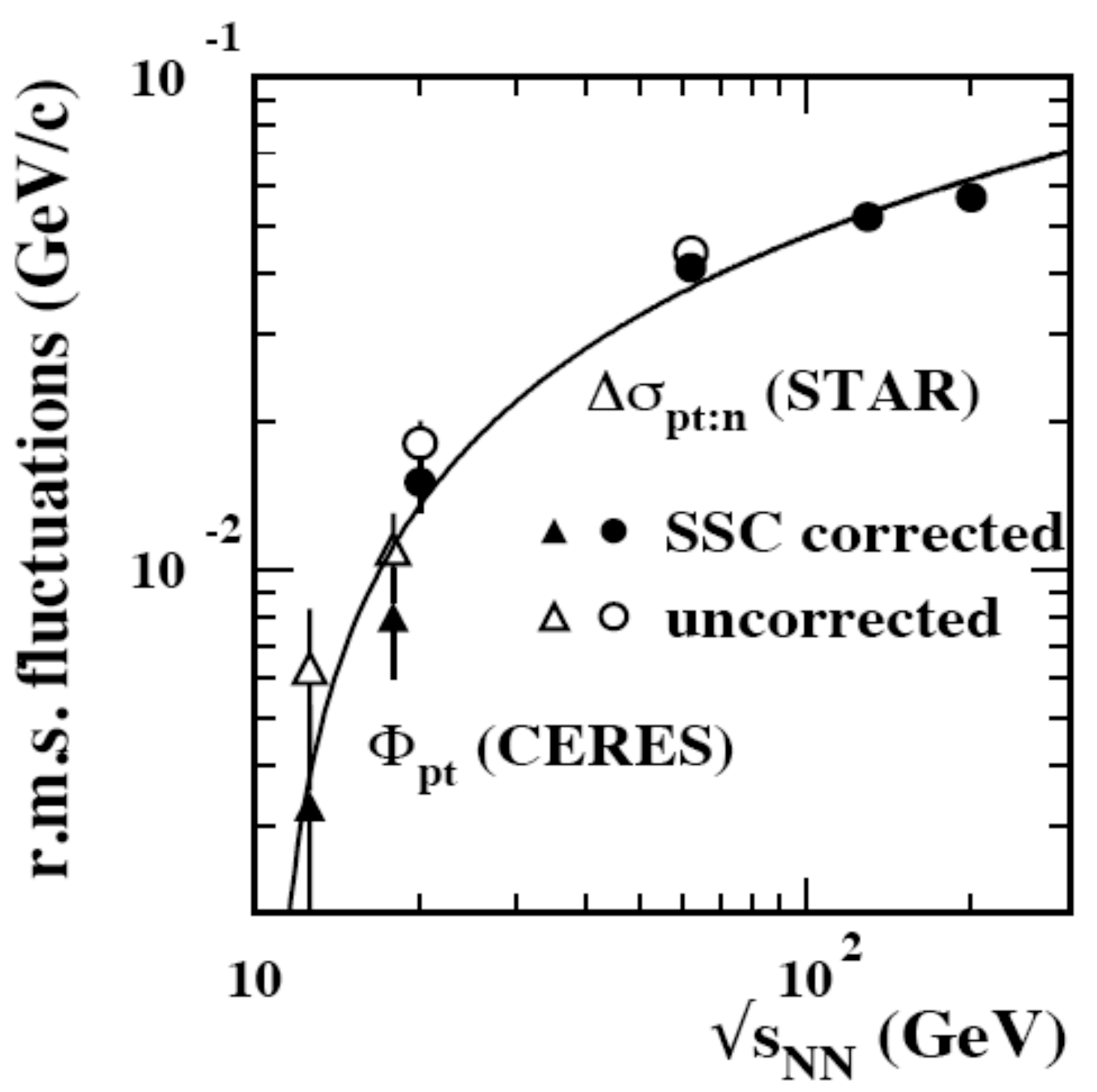}
\caption{The RMS of the $\langle p_{T} \rangle$ fluctuations as a function of \sqrts. The SSC corrected  data have been corrected for small-scale correlations such as HBT and Coulomb effects. The curve is proportional to $\ln [\sqrt{s_{NN}}/10]$. From \cite{Adams:2006sg}.}
\label{Fig:ptfluc}
\end{center}
\end{minipage}
\hspace{1cm}
\begin{minipage}{0.46\linewidth}
\begin{center}
\includegraphics[width=0.9\linewidth]{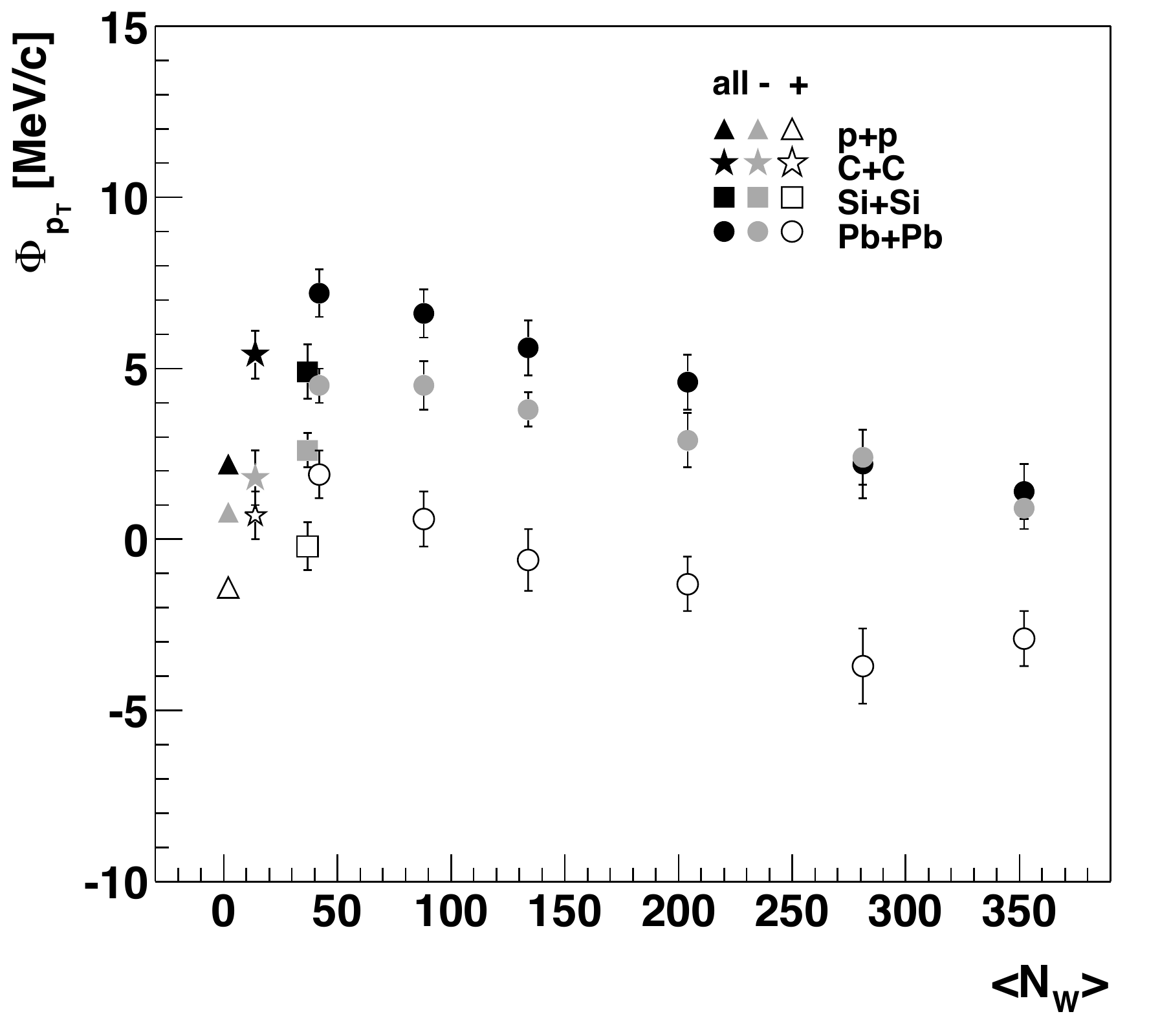}
\caption{ The measured $\Phi_{p_{T}}$ from NA49 at \sqrts=17.3GeV as a function of the  mean number of wounded nucleons $\langle 
N_W \rangle$.  Corrections for limited two track resolution have been applied.
Errors are statistical only. Systematic errors are $<$ 1.6 MeV/c. From \cite{Anticic:2003fd}.}
\label{Fig:PhiPtNa49}
\end{center}
\end{minipage}
\end{figure}

\noindent Fig.~\ref{Fig:ptfluc} shows the RMS of the \pT fluctuations as a function of \sqrts . The RMS grows smoothly   with increasing log(\sqrts)~\cite{Adams:2006sg}. However, there is a relatively large gap in the measurements between  20 and 60 GeV and this region ought to be covered before it is claimed that there are no anomalies in this measure.

~

\noindent An alternative measure of \pT fluctuations is $\Phi$(\pT)~\cite{Gazdzicki:1992ri}. The difference between event-by-event fluctuations of the data and those from mixed events is quantified by $\Phi$(\pT). If the system created in the collisions does not have inter-particle momentum correlations and thus all particles are emitted independently $\Phi$(\pT) will be zero. If instead A+A collisions are an incoherent superposition of N+N events $\Phi$(\pT) will be non-zero but a constant for all centralities in Au+Au and for p+p.  For the most central data, once event-by-event impact parameter fluctuations have been removed NA49 observe no collision energy dependence of $\Phi$(\pT) and its magnitude is consistent with zero~\cite{Grebieszkow:2007xz}. However,  they observe a significant non-monotonic evolution of $\Phi$(\pT) as a function of centrality for highest SPS collision energies (\sqrts=17.3 GeV), Fig.~\ref{Fig:PhiPtNa49}~\cite{Anticic:2003fd}, this result has been confirmed by CERES~\cite{Sako:2004pw}. Further more detailed studies are needed to confirm these results and their potential implications. These will be undertaken by STAR via the BES.

\subsubsection*{K/$\pi$ Fluctuations}\label{SubSubSec:KpiFluc}

\begin{figure}[htb]
	\begin{minipage}{0.46\linewidth}
		\begin{center}
			\includegraphics[width=0.83\linewidth]{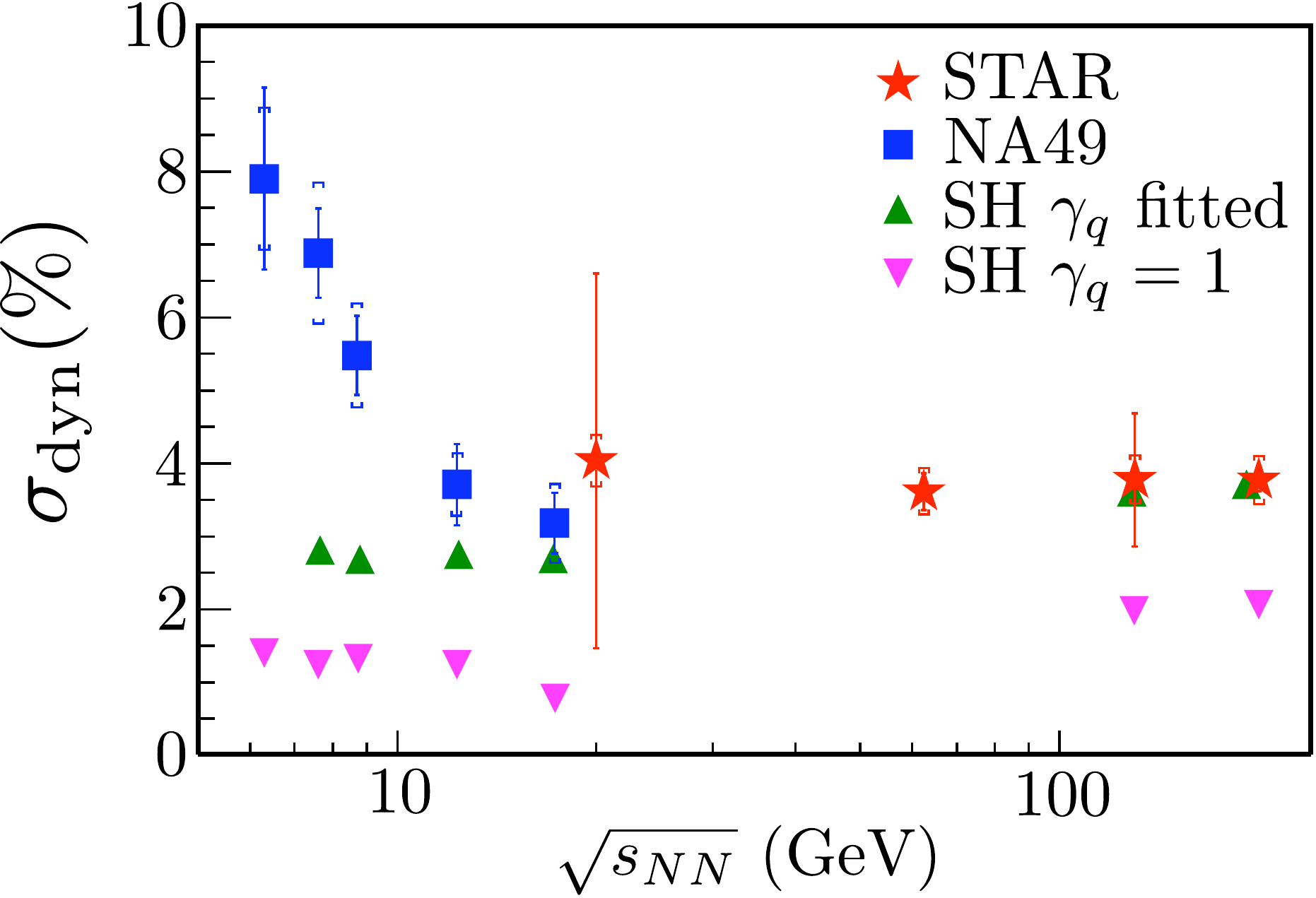}
			\caption{Experimental results for $\sigma_{\rm dyn}$ for $K/\pi$ as a function of \sqrts \cite{Alt:2008ca,Abelev:2009if}.  Also shown are results from the statistical hadronization model of Torrieri~\cite{Torrieri:2007vv}.}
			\label{Fig:kpiSigDyn}
		\end{center}
	\end{minipage}
	\hspace{1cm}
	\begin{minipage}{0.46\linewidth}
		\begin{center}
			\includegraphics[width=0.77\linewidth]{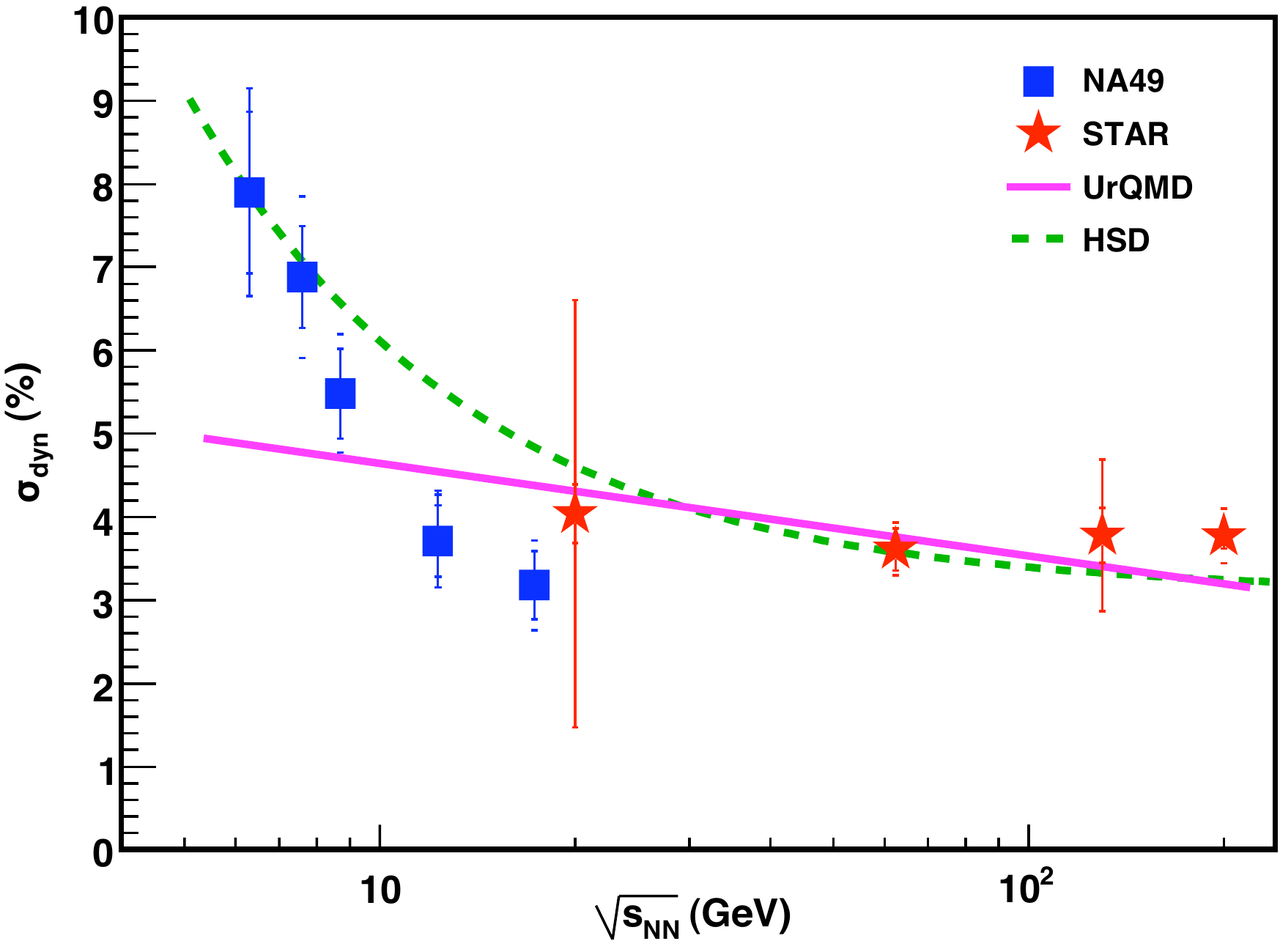}
			\caption{Comparison of the predictions of the HSD and UrQMD models to the experimental data for $\sigma_{\rm dyn}$ for $K/\pi$. Data from \cite{Alt:2008ca,Abelev:2009if}}.
			\label{Fig:kpiSigDynModels}
		\end{center}
	\end{minipage}
\end{figure}

\noindent Current STAR results for $K/\pi$ fluctuations from Au+Au collisions at \sqrts = 19.6, 62.4, 130, and 200 GeV are shown in Fig.~\ref{Fig:kpiSigDyn} along with results observed by NA49 at the SPS in central Pb+Pb collisions at \sqrts = 6.3, 7.6, 8.8, 12.3 and 17.3 GeV \cite{Alt:2008ca}.    The fluctuations are analyzed  using $\sigma_{dyn} =  sign (\sigma^2_{data} - \sigma^{2}_{mixed})\sqrt{|\sigma^2_{data} - \sigma^{2}_{mixed}|}$ where $\sigma_{data}$ is the relative width (standard deviation divided by the mean) of the K/$\pi$ distribution for the data and $\sigma_{mixed}$  is the relative width of the K/$\pi$  distribution for mixed events.  Our results for $K/\pi$ fluctuations in central collisions show little dependence on the incident energies studied and are on the same order as the NA49 measurements at \sqrts = 12.3 and 17.3 GeV.

~

\noindent In Fig.~\ref{Fig:kpiSigDyn}, we compare the statistical hadronization model results of Torrieri \cite{Torrieri:2007vv} to the experimental data.   We see that when the light quark phase space occupancy, $\gamma_{q}$, is one, corresponding to equilibrium, the calculations underestimate the experimental results at all energies.  When $\gamma_{q}$ is varied to reproduce the excitation function of $K^{+}/\pi^{+}$ yield ratios and the excitation function of temperature versus chemical equilibrium over the SPS and RHIC energy ranges \cite{Torrieri:2007vv,Rafelski:2005md}, the statistical hadronization model  correctly predicts the dynamical fluctuations at the higher energies but under-predicts the NA49 data at the lower energies, supporting the conclusion that the lower energy fluctuation data are anomalous \cite{Alt:2008ca}.  In these fits,  $\gamma_{q} > 1$ (chemically over-saturated) for $\sqrt{s_{NN}} < 9$ GeV and $\gamma_{q} < 1$ (chemically under-saturated) for $\sqrt{s_{NN}} >$ 9 GeV.

~

\noindent The changes in susceptibilities as a function of temperature, illustrated in Figs.~\ref{Fig:LatticeQCD1} and \ref{Fig:LatticeQCD2} for $\mu_B$ = 0, are expected to diverge at the critical point.  Such an effect might be observable as deviations of fluctuations from a monotonic dependence on incident energy in central collisions.   However, changes in the underlying physics can also induce changes in the fluctuations as a function of incident energy.  To gain insight into what we might expect from $K/\pi$ fluctuations as a function of energy, we compare the experimental results to predictions from the HSD model~\cite{Gorenstein:2008et} and the UrQMD model \cite{Bleicher:1999xi} in Fig.~\ref{Fig:kpiSigDynModels}.  The NA49 UrQMD results \cite{Alt:2008ca} were carried out using UrQMD version 1.3 with an NA49 acceptance filter while the STAR UrQMD results were carried out using UrQMD version 2.3 with a STAR acceptance filter.   We can see that UrQMD  reproduces the results at RHIC energies, but  under-predicts the fluctuations at low incident energies.  HSD seems to reproduce well the general shape of the measurements, but is slightly above the data in the \sqrts $\sim$ 10 GeV range. The fact that no model completely reproduces the measurements at all \sqrts,  combined with the lack of experimental data in the range of \sqrts =20 - 60 GeV, means that the question of non-monotonic behavior of $K/\pi$ fluctuations must be answered with additional measurements.

\begin{figure}[htb]
	\begin{minipage}{0.48\linewidth}
		\begin{center}
			\includegraphics[width=\linewidth]{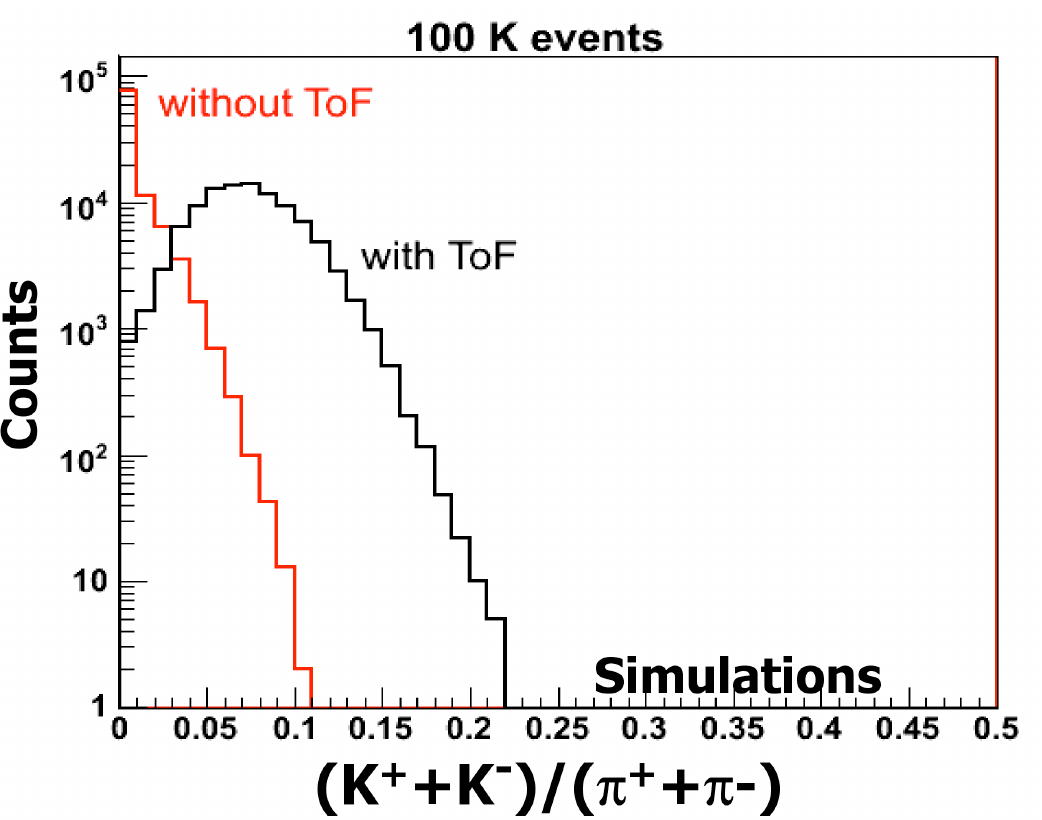}
			\caption{The uncorrected reconstructed K/$\pi$ ratio for 100 k Au-Au central events at \sqrts = 8.8 GeV with and without using the ToF information.}
			\label{Fig:KaonWWOTof}
		\end{center}
	\end{minipage}
	\hspace{0.5cm}
	\begin{minipage}{0.48\linewidth}
		\begin{center}
			\includegraphics[width=\linewidth]{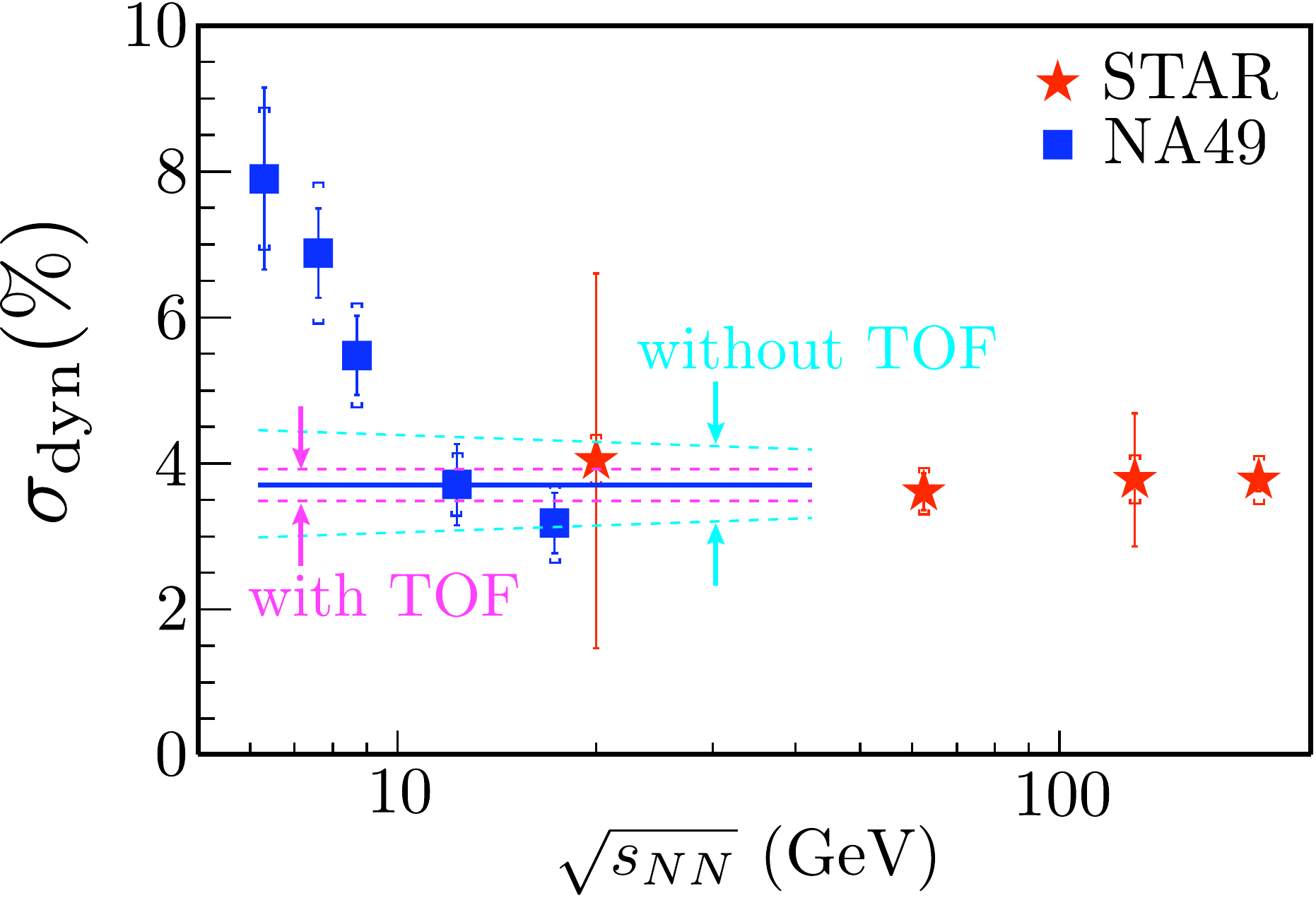}
			\caption{Estimate of the error in $\sigma_{dyn}$ for the charge integrated $K/\pi$ fluctuations as a function of \sqrts with and without the ToF. Shown for comparison are the current measurements from NA49 and STAR. Data from~\cite{Abelev:2009if}.}
			\label{Fig:sigDynToF}
		\end{center}
	\end{minipage}
\end{figure}

~

\noindent To make these measurements, one needs to attempt to measure {\it all}  the kaon and pions.  The $K$ reconstruction efficiency as a function of \pT is rather low, see section~\ref{SubSubSec:STARRates} and Fig.~\ref{Fig:EfficSTAR} for more details,  due to multiple effects. The first is the decay of the $K$, the dominant channel being $K^{\pm} \rightarrow \mu^{\pm}+\nu_{\mu}$ with a $c\tau$ of 3.7 m. This means a significant number of the kaons decay before being measured by the TPC. Secondly the cuts required to select clean kaons, essential for the fluctuation measures, reduce the efficiency further.  By using the ToF we can extend the clean PID range to higher \pT and thus gain essential coverage. As already mentioned not only coverage but clean PID is needed. Again the ToF is useful because it can eliminate much of the electron contamination at low \pT that $dE/dx$ measures alone cannot remove. The most damaging misidentification is that of a $K$ as a $\pi$ or vice-versa, since this effect distorts both the numerator and denominator of the measure, i.e. $K/\pi \rightarrow (K+1)/(\pi-1)$ or $(K-1)/(\pi+1)$. Fig.~\ref{Fig:KaonWWOTof} shows a simulation of the raw STAR charge integrated $K/\pi$ ratio as measured from 100k HIJING simulated central Au+Au collisions at \sqrts=8.8 GeV. The red curve shows the distribution of the $K/\pi$ ratio ratio without using the ToF information and the black curve shows the distribution of the $K/\pi$ ratio using the ToF. It can be seen that the ToF significantly improves our ability to study $K/\pi$ fluctuations.  It has been estimated that 1 M events at each of the proposed collision energies for the BES are sufficient to perform these PID fluctuation studies. Figure~\ref{Fig:sigDynToF} shows the estimated statistical error for STAR's $\sigma_{dyn}$ for the charge integrated K/$\pi$ ratio with and without the ToF information. Also shown for comparison are the current NA49 and STAR measured data points. With the ToF, STAR's relative error is $\pm 5\%$, without the ToF, this doubles to $\pm 10\%$.  Although $K/\pi$ fluctuations have been studied by NA49 and STAR already, there still remains substantial work to fully understand the results. For instance, NA49 only measured the charge integrated K/$\pi$ fluctuations in central collisions. STAR, with its larger acceptance and ToF identification reach, will be able to repeat these measures in much greater detail and also measure the charge separated fluctuations. These improvements will allow us to observe event-by-event if there is truly something special happening in the $K^{+}/\pi^{+}$ ratio as suggested by the integrated measure shown in Fig.~\ref{Fig:kpiSigDyn}.

\subsubsection*{p/$\pi$ Fluctuations}\label{SubSubSec:pPiFlc}

\begin{figure}[htb]
\begin{minipage}{0.46\linewidth}
\begin{center}
\includegraphics[width=0.9\linewidth]{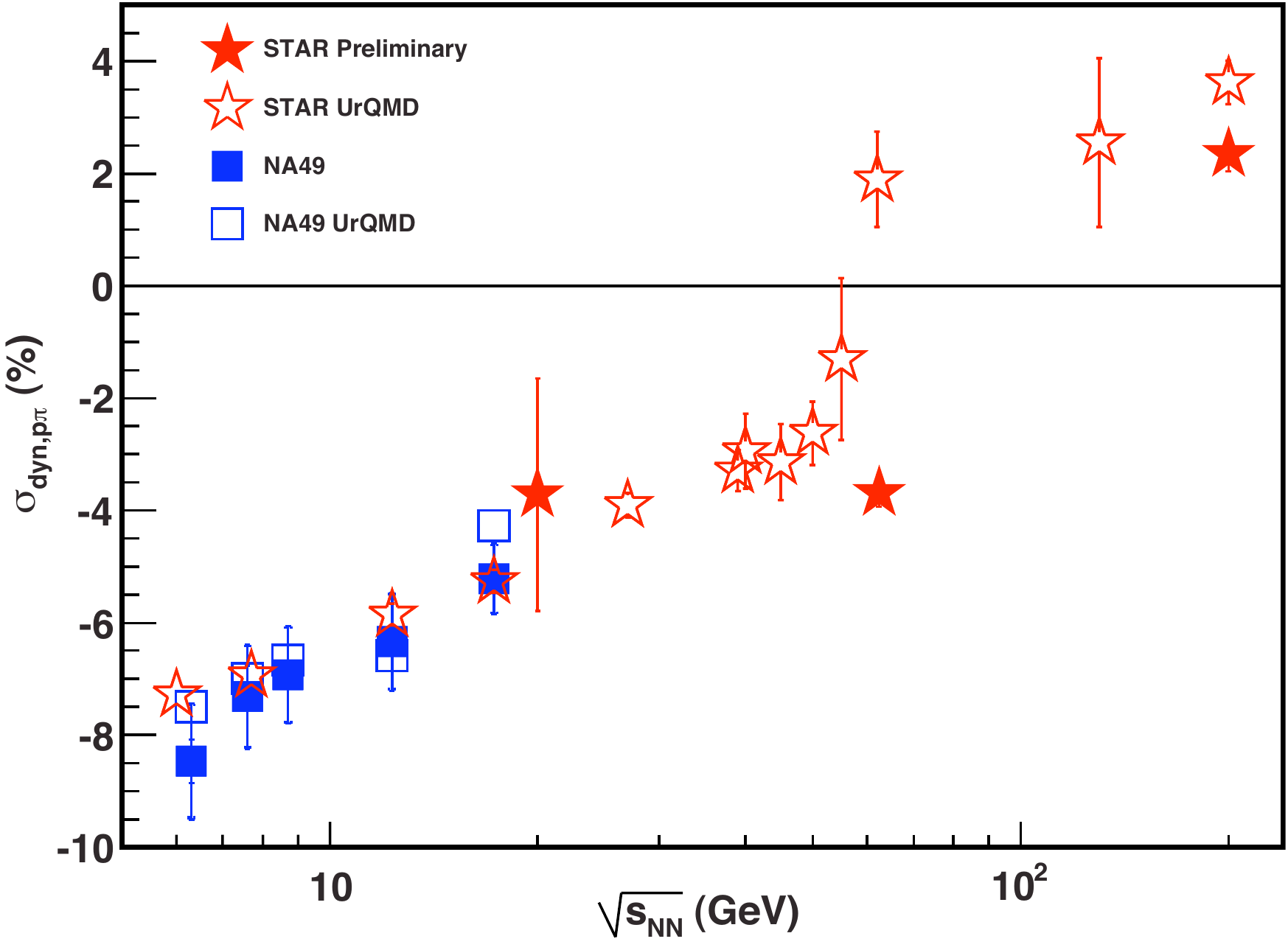}
\caption{Comparison of the predictions of the UrQMD model to the experimental data for $\sigma_{\rm dyn}$ for $p/\pi$~\cite{Alt:2008ca}.}
\label{Fig:sigmadyn_ppi}
\end{center}
\end{minipage}
\hspace{0.3cm}
\begin{minipage}{0.46\linewidth}
\begin{center}
\includegraphics[width=0.9\linewidth]{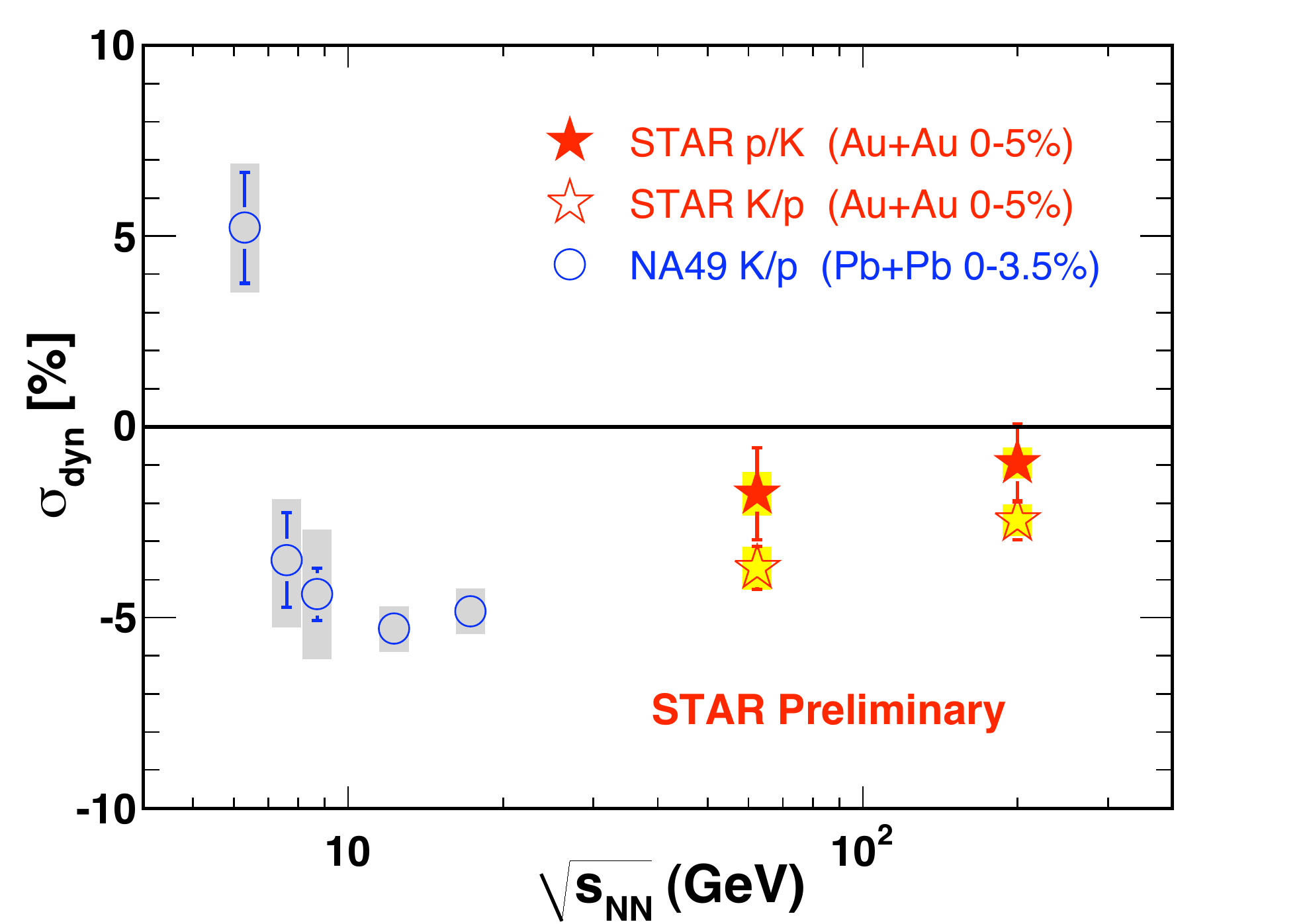}
\end{center}
\caption{Preliminary experimental results for $\sigma_{\rm dyn}$ for K/p (p/K) as a function of \sqrts from NA49  and STAR~\cite{Schuster:2009ak,Tian:2010zz}.}
\label{Fig:k2p}
\end{minipage}
\end{figure}

\noindent The study of $p/\pi$ fluctuations may provide information about baryon fluctuations.  $p/\pi$ fluctuations have been studied as a function of \sqrts by NA49 \cite{Alt:2008ca} and by STAR as shown in Fig.~\ref{Fig:sigmadyn_ppi} using the variable $\sigma_{\rm dyn}$ at the same energies as those used to study $K/\pi$ fluctuations.  The dependence on \sqrts of the NA49 results and the STAR results seem to join smoothly.  Here the fluctuations are negative, which may indicate that the decay of resonances is important for $p/\pi$ fluctuations.  In Fig.~\ref{Fig:sigmadyn_ppi} we compare the predictions of UrQMD  to the measured fluctuations.  The UrQMD calculations were done by NA49 \cite{Alt:2008ca} using the NA49 acceptance filter.  We can see that the predictions of the model is reasonably close to the experimental results.

~

\noindent Studying charge separated $p/\pi$ fluctuations will be crucial to understanding  the energy dependence for several reasons.  One reason is that the $\bar{p}/p$ changes strongly with \sqrts.  Another is that several resonances exist such as $\Lambda \rightarrow p + \pi^{-}$ or $\Delta^{0} \rightarrow \bar{p} + \pi^{+}$ that can have a significant effect on the observed $p/\pi$ fluctuations.  It has been estimated that 1 M events at each of the proposed collision energies for the BES are sufficient to perform these PID fluctuation studies.

\subsubsection*{K/p fluctuations}\label{SubSubSec:kp}

\noindent The correlation between strangeness $S$ and baryon number $B$ is  sensitive to the state of matter created in heavy-ion  collisions~\cite{Koch:2005vg}. In a system consisting of quarks and gluons, strangeness is carried by s and $\bar{s}$ quarks, which also  carry fractional baryon number ( B = 1/3 for s and -1/3 for $\bar{s}$ quarks). In a hadron gas, much of the strangeness is carried by kaons for which B = 0 leading to a weaker correlation between strangeness and baryon number, which has been reproduced in the string hadronic model UrQMD~\cite{Haussler:2005ei}. At the transition from hadron gas to quark gluon-plasma, a rapid fluctuation in the baryon-strangeness is  expected, and recent lattice QCD studies~\cite{Cheng:2008zh} confirm this prediction. As both baryon number and strangeness are conserved quantities, the effect is conveyed to the final state and can be observed via the fluctuations in the event-wise kaon to proton ratio.  The feasibility of measuring K/p fluctuations, characterized by $\sigma_{dyn}$, have recently been shown by STAR and NA49, the preliminary results are reproduced in Fig.~\ref{Fig:k2p}~\cite{Schuster:2009ak,Tian:2010zz}. The same mixing procedure has been used in both of these analyses. There appears to be only a slight increase in the  dynamical fluctuations at RHIC compared to the top SPS energies. An interesting point to note is that $\sigma_{dyn}$  for p/K and for K/p are not exactly the same. The low multiplicity of p and K in each event, and that $\sigma_{dyn}$ itself is small, may account for much of the observed differences. More data are needed to shed light on the RHIC results. The ToF upgrade will facilitate this important measurement in the beam energy scan.  It has been estimated that 1 M events at each of the proposed collision energies for the BES are sufficient to perform these PID fluctuation studies.

\subsubsection*{Net proton Kurtosis}\label{SubSubSec:Kurtosis}

\noindent To date most experimental fluctuation measures have concentrated on the second moments (proportional to the square of the correlation length). However, estimates of the magnitude of correlation length in heavy-ion collisions indicate that they could be small around the critical point (of the order of 2-3 fm)~\cite{Stephanov:2008qz}, making it challenging to be detected in experiments.

~

\noindent Higher moments of event-by-event pion and proton multiplicities might be significantly more sensitive to the existence of the critical point than measures based on second moments. The fourth moment, the kurtosis,  of these multiplicity distributions is expected to be proportional to the seventh power of the correlation length~\cite{Stephanov:2008qz}. In addition, it is expected that the evolution of fluctuations from the critical point to the freeze-out point may lead to a non-Gaussian shape in the event-by-event multiplicity distributions.  The kurtosis of multiplicity distributions might then provide a more sensitive observable for the search of the QCD critical point.

~

\begin{figure}[htb]
\begin{minipage}{0.46\linewidth}
\begin{center}
\includegraphics[width=0.9\linewidth]{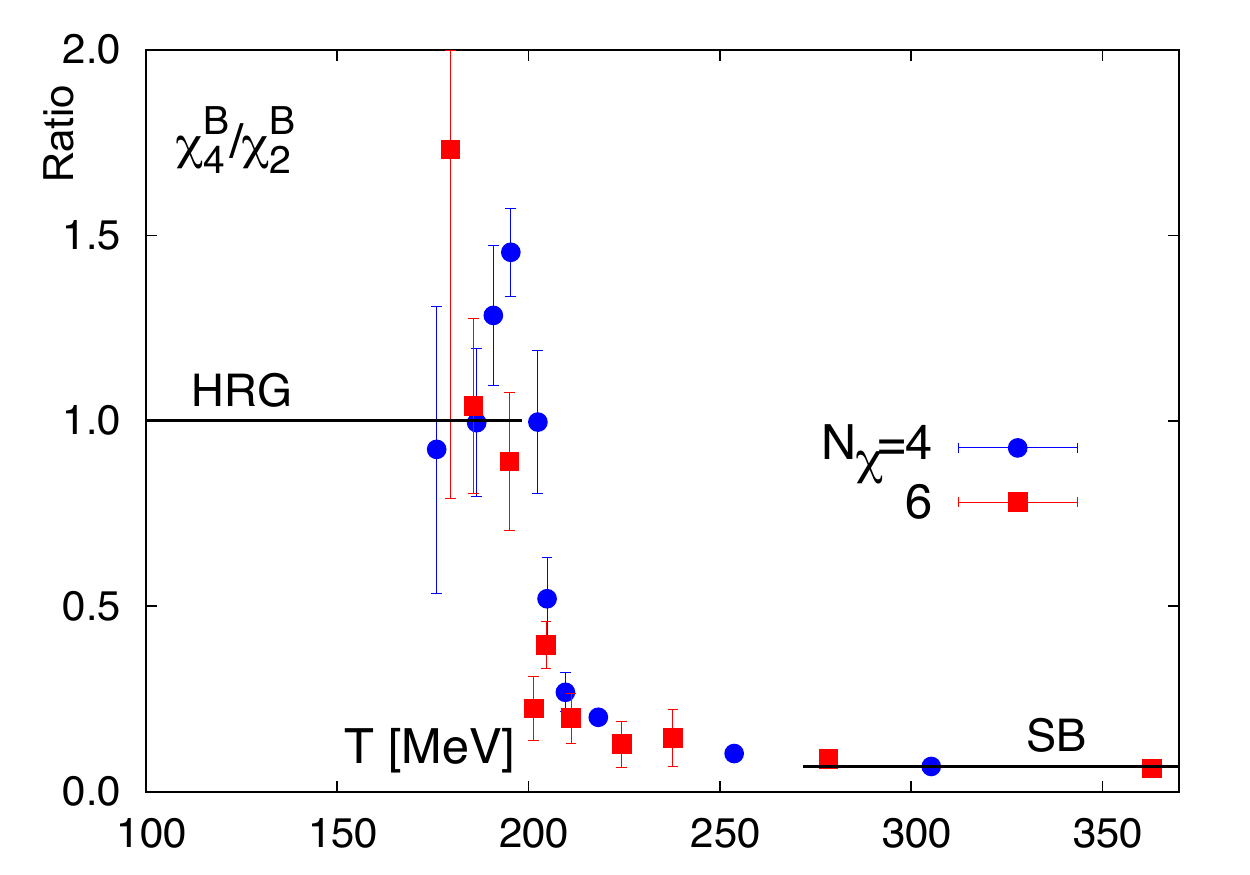}
\caption{The ratio of fourth and second order cumulants of baryon number as a function of temperature \cite{Cheng:2008zh}.  The value from the hadron gas model (HRG) is for the temperature range 100 to 200 MeV.}
\label{Fig:LatticeQCD3}
\end{center}
\end{minipage}
\hspace{1cm}
\begin{minipage}{0.46\linewidth}
\begin{center}
\includegraphics[width=0.9\linewidth]{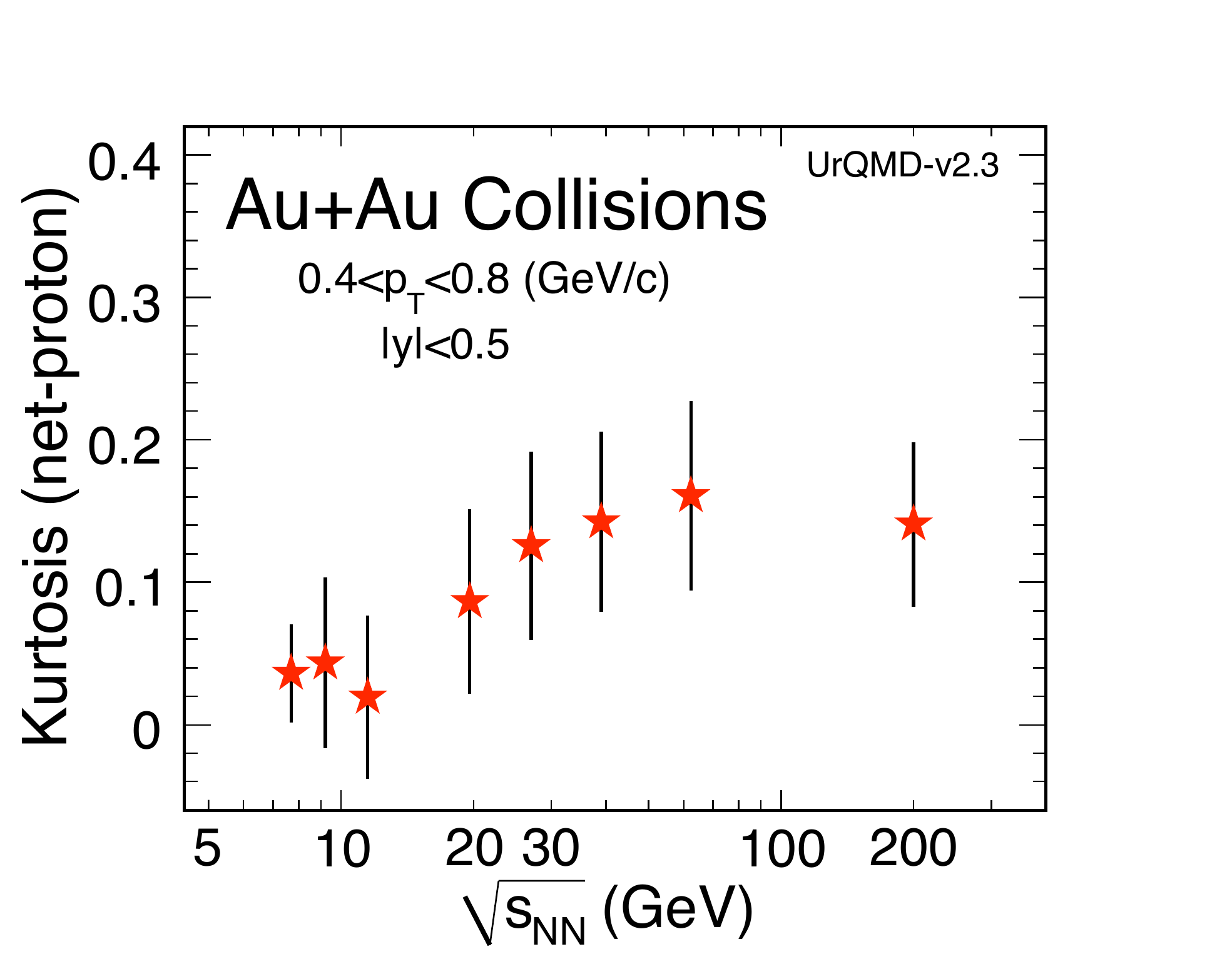}
\end{center}
\caption{The net-proton kurtosis from the UrQMD model, evaluated in the STAR acceptance for 100 k events.}
\label{Fig:KurtosisUrQMD}
\end{minipage}
\end{figure}

\noindent  Further in Lattice calculations, which assume the system to be in thermal equilibrium, the kurtosis of event-by-event net-baryon number, net charge and net strangeness  are related to the respective susceptibilities. These susceptibilities show large values or diverge at the critical temperature~\cite{Gavai:2004sd,Koch:2008ia,Stephanov:2008qz,Cheng:2008zh}, Fig.~\ref{Fig:LatticeQCD3}. The measurement of higher moments of event-by-event identified particle multiplicity distributions and its variation with centrality and beam energy will provide the first direct connection between experimental observables and Lattice Gauge Theory calculations \cite{Cheng:2008zh}.

~

\noindent Using STAR's excellent mid-rapidity PID capabilities,   (anti) protons for \pT =0.2-1.0 GeV/c can be cleanly identified using dE/dx in the TPC. We can therefore carry out a proton kurtosis analysis~\cite{Aggarwal:2010wy}. It is advisable to do this analysis over as large a \pT range as possible to preserve the long-range correlations that the kurtosis is most sensitive to probe.  Therefore the sensitivity of these measurements is greatly increased with the  inclusion of the ToF, and measurements related to isospin susceptibilities and strangeness susceptibilities will be also possible.

~

\noindent  The UrQMD model \cite{Bleicher:1999xi,Bass:1998ca} can serve as a baseline of hadronic processes involved in heavy-ion collisions. To estimate the minimum statistics required to carry out kurtosis analysis in the RHIC beam energy scan, simulations have been performed in the latest UrQMD version 2.3 ~\cite{Petersen:2008kb, Bleicher:2009}. Figure \ref{Fig:KurtosisUrQMD} shows the net-proton kurtosis evaluated within STARÕs central barrel acceptance ($|$y$| <$ 0.5) for central Au+Au collisions in the energy range 5 $<$ \sqrts $<$ 200 GeV. The  analysis shows that a measurement of  the net proton kurtosis can be made with a statistical error of 0.75 when 100 k events are used.  Extrapolating these results, we estimate that 4 M events are needed at each beam energy to obtain a statistically precise measurement of better than 10$\%$.

\subsubsection*{Azimuthal Correlations and v$_{2}$ Fluctuations}\label{SubSubSec:v2Fluc}

\noindent $v_2$ fluctuations $\sigma_{v_{2}}$ have been studied by STAR~\cite{Sorensen:2006nw} and PHOBOS~\cite{Alver:2008hu} as a method to test models of the initial conditions in heavy-ion collisions and to infer the effectiveness of the conversion of eccentricity to momentum-space anisotropy on an event-by-event basis. After further investigation, it has been found that it is not experimentally possible to determine $\sigma_{v_{2}}$ independent of non-flow correlations ($\delta$) (see for instance \cite{Sorensen:2009cz} and references therein for more discussions). The sum of non-flow and elliptic flow correlations ($\sigma_{tot}^{2}$) in the form $\delta+2\sigma_{v_{2}}^{2}$ can be measured via the non-statistical width of the flow vector distribution which is related to the difference between the two- and four-particle cumulants: $\sigma_{tot}^{2} \approx \delta+2\sigma_{v_{2}}^{2} \approx v_{2}\{2\}^2 - v_{2}\{4\}^{2}$. Non-flow due to HBT and resonances can be nearly eliminated by selecting like charge particles and by using rapidity gaps (particles are required to be well separated in rapidity). Those correlations due to jet production should fall off with energy. The energy dependence of $\sigma_{tot}^{2} \approx \delta + 2\sigma_{v_{2}}^2$ should therefore be of considerable interest in an energy scan as it should be increasingly dominated by $\sigma_{v_{2}}^{2}$ as the beam energy is reduced and can be expected to rise if matter is created near the QCD critical point.

~

\noindent Figure~\ref{Fig:v2fluct} shows preliminary STAR results for $\sigma_{tot}^2$ based on 15 million Au+Au collisions at 200 GeV~\cite{Sorensen:2006nw}. The errors are dominated by systematic uncertainties. Measurements  can be carried out using several hundred thousand minimum-bias events. This estimate is supported by the analysis of the two- and four-particle cumulants carried out on the 130 GeV data based on several hundred thousand events. These measurements will therefore be possible at all \sqrts values of the proposed BES, with likely exception of 5 or 6 GeV.

\subsubsection*{Photon Multiplicity Fluctuations}\label{SubSubSec:PhotonFluc}

\noindent In a thermodynamical picture of the system formed in the collision, the fluctuations in particle multiplicities can be related to the matter compressibility \cite{Mrowczynski:1997kz}, which  could aid our understanding of the critical fluctuations at the QCD phase boundary. Through event-by-event photon multiplicity measurements using the Photon Multiplicity Detector (PMD) at forward rapidities, we can access, at the same beam energy, multiplicity fluctuations at a higher baryon chemical potential compared to those measured at mid-rapidity via charged particles in the TPC.  The excellent spatial resolution of the PMD allows one to study these fluctuations in localized regions of phase space. Due to the uncertainty in the location of the critical point we need to investigate as much of the QCD phase diagram as possible. Given the experimental limitations on the number of  beam energy points that can be scanned, the ability to study fluctuations across a wide range in rapidity enhances the sensitivity of our search. 

~

\begin{figure}[htb]
\begin{minipage}{0.46\linewidth}
\begin{center}
\includegraphics[width=\linewidth]{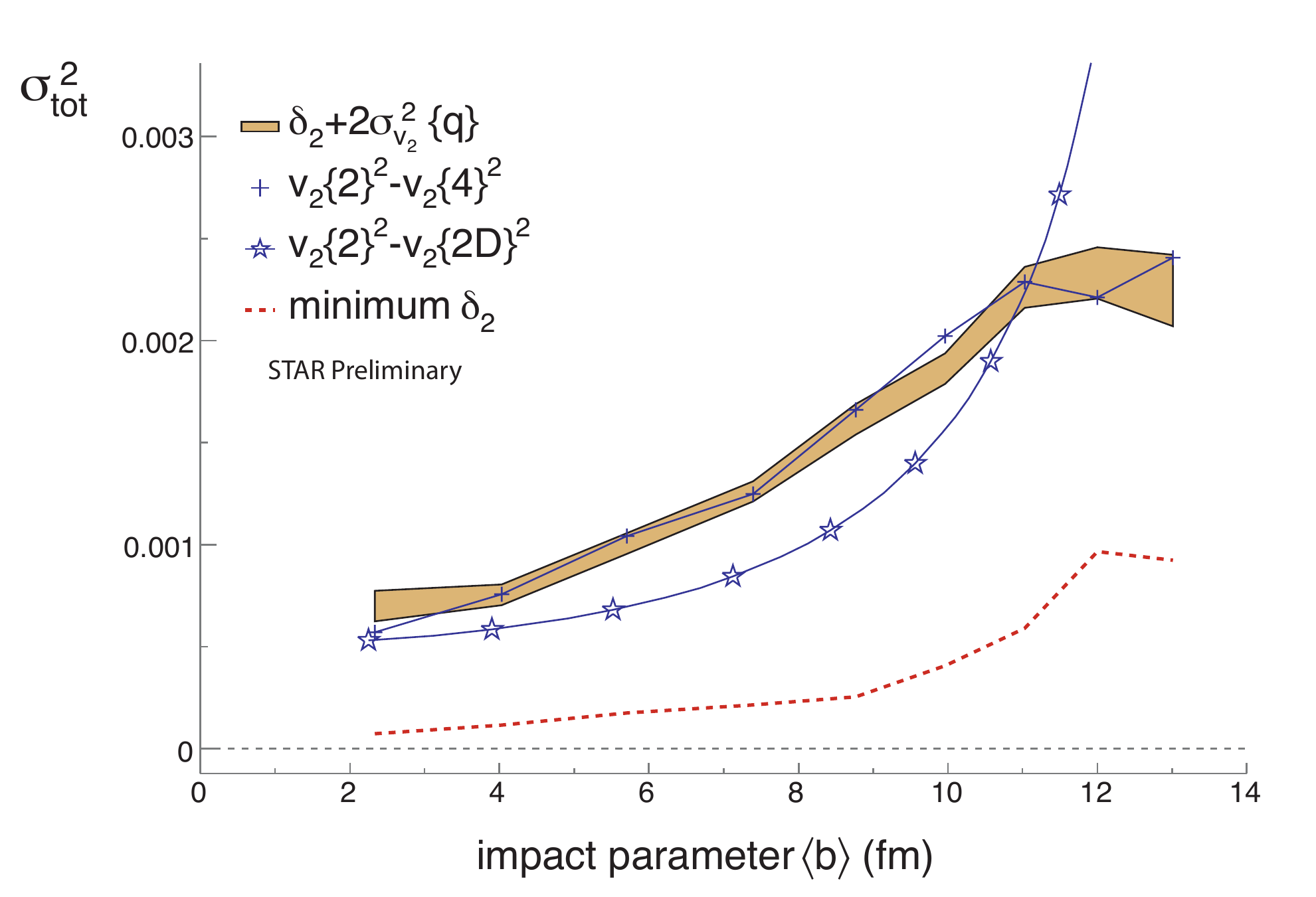}
\caption{Lower bound estimates of the v$_{2}$ fluctuations as a function of centrality for \sqrts=200 GeV Au+Au collisions. From \cite{Sorensen:2006nw}.}
\label{Fig:v2fluct}
\end{center}
\end{minipage}
\hspace{1cm}
\begin{minipage}{0.46\linewidth}
\begin{center}
\includegraphics[width=0.9\linewidth]{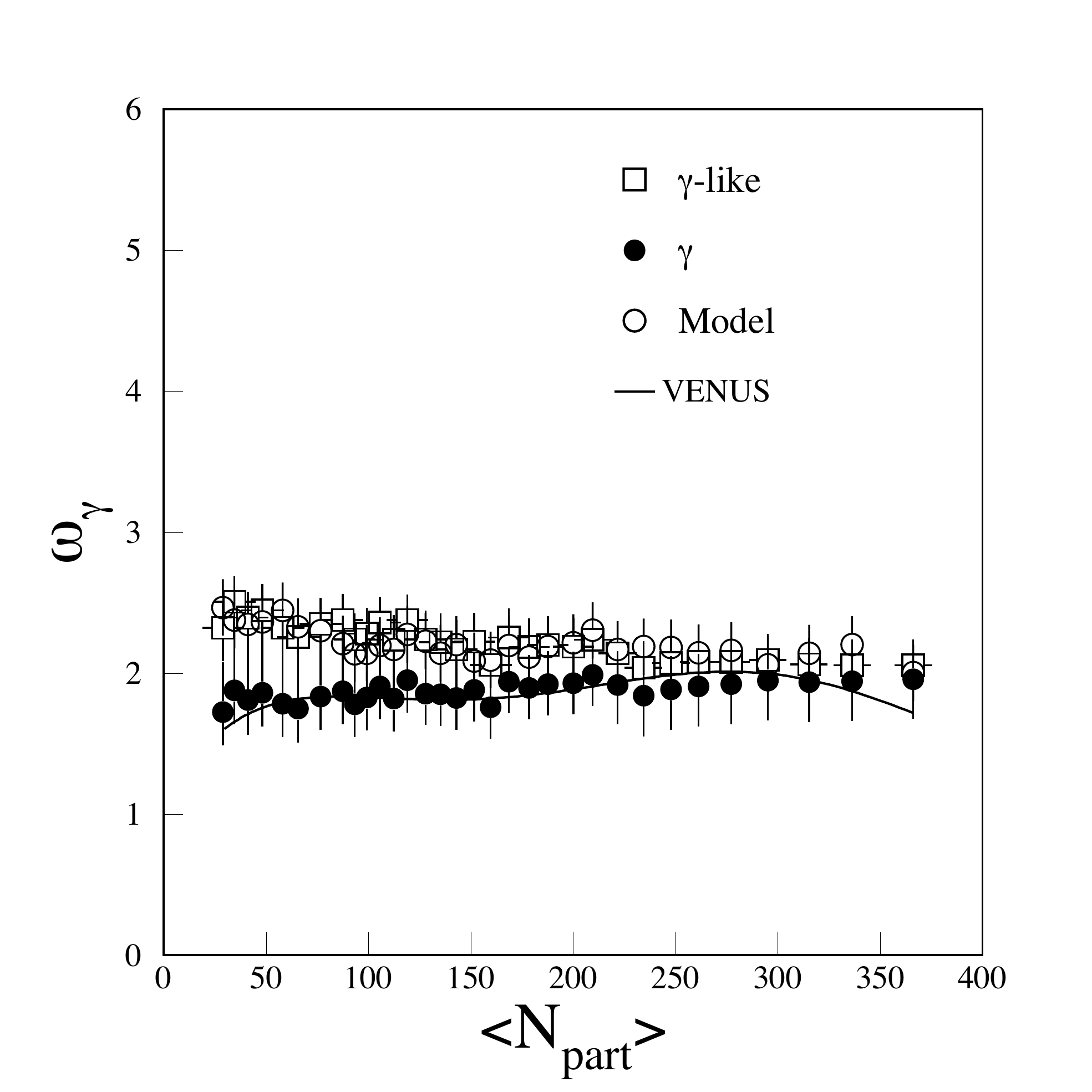}
\caption{The relative fluctuations ($\omega_\gamma$) of photons as a function of number of participants.  The data show the fluctuations in  photon- like clusters and photons after correction for efficiency and purity. These are compared to calculations from a participant model and  those from VENUS event generator \cite{Aggarwal:2001aa}.}
\label{Fig:PhotonFluc}
\end{center}
\end{minipage}
\end{figure}

\noindent The Fig.~\ref{Fig:PhotonFluc} shows the relative fluctuation in  photon multiplicity ($\omega_\gamma$ = $\sigma^2$ /mean) as a function of number of participating nucleons. The results are at a center of mass energy of 17.3 GeV in Pb+Pb collisions within a pseudorapidity coverage of 2.9-4.2 measured using the PMD in WA98 experiment \cite{Aggarwal:2001aa}.

\subsubsection{Forward-Backward Correlations}\label{SubSubSec:ForwardBack}

\noindent Long-range multiplicity correlations have been measured to provide insight on the mechanisms of particle production. Many experiments show strong short-range correlations (SRC) over a region of $\pm$ 1 unit in rapidity \cite{Karsch:2000kv,Fodor:2001au,Allton:2002zi,Braun:1999hv} Correlations that extend over a longer range are observed in hadron-hadron interactions at higher energies \cite{Fodor:2001au, Braun:1999hv}. It has been suggested that long-range correlations (LRC) might be enhanced in hadron-nucleus and nucleus-nucleus interactions due to multiple parton collisions \cite{Braun:2000hd, Braun:2001us}.

~

\noindent The Forward-Backward (FB) correlation strength for multiplicity-multiplicity correlations is defined by the dependence of the average charged particle multiplicity in the backward hemisphere, $\langle N_{b}\rangle$, on the event multiplicity in the forward hemisphere, $\langle N_{f} \rangle$, such that $\langle N_{b} \rangle = a +  \langle N_{f} \rangle$, where a is a constant and the correlation strength is

\begin{eqnarray*}
b = \frac{\langle N_{f}N_{b}\rangle -\langle N_{f}\rangle \langle N_{b}\rangle}{\langle N_{f}^2\rangle-\langle N_{f} \rangle ^2}.
\end{eqnarray*}

\noindent For STAR, the forward-backward intervals are located symmetrically about midrapidity ($\eta$ = 0).

~
 
\begin{figure}[htb]
\begin{center}
\includegraphics[width=0.4\linewidth]{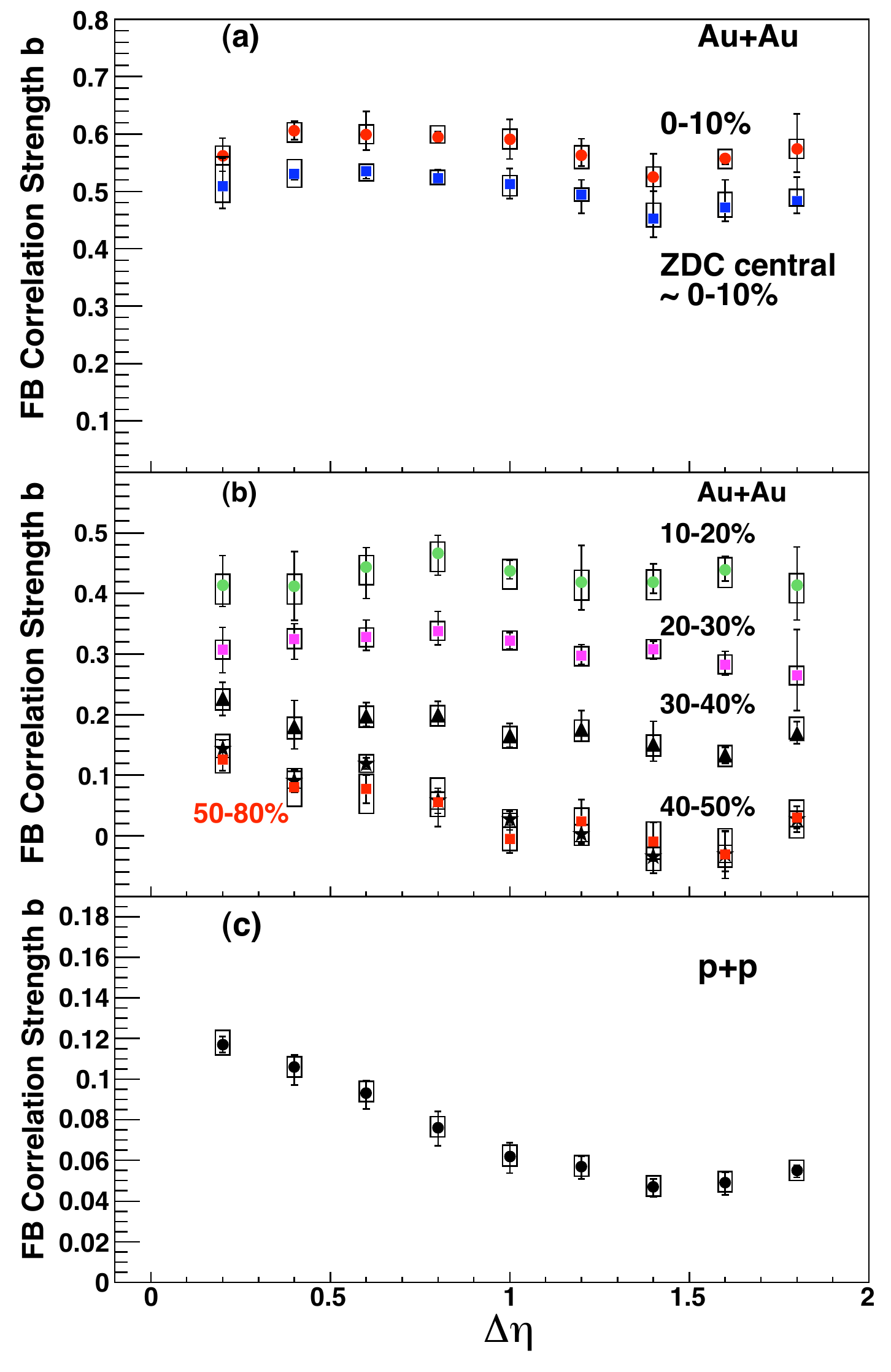}
\end{center}
\caption{FB correlation strength as a function of  $\Delta\eta$ (a) for Au+Au at the 0-10$\%$ most central events,  (b)  more peripheral Au+Au collisions   and c)  p+p collisions at \sqrts = 200 GeV. Figure from \cite{Abelev:2009dqa}. }
\label{Fig:fb2}
\end{figure}

\noindent Previous studies of FB multiplicity correlations indicate that both short and long range correlations are present in Au+Au collisions at \sqrts = 200 GeV, while only short range correlations are present in the more peripheral collisions~\cite{Tarnowsky:Thesis:2008,Braun:2003fn, Brogueira:2006yk}. Measurements for 200 GeV Au+Au collisions are shown in Fig.~\ref{Fig:fb2} along with measurements in p+p for reference~\cite{Abelev:2009dqa}. From the large value of the correlation strength at large $\Delta\eta$, one can infer that a dense partonic system is created with multi-parton interactions that lead to long-range correlations in heavy-ion collisions.

~

\noindent An important extension of this work will be to examine Au+Au collisions as a function of collision energy, particularly in the 5-39 GeV range. By studying the energy dependence of the FB correlation strength, we will 1) be able to test models that have been proposed to describe our data (e.g.  Parton String Model~\cite{Amelin:2001sk}, Color Glass Condensate~\cite{Kovchegov:1999ep}, HIJING~\cite{Wang:1991hta}) and 2) search for non-monotonic trends in shape and magnitude of the FB correlations that might signal fluctuations due to a critical-point or a first order phase transition at freeze-out. These studies can be carried out with approximately 1 million events at each collision energy.

\subsubsection{The Focussing Effect of the QCD Critical Point }\label{SubSubSec:CritPointFocus}

\begin{figure}[htb]
\begin{center}
\includegraphics[width=0.5\linewidth]{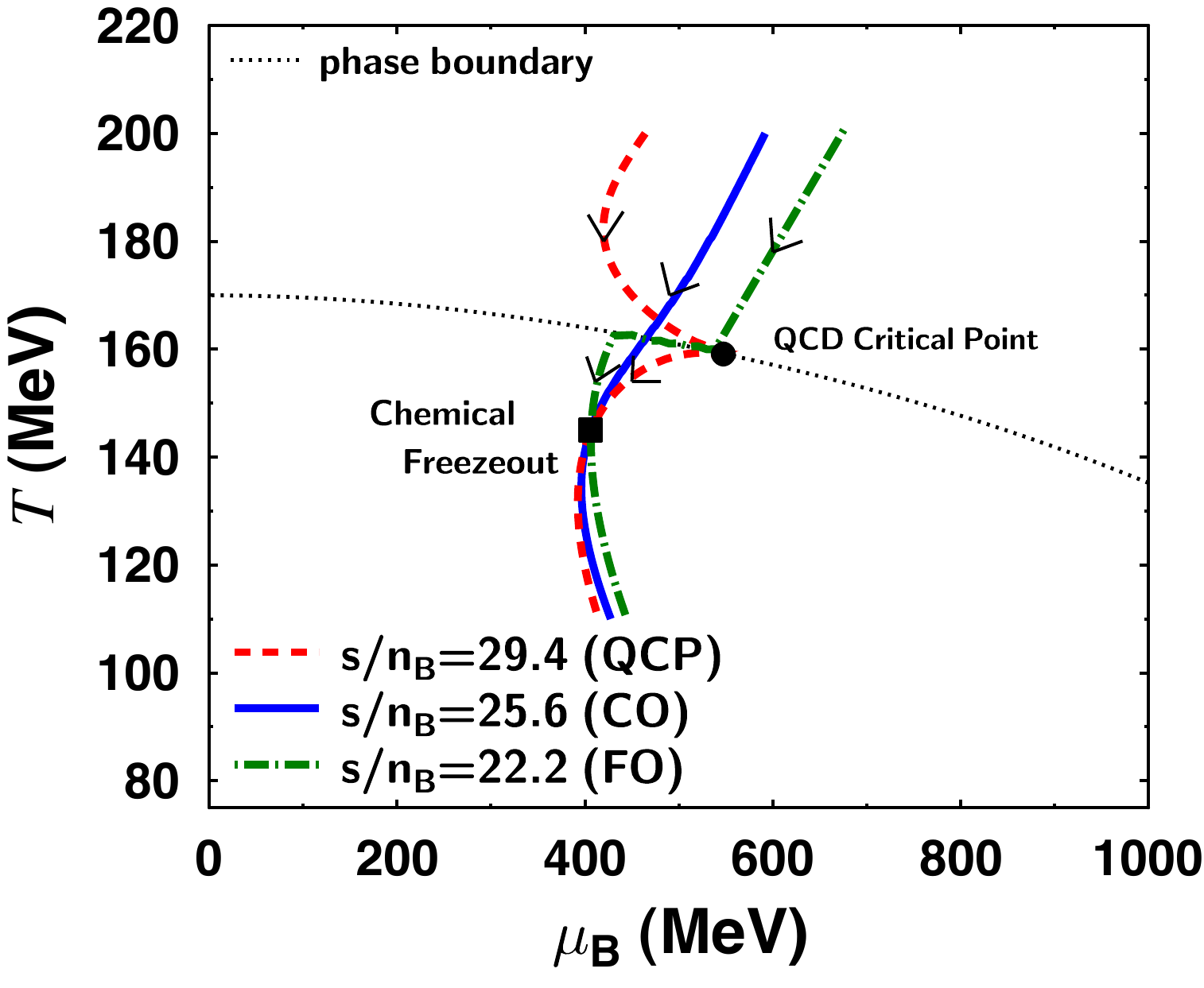}
\caption{The QCD phase diagram showing hydrodynamical trajectories with and without a CP. The Solid line - Crossover transition, dash-dotted - first-order transition. dashed line with CP. From~\cite{Asakawa:2008ti}.}
\label{Fig:Focusing2}
\end{center}
\end{figure}

\noindent Asakawa {\it et al.}~\cite{Asakawa:2008ti} proposed that a critical point would act as an attractor of nearby system expansion trajectories in the ($\mu_{B}$,T) plane. Fig.~\ref{Fig:Focusing2} illustrates the situation for three   system expansion trajectories  which originate from slightly spaced initial conditions. By considering the different options
of a simple crossover transition, a first order phase transition, and a similar transition as displaced by an assumed attraction of the trajectory by an adjacent critical point, the authors demonstrate that different initial system trajectories may end up in the same hadro-chemical freezeout point.

~

\noindent Different initial system trajectories imply different "starting conditions", as governed solely by the the collisional center of mass energy. One expects that such starting points are determined by the primordial conditions, prevailing during the early equilibration phase, just after the end of target-projectile interpenetration. They are thus initially insensitive to the particular conditions, encountered later, in the vicinity of the parton-hadron phase transformation line. The system expansion trajectories, labelled by their initial entropy to baryon number ratios (as specified in  hydrodynamic models underlying such considerations), should therefore be shifted smoothly toward lower baryochemical potential, with increasing collisional energy.

~

\noindent The proposed study rests on the following assumption: if no special singularities arise along the parton-hadron coexistence line (such as a critical point, and/or associated transition from crossover to a first order nature of the phase transformation) the smooth spacing with energy of the initial system expansion trajectories should in turn lead to a similar smooth spacing of the hadronic chemical freeze-out points. These points in the ($\mu_{B}$,T) plane are obtained from the statistical, grand canonical hadronization model~\cite{Alt:2006dk,
Adams:2006ke,Takahashi:2007bh,Stock:1999hm,BraunMunzinger:2001as} which analyzes the observed sets of hadronic production ratios, at various energies. This analysis derives, for each energy, a "hadrochemical freeze-out point" in the ($\mu_{B}$,T)  plane. These points are shown in Fig.~\ref{Fig:PhaseDiagram}. The  expanding system must to go through these points at hadronization. Thus these points, in their succession with incident energy, mark an entry concerning the position of the respective expansion trajectory in this plane. If trajectory re-focusing occurs, owing to the attractor mechanism proposed as a consequence of a critical point, the smooth dependence of the freeze-out points on initial collisional energy should be re-shuffled.

~

\noindent What is required for a search concerning this fascinating conjectured mechanism is a systematic precision measurement of the hadron production yields (including the hyperons and anti-hyperons) at each incident energy. At the relevant low energies of the intended energy scan, the previous data from the SPS low energy runs do not offer adequate event statistics. The resulting, apparently smooth succession of freeze-out points, as illustrated in Fig.~\ref{Fig:PhaseDiagram},  need to be revisited by high statistics STAR runs at these energies. 

\subsection{Locating a 1$^{st}$ Order Transition and/or Changing the Equation of State}\label{SubSec:OrderedTransition}

\subsubsection{Elliptic Flow}\label{SubSubSec:v2}

\begin{figure}[htb]
\begin{minipage}{0.5\linewidth}
\begin{center}
\includegraphics[width=\linewidth]{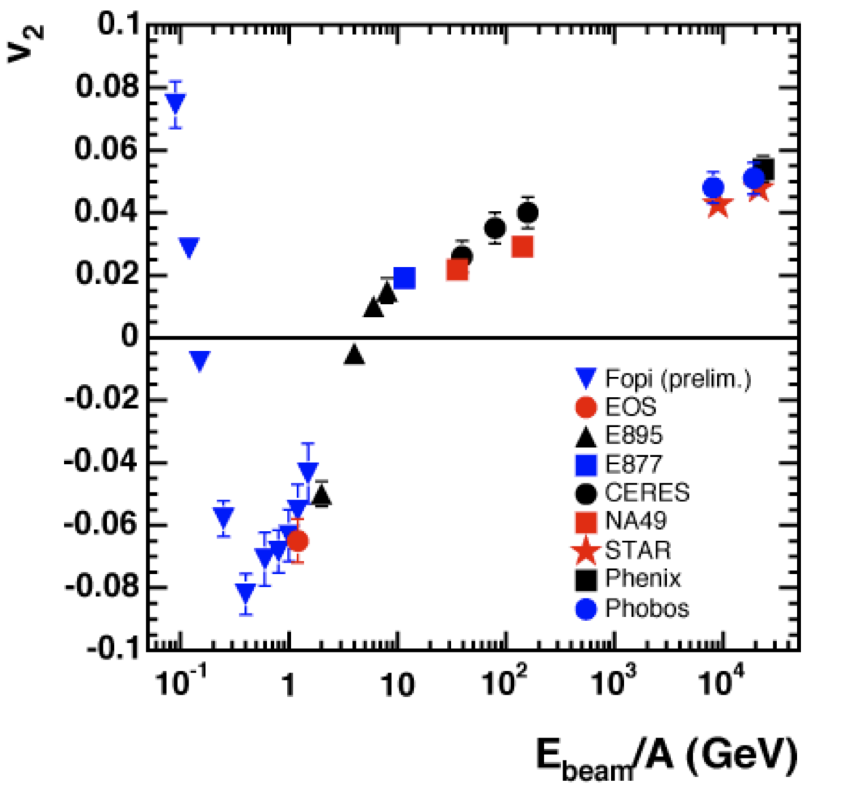}
\end{center}
\caption{The charged particle v$_{2}$ as a function of beam energy.}
\label{Fig:v2sqrts}
\end{minipage}
\hspace{1cm}
\begin{minipage}{0.5\linewidth}
\begin{center}
\includegraphics[width=\linewidth]{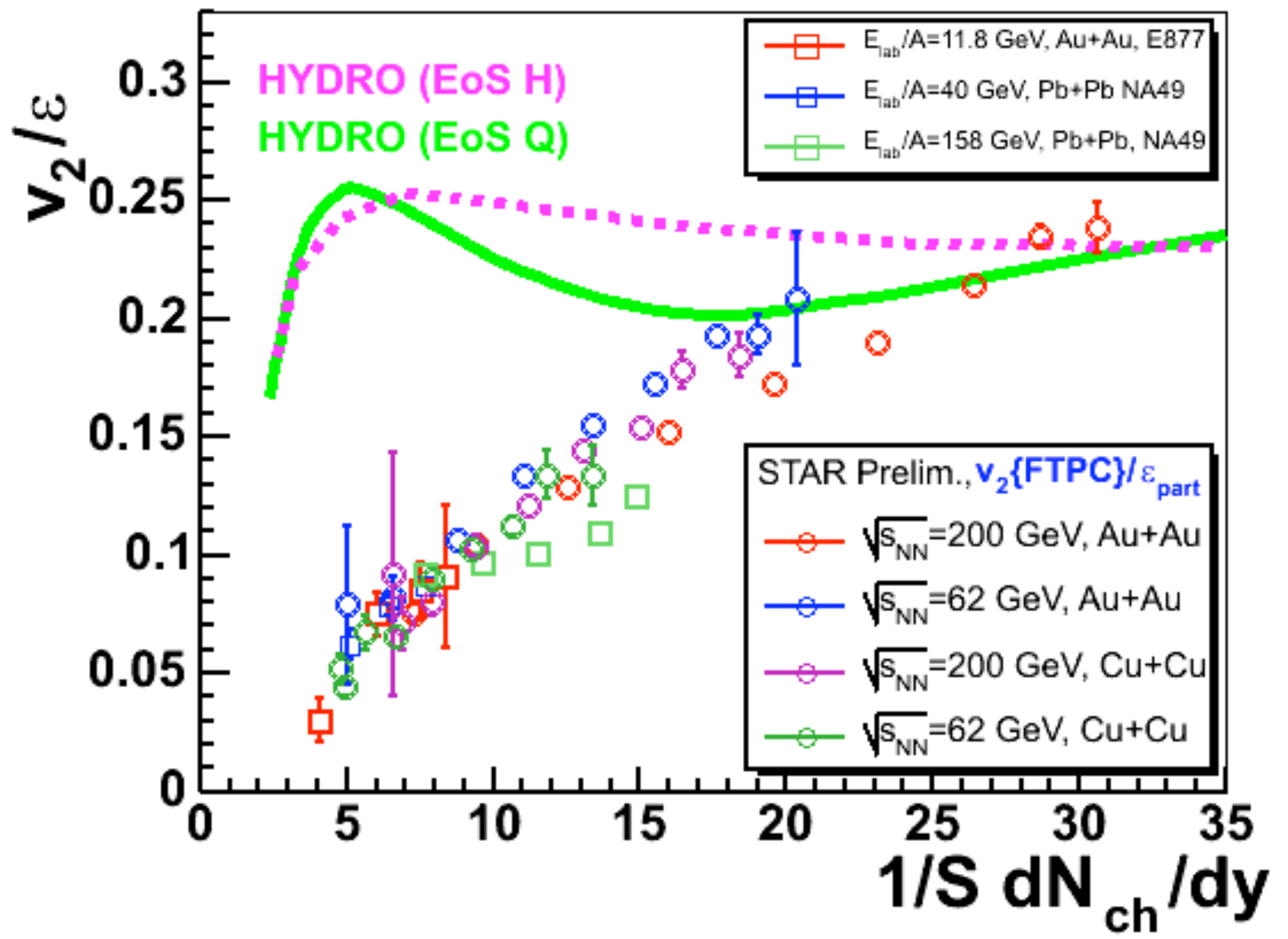}
\caption{The charged particle v$_{2}$/$\epsilon$ as a function of the charged particle mid-rapidity density. From~\cite{Voloshin:2008dg}.}
\label{Fig:v2overEps}
\end{center}
\end{minipage}
\end{figure}

\noindent Elliptic flow (v$_{2}$) measurements have been performed over three orders of magnitude in \sqrts. The most plentiful systematics are available for unidentified particles.  Figure \ref{Fig:v2sqrts} shows the least detailed measure of anisotropy -- \pT- and particle-type-integrated v$_2$-- over a broad range of energies ~\cite{Adler:2004cj}.  The strong non-monotonic behavior observed in the excitation function reflects an evolution in the driving physics.  At the lowest energies, the colliding nuclei orbit each other and may fuse to form a rotating compound nucleus which evaporates particles in-plane (v$_{2}>$0) due to angular momentum considerations.  As the beam energy increases, the crossing timescale becomes much shorter than the thermalization time, and the passing beam nucleons are too fast to be captured by the target nucleus' mean field; at this point, the system is appropriately described in the language of participants and spectators.  At beam energies $\sim$ 200 AMeV, the passing spectators absorb particles emitted by the participant zone in the reaction plane, leading to out-of-plane "squeeze-out" ~\cite{Voloshin:2008dg}, and v$_{2}>$0.  As the energy is increased to top AGS energy (Ebeam $\sim$10 AGeV) and beyond, Lorentz contraction effects and decreasing crossing timescales gradually render the spectators irrelevant.  It is at these energies, \sqrts $>$5 GeV, that the emission anisotropy reflects that from a hot, compressed anisotropic zone, and hydrodynamic expansion is assumed to be the physics driving the evolution.

~

\noindent Above AGS energies any energy dependence is expected to be driven not by changing physics mechanisms, but by characteristics -- equation of state, viscosity, the number of degrees of freedom etc-- of the system determining the hydrodynamic evolution.  Such effects will be more subtle than the obvious structures in Fig.~\ref{Fig:v2sqrts}  described above.  Hence, we will need more detail than integrated v$_{2}$ values. First, if instead of plotting against \sqrts one calculates v$_{2}$/$\epsilon$ (where $\epsilon$ is the initial- state eccentricity)  and plots this quantity  versus the measured charged particle density per unit overlap area, S, one sees that all the data from different energies collapse onto a common curve, Fig.~\ref{Fig:v2overEps}~\cite{Voloshin:2008dg}. Also one can see that it is only at the largest particle densities, corresponding to central collisions at maximum RHIC energies, that the data reach the hydrodynamical limit for elliptic flow. This immediately provokes the question about the validity of using  hydrodynamical calculations for lower collision energies. In light of this and other results much work has been done recently to implement viscous hydrodynamical theories.

~

\noindent The bulk nature of the hypothesized sQGP phase is revealed in strong {\em elliptic} flow, which in central collisions approaches the predictions of ideal hydrodynamics, assuming system thermalization on an extremely short timescale ($\sim 0.5$~fm/c)~\cite{Kolb:2003dz,Huovinen:2006jp}.  However, the mechanism behind such rapid thermalization remains far from clear and is under active theoretical study~\cite{Mrowczynski:2006ad,Muller:2006ad}.  This has a connection to another novel phenomenon that could be relevant at RHIC --- saturation of the gluon distribution, often referred to as the Color Glass Condensate or CGC--- which characterizes the nuclear parton distribution prior to collision~\cite{McLerran:1994vd}. Various theoretical approaches to connect collision geometry, saturated gluon distributions, and the onset of bulk collective behavior are being explored~\cite{Kovchegov:2007pq, Kolb:2003dz,Huovinen:2006jp}; more experimental input is needed to guide these efforts.

~

\begin{figure}[htb]
\begin{minipage}{0.46\linewidth}
\begin{center}
\includegraphics[width=\linewidth]{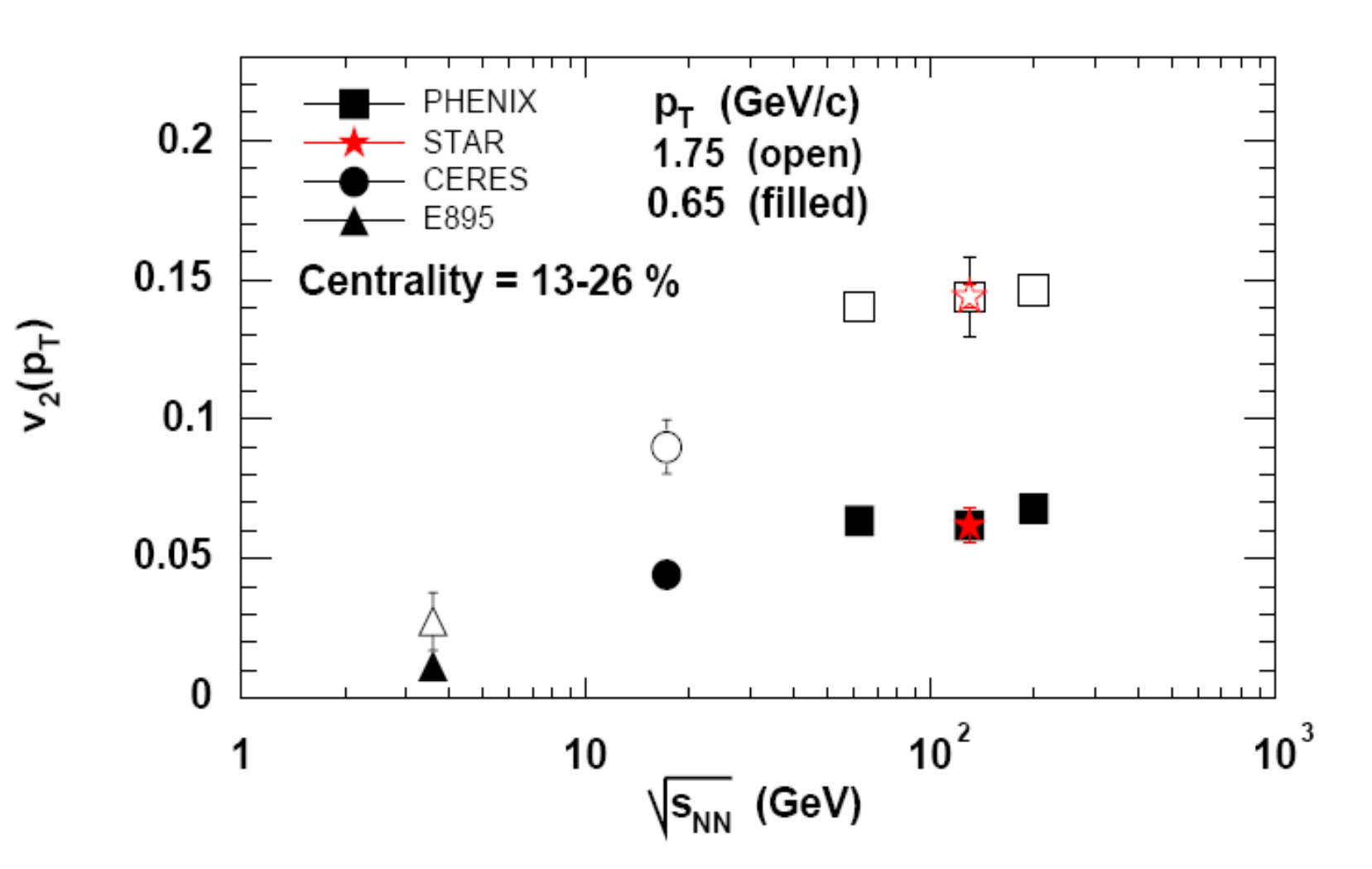}
\end{center}
\caption{The charged particle v$_{2}$ as a function of \sqrts for two different \pT values. From \cite{Adler:2004cj}.}
\label{Fig:v2sqrtspt}
\end{minipage}
\hspace{1cm}
\begin{minipage}{0.46\linewidth}
\begin{center}
\includegraphics[width=\linewidth]{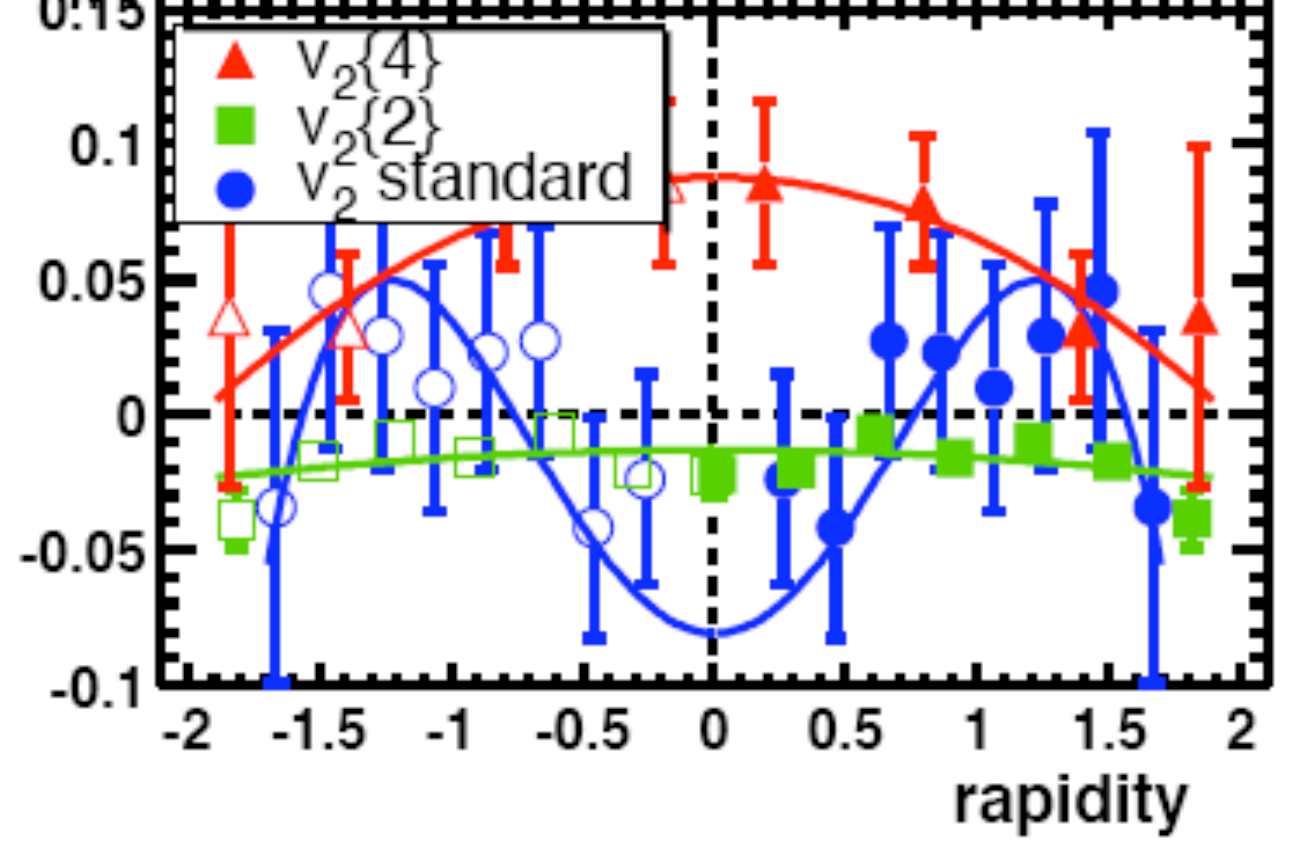}
\caption{Proton v$_{2}$ as a function of rapidity at \sqrts= 8.77 GeV from Pb-Pb collisions for centrality bin b=3.4-5.3 fm. From \cite{Alt:2003ab}.}
\label{Fig:v2ProtonSPS}
\end{center}
\end{minipage}
\hspace{1cm}
\end{figure}

\noindent If one  performs more differential analyses and looks at v$_{2}$(\pT) as a function of \sqrts the results shown in Fig.~\ref{Fig:v2sqrtspt} emerge. For a fixed centrality the v$_{2}$ at fixed \pT values first grows with \sqrts but then reaches a plateau. This leveling-off at similar \sqrts for differing transverse momenta might present evidence of a softening of the equation of state due to a phase transition. It will be extremely enlightening to see if this feature is preserved for differing centralities and using identified particle results.  Moreover, the energy where the leveling-off begins lies  in the unexplored region of the RHIC BES, and ought to be determined with more precision. Table~\ref{Table:v2} shows the number of events  estimated to be needed in order to measure the inclusive particle v$_{2}$ up to \pT $\sim$1.5 GeV/c.

\begin{table}[htb]
\caption{Estimate of events needed to measure the inclusive particle v$_{2}$ up to \pT $\sim$1.5 GeV/c. } 
\begin{center}
\begin{tabular}{|c|c|c|c|c|c|c|}
\hline
\sqrts (GeV) & 5 & 7.7  &11.5  &17.3 & 27 & 39 \\
\hline
 Number of Events v$_{2}$ &  0.3 M & 0.2 M & 0.1 M & 0.1 & 0.1 M &  0.1 M\\
\hline
\end{tabular}
\end{center}
\label{Table:v2}
\end{table}

~

\noindent A potential direct signature of a phase transition is the ``collapse" of the proton v$_{2}$~\cite{Danielewicz:1999vh,Kolb:1999it,Kolb:2000sd}. NA49 first reported observation of such a result for \sqrts= 8.77 GeV. However,  when they repeated this measurement using other techniques to extract v$_{2}$ the results differ, Fig.~\ref{Fig:v2ProtonSPS}~\cite{Alt:2003ab}. The results presented in Fig.~\ref{Fig:v2ProtonSPS}  are from  the ``standard" v$_{2}$ method (blue circles),  cumulants for two-particle correlations, v$_{2}\{2\}$, (green squares) and cumulants for four-particle correlations, v$_{2}\{4\}$, (red triangles). The differences between these   results are beyond the statistical error bars, however, the techniques used are known to have different sensitivities to non-flow effects, such as decays (see ~\cite{Alt:2003ab} and references within for more details).  It is currently unclear if these differences are due to non-flow effects or  physical fluctuations of the v$_{2}$ that come in with differing weights to these different techniques, or represent an indication of the systematics of the measure. With the BES we can perform these measurements with higher statistics and better systematics. Also, recent advances in experimental techniques have developed methods to determine the scale of the v$_{2}$ fluctuations. It is therefore essential that we recalculate the v$_{2}$ and perform the fluctuation measurements with the same apparatus to try and disentangle all these effects.   The feasibility of such a measurement is shown in Section~\ref {Sec:AnalysisOfCurrentTestRuns} where  statistically relevant identified v$_{2}$ measurements have been made with only $\sim$3 k  events. 

~

\begin{figure}[htb]
\begin{minipage}{0.46\linewidth}
\begin{center}
\includegraphics[width=\linewidth]{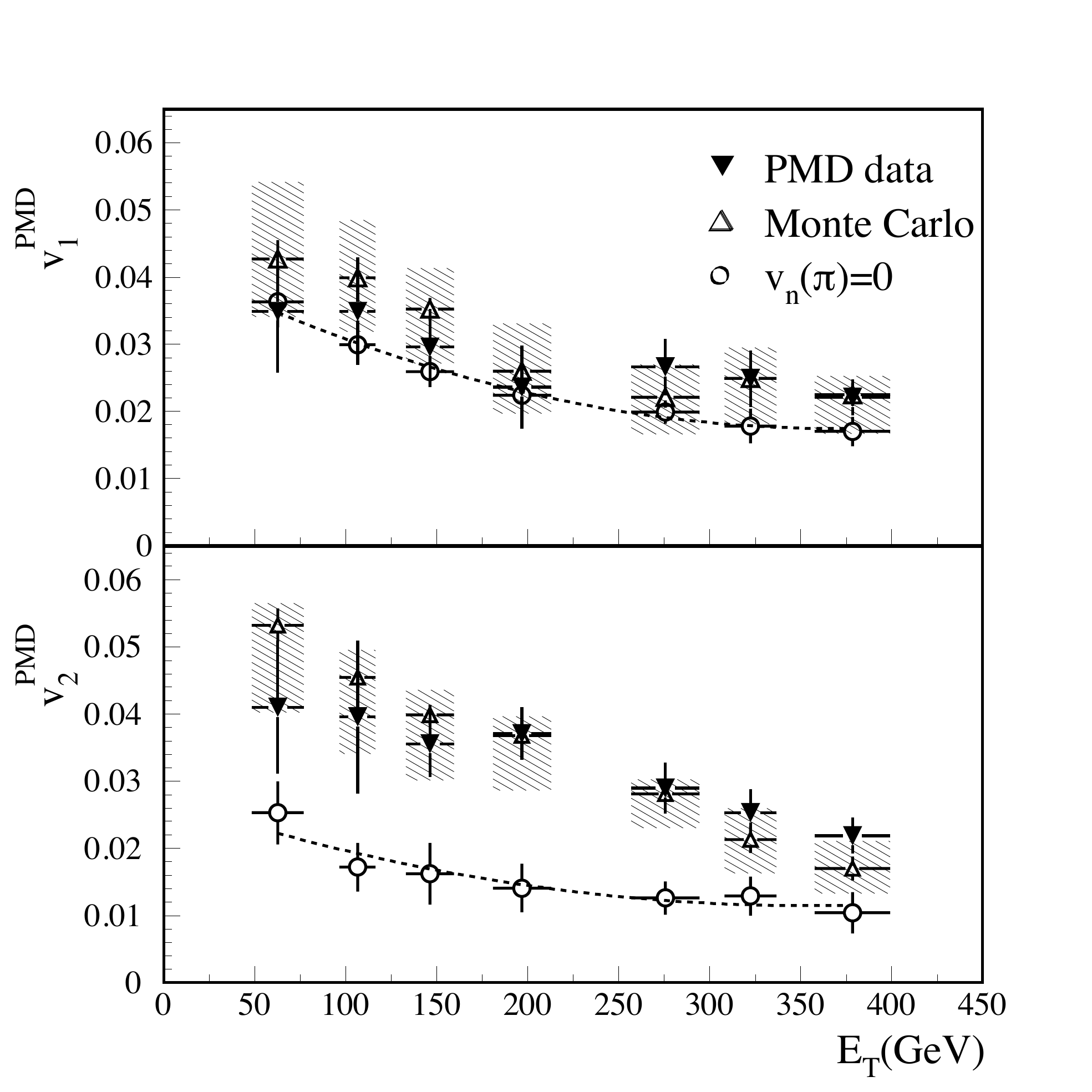}
\end{center}
\caption{Directed and elliptic photon anisotropy coefficients in the pseudorapidity region 3.25-3.75 for different centralities  as measured in PMD for WA98 experiment at SPS energies \cite{Aggarwal:2004zh}. For more details see \cite{Aggarwal:2004zh}.}
\label{Fig:v2photon}
\end{minipage}
\hspace{1cm}
\begin{minipage}{0.46\linewidth}
\begin{center}
\includegraphics[width=\linewidth]{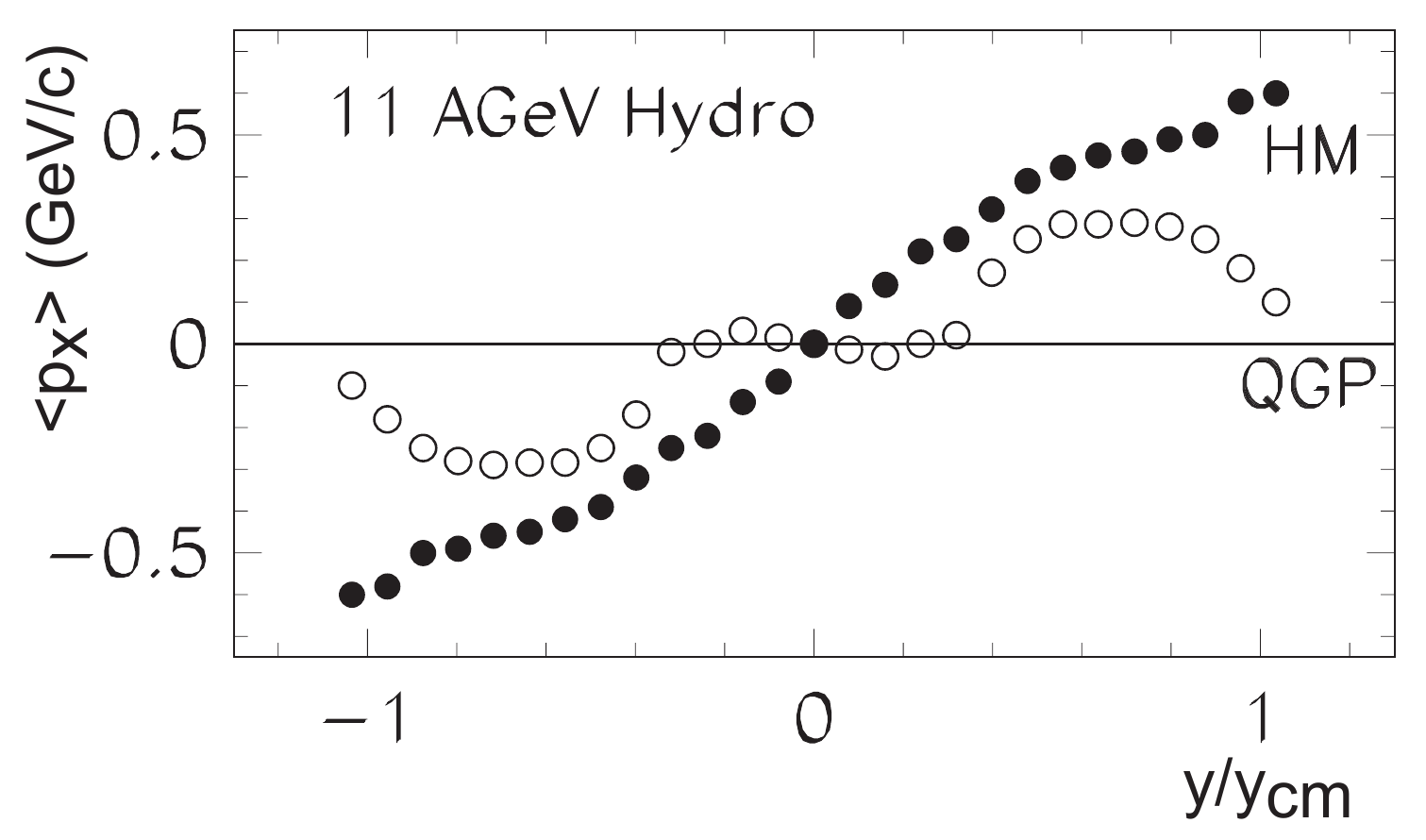}
\caption{ Directed flow of protons from ideal hydrodynamics with a QGP phase (open symbols)
and from the Quark Gluon String Model without QGP phase (full symbols)~\cite{Csernai:1999nf}. Figure from \cite{Stoecker-2005-750}.}
\label{Fig:wiggle}
\end{center}
\end{minipage}
\hspace{1cm}
\end{figure}

\noindent The first evidence of photon azimuthal anisotropy at SPS energies was observed in the distribution of photons from S+Au collisions at SPS energies measured in the pre-shower photon multiplicity detector of the WA93 experiment at CERN \cite{Aggarwal:1997iu}.  Subsequent measurements of both directed and elliptic flow of photons were performed in the WA98 experiment at SPS, these are shown in Fig. \ref{Fig:v2photon} \cite{Aggarwal:2004zh}.  The preliminary measurements at RHIC using the PMD were reported in Ref.~\cite{Raniwala:2008zza}.   These photon measurements will complement the charged particle measurements at mid-rapidity.

~

\subsubsection{Directed Flow}\label{SubSubSec:V1}

\noindent Directed flow is obtained from the first harmonic (v$_{1}$) in the Fourier expansion of the azimuthal anisotropy of emitted particles with respect to the collision reaction plane, $\Psi_r$,~\cite{Poskanzer:1998yz}.  

\begin{eqnarray*}
E\frac{d^3N}{dp^3} = \frac{1}{2\pi}\frac{d^2N}{p_{T}dp_{T}dy}(1+ \sum_{n=0}^{\infty} 2v_{n}\cos[n(\phi-\Psi_r)])
\end{eqnarray*}

\noindent It describes collective sideward motion of the produced particles and nuclear fragments and, even more so than elliptic flow, carries information on the very earliest stage of the collision.  Specifically, at large $\eta$ (in the  fragmentation region), directed flow is believed to be generated during the nuclear passage time ($2R/\gamma \sim 0.1$~fm/c)~\cite{Sorge:1996pc,Herrmann:1999wu}.  It therefore probes the onset of bulk collective dynamics during thermalization, providing valuable experimental guidance to models of the pre-equilibrium stage.

~

\noindent  The shape of $v_1$ vs.~rapidity is of special interest because it has been identified in several theoretical papers as a promising Quark-Gluon Plasma signature~\cite{Brachmann:1999xt,Csernai:1999nf,Stoecker-2005-750}.  At low relativistic energies, $v_1(y)$ is almost directly proportional to rapidity, and relative to protons, the pion $v_1(y)$ is significantly smaller in 
magnitude and opposite in sign.   The sign of $v_1$ is conventionally defined as positive by the direction of nucleon flow in the projectile fragmentation region.  Often, just the slope of $v_1(y)$ at mid-rapidity has been used to define the strength of directed flow.  At RHIC energies, directed flow is predicted to be small near mid-rapidity with very weak dependence on pseudorapidity.  Calculations involving a QGP phase suggest that  $v_1(y)$ may exhibit a characteristic ``wiggle'' \cite{Brachmann:1999xt,Csernai:1999nf,Stoecker-2005-750,Snellings:1999bt,Bleicher:2000sx}, whereby directed flow changes sign three times, not counting a possible sign change near beam rapidities (in contrast to the observed sideward deflection pattern at lower energy, where the sign changes only at 
mid-rapidity), see Fig.~\ref{Fig:wiggle},  if a QGP phase transition is assumed. In these calculations, the wiggle structure is interpreted as a consequence of the expansion of the system, which is initially tilted with respect to the beam direction.  
The expansion leads to the so-called anti-flow or third flow component.  Such a flow can reverse, over a region on either side of mid-rapidity, the normal pattern of sideward deflection as seen at lower energy, and hence can result in either a flatness of $v_1$, or a wiggle structure if the expansion is strong enough.  However, a similar wiggle structure in proton $v_1$ is also predicted if one assumes strong but incomplete baryon stopping together with strong space-momentum correlations caused by transverse radial 
expansion \cite{Snellings:1999bt,Bleicher:2000sx}.  The situation for pion directed flow is less clear in such models. While RQMD model calculations indicate that shadowing by protons causes the pions to flow mostly with opposite sign to the protons, mirroring the proton wiggle, other calculations predict that pions flow with opposite sign only in a limited rapidity range. 

~

\begin{figure}[htb]
\begin{center}
\includegraphics[width=0.5\linewidth]{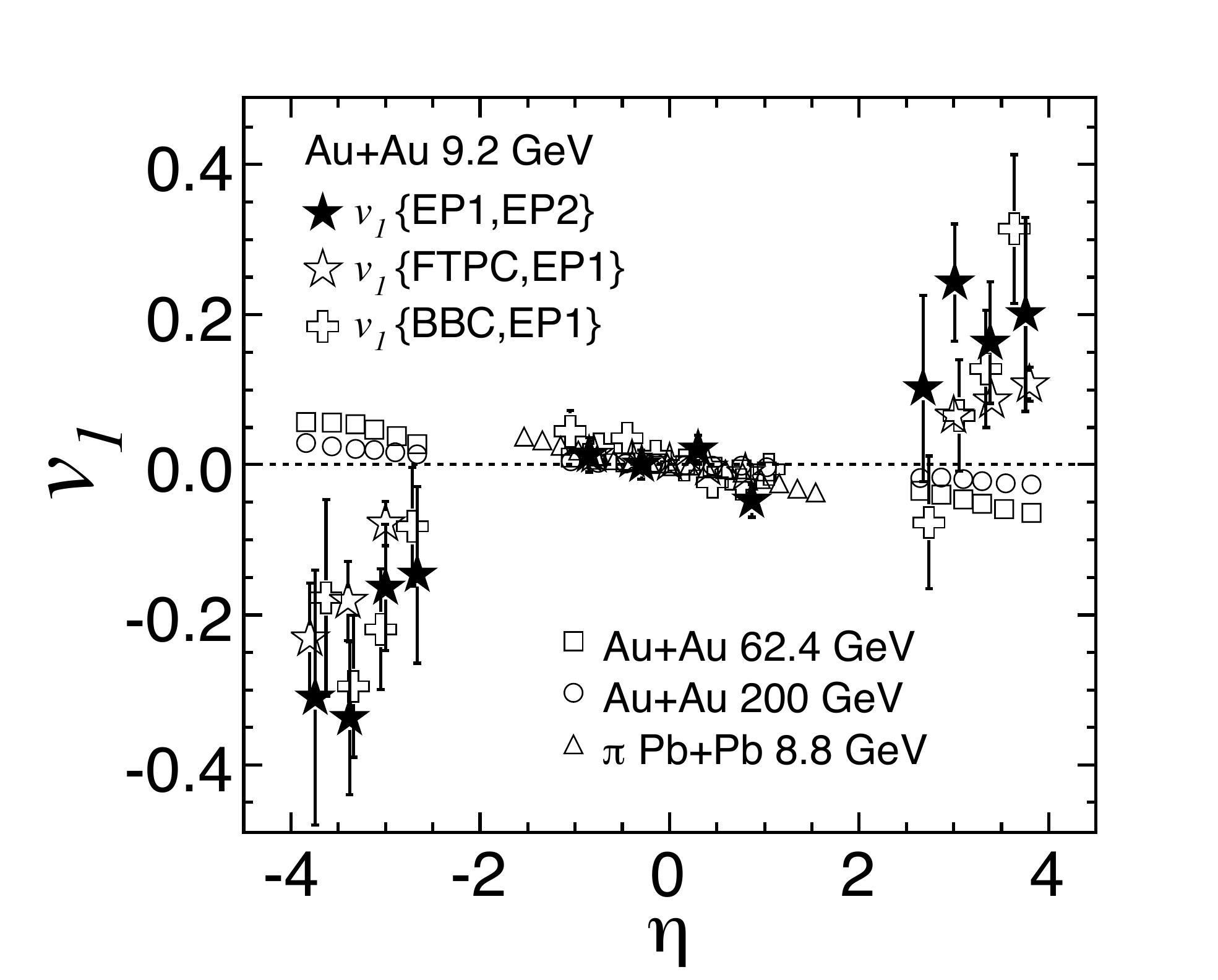}
\caption{The charged particle v$_{1}$ as a function of $\eta$ for Au+Au collisions at \sqrts=9.2, 62.4 and 200 GeV~\cite{Abelev:2009bwa}, and for identified pions in \sqrts=8.8 GeV Pb+Pb collisions~\cite{Alt:2003ab}.}
\label{Fig:v192}
\end{center}
\end{figure}

\noindent It is not until we move down into the collision energy domain of the proposed BES that the acceptance in pseudorapidity of STAR's Forward TPCs comes close enough to the rapidity region of the incoming nuclei to fully map-out all the changes in sign of $v_1$, and therefore, a more thorough understanding of the wiggle phenomenon will then be possible.  Figure~\ref{Fig:v192} shows the measured directed flow for Au+Au collisions at \sqrts = 9.2, 62.4 and 200 GeV for unidentified charged hadrons, and for identified pions in 0-60$\%$ centrality Pb+Pb collisions at \sqrts=8.8 GeV~\cite{Alt:2003ab}. For the 62.4 and 200 GeV the data is from 30-60$\%$ central data, due to a lack of statistics the 9.2 GeV covers 0-60$\%$. This  $v_1$ study at $\sqrt{s_{NN}} =9.2$ GeV, based on just a few hours of sporadic collisions during a machine development test in March 2008 has already corroborated the promising nature of this line of investigation. The predicted sign change near beam rapidities is clearly evident for the lower energy data but there are insufficient statistics currently to attempt to identify any possible ``wiggles" at mid-rapidity.

~

\noindent Based on experimental analyses to date, and on simulations of the performance of the STAR TOF system,  a relatively modest event sample (0.5 M)--- many times smaller than the few million minimum-bias events needed when using elliptic flow of abundant baryons and mesons to study constituent quark scaling --- will be sufficient to unambiguously resolve the pending questions about the ``wiggle" phenomenon,  discussed above.

~

\noindent Important insights into the evolution of the system also can be obtained from flow of identified particles.  At low \sqrts most of the  protons are transported to mid-rapidity by baryon stopping while pions are created from the collisions, and the difference of their flow pattern will shed light on the evolution of the system.  In particular, if the $v_1$ of protons and pions would have the same sign near mid-rapidity, then that would be a signature for a tilted source and would be consistent with one-fluid hydrodynamic models incorporating a first-order phase transition.  It has been argued by St\"ocker that the so-called 
``collapse of proton flow" reported at CERN may be evidence for such a transition~\cite{Stoecker-2005-750}.  All of the above considerations stimulate much interest in directed flow over the proposed range of the BES.      

~

\noindent A separate but equally curious sign-change phenomenon occurs for charged-particle $v_1(p_T)$ in the vicinity of $p_T \sim 1$ GeV$/c$ near mid-rapidity at higher RHIC energies (i.e. it is observed in the main TPC of STAR but not in the FTPCs).  If we assume that $v_1$ for pions and protons has opposite sign, then the steeply increasing relative abundance of protons above 1 GeV$/c$ has the potential to explain the change of sign, but more data are needed over a wider range of beam 
energies to determine if this phenomenon is properly understood.

\subsubsection{Identified Particle \pT Spectra}\label{SubSubSec:pTSpectra}

\begin{figure}[htb]
\begin{minipage}{0.46\linewidth}
\begin{center}
\includegraphics[width=0.8\linewidth]{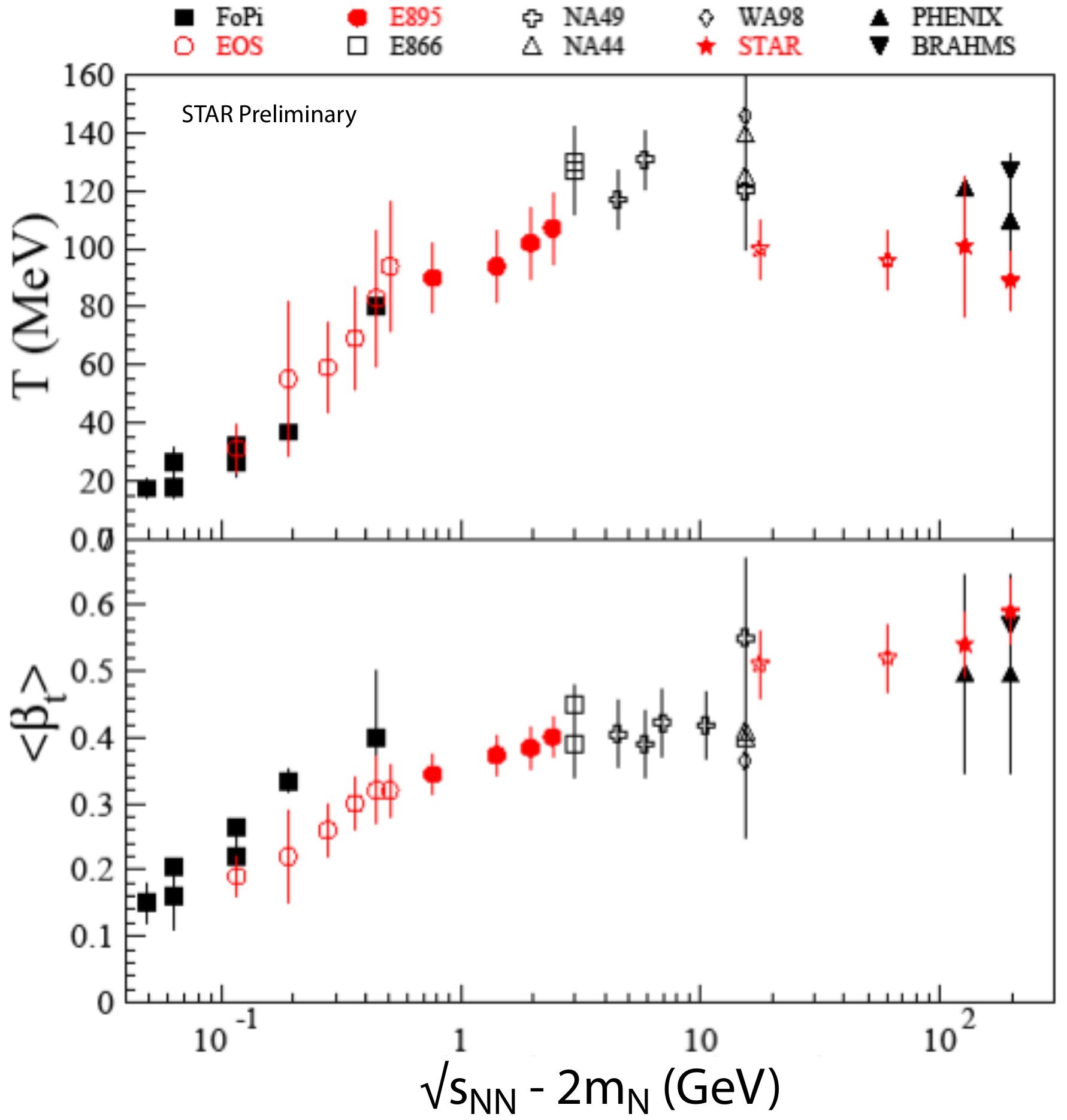}
\caption{The calculated thermal/kinetic freeze-out temperature and mean transverse radial flow as a function of \sqrts~\cite{Cebra:2009fx,Kumar:2008ek}.}
\label{Fig:TBeta}
\end{center}
\end{minipage}
\hspace{1cm}
\begin{minipage}{0.46\linewidth}
\begin{center}
\includegraphics[width=\linewidth]{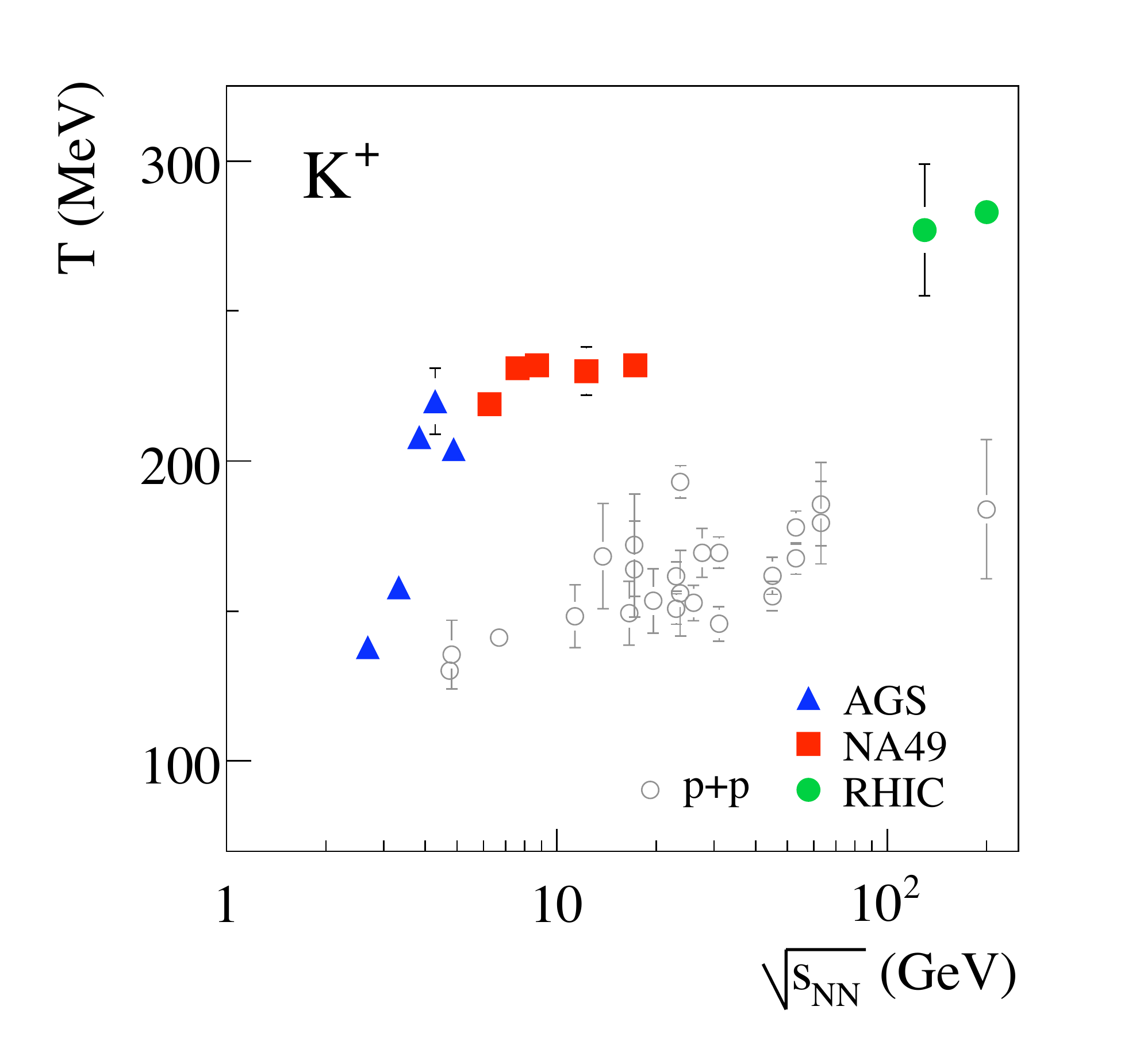}
\end{center}
\caption{The extracted inverse slope parameter of the \pT distributions of  K$^{+}$ as a function of \sqrts for A+A and p+p collisions. From \cite{Alt:2007fe}.}
\label{Fig:TKsqrts}
\end{minipage}
\end{figure}

\noindent Changes in shape of the particle \pT spectra as a function of \sqrts and centrality can be studied to reveal changes in the  kinetic freeze-out temperature of the hadrons from the medium and their mean transverse velocity. Typically a Blast-Wave function is used to extract these variables from fits to the \pT spectra~\cite{Schnedermann:1993ws}. The fits are usually restricted to  \pT$<$ 2 GeV/c so that the bulk of the particle kinematics are not dominated by hard processes. Fig.~\ref{Fig:TBeta} shows the extracted kinetic freeze-out temperature, T$_{kin}$,   and mean radial velocity, $\beta_{t}$, for identified pions as a function of \sqrts for the most central events~\cite{Cebra:2009fx,Kumar:2008ek}. It can be seen that at very low \sqrts there is a steep rise in the extracted T$_{kin}$ and $\beta_{t}$. For T$_{kin}$ a plateau appears at higher \sqrts, although the radial velocity continues to increase.

~

\begin{figure}[htb]
\begin{minipage}{0.46\linewidth}
\begin{center}
\includegraphics[width=0.7\linewidth]{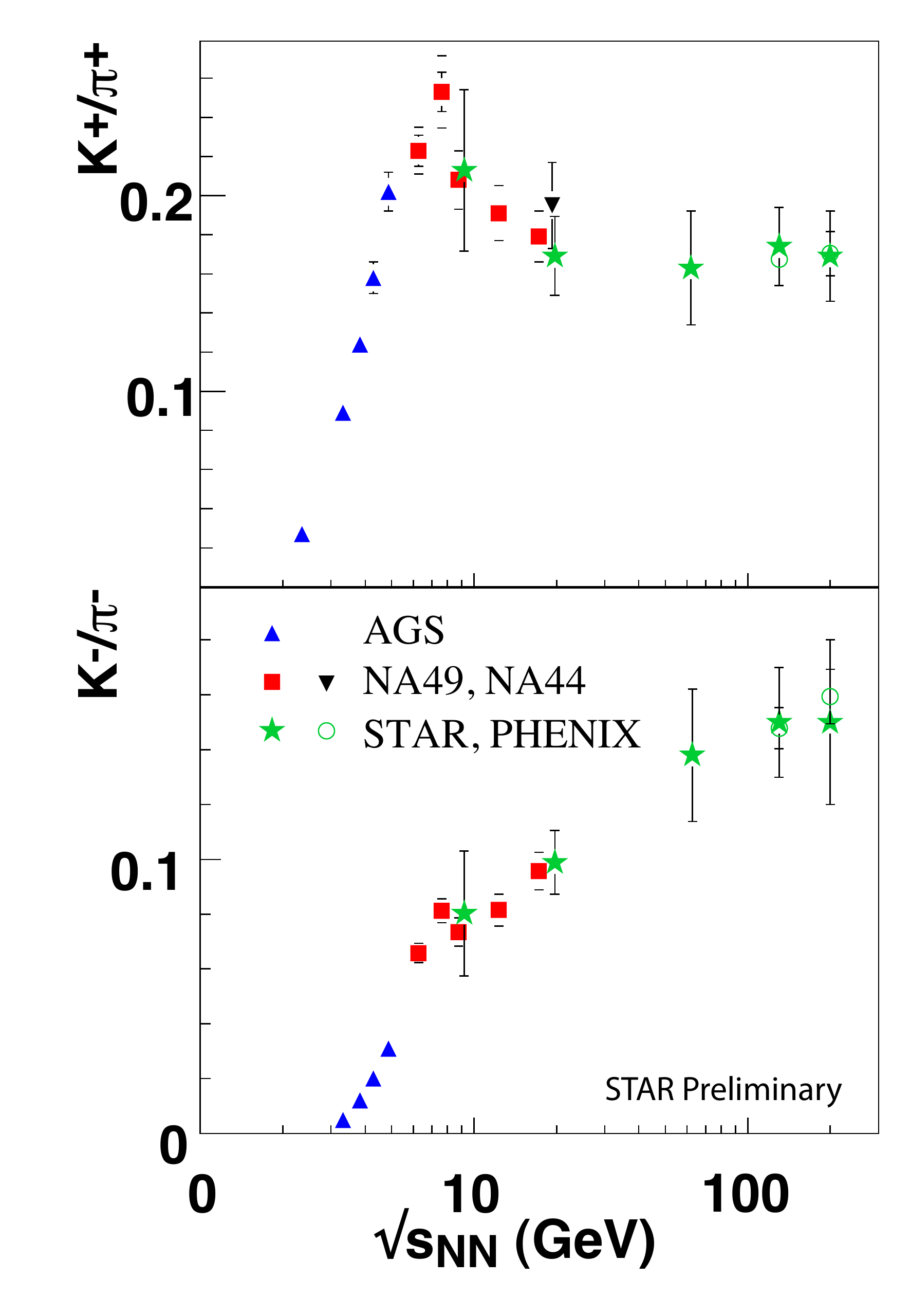}
\caption{The K$^{+}$/$\pi^{+}$ and K$^{-}$/$\pi^{-}$ ratio as a function of \sqrts~\cite{Odyniec:2008zz}. Low energy data added to plot from  \cite{Alt:2007fe}.}
\label{Fig:Kpi}
\end{center}
\end{minipage}
\hspace{1cm}
\begin{minipage}{0.46\linewidth}
\begin{center}
\includegraphics[width=0.8\linewidth]{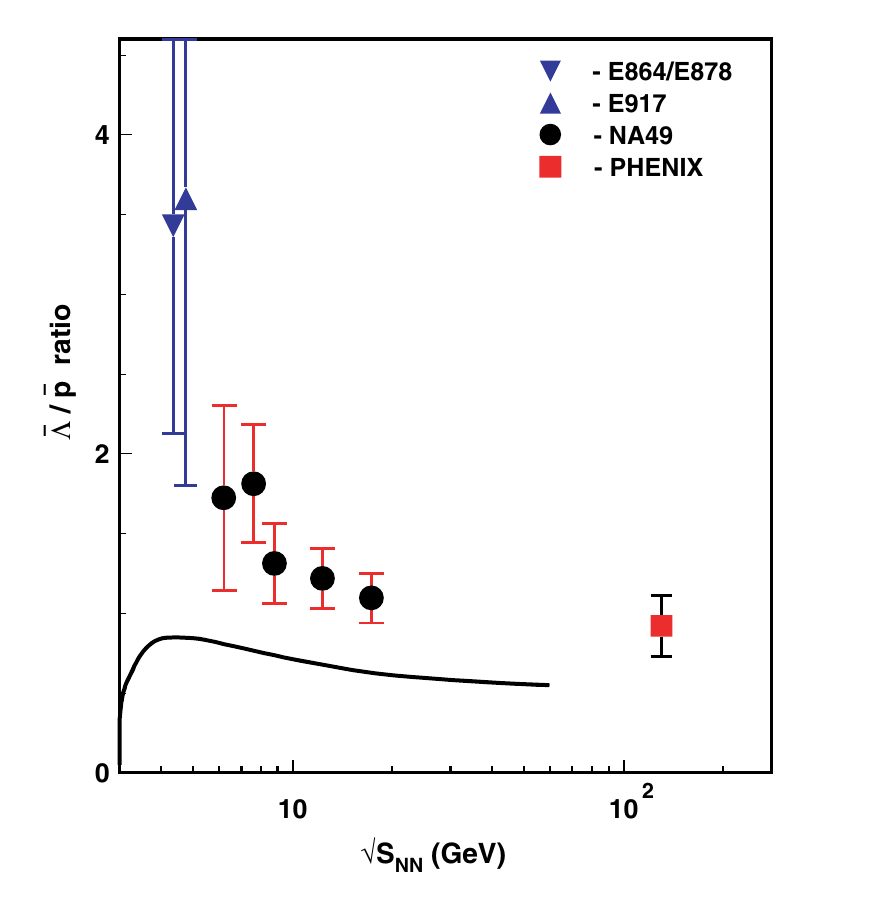}
\end{center}
\caption{The $\bar{\Lambda}$/$\bar{p}$ ratio as a function of \sqrts.  From ~\cite{Alt:2006dk}. The solid curve is a prediction from the statistical hadron gas model described in \cite{Cleymans:1998fq}. }
\label{Fig:lambdabarpbar}
\end{minipage}
\end{figure}

\noindent If instead of $\pi$ one looks at $K^{+}$, Fig.~\ref{Fig:TKsqrts},  a more interesting picture emerges. Instead of the extracted T$_{kin}$ rising steadily to the topmost collision energies, the measured inverse slope of the \pT distribution  appears to form a plateau for \sqrts $\sim$ 8-12 GeV before rising to significantly higher values  for \sqrts=130 and 200 GeV~\cite{Alt:2007fe}.  Such behavior is seen  for A+A collisions but not for p+p. If at intermediate \sqrts the system instead forms a mixed phase region the early stage pressure and temperature are predicted to become independent of the energy density~\cite{Gazdzicki:1998vd}.  This effect creates a step like dependence of the pressure and temperature on the collision energy.  This leads, in turn, to a  weakening of the increase of the inverse slope parameter with \sqrts as seen in Fig.~\ref{Fig:TKsqrts}. Measurements need to be made to see how far this possible plateau extends as this may signal the transition to a single phase QGP system. The K$^{+}$/$\pi^{+}$ ratio also exhibits interesting behavior around these collision energies (see subsection~\ref{SubSubSec:Horn} and Fig.~\ref{Fig:Kpi}).  

\subsubsection{The K/$\pi$ and $\bar{\Lambda}/\rm{\bar{p}}$ Ratios}\label{SubSubSec:Horn}

\noindent One of the most challenging set of results to model from the SPS are the K$^{+
}$/$\pi^{+}$ ratios as a function of \sqrts, shown in Fig,~\ref{Fig:Kpi} with data from \cite{Afanasiev:2002mx,Alt:2007fe,Abelev:2008ez,Ahle:1998gv,Ahle:1999va,Klay:2003zf}.  Given this result, and that of the fitted inverse slopes of the K$^{+}$ (Fig.~\ref{Fig:TKsqrts}), it is  essential to go back and study this region in greater detail, as it suggests novel physics occurring. Smaller error bars will also prove if this peak is merely a statistical fluctuation. With our increased precision we will also be able to  study this ratio differentially and see if this peak persists as a function of \pT and centrality.

~

\noindent Another ratio that is equally hard to explain with statistical hadronization models is the  $\bar{\Lambda}/\rm{\bar{p}}$ ratio.  This ratio becomes extremely high at smaller \sqrts, Fig.~\ref{Fig:lambdabarpbar} ~\cite{Alt:2006dk}. These measurements, especially those at the AGS, were an experimental tour-de-force, but lack statistical precision. STAR's large acceptance and high reconstruction efficiency of the $\bar{\Lambda}$ will drastically reduce the systematic and statistical errors. It is estimated that 1 Million events at \sqrts= 6 GeV will result in several thousand reconstructed $\bar{\Lambda}$. 

~

\noindent It is possible that  the dramatic rise at low \sqrts is due to p-$\rm{\bar{p}}$ annihilation in the baryon-rich medium produced at these energies~\cite{Koch:1989zt,Wang:1998sh}. However, it remains to be seen if this can account for all of the observed behavior since  K$^{+}$/$\pi^{+}$ and $\bar{\Lambda}/\rm{\bar{p}}$ both represent $\bar{s}/\bar{d}$ and should therefore carry some of the same physics. It is also evident that there is a gap in these measurements between \sqrts=20-60, under our proposed run plan this region will be filled in.

\subsubsection{Interferometry}

\noindent By studying interferometry as a function of beam energy we can infer the energy density of the medium produced at the last re-scattering of the hadrons. The source dimensions, or homogeneity regions, as determined via HBT encode different information. R$_{side}$  only contains information about the spatial extension while R$_{out}$ holds spatial and temporal data. The ratio R$_{out}$/R$_{side}$ can therefore reveal the emission duration of the source. It is predicted that for a first order phase transition  this ratio should get much greater than unity due to a stalling in the emission during the phase transition~\cite{Rischke:1996em}.  Measurements of R$_{out}$, R$_{side}$ and R$_{long}$ as a function of collision energy have been studied in great detail and are shown, including new results at \sqrts = 9.2 GeV, later  in Fig.~\ref{Fig:HBTsqrts}. No major jumps in  R$_{out}$/R$_{side}$ are observed.

~

\begin{figure}[htb]
\begin{minipage}{0.7\linewidth}
\begin{center}
\includegraphics[width=\linewidth]{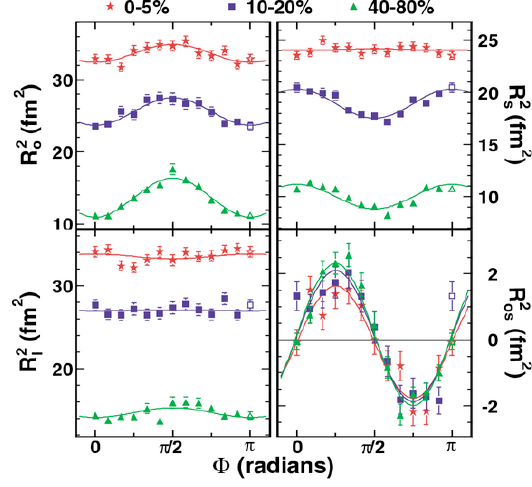}
\end{center}
\caption{HBT radii relative to 2nd-order event plane, in Au+Au at 200 GeV \cite{Adams:2003ra}. }
\label{Fig:HBTAzimuth}
\end{minipage}
\hspace{1cm}
\begin{minipage}{0.7\linewidth}
\begin{center}
\includegraphics[width=\linewidth]{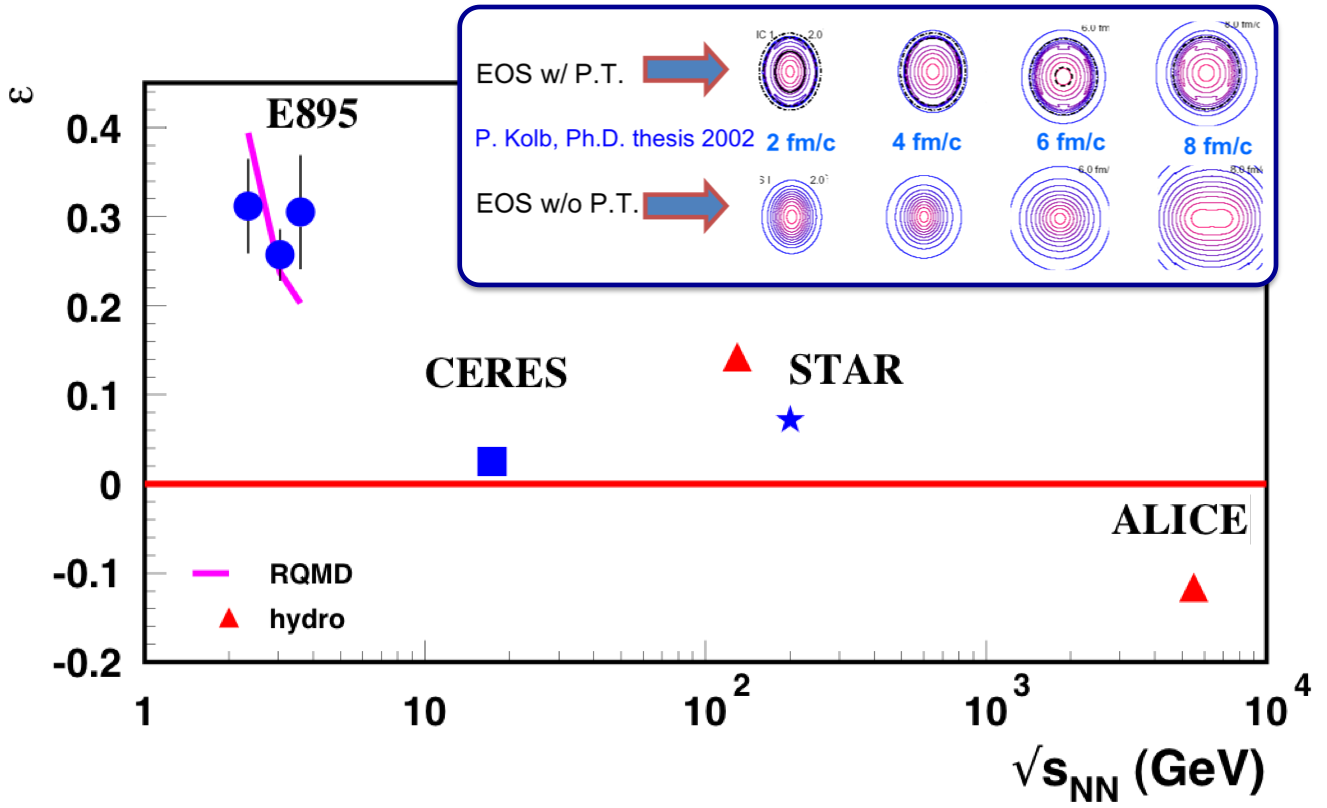}
\caption{Freeze-out anisotropy from 2$^{nd}$-order oscillations of HBT radii. Inset shows hydro. evolution of the source shape for an equation of state with (upper) and without (lower) softening due to finite latent heat \cite{Lisa:2008gf}.}
\label{Fig:FreezeoutAnis}
\end{center}
\end{minipage}
\end{figure}

\noindent Momentum spectra and anisotropy tell only half of the story of collective flow.  The bulk response of the system has a non-trivial structure in both space and time.  Just as the \pT-dependence of azimuthally-integrated HBT radii gives access to the geometric substructure generated by radial flow (e.g. \cite{Retiere:2003kf}), HBT measured relative to the first- and second-order event plane are the spatial analogs of directed and elliptic flow  \cite{Retiere:2003kf, Lisa:2000ip}, respectively, and contain important information not accessible in momentum space alone.  As explained below, these measurements can be sensitive to a ÒsofteningÓ in the equation of state, related to a first-order phase transition, or even rapid crossover.

~

\noindent STAR has measured oscillations of pion HBT radii relative to the second-order event plane (Fig.~\ref{Fig:HBTAzimuth}) \cite{Adams:2003ra}.  In addition to the overall size of the source, these reveal that the transverse shape is  extended out of the reaction plane at freeze-out (the stage probed by HBT).  It is, however, less anisotropic (more round) than the initial source defined by the overlap of the colliding nuclei at finite impact parameter, reflecting the evolution over time of preferential in-plane expansion.  
The anisotropy has been measured at a few lower energies, as well.  Since the lifetime of the system and the elliptic flow increase with collision energy, one naively expects that, for a fixed initial anisotropy, the freeze-out anisotropy becomes less and less out-of-plane extended, and may even become in-plane extended (as predicted \cite{Heinz:2002un}  for example at the LHC).  Figure \ref{Fig:FreezeoutAnis} shows the  freeze-out anisotropy calculated from the  second-order oscillations of these HBT radii as a function of collision energy. The results show an intriguing non-monotonic behavior.  A possible explanation may be as follows: at low energies (say 3-10 GeV) the stiff equation of state of a hadronic system generates a large pressure, pushing the system quickly towards a round shape.  But at some energy (say 20 GeV), a threshold to generate a phase transition is crossed, characterized by a finite latent heat.  This generates a Òsoft pointÓ in the equation of state, and the push towards a round, $\epsilon$ = 0, state stalls briefly.  As the energy increases beyond this threshold, the time spent in the Òsoft stateÓ grows, and the Òout-of-plane-nessÓ at freeze-out grows with energy, until some point (say 70 GeV).  Then, at even higher energy, the system spends most of its time in the (stiff) QGP phase, and the Òout-of-plane-nessÓ again decreases with energy with no further non-monotonic behaviour.  This would be the direct analog of the non-monotonic excitation function of v$_{2}$ originally predicted by ideal hydro models with a softening due to a phase transition \cite{Kolb:1999it}.  The signal in v$_{2}$ has not been observed, perhaps because increasing viscous effects at lower energies smears the structure.  However, the spatial anisotropy probed by HBT is weighted in the time evolution differently, so may retain sensitivity to the softest point.  Figure~\ref{Fig:FreezeoutAnis} represents one of the very rare bulk-sector probes with a non-monotonic excitation function.  Especially given its potential to probe the long-sought ``soft point,Ó this excitation function must be mapped.

~

\noindent Measuring HBT correlations relative to the first-order event plane yields even more unique geometrical information.  In particular, if one approximates the spatial configuration of the freeze-out system as an ellipse, one can extract the angle between its major axis and the beam direction.  This Òtilt angleÓ \cite{Lisa:2000ip} is the spatial analog of the so-called Òflow angleÓ \cite{Reisdorf:1997fx} formerly used to characterize directed flow.  It is, however, much larger ($\sim$40$^{0}$ at AGS energies, compared for flow angles of $<2^{0}$) and may even have opposite sign.  Simultaneous measurement of both the tilt and flow angles provides unique insight on the nature and physics behind directed flow at lower energies.  At RHIC energies, the directed flow becomes even more important, since here we are studying bulk response of the system at the very earliest stages of the collision.  As discussed above, crossing a threshold to a phase transition will generate a ÒwiggleÓ in the directed flow at midrapidity, as the system spends most of the relevant time (keeping in mind that the very small time window to generate directed flow is of order of the crossing time) in the ÒsoftÓ state. This same physical scenario is predicted to generate a non-trivial fingerprint on the coordinate-space configuration (of which the tilt angle is the dominant component).  The geometry will probe the physics behind the Òthird componentÓ of flow generating the v$_{1}$ wiggle. Table~\ref{Table:HBT} shows the estimated number of events required to perform these measurements.

\begin{table}[htb]
\caption{Estimate of number of events needed to measure azimuthally sensitive HBT.} 
\begin{center}
\begin{tabular}{|c|c|c|c|c|c|c|}
\hline
\sqrts &  5 & 7.7 & 11.5 & 17.3 & 27 & 39 \\
\hline
Number of Events  & 4M & 4 M& 3.5 M & 3.5 M & 3 M & 3 M \\
\hline
\end{tabular}
\end{center}
\label{Table:HBT}
\end{table}

\begin{figure}[htb]
\begin{minipage}{0.4\linewidth}
\begin{center}
\includegraphics[width=\linewidth]{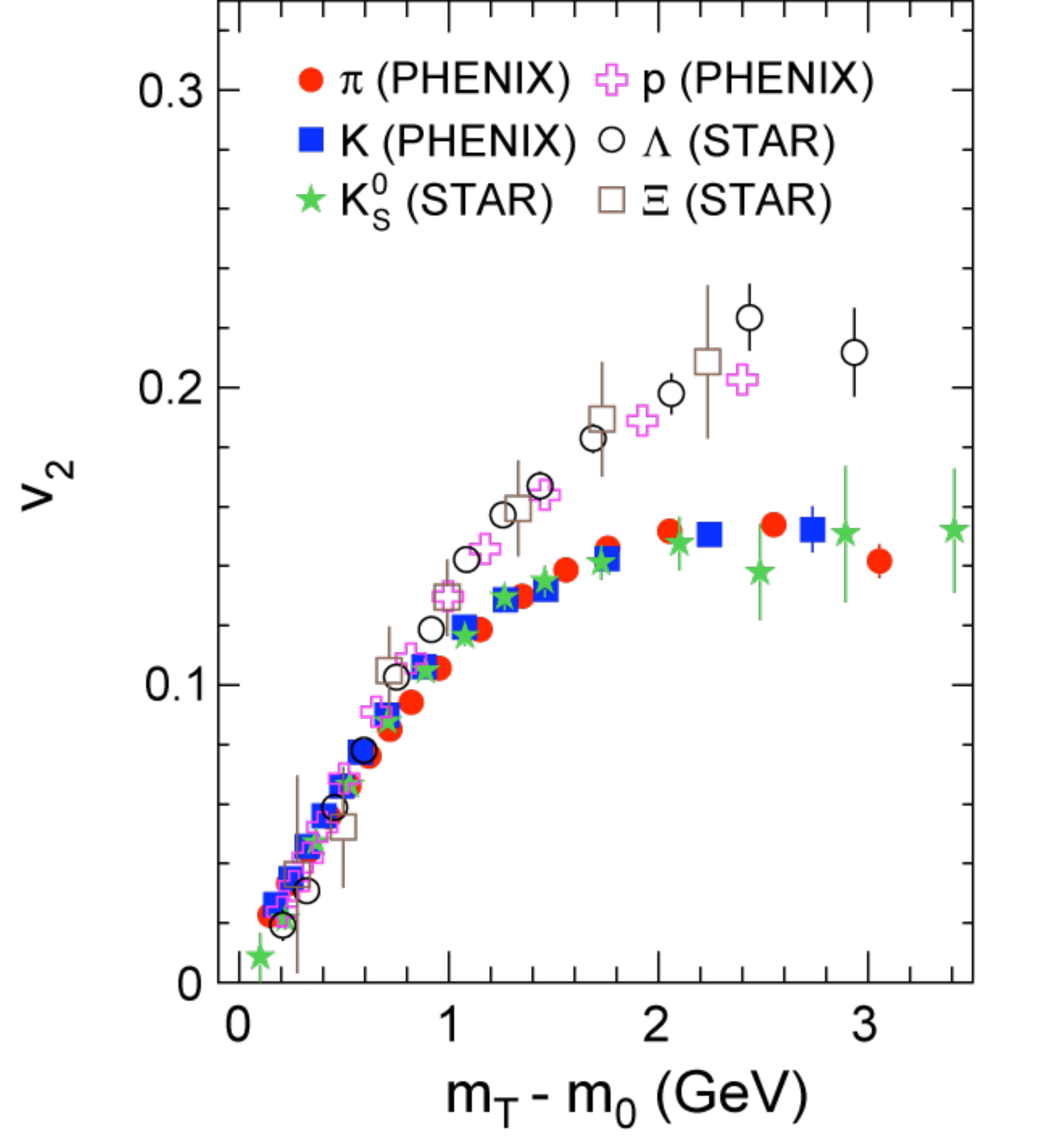}
\caption{Identified particle v$_{2}$ as a function of transverse kinetic energy  for Au+Au collisions at \sqrts=200 GeV.}
\label{FFig:v2KET}
\end{center}
\end{minipage}
\hspace{1cm}
\begin{minipage}{0.55\linewidth}
\begin{center}
\includegraphics[width=\linewidth]{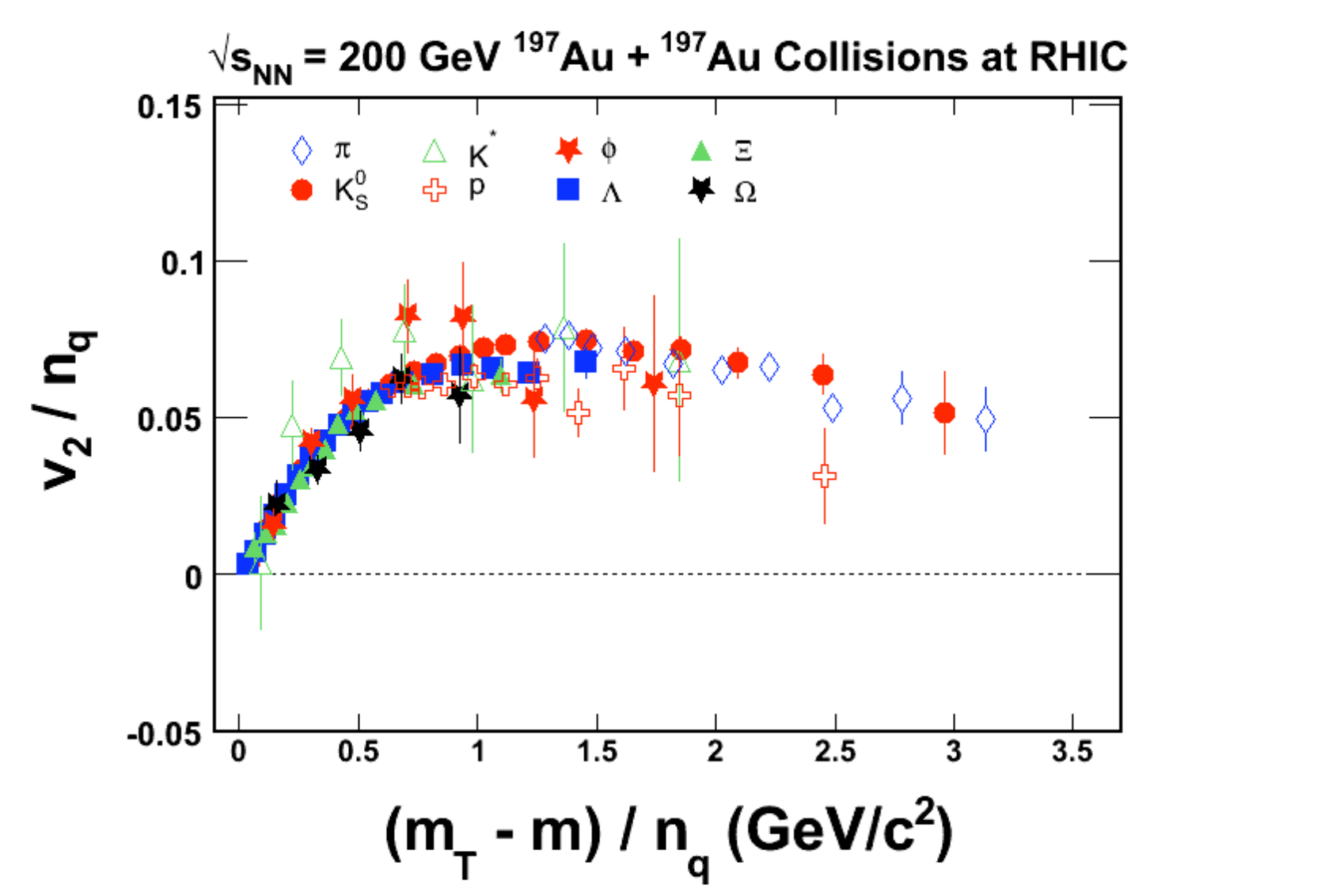}
\caption{Identified particle v$_{2}$ scaled by the number of constituent quarks in the hadron as a function of transverse kinetic energy scaled by the number of constituent quarks in the hadron  for Au+Au collisions at \sqrts=200 GeV.}
\label{Fig:v2KET2}
\end{center}
\end{minipage}
\end{figure}

\subsection{Turning off the Quark Gluon Plasma Signatures}\label{SubSec:QGP}

\noindent RHIC has reported several measures that  indicate that the medium created passes through a partonic phase at top energies~\cite{Adams:2005dq,Adcox:2004mh,Back:2004je,Arsene:2004fa}. We are interested in studying how these signals evolve as a function of \sqrts. As well as these partonic signatures, there are interesting results as a function of centrality that suggest different physical mechanisms  dominate when transverse particle densities change from low to high, and the switch from one to the other is both dramatic in appearance and  sudden. Many studies of these results can be performed with the energy scan. Some of the most promising are discussed below.

\subsubsection{Constituent Quark Scaling of Elliptic Flow and Elliptic Flow of Multi-Strange Hadrons }\label{SubSubSec:NCQ}

\noindent As already stated above,  extensive identified particle v$_{2}$ measurements have been made. Some of these are shown in Fig.~\ref{Fig:v2KET} for various particle species as a function of transverse kinetic energy for Au+Au collisions at \sqrts= 200 GeV at RHIC. Baryons and mesons fall onto two separate curves. At low m$_{T}$-m$_{0}$ the v$_{2}$ follows hydrodynamical type behavior~\cite{Csanad:2005gv}. However, at intermediate values of m$_{T}$-m$_{0}$ the magnitude of the v$_{2}$ reaches a plateau with baryon v$_{2}>$ meson v$_{2}$, Fig.~\ref{Fig:v2KET}.  If however one scales the v$_{2}$ and transverse kinetic energy, or \pT,  by the number of constituent quarks in the hadron, {\it all} particles now fall on a common curve ~\cite{Adams:2003am,Adams:2004bi,Adams:2005zg,Adams:2003xp}, even in the region where two curves are seen in Fig.~\ref{Fig:v2KET2}.   Such scaling can be explained by quark coalescence, or recombination models ~\cite{Hwa:2002zu, Greco:2003xt, Fries:2003vb, Molnar:2003ff}, which provide an intriguing framework for hadronization of bulk  partonic matter at RHIC. The essential degrees of freedom  at the hadronization seem to be effective constituent quarks which have developed a collective elliptic flow during the partonic evolution. Further evidence is the sizable magnitude of v$_{2}$ of the $\phi$ and $\Omega$. Both of these hadrons have small hadronic cross-sections so it is unlikely that they can develop an elliptic flow during the hadronic phase. Since v$_{2}$ is self quenching this suggests that the early stages of the collision are partonic~\cite{Adams:2003xp,Adams:2005zg}.  

~

\begin{figure}[htb]
\begin{center}
\includegraphics[width=0.9\linewidth]{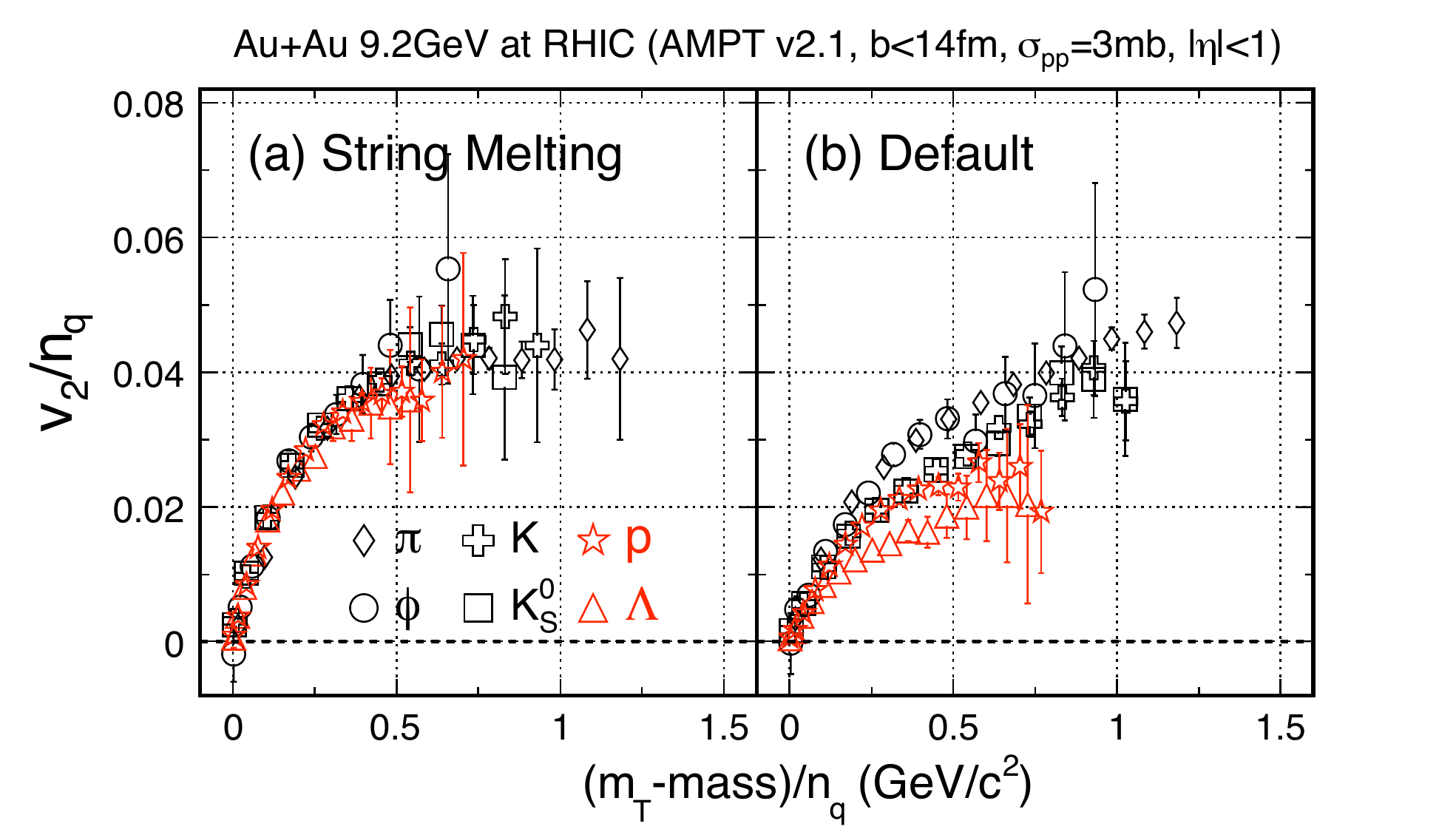}
\caption{The identified particle v$_{2}$ parameter from AMPT model at \sqrts = 9.2 GeV, with and without string melting. See text for more details.}
\label{Fig:AMPTv2}
\end{center}
\end{figure}

\noindent If this is indeed a QGP signature, one would expect  these effects to turn off at lower \sqrts, especially if we drop below the transition temperature at the lowest \sqrts. At these energies, the elliptic flow would develop  when the degrees of freedom of the system are dominated by hadronic interactions.  To examine if the number of constituent quark scaling would still be observed when in the hadronic stage, we looked at simulations from the AMPT model~\cite{lin-2005-72}. This model can be run in two modes, with and without a partonic stage. Mode one is without  string melting and the simulation involves purely hadronic interactions; this is the default setting. In mode two, string melting  is included, resulting in a fully partonic stage at early times.  Full details of the model can be found in reference \cite{lin-2005-72}. AMPT with string melting has been shown to reproduce the trends of the RHIC v$_{2}$ data at \sqrts= 200 GeV, but it cannot when string melting is turned off.  The results for identified particle v$_{2}$ for both these modes at \sqrts= 9.2 GeV are shown in Fig.~\ref{Fig:AMPTv2}. Constituent quark scaling is suggested in the string melting scenario whereas it is clearly absent in the case where only fragmentation via the Lund string model occurs.

\begin{figure}
\begin{center}
\includegraphics[width=0.6\linewidth]{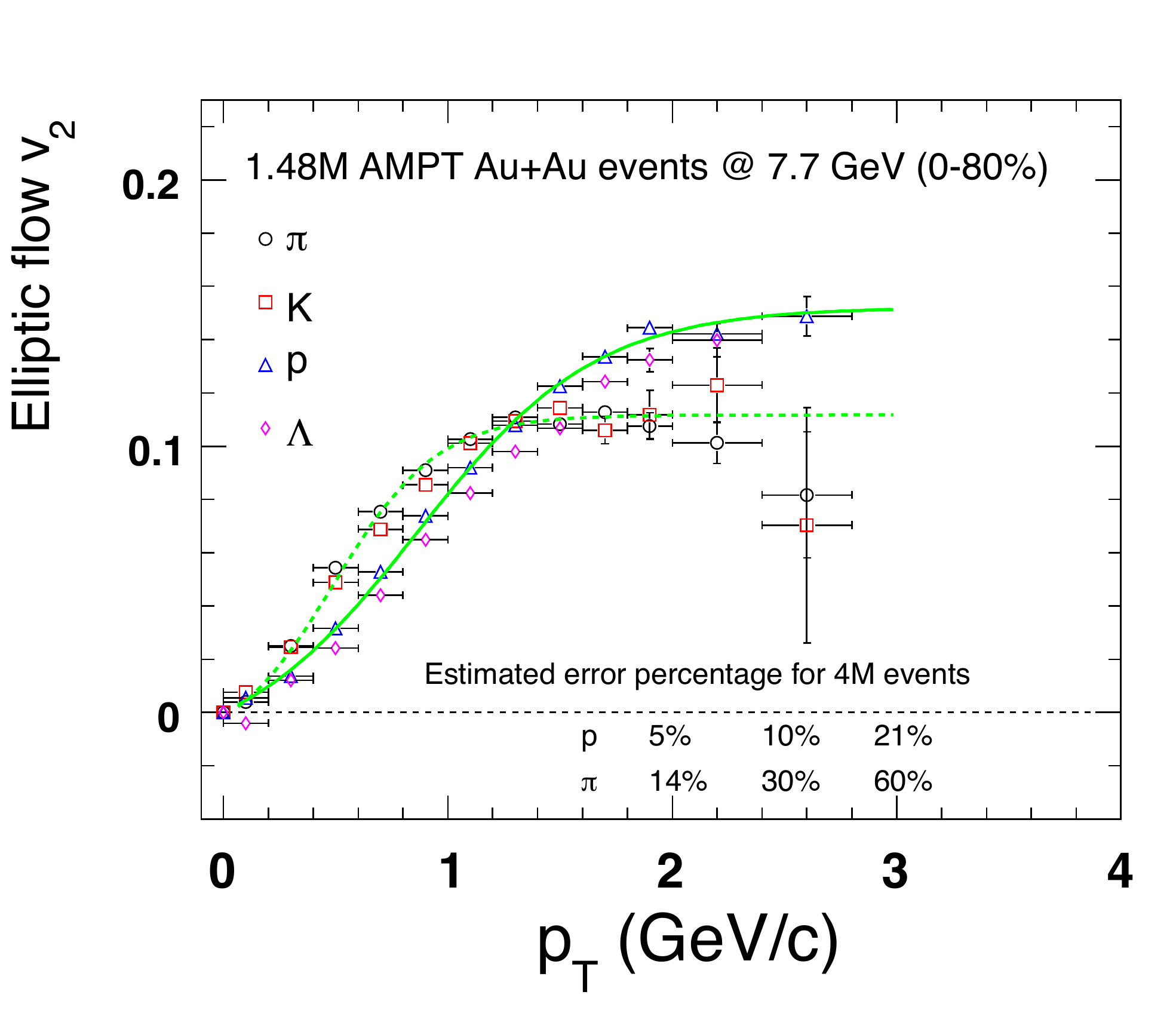}
\end{center}
\caption{AMPT predictions of identified particle v$_{2}$ as a function of \pT  for 1.48 M Au+Au collisions  at \sqrts=7.7 GeV. Also shown is an estimation of the statistical errors for identified proton and $\pi$ v$_{2}$ for 4 M events at 2, 2.5 and 3  GeV/c. The solid/dashed green curves are the predictions from NCQ scaling of v$_{2}$ for baryons/mesons. }
\label{Fig:v2KETnq}
\end{figure}
~

\noindent This measurement requires high statistics for identified  hadrons at intermediate \pT. As the collision energy drops, such measurements therefore become very challenging. There are insufficient statistics to make this measurement with the current SPS data. Figure~\ref{Fig:v2KETnq} shows an estimate of the statistical errors of the v$_{2}$ for identified protons and $\pi$ at \sqrts= 7.7 GeV from 1.4 M  0-80$\%$ centrality events from AMPT predictions. The green curves show the estimated v$_{2}$ for baryons and mesons should Number of Constituent Quark (NCQ) scaling exist at \sqrts=7.7 GeV. Also shown in this figure are estimates of the relative error on the proton and $\pi$ v$_{2}$ at \pT = 2, 2.5, and 3 GeV/c for 4 M events.  Extrapolating from this simulations we conclude that for \sqrts$>$ 12  GeV and higher a significant measurement of potential constituent quark scaling can be made for $\pi$, K (K$^{0}_{s}$), p and, $\Lambda$ up to (m$_{T}$-m$_{0}$)/NCQ $\approx$ 1.5 GeV with ~$\sim$5 M minimum bias events. For the $\phi$ and $\Omega$ at least 25 M events are required to reach  \pT =2 GeV with a 10$\%$ statistical error due to their low production rates and the poor signal to background ratios ~\cite{Alt:2004kq, Alt:2008iv}.

~

\noindent Even if a QGP is always formed at collision energies available at RHIC, the fraction of  time spent in the hadronic phase should increase with decreasing \sqrts. If constituent quark scaling is a pure partonic signal this may become washed out via hadronic interactions in this later stage. Thus a lack of constituent quark scaling could indicate either failure to push the system above the critical temperature or domination by the hadronic stage of the collision on the measured signals.

\begin{figure}[htb]
\begin{center}
\includegraphics[width=0.8\linewidth]{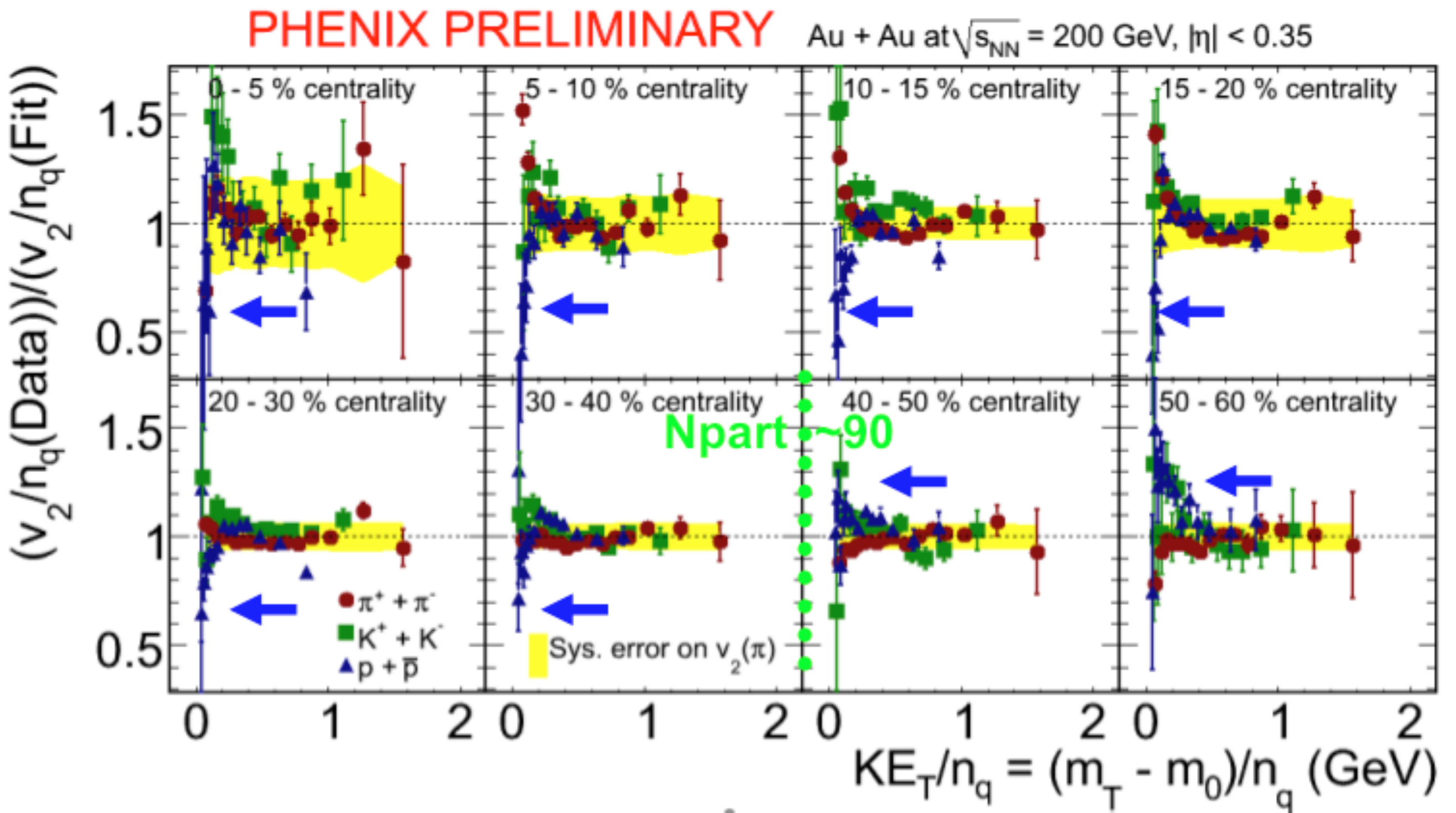}
\caption{Identified particle v$_{2}$ per constituent quark divided by a common fit function as a function of transverse kinetic energy per constituent quarks  for Au+Au collisions at \sqrts= 200 GeV for different centralities.}
\label{Fig:v2NCQDev}
\end{center}
\end{figure}

~

\noindent The PHENIX collaboration has reported an interesting systematic trend in the deviations of the v$_{2}$ of protons
from constituent quark scaling as a function of centrality~\cite{Issah:2008rk}. STAR have reported similar observations for protons and $\Lambda$~\cite{Shi:2009jg}.  To measure the accuracy of the constituent quark scaling with scaled transverse kinetic energy, PHENIX perform a fit to all the scaled measured data and then plot the deviations of each hadronic species to this common fit. This fit is performed as a function of centrality for \sqrts = 200 GeV Au+Au collisions and shown in Fig.~\ref{Fig:v2NCQDev}. If one focuses on the (anti)proton data (the blue triangles) one sees that for the more central collisions, the low \pT data fall below the common fit. As the centrality decreases this deviation from the scaling diminishes  smoothly until the 40-50$\%$ data is reached at which point the data changes dramatically, to show a positive deviation. The 40-50$\%$ data is at N$_{{\rm part}}\sim$ 90 matching the centrality where the low \pT ridge correlation measured by STAR also starts to show dramatic deviations from a \pp linear superposition model~\cite{Prindle:2009aa}, and this result is discussed in the sub-section~\ref{SubSubSec:Jets}.   This feature of the low \pT identified particle  v$_{2}$ measurements can be extremely well mapped during the energy scan, to see if it is a coincidence or whether possible dramatic changes in the underlying physics mechanism affect both measures.

\subsubsection{Nuclear Modification Factors  and Baryon/Meson Ratios}\label{SubSubSec:NMF}

\begin{figure}[htb]
\begin{minipage}{0.46\linewidth}
\begin{center}
\includegraphics[width=0.8\linewidth]{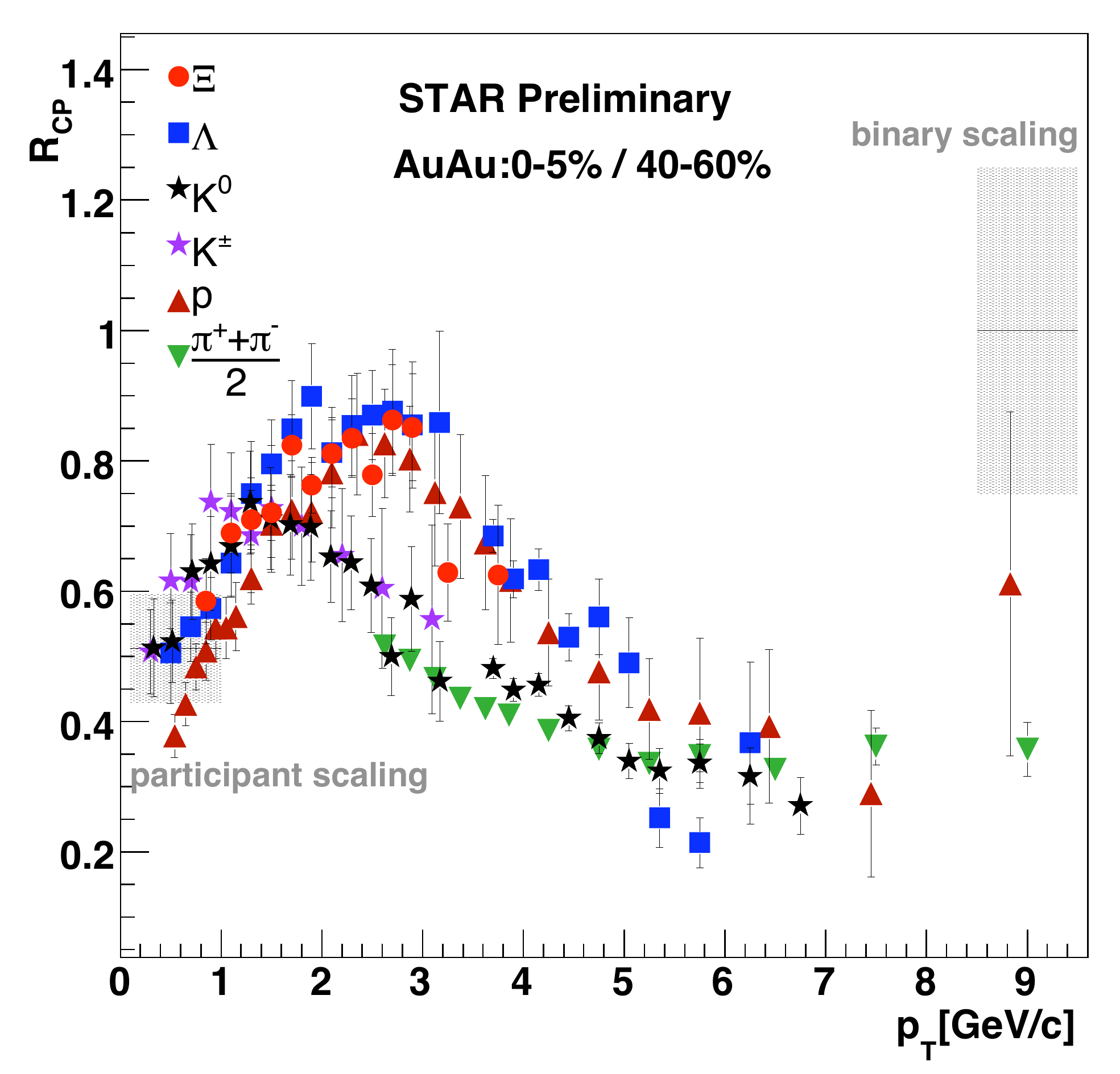}
\caption{Identified particle R$_{CP}$ for Au+Au collisions at \sqrts=200 GeV. From \cite{Lamont:2006rc}.}
\label{Fig:Rcp}
\end{center}
\end{minipage}
\hspace{1cm}
\begin{minipage}{0.46\linewidth}
\begin{center}
\includegraphics[width=\linewidth]{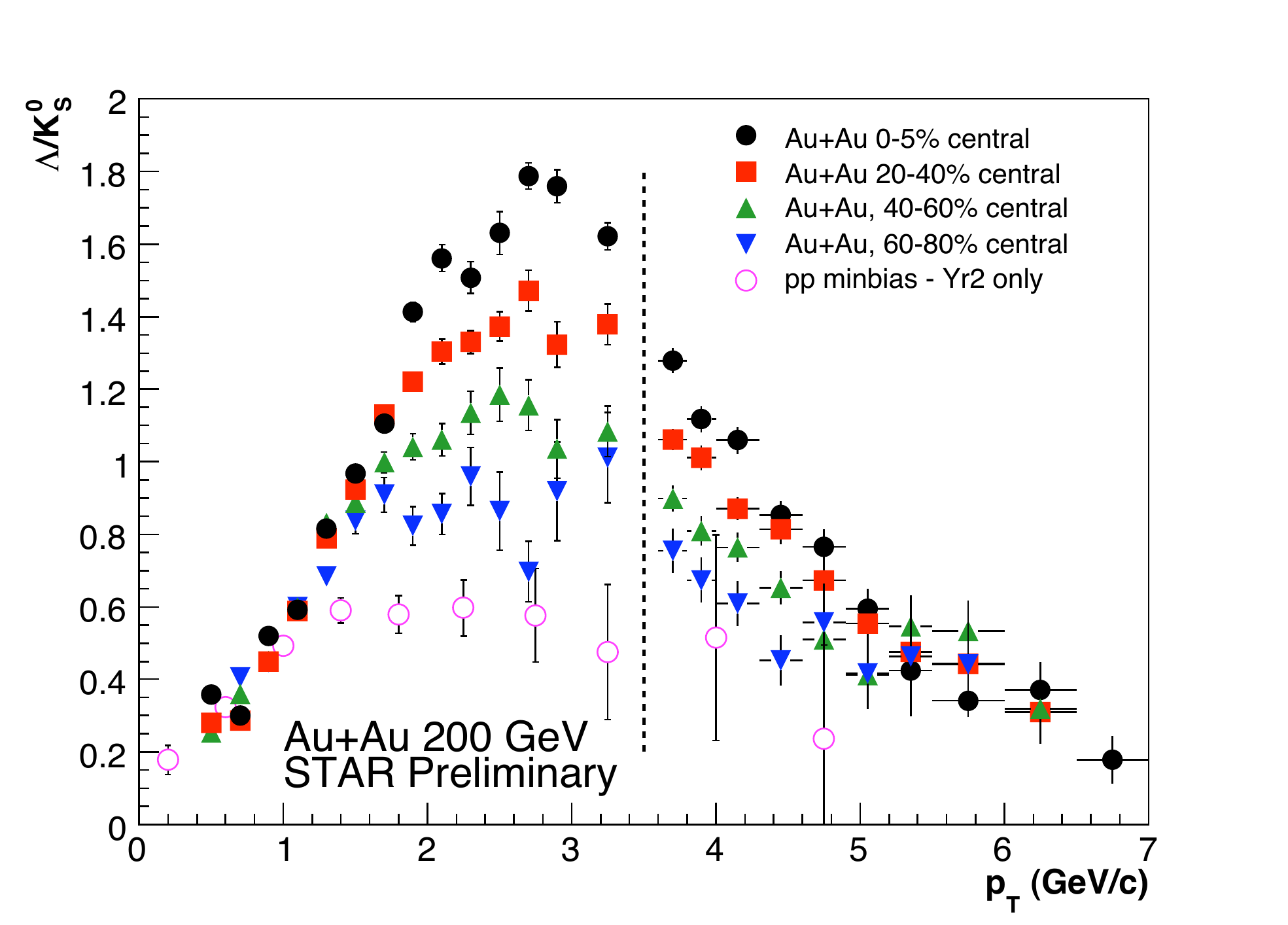}
\end{center}
\caption{The $\Lambda/K^{0}_{s}$ ratio as a function of \pT for Au+Au and p-p collisions at 200 GeV. From \cite{Lamont:2006rc}.}
\label{Fig:LamK0s}
\end{minipage}
\end{figure}

\noindent At intermediate to high \pT, it was initially expected that hard processes, that could be calculated via pQCD, would dominate. The nuclear modification factor, R$_{CP}$, is the N$_{bin}$ scaled ratio of central to peripheral \pT distributions. If  hard processes were not affected by the presence of the medium, and should scale with the number of binary collisions and R$_{CP}$ would equal unity. As shown in Fig.~\ref{Fig:Rcp}, high \pT particles are strongly suppressed~\cite{Lamont:2006rc}. Above 4-5 GeV/c all particle species, including  non-photonic electrons from the decay of heavy-flavored hadrons, are  suppressed by a factor of 5 in central Au-Au collisions.  Attempts have been made to measure the R$_{CP}$ of hadrons at the SPS~\cite{Alt:2007cd,Aggarwal:2007gw}. Unfortunately the reach of their data is not sufficient to make a firm statement as to whether the suppression is the same as that at top RHIC energies.  Since all species measured to date show significant suppression one can measure the charged hadron nuclear suppression factor to determine that "jet suppression" has started. At lower collision energies initial state effects such as the Cronin effect~\cite{Cronin:1974zm} -- the enhancement of particle yield at intermediate \pT with respect to binary collision scaling -- become more prominent. By measuring the  nuclear modification factor, R$_{CP}$, instead of R$_{AA}$,  the N$_{bin}$ scaled ratio of A+A to p+p  \pT distributions, the initial state effects should be minimized.  Finally. since the suppression is also large, one can have a sizable statistical error and still make a meaningful estimate. Table~\ref{Table:Rcp} shows estimates of the number of events needed to determine R$_{CP}$ of charged hadrons at each beam energy, in order to locate the beam energy at which interactions with the medium begin to affect hard partons. It has been estimated that too many events are needed to perform this measurement at lower energies.

\begin{table}[htb]
\caption{Estimates of events needed to measure R$_{CP}$ up to various \pT values at three different beam energies. It has been estimated that too many events are needed to perform this measurement at lower energies so they are not shown here.} 
\begin{center}
\begin{tabular}{|c|c|c|c|}
\hline
\sqrts (GeV) & 18 & 27 & 39 \\
\hline
\pT reach (GeV/c) & 4.5 & 5.5 & 6.0 \\
\hline
Number of Events  & 15 M & 33 M & 24 M  \\
\hline
\end{tabular}
\end{center}
\label{Table:Rcp}
\end{table}

~

~

\noindent   One of the first indications that particle production at \pT above 2 GeV/c was not purely from modified jet fragmentation came from the measurement of baryon/meson ratios.  The $p/\pi$, and $\Lambda/K^{0}_{s}$  ratios surpassed unity for  intermediate \pT  in the more central events, far higher than that observed in p+p collisions, Fig.~\ref{Fig:LamK0s}~\cite{Lamont:2006rc,Abelev:2007ra,Abelev:2006jr}.  At higher \pT the ratios turn over and decrease steadily until reaching  values close to that observed in elementary collisions at \pT around 6 GeV/c.  This intermediate \pT range is also the region where the v$_{2}$ of baryons is larger than that of mesons. Novel hadronization mechanisms are required to reproduce these data.  The majority of these models use a recombination/coalescence (ReCo) mechanism to form hadrons, as with the quark scaling  of v$_{2}$, described above, these models only combine constituent quarks~\cite{Hwa:2002zu, Greco:2003xt, Fries:2003vb, Molnar:2003ff}.  This mechanism  naturally leads to an enhanced  baryon-to-meson ratio  when the parton \pT  distribution is an exponential, i.e. when hadron production is not dominated by fragmentation which gives a power-law-like distribution.  Such models require quarks to coalesce and hence the observation of ReCo-like behavior is one of the corner-stone pieces of evidence of the formation of the sQGP.

~

\noindent One of the interesting topics of the energy scan is to investigate in which energies these phenomena are prevalent.  Although the cross-section for high \pT processes is smaller at lower \sqrts, the push to high \pT due to radial flow is also lower so the ReCo regime may move to lower \pT ranges, counterbalancing somewhat the loss in the hard scattering production rate.

\subsubsection{Jet Correlations and The {\it Ridge}}\label{SubSubSec:Jets}

\noindent   Hard scattered partons fragment into a spray of collimated hadrons known as a ``jet".  By examining di-hadron  $\Delta\eta$ -$\Delta\phi$ ($\Delta\eta$ =$\eta_1$-$\eta_2$ and $\Delta\phi$=$\phi_1$-$\phi_2$)  correlations, these jets can be identified in the presence of a large background. This background is subtracted statistically. In p-p collisions, clear back-to-back peaks from the high Q$^{2}$ interactions are apparent. However,  in more central A+A collisions, the away-side correlation disappears. This is believed to be  due to re-scattering of the away-side parton as it traverses the sQGP. The near-side correlation remains, showing evidence of vacuum-like fragmentation. While making these measurements, an interesting feature emerged in more central A+A collisions, namely a long range $\Delta\eta$  correlation at small  $\Delta\phi$. This long range correlation sits under the jet peak and extends to at least $\Delta\eta$=1. This phenomenon, called the ridge, appears to be correlated with jet-like triggers but the particles within the correlation have features reminiscent of the bulk rather than jet fragmentation, i.e. the p/$\pi$ ratio is close to that of the bulk, not a jet, and the \pT spectra are softer than those from fragmentation\cite{Abelev:2009qa}. This ridge correlation appears in both high p$_{T}$ triggered and un-triggered  correlation studies.

\subsubsection*{UnTriggered Correlations}\label{SubSubSec:UntriggeredRidge}

\noindent At RHIC,  differential analyses have been developed to allow more detailed investigations of the observed fluctuations. Instead of looking at the event-wise quantities, we instead study correlations:
\begin{eqnarray*}
\frac{\Delta\rho}{\sqrt{\rho_{\rm{ref}}}}
\end{eqnarray*}

\begin{figure}[htb]
\begin{center}
\includegraphics[width=\linewidth]{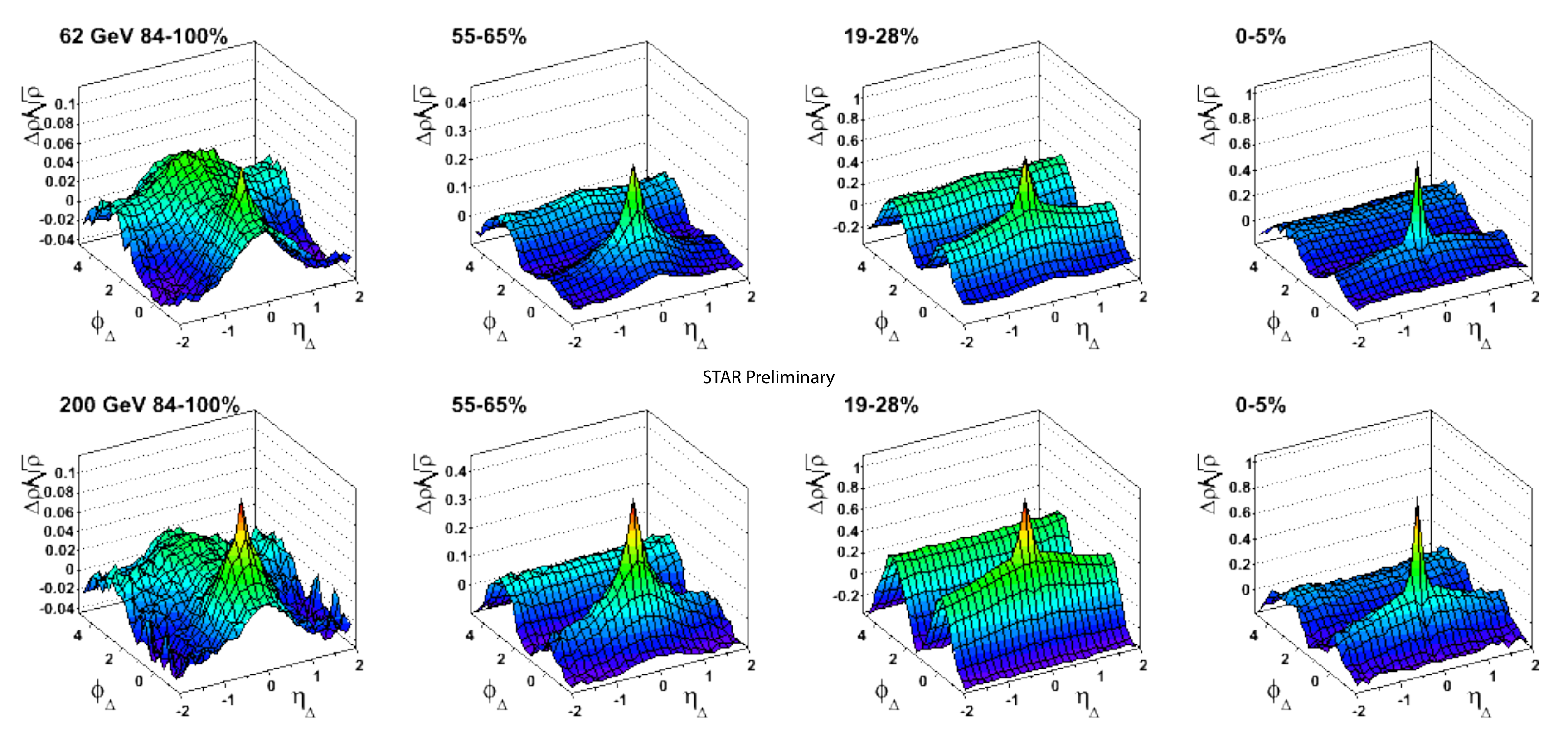}
\caption{Pair density correlations for Au+Au collisions at \sqrts = 62 and 200 GeV as a function of centrality. From ~\cite{Daugherity:2008su}.}
\label{Fig:Autocorr}
\end{center}
\end{figure}

\noindent where  $\rho_{{\rm ref}}$ is the number of mixed event pairs in the distribution being studied and $\Delta\rho$ is the number of pairs in the real event minus $\rho_{{\rm ref}}$.  These correlations are then plotted  as a function of relative angles $\phi$ and $\eta$. The resulting distributions for Au+Au collisions at \sqrts= 62 and 200 GeV are shown in Fig.~\ref{Fig:Autocorr}~\cite{Daugherity:2008su}. Clear structures, including the near-side jet peak can be seen for all centralities. In p-p collisions,  these distributions can be well described by a fit function derived from four simple components. These are a: 

\begin{enumerate}
\item 1-D Gaussian in $\phi_{\Delta}$ accounting for longitudinal fragmentation.
\item 2-D exponential at very small $\Delta\eta$, $\Delta\phi$ for HBT, resonances and e$^{+}$e$^{-}$ pairs.
\item  2-D Gaussian accounting for a mini-jet peak at  small $\Delta\eta$, $\Delta\phi$ 
\item  cos($\Delta\phi$) term for the dipole on the away-side
\end{enumerate}

\begin{figure} [htb]
\begin{center}
\includegraphics[width=0.8\linewidth]{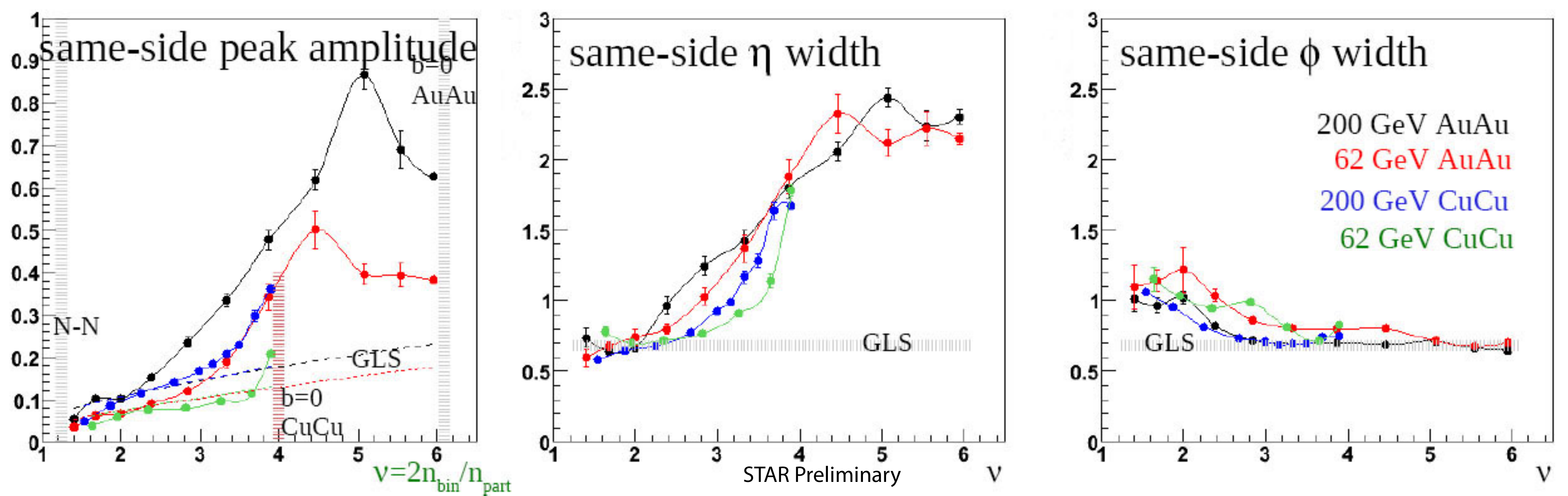}
\caption{Preliminary fit parameters of the 2-D mini-jet Gaussian to the pair density correlations for Au+Au and Cu+Cu collisions at \sqrts = 62 and 200 GeV as a function of $\nu$. From \cite{Prindle:2009aa}. The vertical lines in the left figure indicate the value of $\nu$ for collisions with impact parameters of b=0. The curves/bands labeled "GLS" indicate the values of the parameters if the data scaled with the number of binary collisions.}
\label{Fig:MiniJetFit}
\end{center}
\end{figure}

In A+A events, this fit function must be augmented with a cos($2\Delta\phi$) term to account for an additional  component that is usually associated with elliptic flow. This simple functional form gives a reasonable description of the main features of the data.  In Fig.~\ref{Fig:Autocorr}, it can be seen that in the more central bins, the long range $\Delta\eta$ correlation emerges. The parameters of the fit to the mini-jet 2-D Gaussian are shown in Fig.~\ref{Fig:MiniJetFit} as a function of $\nu$=2N$_{{\rm bin}}/N_{{\rm part}}$. An increase in both the peak amplitude and $\eta$ width, but not the $\phi$ width, above binary scaling (indicated by the dashed curves in Fig.~\ref{Fig:MiniJetFit}), occurs for both collision systems and energies at large values of $\nu$.  It is not yet clear what causes the ridge correlation or why there is a rapid onset.  This is a particle number correlation, and there is no such rapid transition occurring in the equivalent \pT correlation studies.  It would also be of great interest to attempt these studies using PID to see if these phenomenon occurs for all baryons and mesons. Table~\ref{Table:ridge} shows our estimate of the number of events required to perform these studies using non-identified particles.

\begin{table}[htb]
\caption{Estimate of events needed to measure the un-triggered ridge correlations It has been estimated that too many events are needed to perform this measurement at \sqrts  = 5GeV.} 
\begin{center}
\begin{tabular}{|c|c|c|c|c|}
\hline
\sqrts &  11.5 & 17.3 & 27 & 39 \\
\hline
Number of Events  & 13 M & 8 M & 6 M & 6 M  \\
\hline
\end{tabular}
\end{center}
\label{Table:ridge}
\end{table}

~

\noindent One possible explanation is that this is caused by Glasma flux tube radiation in combination with radial and elliptic flow and a flowing sQGP~\cite{Dumitru:2008wn}. A plausible interpretation of the onset of the un-triggered   ridge relates the onset to collisions passing through  the transition scale Q$_{s}/\alpha_{s}(Q_{s})$~\cite{Sorensen:2008bf}.  The saturation momentum of partons in the nuclear wavefunction, Q$_{s}$, grows rapidly with nuclear size and energy. Hence $\alpha_{s}( Q_{s})$, the strong interaction strength, is  correspondingly weak for  high energy hadron-hadron collisions  and nuclear collisions.

~

\begin{figure}[hb]
\begin{minipage}{0.46\linewidth}
\begin{center}
\includegraphics[width=0.8\linewidth]{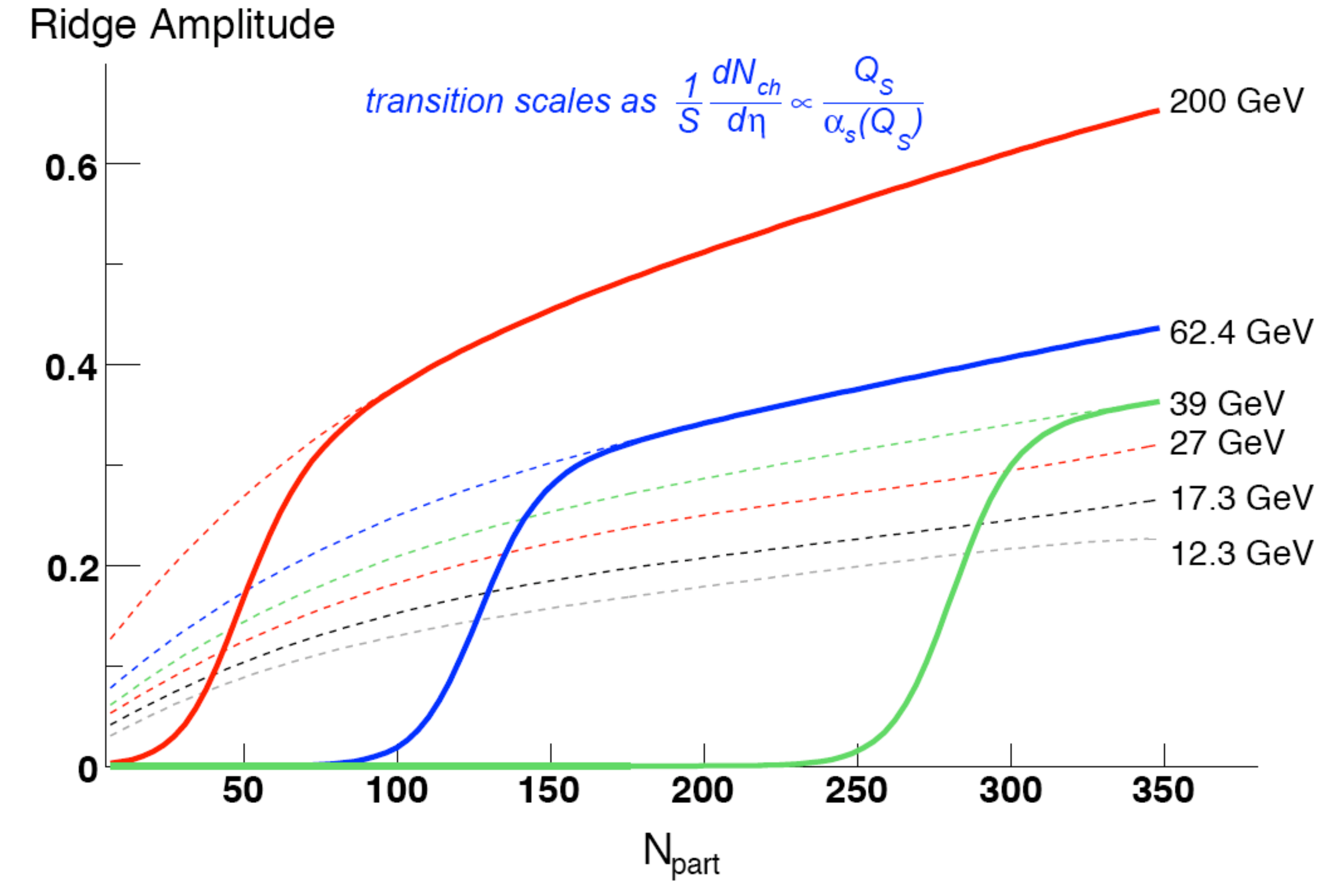}
\caption{The un-triggered ridge amplitude as a function of N$_{{\rm part}}$ when the transition scale is related to the energy density. The different colored curves represent different \sqrts.}
\label{Fig:Glasma1}
\end{center}
\end{minipage}
\hspace{1cm}
\begin{minipage}{0.46\linewidth}
\begin{center}
\includegraphics[width=\linewidth]{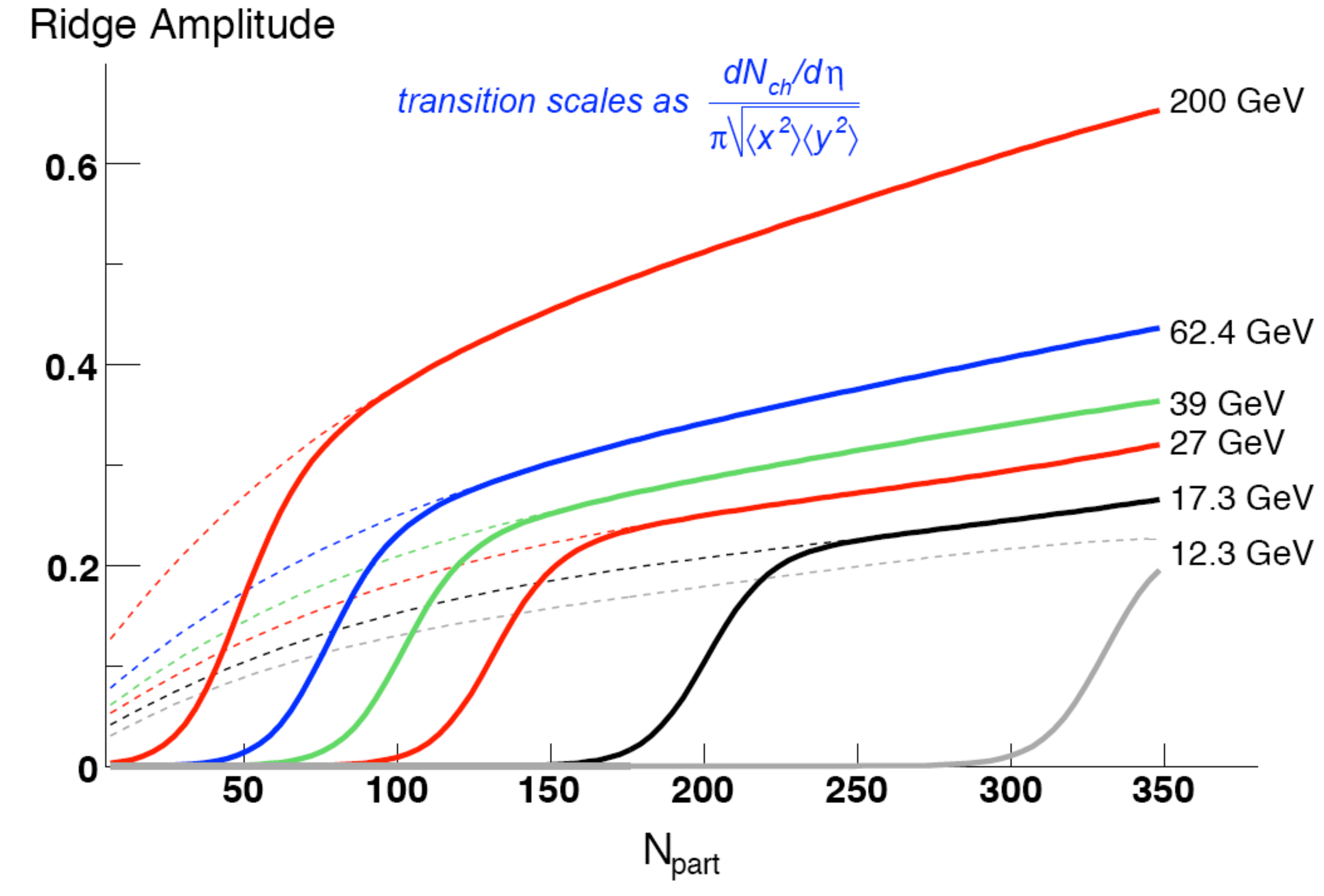}
\end{center}
\caption{The un-triggered ridge amplitude as a function of N$_{{\rm part}}$ when the transition scale is related to the particle density. The different colored curves represent different \sqrts.}
\label{Fig:Glasma2}
\end{minipage}
\end{figure}

\noindent This could be tested via the energy scan as different \sqrts would produce the ridge at different centralities but the same transition scale. Fig.~\ref{Fig:Glasma1} and Fig.~\ref{Fig:Glasma2} represent the two cases when the transition scale scales with either energy density or particle density. The colored curves represent differing collision energies. Below the transition scale, no ridge is produced and the curves are interpolated smoothly between the two cases. It can be seen that as  the \sqrts of the collision is lowered, the number of participants required to pass above the transition point rises. In the case of energy density scaling no ridge is produced below \sqrts= 35 GeV, in the particle density case the ridge persists down to \sqrts= 13 GeV. These predictions could be tested via the energy scan.

~

\noindent

\subsubsection*{Triggered Correlations}\label{SubSubSec:Ridge}

\begin{figure}[htb]
\begin{minipage}{0.46\linewidth}
\begin{center}
\includegraphics[width=\linewidth]{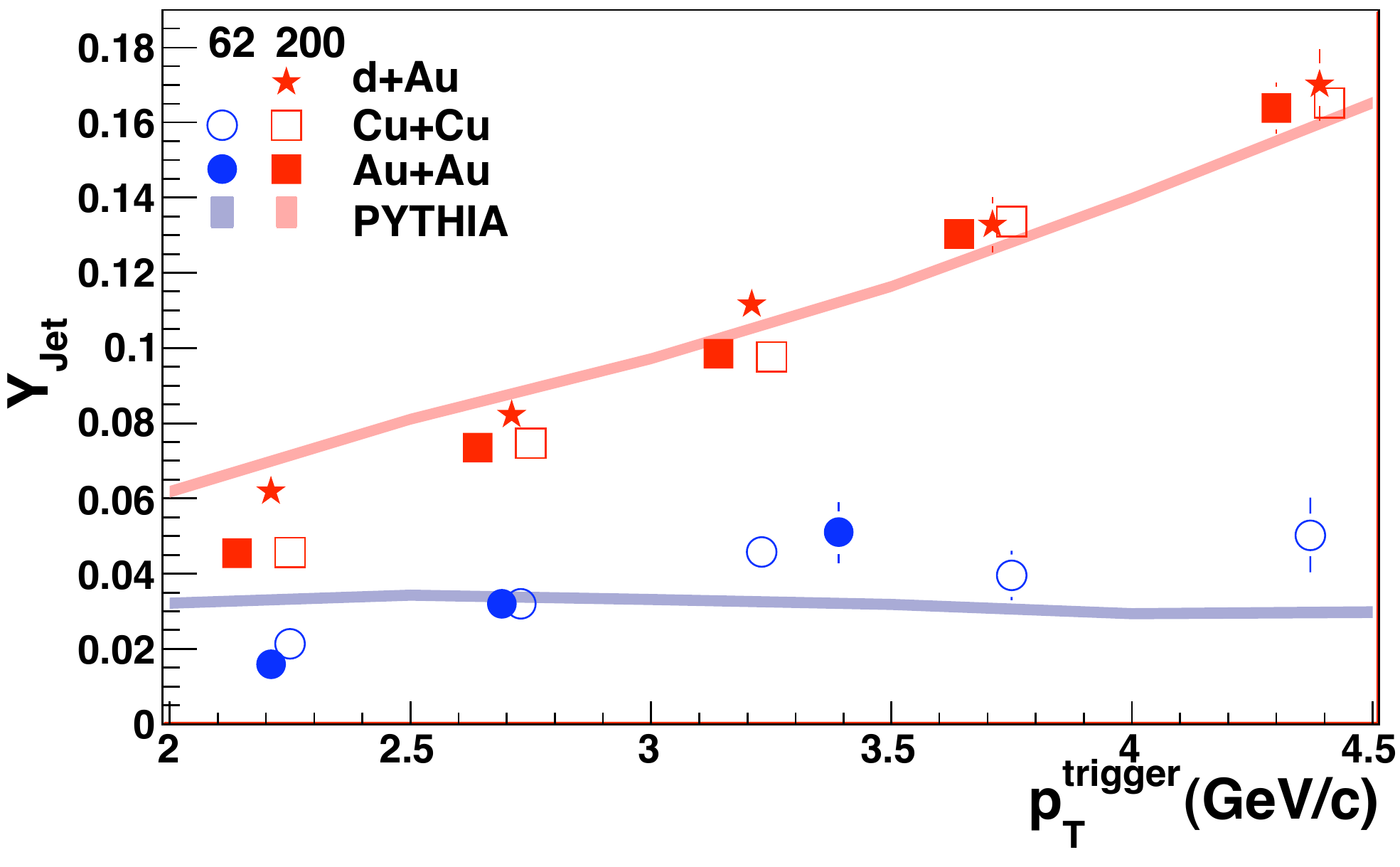}
\caption{The jet correlation yield per trigger on the near-side as a function of trigger particle's \pT for \sqrts= 62 and 200 GeV Au+Au, Cu+Cu and d-Au collisions. The colored curves are predictions from PYTHIA.}
\label{Fig:JetYield}
\end{center}
\end{minipage}
\hspace{1cm}
\begin{minipage}{0.46\linewidth}
\begin{center}
\includegraphics[width=\linewidth]{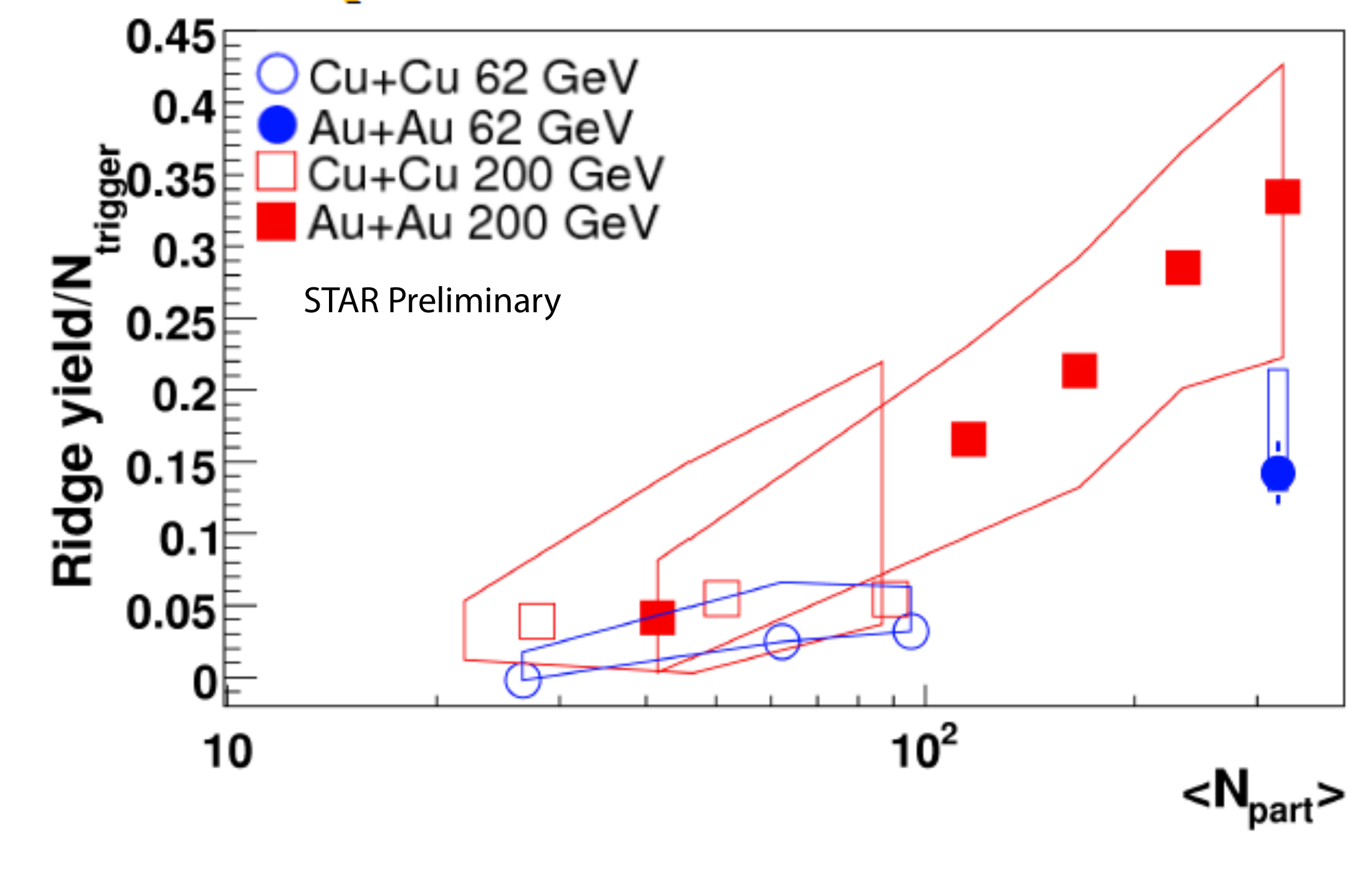}
\end{center}
\caption{The number of particles in the near side ridge correlation as a function of N$_{{\rm part}}$ for \sqrts = 200 and 62 GeV Cu+Cu and Au+Au collisions.}
\label{Fig:RidgeYield}
\end{minipage}
\end{figure}

\begin{figure}[htb]
\begin{minipage}{0.46\linewidth}
\begin{center}
\includegraphics[width=\linewidth]{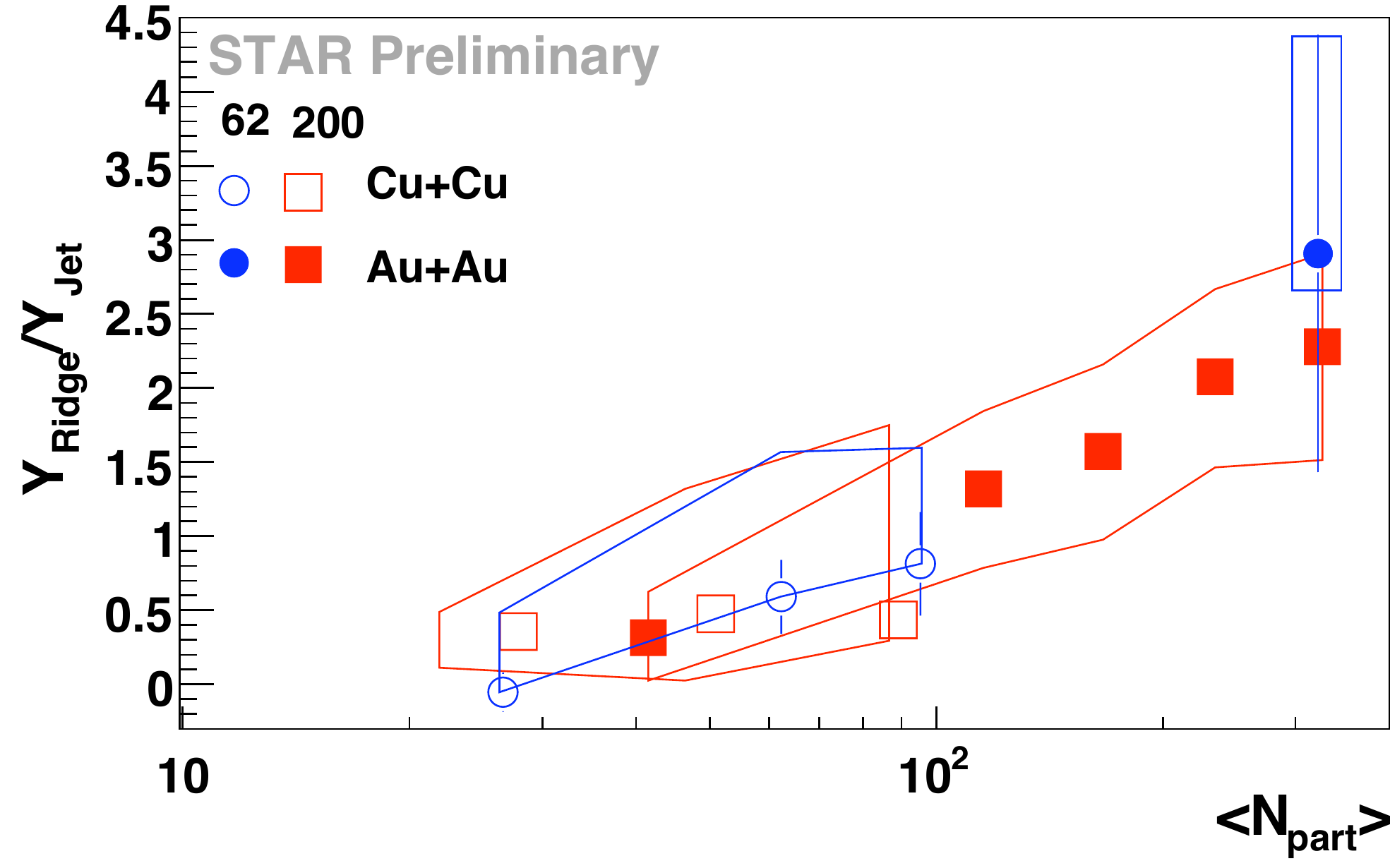}
\caption{The Ridge/Jet correlation yield ratio as a function of N$_{{\rm part}}$ for Cu+Cu and Au+Au collisions at \sqrts = 62 and 200 GeV. From \cite{Nattrass:2008tw}. }
\label{Fig:Ridge2JetRatio}
\end{center}
\end{minipage}
\hspace{1cm}
\begin{minipage}{0.46\linewidth}
\begin{center}
\includegraphics[width=\linewidth]{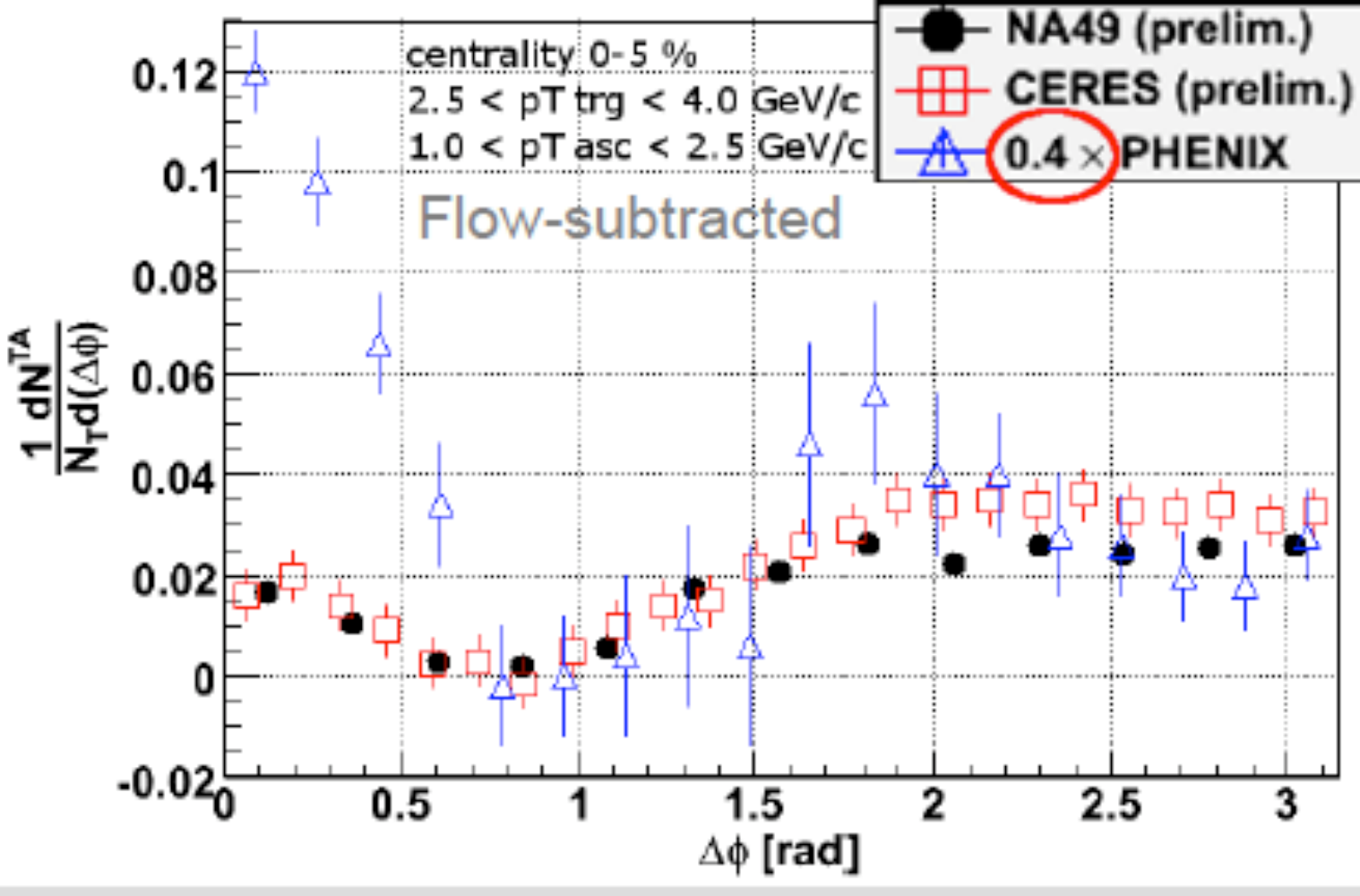}
\end{center}
\caption{The triggered correlations from SPS compared to PHENIX. The away-side distribution does not change. Note the scaling of PHENIX by a factor 0.4.}
\label{Fig:SPSRecoil}
\end{minipage}
\end{figure}

\noindent   By triggering on high \pT particles and associating  them only with the particles in intermediate  \pT ranges, the jet component of the two particle correlations can be selected, while minimizing the soft particle background. In these analyses, clear near-side jet peaks are observed~\cite{Adams:2006yt} as well as the long range $\Delta\eta$ correlation~\cite{Putschke:2007mi}. As expected from jet fragmentation, the near-side  jet correlation yield rises as a function of trigger \pT, Fig.~\ref{Fig:JetYield}, but stays approximately constant as a function of centrality~\cite{Nattrass:2008tw}. The ridge yield on the other hand increases with N$_{{\rm part}}$, Fig.~\ref{Fig:RidgeYield}.  Interestingly, while both the jet and ridge yields rise with  \sqrts, the ratio Ridge/Jet is  constant, for \sqrts= 62 and 200 GeV, Fig.~\ref{Fig:Ridge2JetRatio} \cite{Nattrass:2008tw} . It needs to be tested if this is true at other collision energies or merely a coincidence.

~

\noindent Both CERES~\cite{Kniege:2007mp} and NA49~\cite{Szuba:2009tu} have attempted this analysis at the SPS. While they see evidence for a near-side jet correlation their away-side peaks are consistent with that arising from momentum conservation only, Fig.~\ref{Fig:SPSRecoil}. Also, they do not have the acceptance or statistics to attempt a two dimensional analysis so they can make no statements about the presence of the ridge. We therefore propose to pursue this analysis with STAR to further our understanding of this phenomenon.  

~

\noindent To estimate the feasibility of these studies we have performed PYTHIA simulations. While PYTHIA does not produce the ridge correlation, it is a good representation of fragmentation so we focus on the near-side jet correlation. Since the ridge correlation is similar in magnitude to the jet correlation at \sqrts= 62 and 200 GeV, we determine the event rate estimates by assuming that if there are sufficient statistics to see the jet, the ridge should also be measurable, i.e. we assume that the ridge/jet ratio remains a constant.
 PYTHIA simulations were run at various energies and the correlations produced via the same techniques used on the real data. Two trigger and associated thresholds were used following the cuts of STAR and the lower values of NA49. The cuts used were 1) 3 $<$\pT$^{trig} < 6$ GeV/c and 1.5 $<$\pT$^{{\rm assoc}} <$ \pT$^{{\rm trig}}$ GeV/c  and 2) $2 <$\pT$^{{\rm trig}} < 4$ GeV/c and $1 <$\pT$^{{\rm assoc}} <$ \pT$^{{\rm trig}}$ GeV/c. The NA49 studies are performed with lower thresholds to improve their statistics. This is not possible at the current higher energies of STAR because of the increased reach in \pT of the soft physics background.  The statistics are estimated by requiring a better than 50$\%$ relative error on the extracted jet correlation yield. The required number of events is shown in Table.~\ref{Table:JetRates}. This is likely a slight underestimate as the error due to the background subtraction is not included. It can be seen that this study is possible down to \sqrts = 27 GeV when the NA49 approach is used. The ridge yield is strongly affected by the v$_{2}$ subtraction, however as the \sqrts of the collision reduces, so does the measured elliptic flow, so this systematic error is likely to reduce faster than the signal.

\begin{table}[htb]
\caption{Estimate of events needed to measure high \pT jet correlations as a function of \sqrts. } 
\begin{center}
\begin{tabular}{|c|c|c|}
\hline
&  Number of Events  & Number of Events  \\
$\sqrt{s_{NN}}$ (GeV)  &   $2<p_{T}^{trig}<$ 4 GeV  & $3<p_{T}^{trig}<$ 6 GeV  \\
  &   1.0$<p_{T}^{assoc}<p_{T}^{trig}$ &  1.5$<p_{T}^{assoc}< p_{T}^{trig}$ \\
\hline
\hline
200 & 0.8 M & 1.6 M\\
62  & 2 M & 8 M\\
39 &4.5 M & 24 M\\
27  & 8.8 M &53 M \\
20  & 23 M& 240 M\\
\hline
\end{tabular}
\end{center}
\label{Table:JetRates}
\end{table}

\subsection{Particle Production Studies}\label{SubSec:Trends}

\subsubsection{Identified Particle Yields, Spectra and Statistical Model Fits}\label{SubSubSec:StatModel}

\noindent Since phase diagrams describe only thermalized media, if we are to place a point on the phase diagram we  need to ascertain that the medium created in heavy-ion collisions is, at least temporarily, in thermal equilibrium. This goal continues to be a  challenge for the whole heavy-ion community.  To this end the bulk/soft physics of integrated yields, transverse momentum spectra and particle ratios need to be studied.

~

\noindent Statistical models have had great success in describing the mid-rapidity, and integrated, particle yields as a function of \sqrts. While successful fits do not prove the validity of the model's interpretation of the parameters as the thermal parameters of chemical freeze-out temperature, T$_{ch}$, and baryon chemical potential, $\mu_{B}$, should the fits not work the system {\it cannot} be in thermal equilibrium at hadronization. It is therefore useful to perform these fits before embarking on other measures that require the assumption of thermal equilibration to be interpreted. With STAR's large acceptance and good PID these measures are achieved to high statistical precision with only a very small event count. Previous fits to the data, for example Fig.~\ref{Fig:TMuB}, have shown that from SPS energies upwards the extracted T$_{ch}$  is very close to the cross-over temperature, T$_{c}$, for a phase transition predicted from lattice QCD  of  $\sim$  170-195 MeV~\cite{Fodor:2009ax,Bazavov:2009zn}. This is one of the indications that the QGP is formed in these collisions.

\subsubsection{Strange Particle Yields}\label{SubSubSec:StrangeYields}

\noindent For more than 25 years, it was expected that the production of strange particles in a QGP phase would be enhanced with respect to a hadron gas~\cite{Rafelski:1982pu, Koch:1986ud}. Also, during hadronisation, the $s$ and $bar{s}$ quarks from the plasma may coalesce to form $\phi$ mesons. Production by this process is not suppressed as per the OZI (Okubo-Zweig-Izuka) rule~\cite{Okubo:1963fa,Zweig:1981pd,Zweig:1964jf,Iizuka:1966wu}. This, coupled with large abundances of strange quarks produced in the medium, would again lead to a dramatic increase in the production of  $\phi$ mesons relative to non-QGP production~\cite{Rafelski:1982pu, Koch:1986ud}.  The NA35 and NA49 experiments reported~\cite{Afanasev:2002qp,Bormann:1997qa} that in central S+S (Pb+Pb) collisions at \sqrts = 19.3 (17.3) GeV the kaon to pion ratio is approximately two times higher than in N+N interactions at the same energy per nucleon. However, it was later discovered that an even greater  enhancement occurs at lower energies, where no QGP is expected~\cite{Ahle:1999uy, Ahle:1998jc, Ahle:1998gv, Ahle:2000wq, Vincent:1989jz, Afanasiev:2002mx, :2007fe, Bearden:2004yx, Lee, Gazdzicki:1995zs, Gazdzicki:1996pk}. Interpretation of the experimental data is therefore difficult and additional high precision measurements at lower collision energies would be illuminating.

~

\begin{figure}
\includegraphics[width=\linewidth]{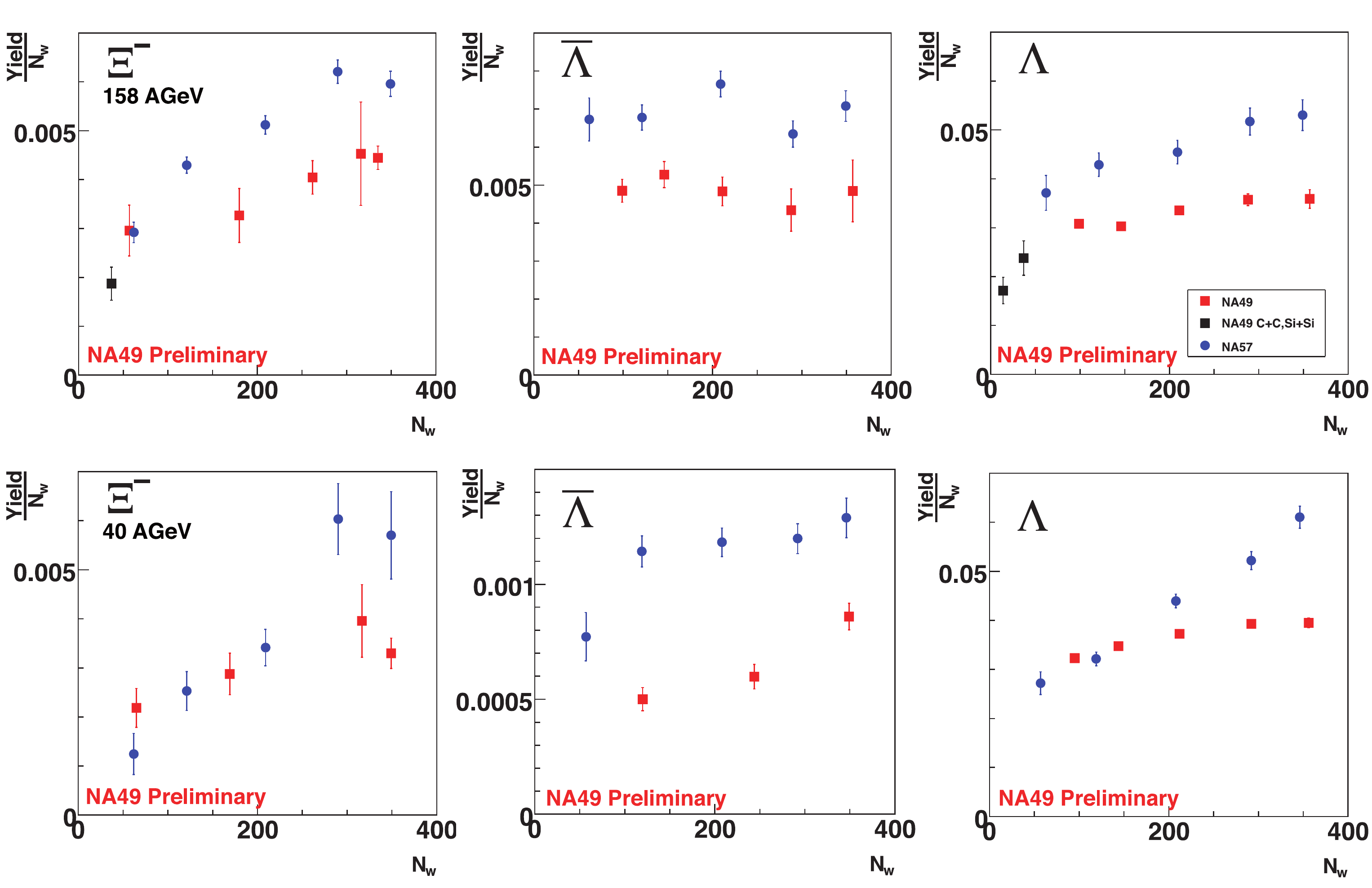}
\caption{\label{Fig:Discrepancy_Centrality_Dependence}Comparison of the centrality dependence of $\Xi^{-}$, $\Lambda$ and $\bar{\Lambda}$ yields ($\mid$y$\mid$ $\leq$ 0.5) at \sqrts = 17.3 (top panel) and \sqrts=8.8 GeV (bottom panel) between NA57 and NA49.}
\end{figure}

\noindent Another crucial point is the centrality dependence of strangeness production. It was observed by the NA57 and NA49 collaborations~\cite{Antinori:2006ij, Antinori:2004ee, NA57_Webpage, Mitrovski:2006js, Blume:2007kw} that an enhancement is also visible for other strange particle species and that it depends on the strange quark content. Fig.~\ref{Fig:Discrepancy_Centrality_Dependence} shows the comparison of the centrality dependence of $\Xi^{-}$, $\Lambda$ and $\bar{\Lambda}$ yields ($\mid$y$\mid$ $\leq$ 0.5) between NA57 and NA49 at \sqrts = 17.3 (top panel) and \sqrts=8.8 GeV (bottom panel). For all particle species and energies a discrepancy is observed for the more central collisions except for $\bar{\Lambda}$ where a discrepancy is also visible in peripheral collisions. Furthermore, the shape of the $\Lambda$ centrality dependence at \sqrts = 17.3 GeV is different between NA57 and NA49. 

\begin{figure}[ht]
\includegraphics[scale=0.6]{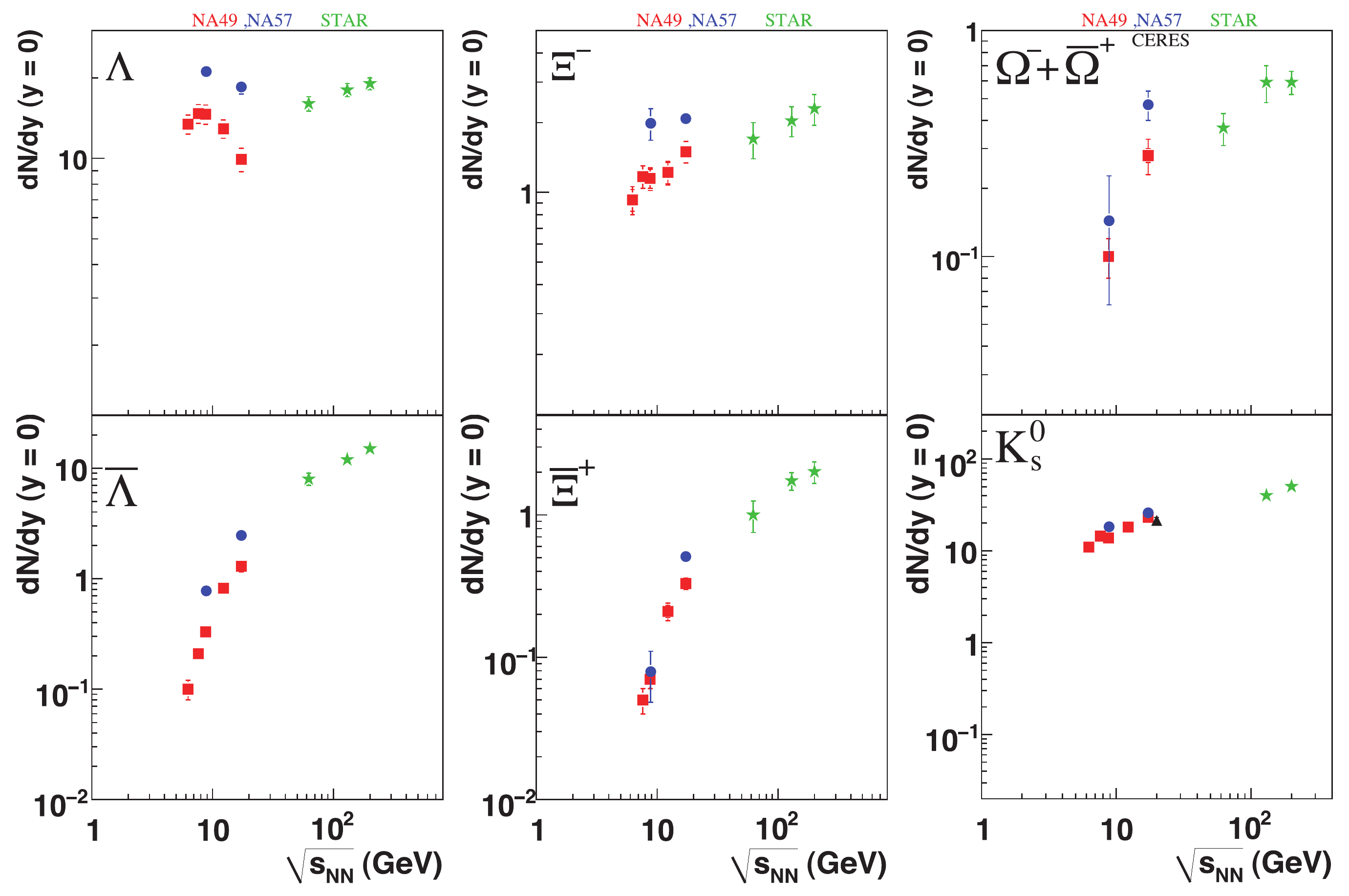}
\caption{\label{fig:Discrepancy_Energy_Dependence}The energy dependence of strange particle and anti-strange particle yields at midrapidity ($\mid$y$\mid$ $\leq$ 0.5) from SPS to RHIC energies.}
\end{figure} 

~

\noindent Fig.~\ref{fig:Discrepancy_Energy_Dependence} shows the collision energy dependence of mid-rapidity strange particle production from SPS to RHIC measured by the NA57~\cite{Antinori:2006ij, Antinori:2004ee, NA57_Webpage}, NA49~\cite{Alt:2004kq, Alt:2008qm}, CERES~\cite{Milosevic:2006gw} and STAR~\cite{Adler:2002uv, Adams:2003fy, Adams:2006ke, Takahashi:2007bh} collaborations. A clear discrepancy between NA57 and NA49 is visible for all particles. For $K^{0}_{s}$ CERES and NA49 are in good agreement whereas NA57 measure a slightly higher $K^{0}_{s}$ yield.

~

\begin{figure}[ht]
\includegraphics[scale=0.43]{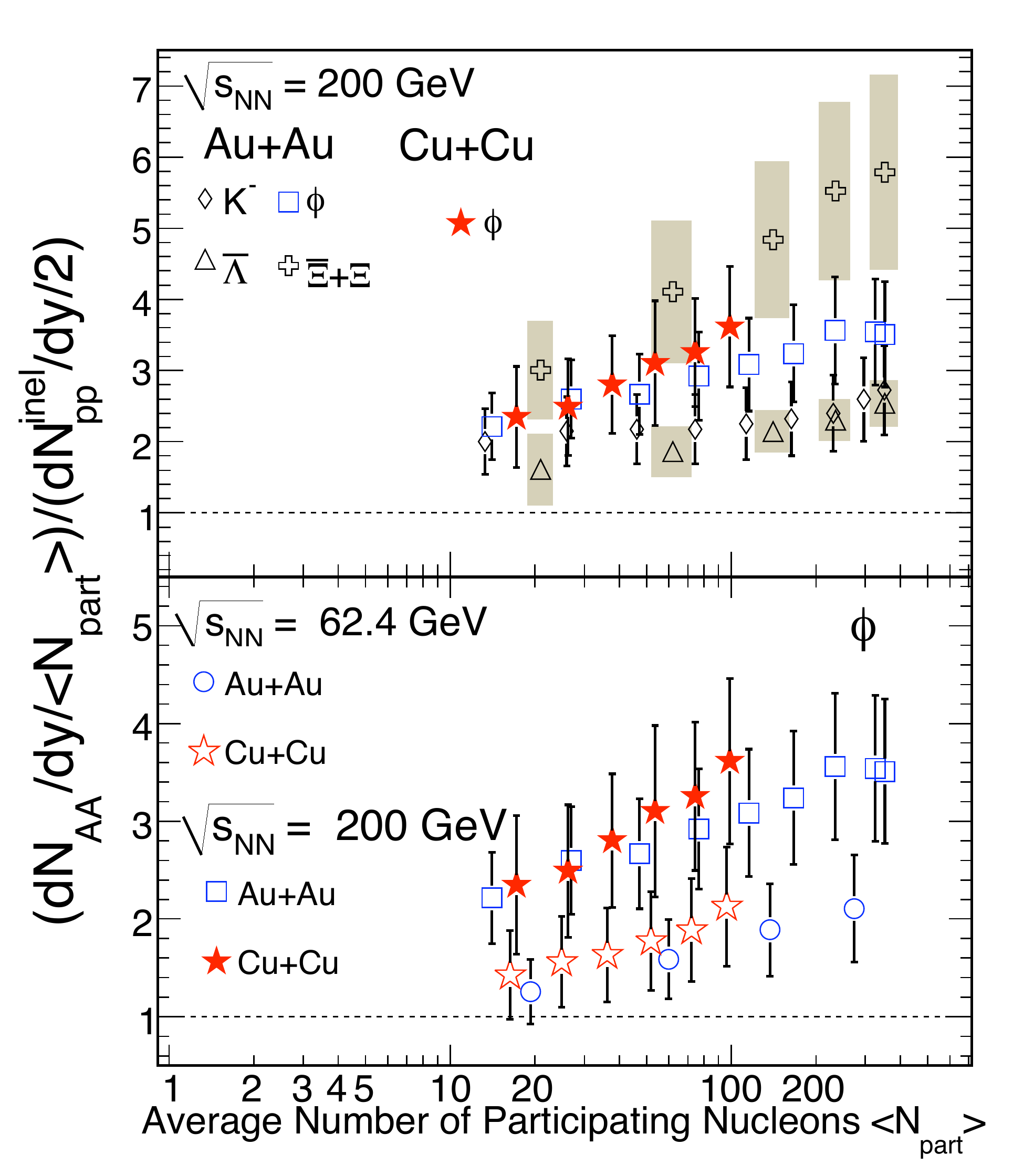}
\caption{\label{Fig:phiEnhance}The measured enhancement of various strange particles in A+A compared to \pp at the same beam energy as a function of the number of participants.  Top panel: For K, $\phi$, $\bar{\Lambda}$ and $\Xi$ in Au+Au and Cu+Cu collisions at \sqrts=200GeV. Bottom panel: For the $\phi$ in Au+Au and Cu+Cu collisions at \sqrts=200 and 62.4 GeV. From \cite{Abelev:2008zk}.}
\end{figure}

\noindent In order to describe the shape and the enhancement in the centrality dependence it is important to consider different types of models. Statistical models are quite successful at fitting particle yields in central Pb+Pb/Au+Au collisions~\cite{Becattini:2005xt, Andronic:2005yp}. The centrality dependence of strangeness production is described in terms of a transformation from canonical to a grand canonical ensemble in the statistical model~\cite{Redlich:2001kb}. This model uses a correlation volume in order to make a comparison to experimental data. A different approach to determine the correlation volume is used in the percolation model~\cite{Hohne:2005ks}. In this model, the fireball volume is segmented into smaller sub-volumes (clusters). A similar approach is used by the so-called core-corona model~\cite{Werner:2007bf}, where an A+A collision is assumed to consist of a central core in  full chemical  equilibrium ($\gamma_s$=1)  surrounded by a corona  produced by \pp type of interactions where strangeness is under-saturated. In order to use a hadronic baseline, microscopic models like UrQMD~\cite{Petersen:2008kb} can be used  to distinguish between a hadronic and partonic world.  So far, discrimination between strange hadron enhancement  due to dense partonic medium formed in heavy-ion
collisions or  canonical supression  in \pp collisions has been highly ambiguous,  when using the available experimental data on strange hadrons with non-zero net-strangeness. However,  the measured enhancement of the $\phi$ ($s\bar{s}$) production (zero net strangeness) in heavy-ion collisions relative to \pp collisions at RHIC, Fig.~\ref{Fig:phiEnhance}, seems to clearly indicate the formation of a dense partonic medium~\cite{Abelev:2008fd,Abelev:2008zk}. Furthermore $\phi$ mesons do not follow the strange quark ordering as expected in the canonical picture for the production of other strange hadrons~\cite{Abelev:2008fd,Abelev:2008zk}. This seems to rule out canonical suppression effects being the predominant cause of the observed enhancement in other strange hadrons in high energy heavy-ion collisions. The BES will allow the  further testing of the centrality and  energy dependence of strangeness enhancement predicted from the various models.

\subsubsection{Mechanisms of particle production at forward rapidity}\label{SubSubSec:Forward}

\begin{figure}[htb]
\begin{minipage}{0.46\linewidth}
\begin{center}
\includegraphics[width=0.8\linewidth]{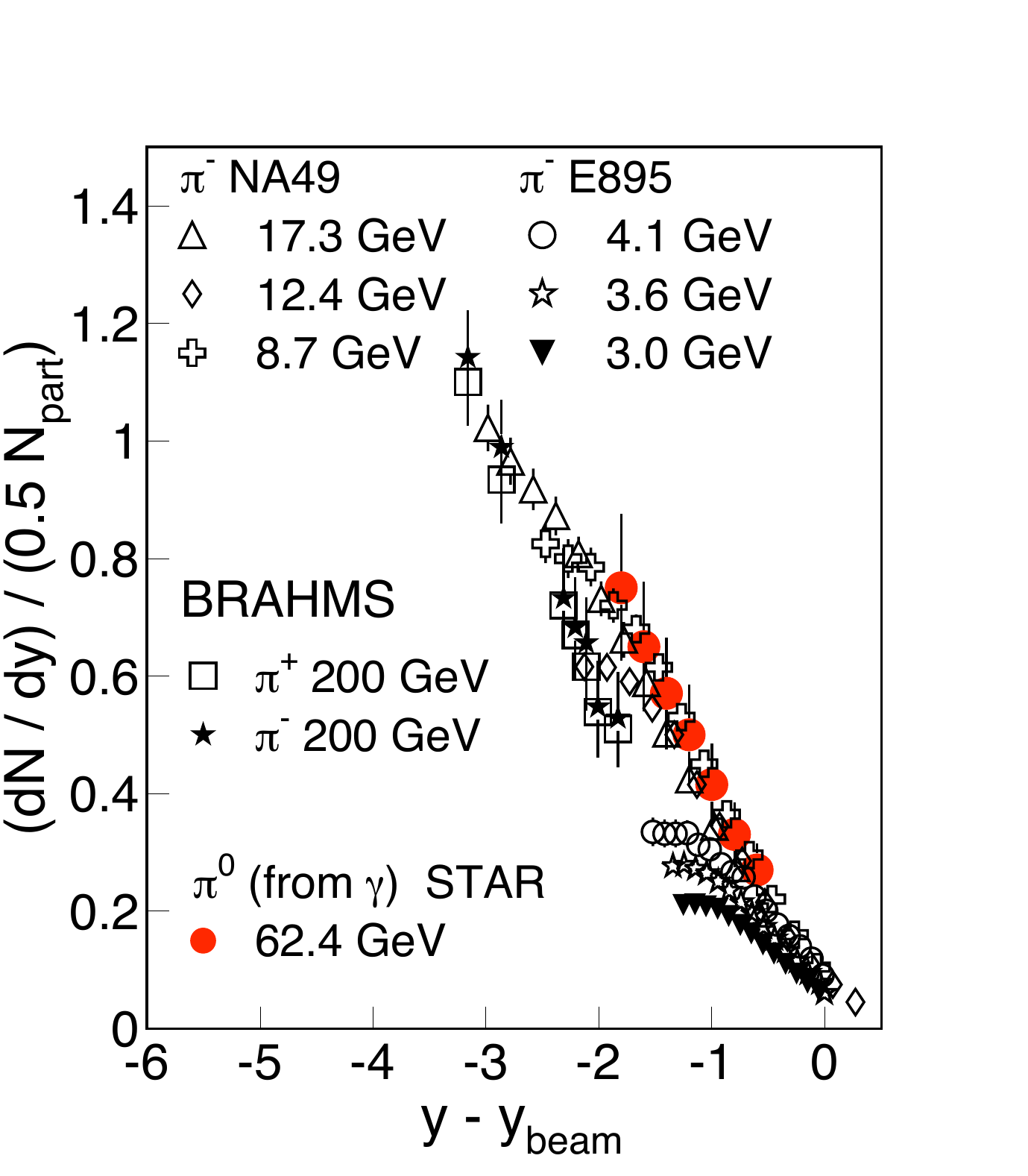}
\end{center}
\end{minipage}
\hspace{1cm}
\begin{minipage}{0.46\linewidth}
\begin{center}
\includegraphics[width=0.8\linewidth]{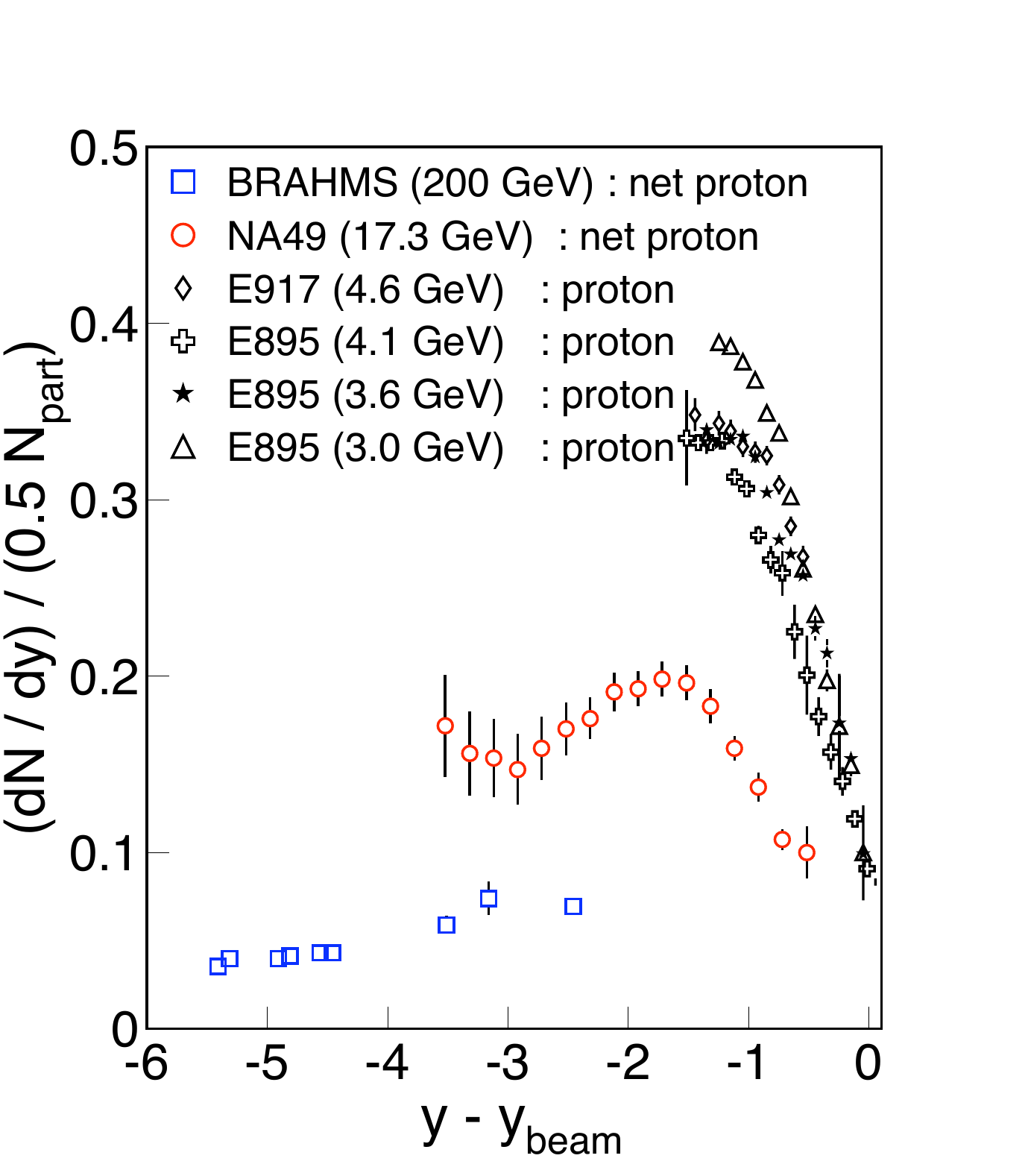}
\end{center}
\end{minipage}
\caption{Longitudinal scaling of $\pi$ and protons rapidity distributions for various collision beam energies \cite{Adams:2005cy}. Rapidity distribution normalized to number of participating nucleon pair is plotted as a function of rapidity shifted by beam rapidity.}
\label{Fig:ForwardScaling}
\end{figure}

\noindent At top RHIC energies, two interesting features have been observed at forward rapidities:
(a) While at forward rapidities particle production scales purely on  the number of participants \cite{Adams:2006ke}, at mid-rapidity the production  follows a dependence on both  the number of participants and the number of binary collisions \cite{Adcox:2000sp}.  This suggests that there are different particle production mechanisms at work as a function of rapidity. The beam energy dependence of these scalings needs to be studied to fully understand how particles are produced in heavy-ion collisions. 
(b)	 The longitudinal scaling of charged particles and $\pi$ at near-beam rapidities observed at RHIC, left plot in Fig.~\ref{Fig:ForwardScaling},  has been found to be beam energy independent, but centrality dependent \cite{Back:2003hx}.  The reason for this centrality dependence may be attributed to fact that the measured charged particle yields contain  significant contributions from baryons which  vary strongly with the number of participants in the collision.  This changing baryon transport plus direct contribution from beam protons could be reason for violation of the scaling (as shown in Fig.~\ref{Fig:ForwardScaling}). Photon production being mostly from $\pi^{0}$ decays provides a clean test of the energy and centrality dependence of longitudinal scaling.

~

\noindent Not many multiplicity detectors will be available at forward rapidity at RHIC during the proposed beam energy scan program; data from the PMD will provide crucial information regarding the rapidity dependence of particle production.

\subsubsection{Light Nuclei Production \label{SubSubSec:lightnuclei}}

\noindent 
The very small binding energies of nuclei of mass number A imply that these composite objects are formed via the coalescence of A nucleons, that are close together in phase space at the time of chemical freeze-out. Their production rates and spectra depend sensitively on the interplay of several dynamical aspects of the collisions such as nucleon density profiles, temperatures, and collective flow. Volume expansion due to secondary interactions tends to diminish cluster yields as the overall
particle production increases with the total energy \sqrts of the collision. Counterbalancing effects arise from collective flow which tends to focus nucleons and nuclei in phase space and hence increases the production rates of clusters and decreases the apparent source sizes with respect to the total volume of the system. Collective motion, temperatures, and densities are also related to entropy production and pressures, which are important quantities in the study of phase diagrams. Thus, the study of the rates and spectra of light nuclei remains a useful tool for understanding the expansion and correlations of the source.

~

\noindent A wide variety of physics can be extracted from the study of light fragment production.
\begin{itemize}
\item
A specific ratio of the invariant cross-sections for light
nuclei and nucleons, called ``B$_{\rm A}$", can be converted
into effective source volumes (V$_{\rm A}$$\sim$1/B$_{\rm A}$)
in a number of model-dependent ways. These effective volumes
can be compared directly to those obtained from intensity
interferometry, {\it e.g.} the product of the ```side" and
``long" radii $(R_s2 R_l)^{1/3}$. The discrepancies in these
comparisons are often interesting as they can result from the
specific model assumptions used to convert the coalescence
B$_{\rm A}$ values into volumes in the presence of strong
collective flow and the temporal development of the source size
(higher mass nuclei freeze-out earlier than lighter nuclei).
\item
Light nucleus production rates can be used to estimate particle
phase-space densities at freeze-out \cite{Murray:2000cw}. Such
studies have indicated complex relationships between the
freeze-out surfaces for pions and nucleons that depends on
\sqrts via the overall production rates of pions and nucleons.
At the lowest (highest) energies studied at the SPS, freezeout
is governed by the nucleons (pions) \cite{Murray:2000cw}.
\item
The ratio of invariant cross-sections for particles and
antiparticles of mass, A, can be used to deduce values of the
chemical potentials to the temperature (and the square of the
``Fugacity"), for example: $\rm{\bar{d}/d} = $exp(-2$\mu_B^A$/T).
As the proton chemical potential can be independently measured,
the measurement of $\rm{\bar{d}/d}$ (and the ratios for higher
mass nuclei) can provide information on the neutron chemical
potential and the $\rm{\bar{n}/n}$ ratios, which are not
generally directly measurable \cite{Belmont:2008zz,Adler:2004uy}.
\item
At the SPS \cite{Kolesnikov:2007ps,Melkumov:2007su}, the
initial n/p ratio of 1.54:1 for $^{208}$Pb evolves towards the
equilibration value of $\sim$1:1 in the fireball due to strong
resonance production. Thus, the measurement of the
triton/$^3$He ratio, for example, can be converted using a
simple additive procedure to the freeze-out n/p ratio in the
source, and hence determine the degree of chemical
equilibration achieved there.
\item
The deuteron to proton ratios can be used to determine the
entropy per baryon \cite{Stoecker:1984py,SimonGillo:1995dh} via
S/A = 3.95-ln(d/p) + (N$_\pi$/N$_{\rm n}$).
\item
Recent studies from the lattice indicate strong effects in the
net-baryon kurtosis (measured via net-protons in an experiment)
near the critical point (see Section:\ref{SubSubSec:Kurtosis}).
Similar effects may be visible in the study of the fluctuations
in the deuteron production rates in the same events.
\end{itemize}

~

\noindent The EOS, E877, and E878 experiments at the AGS, the
NA44, NA49, and NA52 experiments at the CERN-SPS, and the RHIC
experiments, have provided a wealth of information on the
\sqrts dependence of light nuclei production in A$+$A
collisions. However, a systematic study of the light nucleus
production rates and spectra in a RHIC energy \sqrts scan may
clear up some unresolved aspects of the data from the SPS
experiments. These aspects are as follows.
\begin{itemize}
\item
The SPS experiments do not in general measure the same
kinematic regions near mid-rapidity. The comparison of the
results thus requires an approximate scaling from one
experiment's acceptance to another's, and the production rates
so compared are not always in perfect agreement
\cite{Afanasiev:2000,Anticic:2004yj}. The B$_{\rm A}$ ratios
measured by NA44 and NA52 at the top SPS energy differ by a
factor of two (a $\sim$3-$\sigma$ discrepancy) while the NA49
result is in between. Only NA49 measured light fragments at SPS
beam energies lower than \sqrts = 17.3 GeV.
\item
In the NA49 data, the rapidity distributions of $^3$He nuclei
are ``concave" versus the rapidity ({\it i.e.} indicate a
shallow minimum at y$_{\rm CM}$ = 0), while the nucleon
rapidity spectra are essentially flat in the same rapidity
region. The observed increase in the $^3$He formation rates for
increasing $|$y$_{\rm CM}$$|$ is not yet understood
\cite{Kolesnikov:2007ps,Melkumov:2007su}. The observed
dependence of the light nucleus production rates on the
rapidity in NA44 \cite{Bearden:2002ta} is a feature that
disagrees with the model of Ref. \cite{Scheibl:1998tk}, which
is otherwise successful in describing direct comparisons of the
source dimensions inferred from HBT intensity interferometry to
those obtained from light nucleus coalescence prescriptions.
\item
The distinction \cite{Llope:1995zz} between the coalescence of
A nucleons versus the coalescence of nucleons and lighter
fragments for the formation of a cluster of mass A is not
completely clear. Some analyses imply that deuteron formation
as an intermediate step to triton formation is required to
match the data \cite{Polleri:Thesis:2003} \cite{Bearden:1999iq}
while others disagree \cite{Bearden:2002ta}.
\end{itemize}

\subsubsection{(Anti-)Hypernuclei}\label{SubSubSec:Hypernuclei}

\noindent A hyper-nucleus is a nucleus which contains at least one hyperon in addition to nucleons. The first hyper-nucleus was discovered by Marian Danysz and Jerzy Pniewski in 1952 in a cosmic ray experiment~\cite{Danysz:1953aa}. The smallest and simplest hyper-nucleus is hypertriton (3$\Lambda$H), consisting of a $\Lambda$, a proton and a neutron. Hyperons inside a hypernucleus contain strangeness quantum number, and hence, provide one more degree of freedom for nuclear spectroscopy than the normal nucleus, which consists of protons and neutrons. These hypernuclei also provide an ideal laboratory for studying the force between hyperon and nucleon (Y-N interaction), which was otherwise not possible with normal nuclei or the traditional hadron-hadron or electron-positron beams. This information is much needed to understand the configuration of a neutron star, which depending on the strength of the interaction can alternatively be an object of strange quark matter, hyperon star, or a kaon condensate at the core~\cite{Lattimer:2004pg}. 

~

\noindent The conventional methods of providing hypernuclei were by cosmic ray interaction, kaon capture or strangeness exchange~\cite{Davis:1992dt}.  There are currently several hypernucleus experiments in the major nuclear facilities: MAMIC at Mainz and JLab (photo-production), FINUDA at DA$\phi$NE (stopped kaon beam from e$^{+}$e$^{-}$ collider), J-PARC (stopped kaon from hadron beam), PANDA at FAIR (stopped anti-proton annihilation), and HypHI at FAIR and SPHERE at JINR (heavy ion beam). All these experiments require a target of baryon-rich dense nuclear matter to provide the nucleons necessary for hypernucleus production, and are incapable of producing anti-hypernuclei: the anti-matter of hypernuclei. 

~

\noindent RHIC collisions produce an abundance of particles with a high phase-space density. This environment is thus uniquely suited for the production of exotic nuclei and anti-nuclei and anti-hypernuclei via coalescence at the late stages of the evolution of the medium produced. The abundances of nuclei and anti-nuclei are similar at the core of the reaction zone since the anti-matter and matter are more or less equal at top RHIC energies.  This offers the first opportunity for discovery of anti-hypernuclei as well as anti-nuclei (A $>$ 3). Since the coalescence process for formation of hypernuclei (anti-hypernuclei) requires that nucleons (anti-nucleons) and hyperons (anti-hyperons)  are produced in proximity, the hypernucleus (anti-hypernucleus) production are sensitive to correlations of the coordinate and momentum phase space distributions of nucleons and hyperons.  Similarity or equilibrium between these two species is one of the signatures of the QGP formation, which would also result in higher hypernucleus (anti-hypernucleus) yields. The hypertriton yields can be compared to the yields of $^{3}$He and triton which have the same atomic mass number.

\noindent We have found clear evidence for the first-ever observation of a anti-hypernucleus, the anti-hypertriton, as reported in ~\cite{Abelev:2010rv}. The anti-hypertriton and the hypertriton were reconstructed in the TPC and identified via the secondary vertex of hypertriton to ($^{3}$He + $\pi$).  We have also been able to measure the anti-hypertriton lifetime, though with  a large statistical uncertainty (20 to 30\%).  

~

\noindent We plan to increase statistics for the signals currently under study, which for example, will result in a more precise lifetime measurement for the anti-hypertriton.  We plan to carry out a few key measurements of hypernucleus production  which will provide crucial information about the hyperon --  nucleon correlation. The BES will provide data points between AGS and top RHIC energies to establish the trend of the 
hypertriton/$^{3}$He ratio. This can then be compared to calculations at a quantitative level.  For \sqrts = 17 (5) GeV, with a penalty factor of 368 (48)~\cite{Liu:2006aa} and a $^{3}$He yield of 2$\times$10$^{-4}$ ( 0.01) and a hypertriton/$^{3}$He ratio of 0.3 (0.05) the estimated number of min-bias Au+Au events needed to extract a hypertriton signal at the  5$\sigma$ level is 10-100 M (1-10 M).  The above estimates assume that the reconstruction efficiency is the same at all energies. In reality, the background and hyperon yields decrease with decreasing beam energy while reconstruction efficiency increases due to lower TPC occupancy.

\subsection{Local Parity Violation in the Strong Interaction}\label{SubSubSec:ParityViolation}

\noindent The observation of local parity violation in the strong interaction in heavy-ion collisions would be an extremely important result. The concept that non-central collisions may result in such a violation was first postulated over a decade ago~\cite{Kharzeev:1998kz,Kharzeev:1999cz}.  More recently it was suggested that it can result in charge separation relative to the reaction plane~\cite{Kharzeev:2004ey}. Non-central events should produce media with  large orbital angular momenta. Since the system has a net charge this should result in strong magnetic fields. If the system is also de-confined this may result in P violating domains and different number of left and right handed quarks.
In the strong magnetic field it will result in preferential emission of like sign charged particles from these domains along the angular momentum vector, i.e. in the direction normal to the reaction plane.  If one could measure the number of positively or negatively  charged particles with respect to this plane, the P violating domains should result in 

\begin{eqnarray*}
\frac{dN_{\pm}}{d\phi} \propto 1 + 2a_{\pm} \sin(\phi - \Psi_{RP}) + 2v_{1}\cos(\phi - \Psi_{RP}) + 2v_{2}\cos(2(\phi - \Psi_{RP})) + ...
\end{eqnarray*}

\noindent where a$_{\pm}$ is the asymmetry and thus a measure of the scale of the parity violation.  
The problem is that while the emission direction is preferred to be along the angular momentum vector, the sign of this emission vector is random for each domain.  This means that $\langle a_\pm \rangle$ = 0 and since dN$_{\pm}$/d$\phi$ can only be measured statistically by experiments the proposed signal is lost. What can be measured is the mean pair asymmetry, $\langle a_{\alpha}a_{\beta}  \rangle $, where $\alpha$ and $\beta$ indicate all possible particle charge sign combinations.

\begin{eqnarray*}
\langle \cos( \phi{_\alpha} + \phi_{\beta} -2\Psi_{RP}) \rangle \approx (v_{1,\alpha}v_{1,\beta} - a_{\alpha}a_{\beta})
\end{eqnarray*}

\begin{figure}[htb]
\begin{center}
\includegraphics[width=0.5\linewidth]{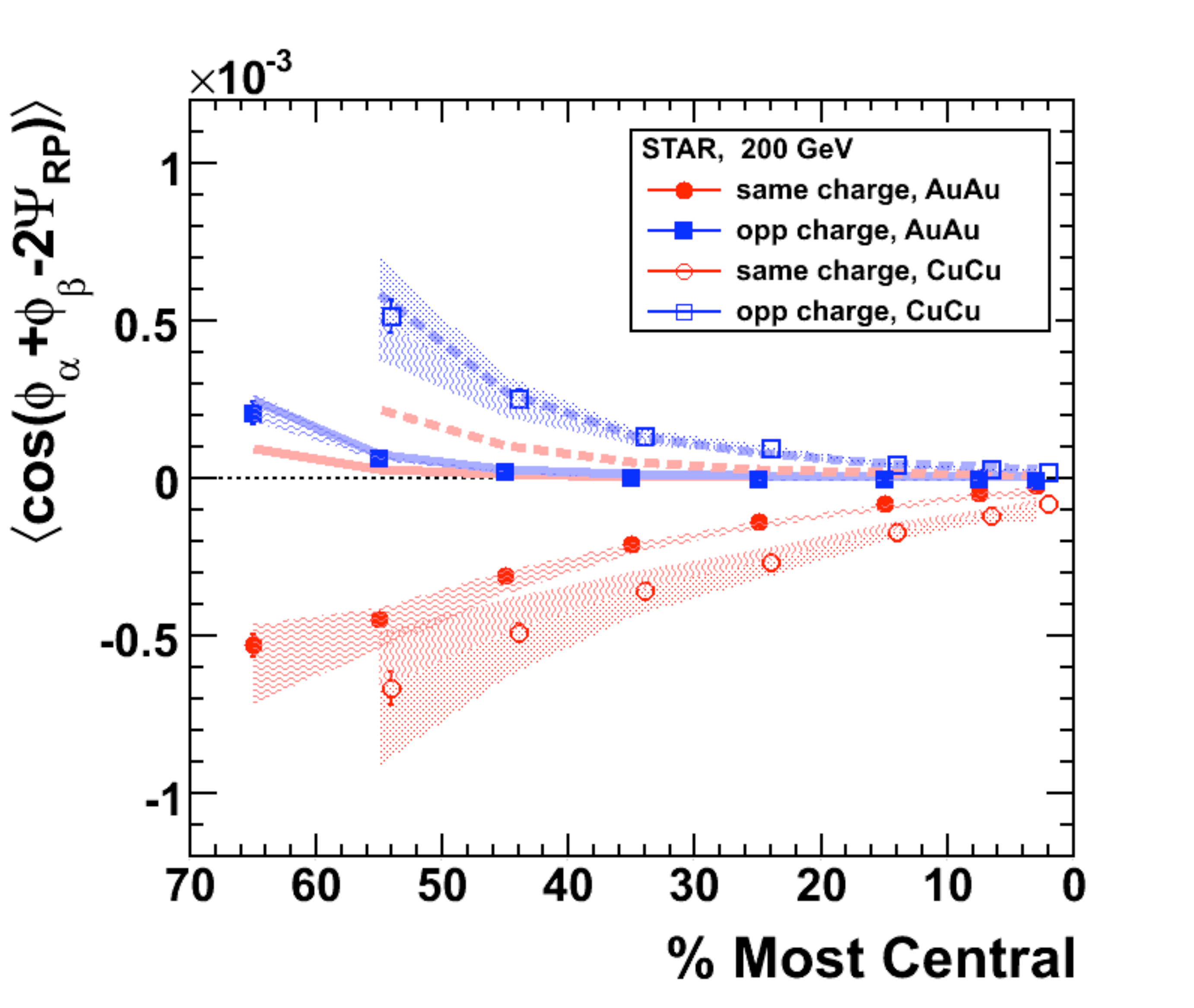}
\end{center}
\caption{$\langle \cos( \phi{_\alpha} + \phi_{\beta} -2\Psi_{RP})\rangle$ as a function of centrality for Au+Au and Cu+Cu events at \sqrts= 200 GeV. The shaded regions indicate the systematic uncertainties. The thick and dashed curves are HIJING 3-particle correlation calculations for Au+Au and Cu+Cu collisions respectively. Figure from \cite{Abelev:2009uh}. The uncertainties will be smaller at lower collision energies.}
\label{Fig:ParityViolation}
\end{figure}

\noindent $ \langle a_{+}a_{-} \rangle $ should show a positive correlation for like-sign pairs and a negative one for unlike-sign pairs~\cite{Voloshin:2004vk}. Such signals, Fig.~\ref{Fig:ParityViolation}, have recently been reported by STAR in non-central Au+Au and Cu+Cu events at \sqrts=200 GeV where the magnitude of the effect grows with decreasing centrality as predicted~\cite{Abelev:2009uh}.  Such a parity violating signal only results from the combination of a strong B-field (peripheral events) and de-confinement. This means there is likely to be a strong threshold effect that can be sought via the energy scan. Due to the predicted  increase in  the signal as the collision energy decreases (unless the sQGP is no longer formed) it  has been estimated that 4 M events are sufficient to resolve the signal for \sqrts = 5-39 GeV.

~

\noindent Care must be taken however before parity violation is declared in these events because $\langle a_{\alpha}a_{\beta} \rangle$ is P-even not P-odd and so may contain significant contributions from other sources such as resonance decays. Several of these contaminating sources have been investigated and none have been able to consistently create signals of the right magnitudes and centrality dependence~\cite{Abelev:2009txa}.  The shaded areas in Fig.~\ref{Fig:ParityViolation} indicate the  magnitudes of the background contributions, these effects should be smaller at lower collision energies. One way to lend credence to this P-violation measure is to observe the same result at different collision energies.  By varying \sqrts the contributions from the other contaminating sources will occur with different weights. Thus the observation of this potentially P violating signature at other energies will help rule out other more traditional explanations of the results.

\section{Analysis of Current Test Runs - Au+Au at \sqrts = 9.2 and 19.6 GeV }\label{Sec:AnalysisOfCurrentTestRuns}

\noindent A number of short low energy  test  runs have been performed at RHIC, the data recorded are used below to examine STAR's readiness for the BES. In particular data have been taken at \sqrts = 19.6 and 9.2 GeV in 2001 and 2008 respectively.  The  2008 9.2 GeV  run was below injection energy. STAR recorded $\sim$175 k Au+Au collisions at the injection energy of \sqrts= 19.6 GeV. From these data 5k events belongs to the top 10$\%$ collision centrality and had a well reconstructed primary vertex.   During the two days of \sqrts = 9.2 GeV tests 200 k triggers were recorded resulting in $\sim$3.5 k good Au+Au events. A good event is defined as having a reconstructed primary vertex that is well within the TPCs acceptance ($|z|<$30 cm) and consistent with coming from along the beamline. The results have been published in~\cite{Abelev:2009bwa}.

~

\begin{figure}[htb]
\begin{minipage}{0.46\linewidth}
\begin{center}
\includegraphics[width=0.8\linewidth]{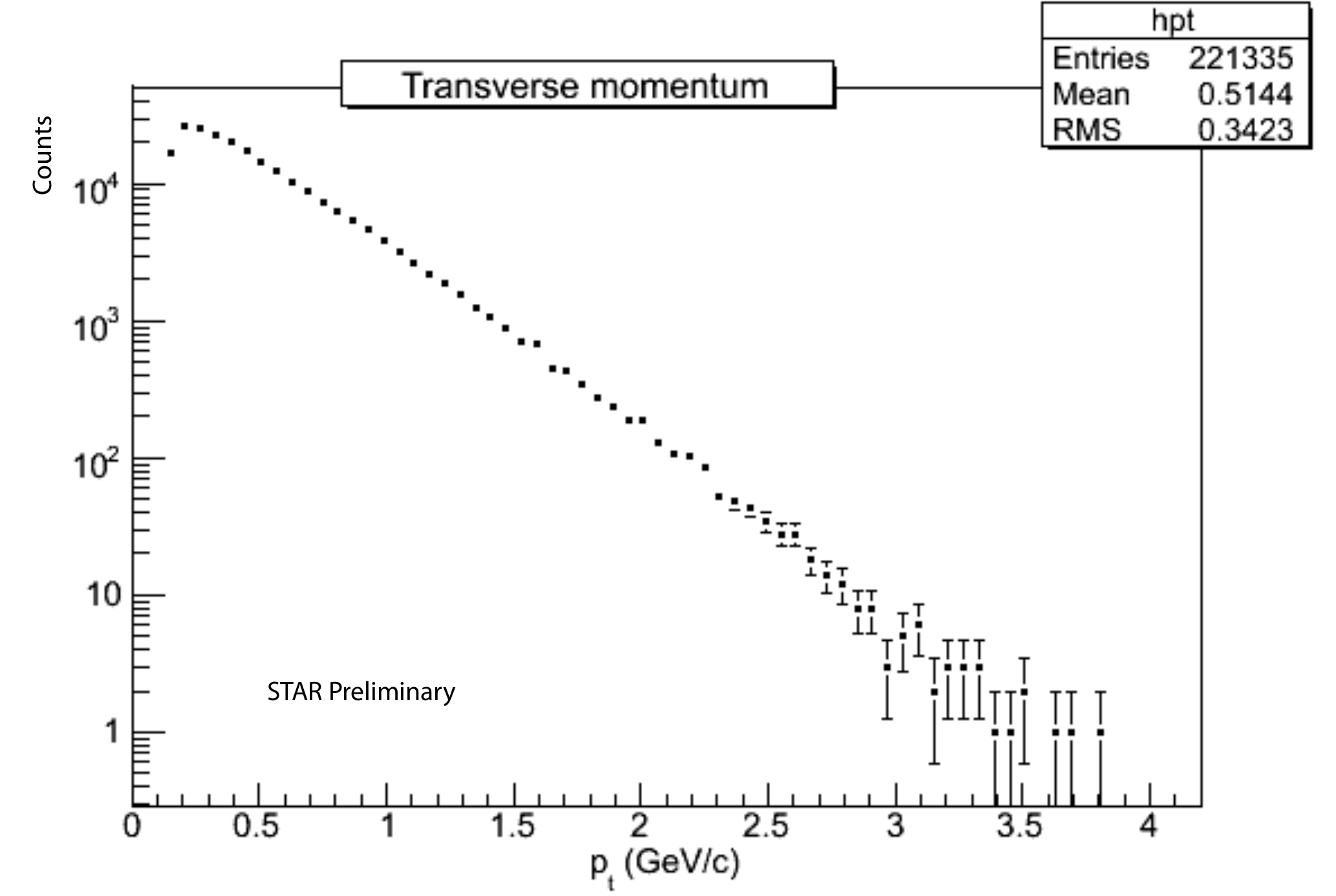}
\caption{The uncorrected \pT spectrum of charged particles for \sqrts= 9.2 GeV minimumbias Au+Au data.}
\label{Fig:UncorrectedSpectrum9GeV}
\end{center}
\end{minipage}
\hspace{1cm}
\begin{minipage}{0.46\linewidth}
\begin{center}
\includegraphics[width=0.8\linewidth]{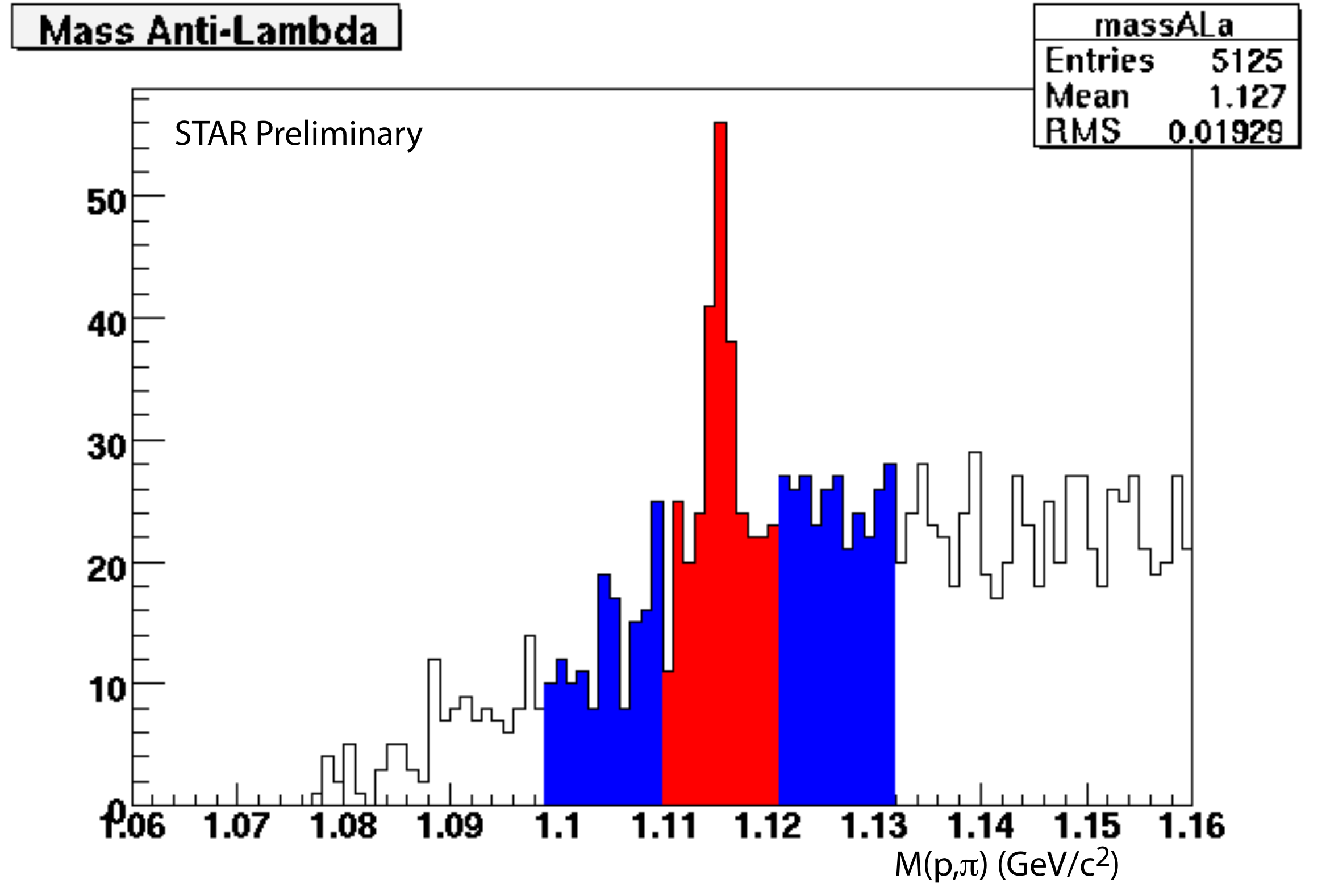}
\end{center}
\caption{The invariant mass distribution of the $\bar{\Lambda}$ for \sqrts= 9.2 GeV minimumbias Au+Au data.}
\label{Fig:AntiLambda9GeV}
\end{minipage}
\end{figure}

\noindent Even with these few events the uncorrected minimum bias charge particle transverse momentum  spectrum at 9.2 GeV extends to $>$ 3 GeV/c, Fig.~\ref{Fig:UncorrectedSpectrum9GeV}, and there are indications of an $\bar{\Lambda}$ peak, Fig.~\ref{Fig:AntiLambda9GeV}.

\begin{figure}[htb]
\begin{center}
\includegraphics[width=\linewidth]{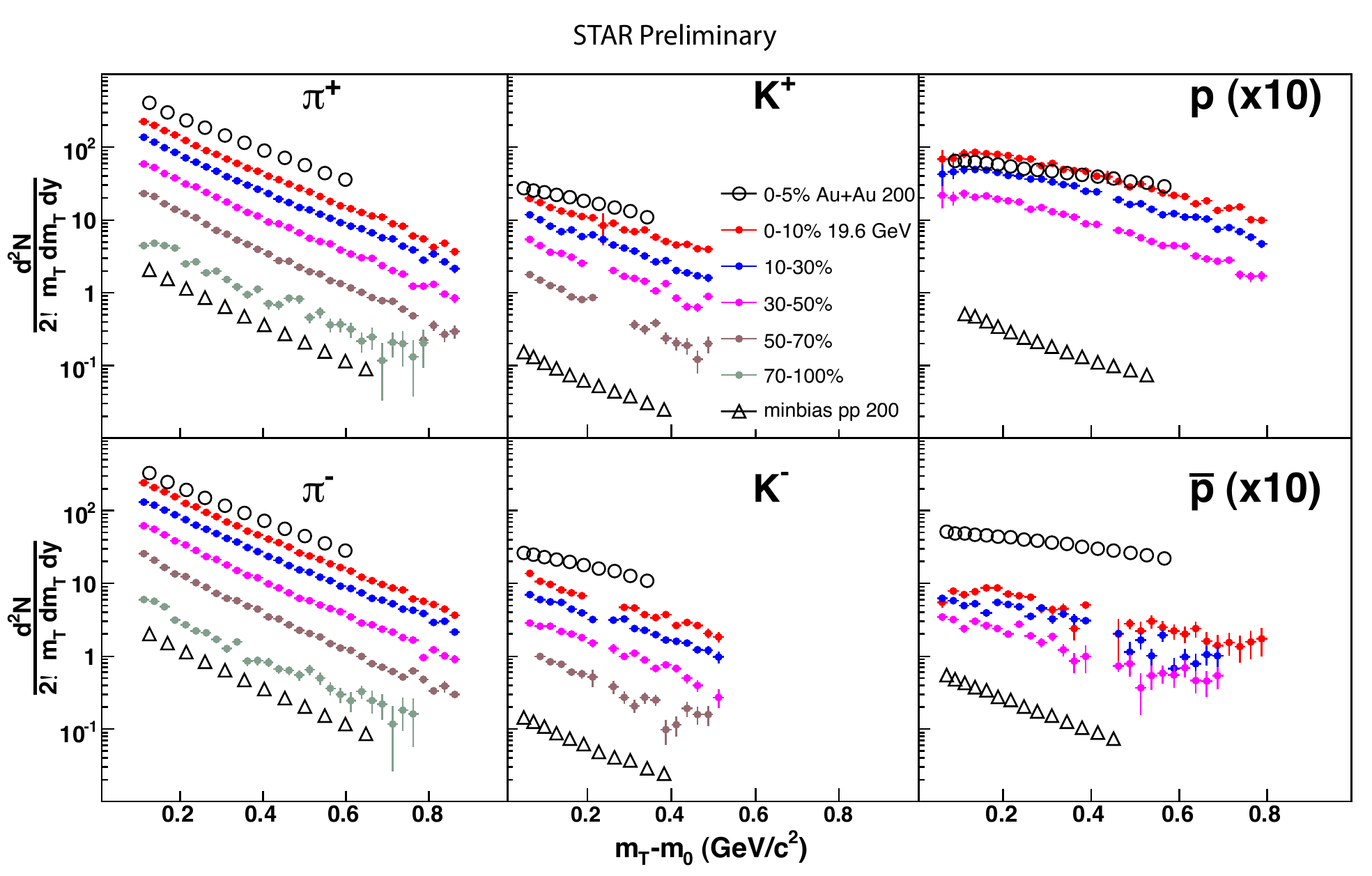}
\caption{The identified particle spectra as a function of centrality for Au+Au collisions at \sqrts=19.6 GeV. From ~\cite{Cebra:2009fx}.}
\label{Fig:Spectra19}
\end{center}
\end{figure}

\begin{figure}[htb]
\begin{minipage}{0.46\linewidth}
\begin{center}
\includegraphics[width=0.8\linewidth]{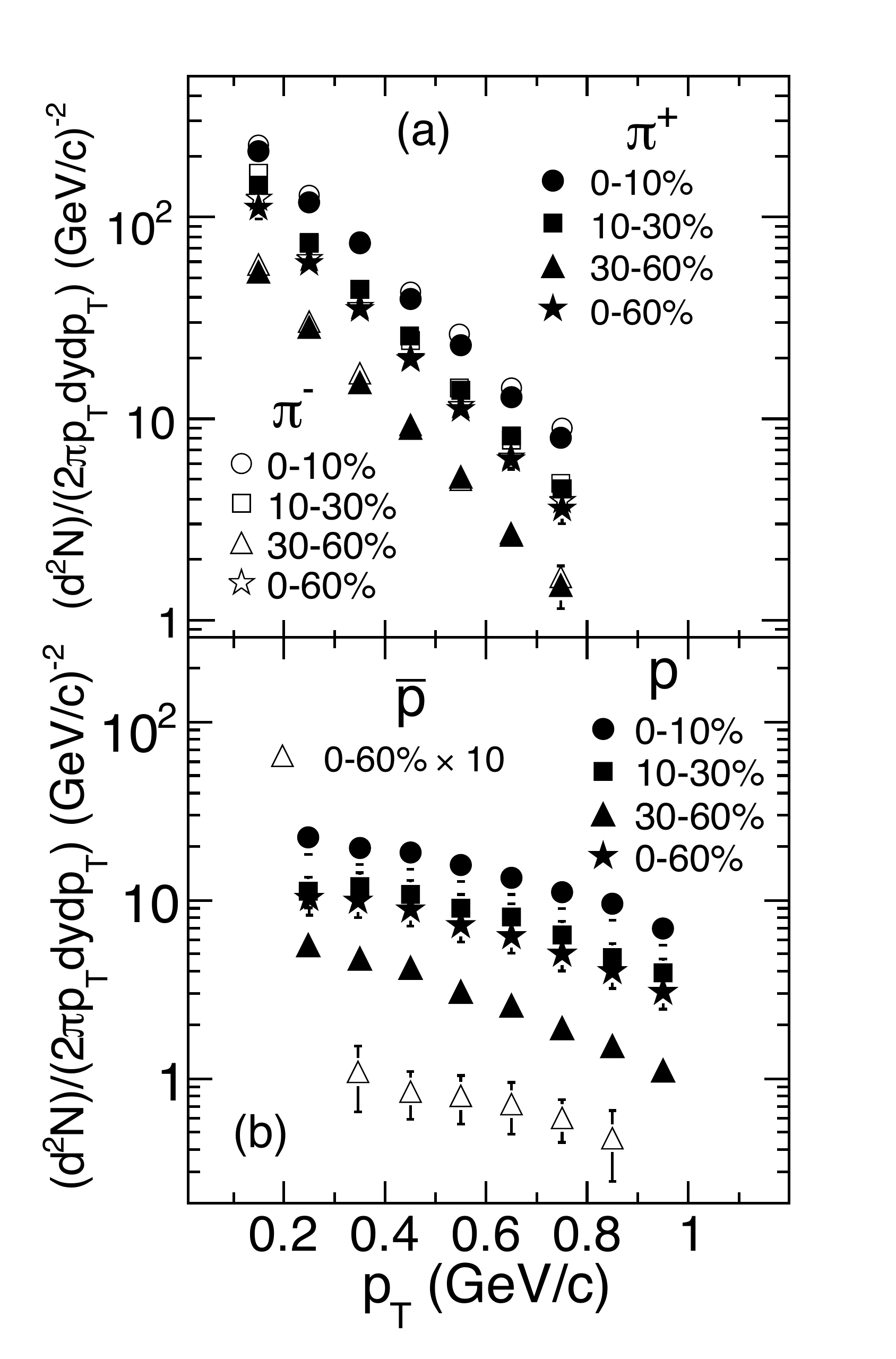}
\caption{The corrected a) $\pi$  and b) p \pT spectrum  for \sqrts= 9.2 GeV  Au+Au data as a function of centrality at midrapidity ($|y|<$0.5). From ~\cite{Abelev:2009bwa}.}
\label{Fig:pippT9.2GeV}
\end{center}
\end{minipage}
\hspace{1cm}
\begin{minipage}{0.46\linewidth}
\begin{center}
\includegraphics[width=\linewidth]{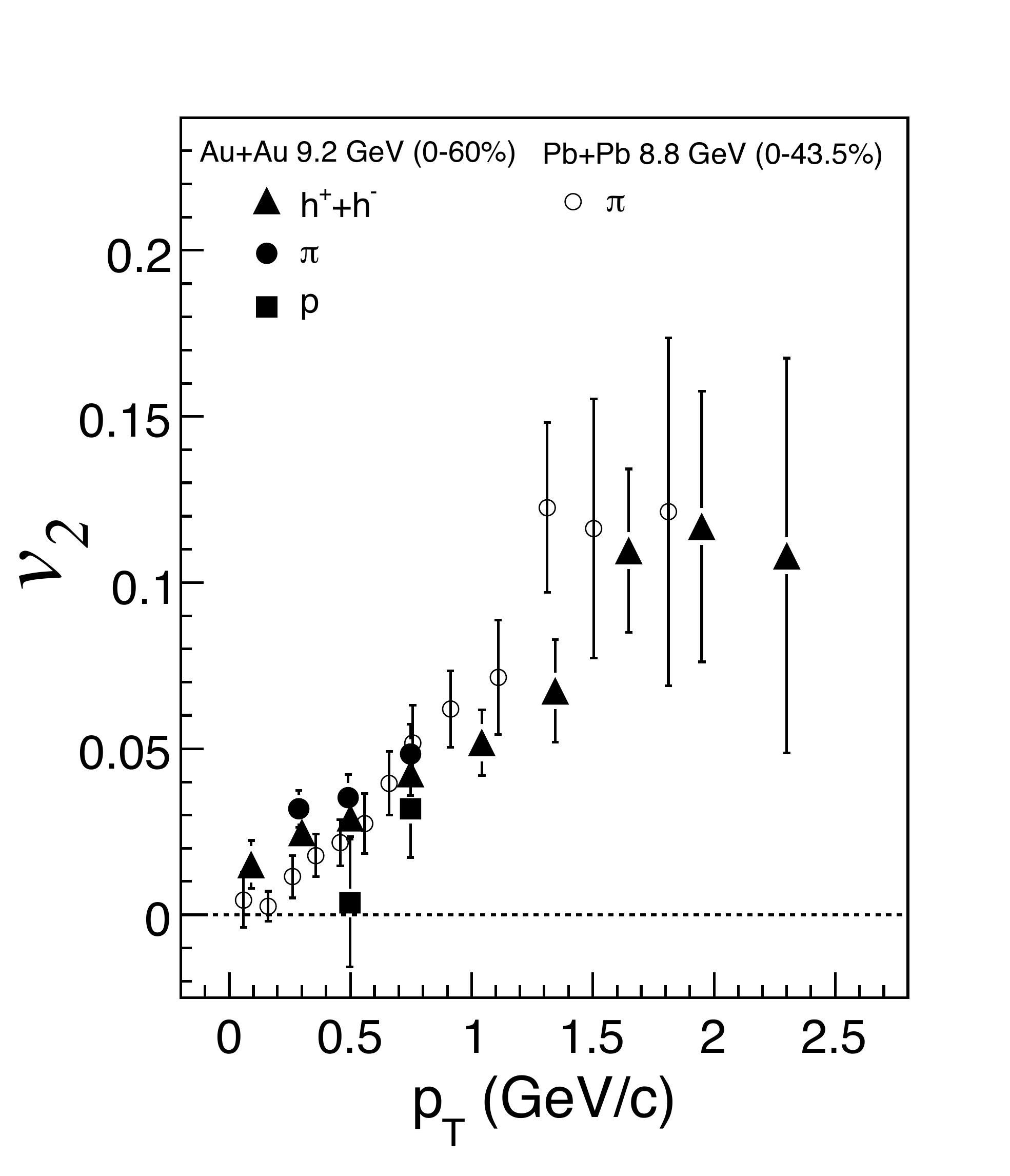}
\end{center}
\caption{The charged particle v$_{2}$ at \sqrts=9.2 GeV as a function of \pT. From \cite{Das:2009ni,Abelev:2009bwa}.}
\label{Fig:v2Pt9.2}
\end{minipage}
\end{figure}

\begin{figure}[htb]
\begin{minipage}{0.46\linewidth}
\begin{center}
\includegraphics[width=0.7\linewidth]{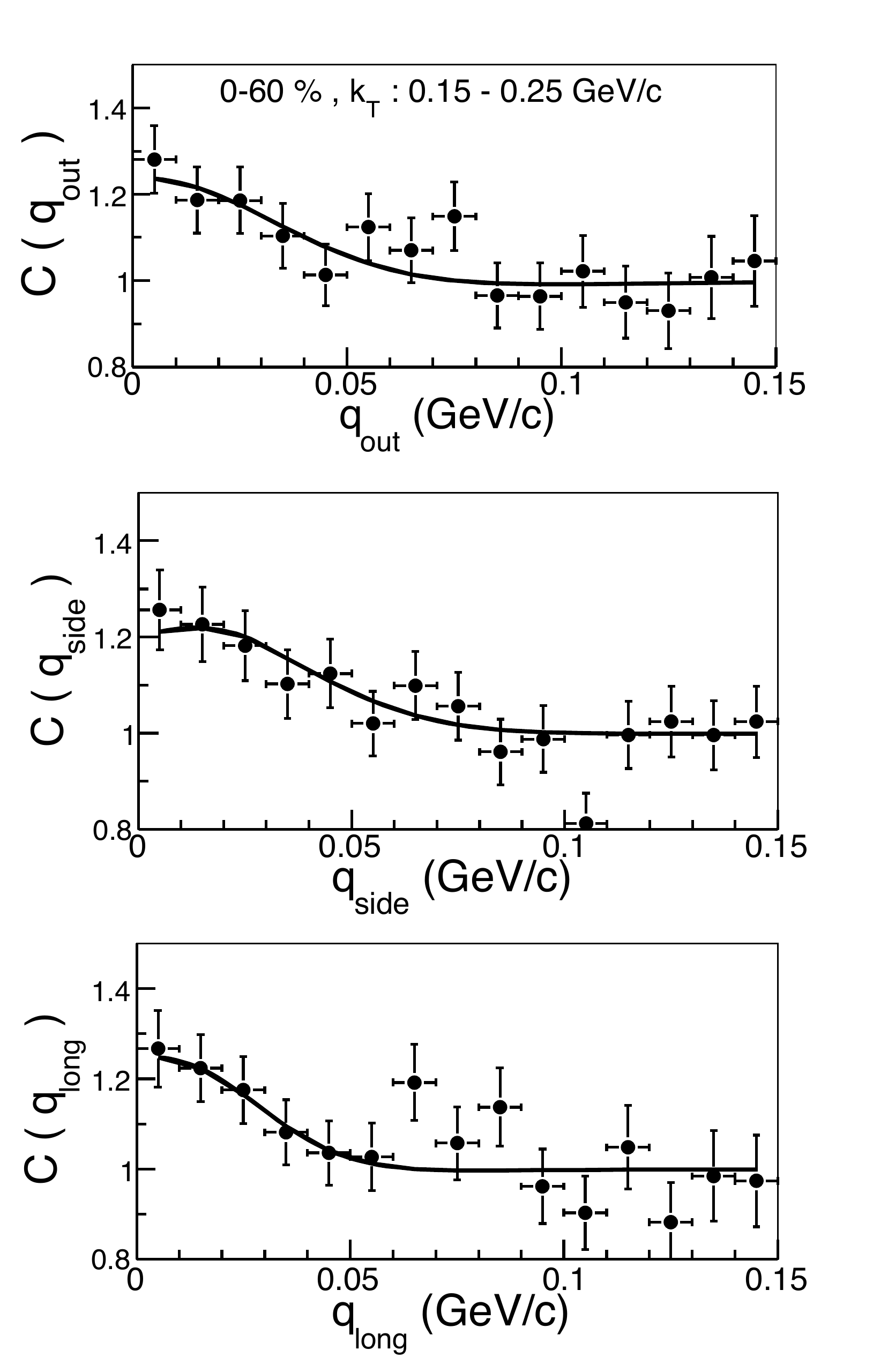}
\end{center}
\caption{The HBT projections of the 3-dimensional correlation functions for negative $\pi$ data  from 0-60$\%$ central Au+Au collisions at \sqrts= 9.2 GeV.  Fits shown are the Bowler-Sinyukov functions~\cite{Bowler:1991vx,Sinyukov:1998fc}. From~\cite{Das:2009ni,Abelev:2009bwa}.}
\label{Fig:HBT9.2}
\end{minipage}
\hspace{1cm}
\begin{minipage}{0.46\linewidth}
\begin{center}
\includegraphics[width=0.8\linewidth]{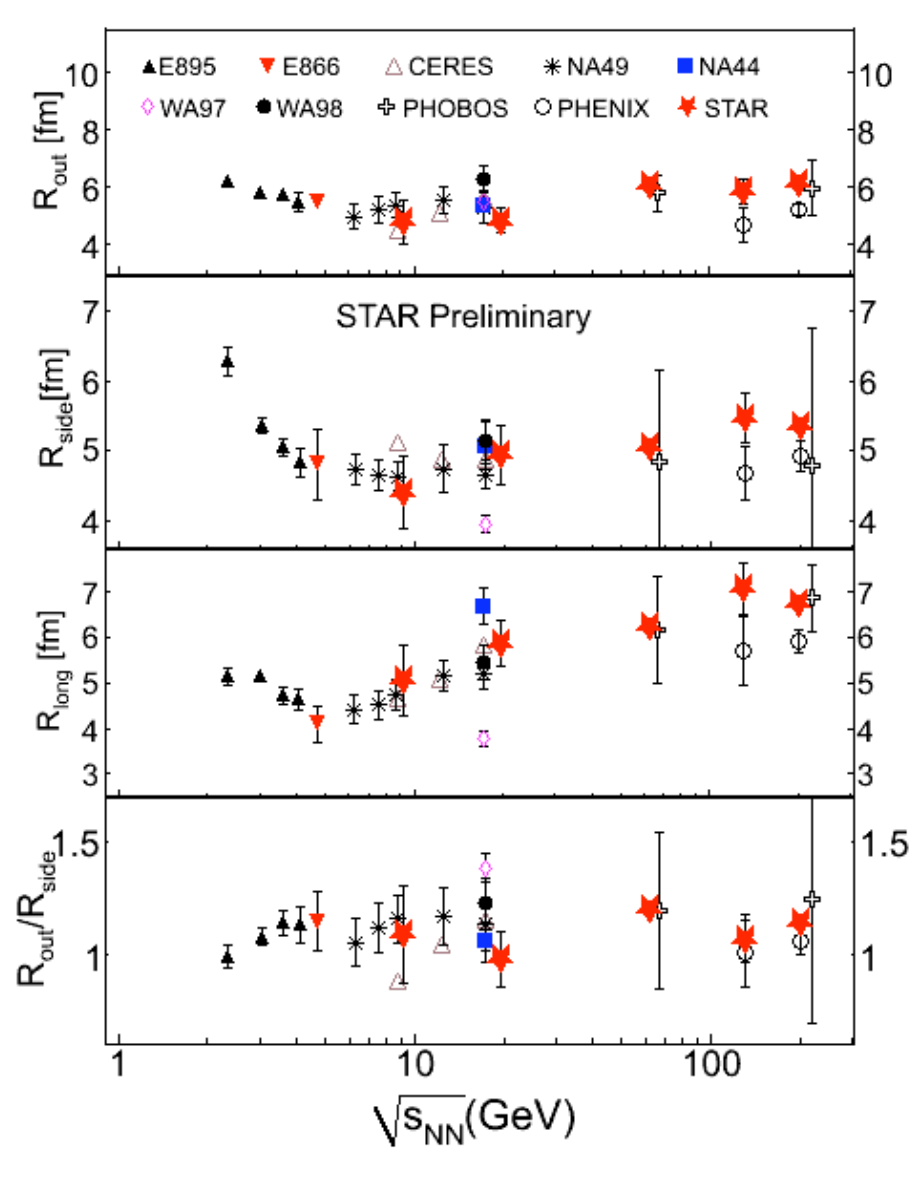}
\caption{The HBT systematics as a function of \sqrts.  Low energy data from STAR added to plot from \cite{Adler:2001zd}.}
\label{Fig:HBTsqrts}
\end{center}
\end{minipage}
\end{figure}

\noindent While neither of these runs were long enough to improve on the physics measurements reported by the SPS experiments many checks have been performed.  Some of the results of the analyses of these data are shown briefly below, or have been included in the figures and discussions in the previous sections.  The resulting \pT spectra are shown for $\pi$, K and (anti)protons in Fig.~\ref{Fig:Spectra19} for the 19.6 GeV  and for  pions and protons  at 9.2 GeV in Fig.~\ref{Fig:pippT9.2GeV}  \cite{Kumar:2008ek,Abelev:2009bwa,Chen:2009aj}. Similar results exist for the charged kaons. We measure 82$\%$ of the produced $\pi$ at mid-rapidity and 75$\%$ of the protons, the kaon reach is not as complete, covering only 47$\%$ of the yield, however this will be greatly improved  with the inclusion of the ToF. HBT radii have been extracted from the \sqrts= 9.2 GeV Au+Au collisions  for $\pi$ in the 0-30$\%$ centrality bin, Fig.~\ref{Fig:HBT9.2}. The fits yield results that correspond with  the \sqrts systematics from AGS to RHIC shown in Fig.~\ref{Fig:HBTsqrts}~\cite{Adler:2001zd}.  No anomalous jump is seen in the R$_{out}$/R$_{side}$ ratio, although the peak could be narrow and located between \sqrts=20 and 60 GeV.   The feasibility of such a measurement is indicated in Fig.~\ref{Fig:v2Pt9.2} where  statistically relevant identified v$_{2}$ measurements have been made with only $\sim$3 k  events. To date all of STARs results fit well into the measured systematics reported by the SPS, indicating that STAR is ready for full length data taking.

~

\noindent Running at sub-injection energies requires a different harmonic to top energy running and is challenging to both the experiments and the collider. For instance the particular choice of \sqrts = 9.2 GeV did not allow the beams to be cogged at STAR and PHENIX simultaneously. However, data were recorded and this cogging issue will be resolved for the BES by an appropriate selection of collision energies. The data taken during this short run are used below to demonstrate that  the event rate projections for the BES are correct, and that STAR's triggering, and event and vertex reconstruction capabilities, designed for top energies, are applicable at these low \sqrts.

~

\noindent During this test beam an event rate at STAR of $\sim$ 1 Hz was observed. This is a much lower rate than that expected during the actual energy scan due to a number of factors. The two major ones were: a) since this was a test run only 56 bunches, out of a maximum of 120, were used and b) the intensity of each bunch was less than the maximum possible by a factor 3-6. A further gain in measured luminosity could be obtained by running in continuous injection mode. All these factors combined suggest that during the actual energy scan  an event rate of $>$5 Hz can be expected.

~

\begin{figure}[htb]
\begin{minipage}{0.46\linewidth}
\begin{center}
\includegraphics[width=\linewidth]{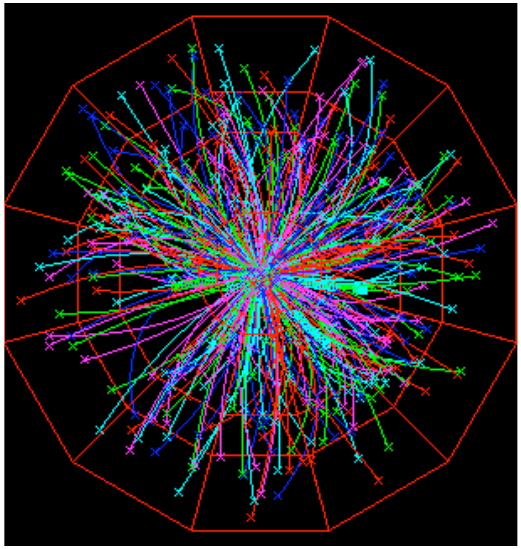}
\end{center}
\end{minipage}
\begin{minipage}{0.46\linewidth}
\begin{center}
\includegraphics[width=\linewidth]{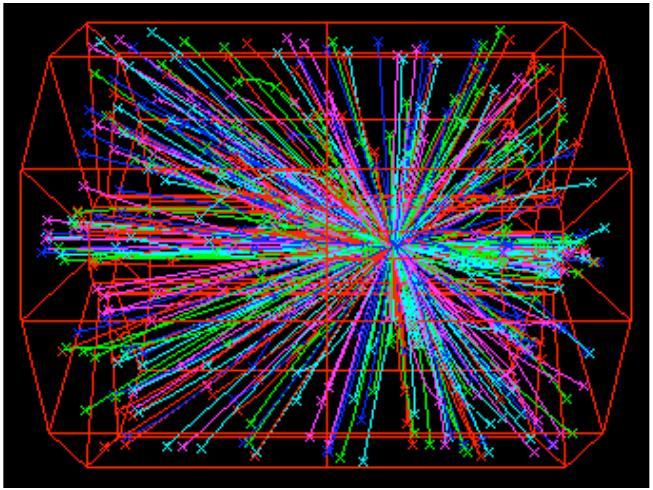}
\end{center}
\end{minipage}
\caption{A more central Au+Au event at \sqrts= 9.2 GeV, shown end on and from the side. The red frame is the TPC.}
\label{Fig:AuAu91}
\end{figure}

\begin{figure}[htb]
\begin{minipage}{0.46\linewidth}
\begin{center}
\includegraphics[width=\linewidth]{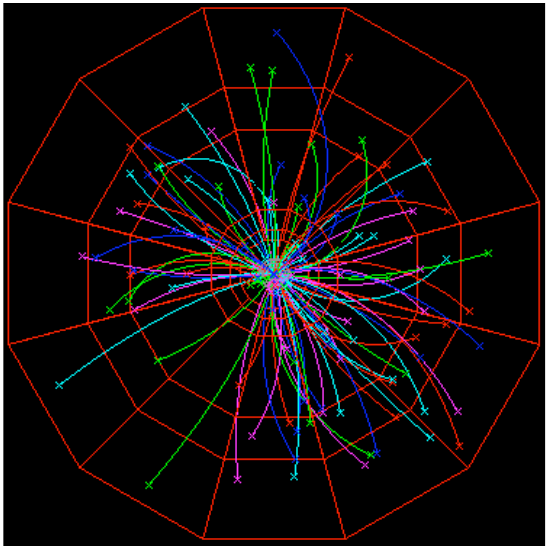}
\end{center}
\end{minipage}
\begin{minipage}{0.46\linewidth}
\begin{center}
\includegraphics[width=\linewidth]{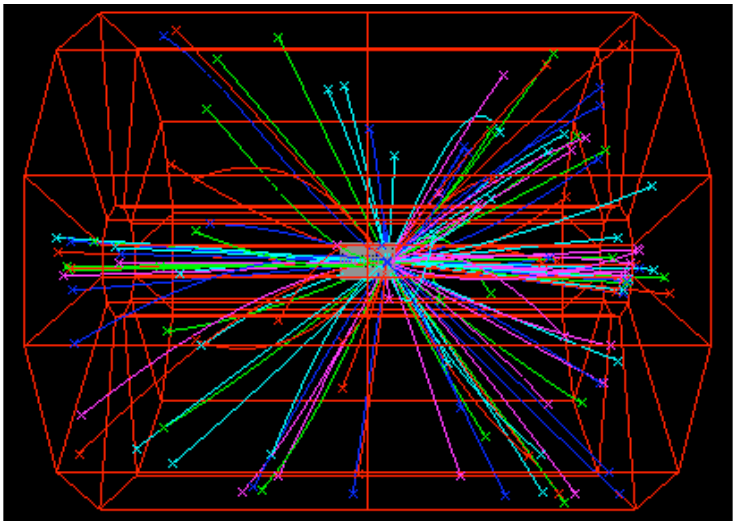}
\end{center}
\end{minipage}
\caption{A more peripheral Au+Au event at \sqrts= 9.2 GeV, shown end on and from the side. The red frame is the TPC.}
\label{Fig:AuAu92}
\end{figure}

\noindent   Two \sqrts = 9.2 GeV Au+Au events are shown in side and end view in Fig.~\ref{Fig:AuAu91} and Fig.~\ref{Fig:AuAu92}. The collision vertex is clearly defined and the mean $\eta$ of the charged tracks is approximately zero in both cases, strongly indicating that these are beam-beam collisions and not beam-gas or beam-beampipe.

~

\newpage

\newpage

\begin{figure}[htb]
\begin{minipage}{0.46\linewidth}
\begin{center}
\includegraphics[width=0.8\linewidth]{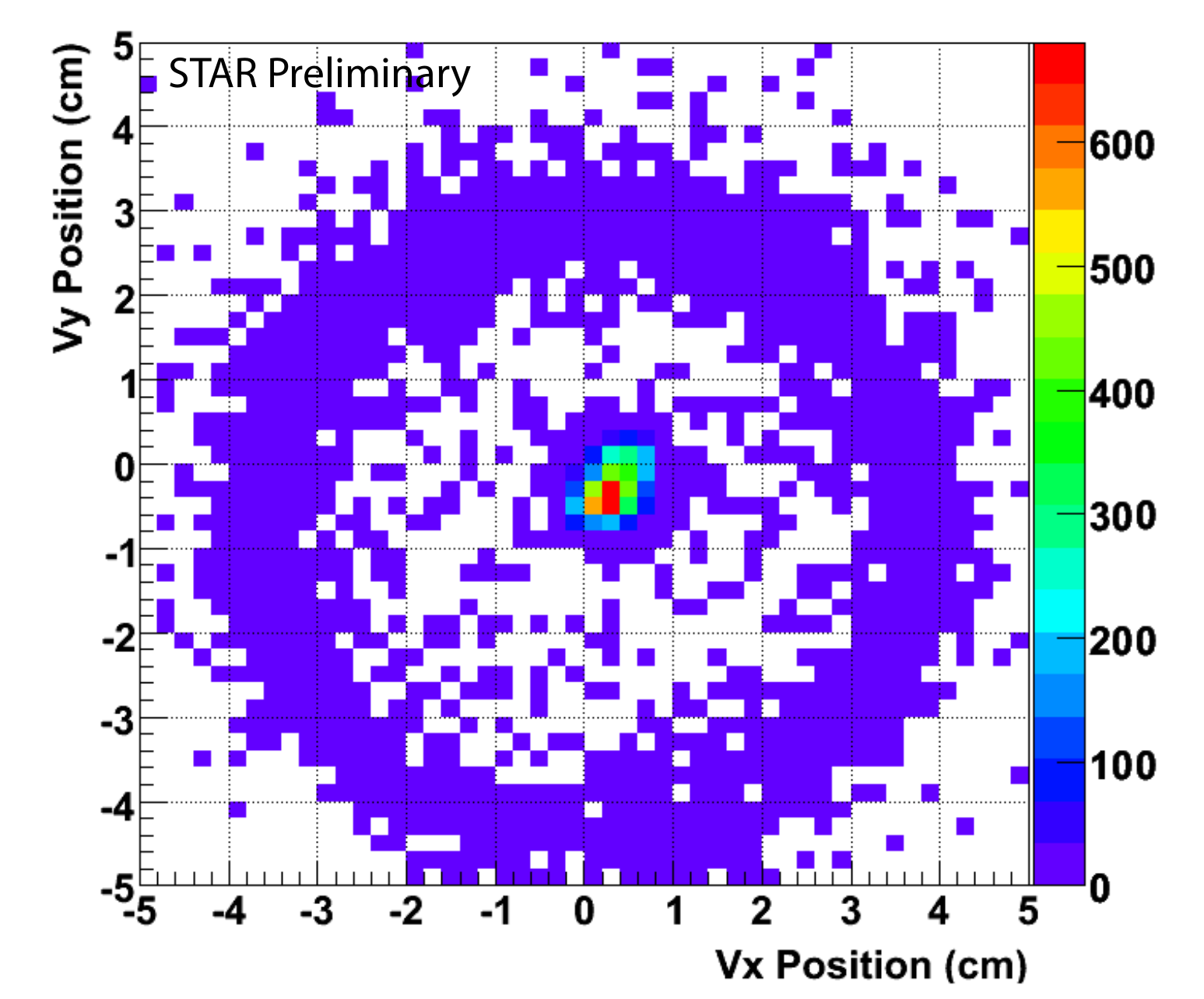}
\end{center}
\end{minipage}
\begin{minipage}{0.46\linewidth}
\begin{center}
\includegraphics[width=0.8\linewidth]{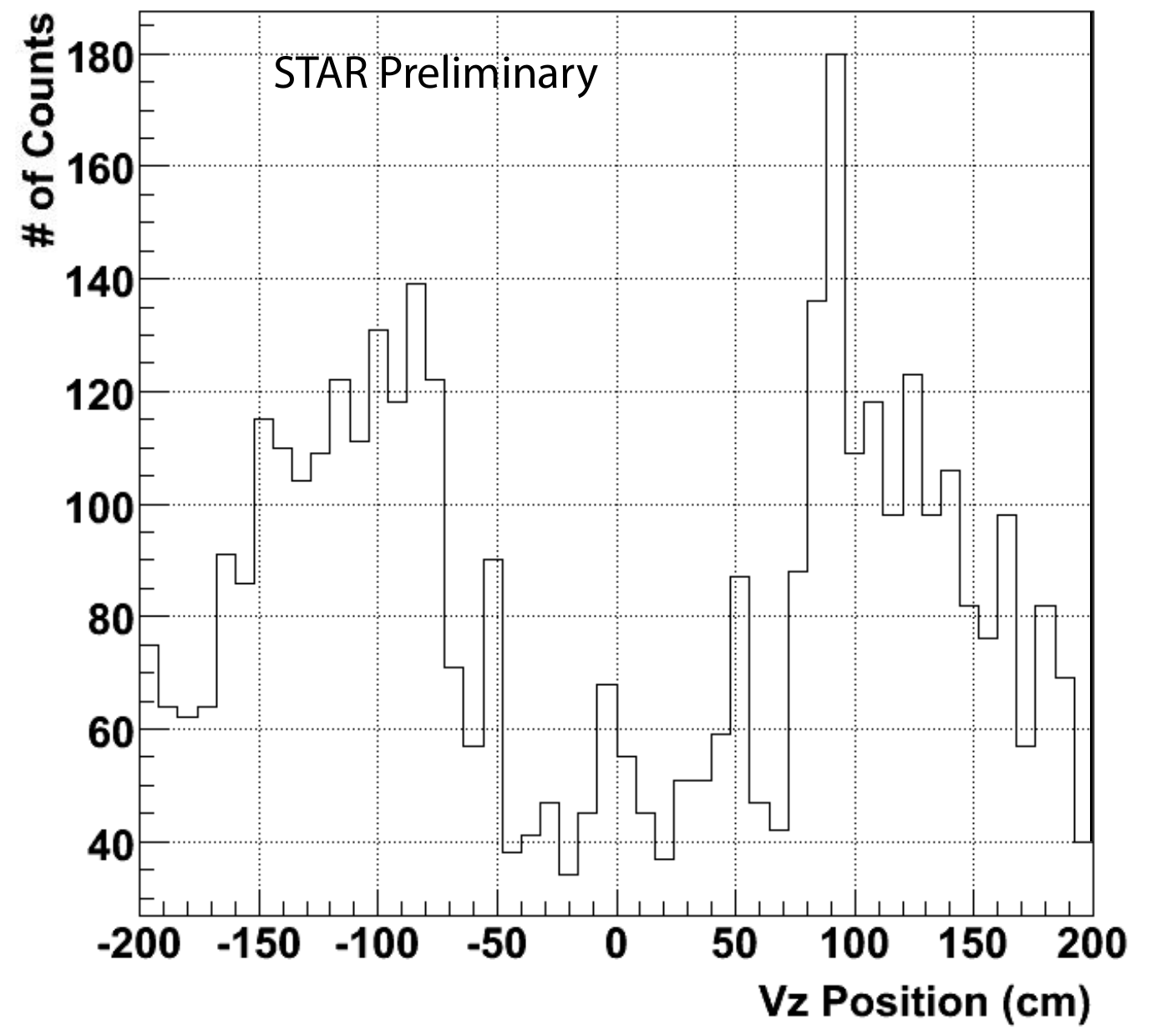}
\end{center}
\end{minipage}
\caption{Left: The radial position of the reconstructed primary vertex in the  Au+Au \sqrts= 9.2 GeV test run. Right: The z position of the primary vertex for those events with a radial primary vertex position indicating they occurred in the beam pipe, R $>$ 2 cm.  The beampipe is constructed from beryllium for $|z| < $50 cm and aluminum outside of this range. }
\label{Fig:beambeampipe}
\end{figure}

\noindent Many of the events recorded were beam-beampipe collisions due to the large diameter of the 9.2 GeV beam. These events, Fig.~\ref{Fig:beambeampipe},  are easily rejected off-line via cuts on the radial position of the primary vertex. In the left plot the beam spot due to beam-beam collisions is clearly seen at x $\sim$ 0.5 cm  and y $\sim$ -0.5 cm. The beampipe, radius 5 cm, is also illuminated due to beam-beampipe events. The right hand plot of Fig.~\ref{Fig:beambeampipe} shows the location in z of the beam-beampipe collisions. STAR's beampipe is made of beryllium in the $|z|< $ 50 cm region and changes to aluminum in the  $|z| >$ 50 cm,  outside of the nominal collision diamond. The increase in the number of events occurring in the large z range is due to the higher cross-section of Au on aluminum. Since the event rate will be low it has been decided that no attempt will be made to reject these online.

\section{The p+p baseline}

\noindent For certain measurements in the A+A energy scan (\sqrts$=5-60$ GeV) it is important to
know well the baseline NN cross section and invariant spectra to use as references. Previous relevant measurements come from many different accelerators, AGS, PS, SPS, FNAL and ISR spanning 40 years of
experiment.  The quality of these measurements are in general poor, particularly at higher \pT due to limited statistics, but this may be all that we will have access to, since performing these measurements at RHIC is not feasible in the
first years of BES program. In addition, agreement between different experiments is not perfect in
several cases with differences up to $50\%$ at high \pT. A number of articles have discussed the \pp systematics from available data in particular for high-\pT\ particles in regard to the RHIC data. Most data are for pions, both charged and neutral. Data are much sparser for other hadrons. 

~

\noindent For fixed target energies relative to \sqrts~ we tabulate the relation between the beam momentum and the \sqrts~ for a number of energies where experiments have been performed in fixed target experiments, Table~\ref{Table:ptos}.  In the following sections we discuss briefly relevant pion data from \pp  collisions at lower energies, particularly related to the higher \pT ranges.

\begin{table}[h]
  \centering
  \begin{tabular}{|c|c|c|}
\hline
$P_{lab}$ (GeV) &$\sqrt{s}$ (GeV) &Facility\\\hline
12& 4.55&CERN\\\hline
24& 6.83&CERN\\\hline   
158& 17.3&SPS\\\hline
200& 19.32&CERN,FNAL\\\hline
400& 27.36&CERN,FNAL\\\hline
800& 38.8&FNAL\\\hline
  \end{tabular}
  \caption{Relation between P$_{lab}$ and c.m. energy in \pp  collisions, where\ pp data exist.}
  \label{Table:ptos}
\end{table}

\subsection{Energies below top SPS}\label{SubSec:ppSPS}

\noindent The very low energies were studies at AGS and CERN in the 1970's. In particular
data were obtained at 12 and 24 GeV incident energy. Production at these low energies has no hard components, but arises essentially from various resonance productions resulting in an approximate $m_T$-scaled
spectra up to the kinematic limit. This same feature is seen in A+A collisions, see e.g. Ref.\cite{Cole:1991pr} with preliminary data from Si+A at 14.6 A~GeV. The main \pp data comes from Ref.\cite{Blobel:1973jc} and for higher
\pT data from Ref.\cite{Beier:1978ux}. In the energy range of 24-158 GeV fixed target there is really no available data and, as will be discussed, extrapolating between the measured data may be difficult and imprecise.

\subsection{SPS, FNAL and ISR  energies}\label{SubSubSec:ppISR}

\noindent A fairly extensive data-set for pion production from NA49~\cite{Alt:2005zq} covers a large $x_F$~range.  The data at mid-rapidity extend to \pT of 2.1 GeV/c. A summary of data around \sqrts = 22 GeV was made by Arleo and d'Enterria~\cite{Arleo:2008zd} in order to parametrize the cross section for use in the comparison to RHIC data at \sqrts = 22.4 GeV (Cu+Cu). They made small energy dependent corrections based on pQCD to the
available data between 21 and 23 GeV and gave a parameterization that will be used below. The parameterization is given by
$$f(p_T) = p_0.[1+(p_{T}/p_1)]^{p_2}*[1-x_T]^{p_3},$$
with $p_0=176.3$, $p_1$=2.38, $p_2$=-16.13,  $p_3=6.94$, and $\chi^2$/ndf = 208.2/190.

~

\noindent At FNAL high momentum \piplus and \piminus data from \pp collisions were obtained by the E605 experiment at fixed target energies of 400 and 800 GeV with significant integrated luminosity, 436 $pb^{-1}$ and 615 $pb^{-1}$, respectively~\cite{Jaffe:1989bb}. The
\sqrts=38.8 GeV data (800 GeV fixed) probes \pT out to 10 GeV/c, i.e. x$_T$ of $\sim$ 0.5. The errors
on the data are typically $10\%$ statistical and a similar value for the systematic uncertainty. This data-set is fairly consistent with a measurement by the earlier data from the Chicago-Princeton collaboration~\cite{Antreasyan:1978cw}, albeit there are differences in the order of 30$\%$. The data at \sqrts=27.4 GeV has smaller \pT coverage. 

~

\noindent The ISR \pinull~data has often been summarized  emphasizing the $x_T$-scaling properties, as has the fact that  while the within experiment energy dependence (like CCOR) are well established,  overall 30$\%$
disagreements on cross sections are quite common (see e.g. Ref.~\cite{Tannenbaum:2005by}). This kind of disagreement also exists
between the  ISR and FNAL experiments, even though the general trends are in good agreement.

\begin{figure}[htb]
\begin{minipage}{0.47\linewidth}
\begin{center}
\includegraphics[width=\linewidth]{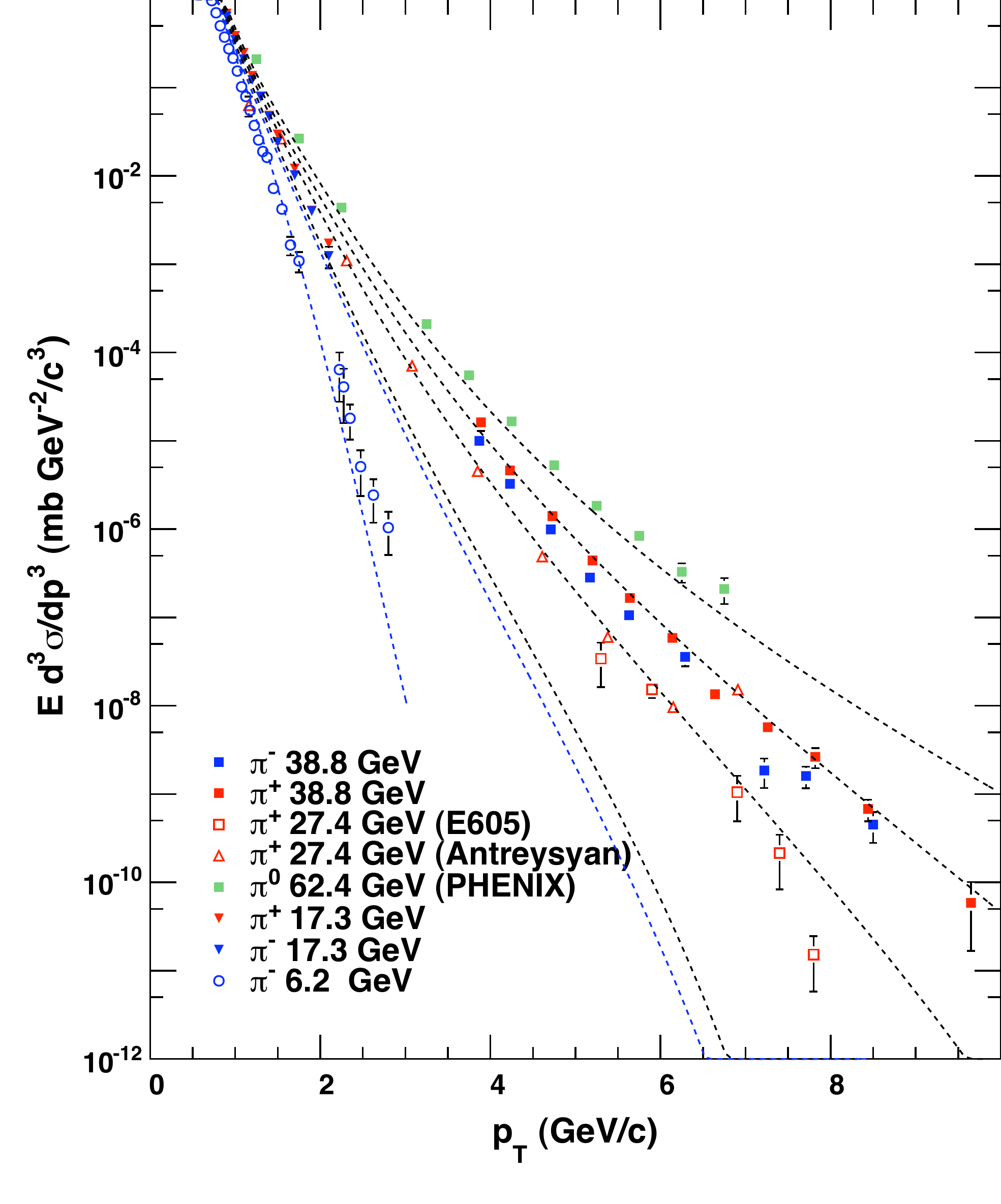}
\caption{Invariant cross-sections for \piplus, \piminus, and \pinull  for various c.m. energies as indicated in the legend. The data are from reference given in the text.}
\label{Fig:pion_pt}
\end{center}
\end{minipage}
\hspace{1cm}
\begin{minipage}{0.46\linewidth}
\begin{center}
\includegraphics[width=\linewidth]{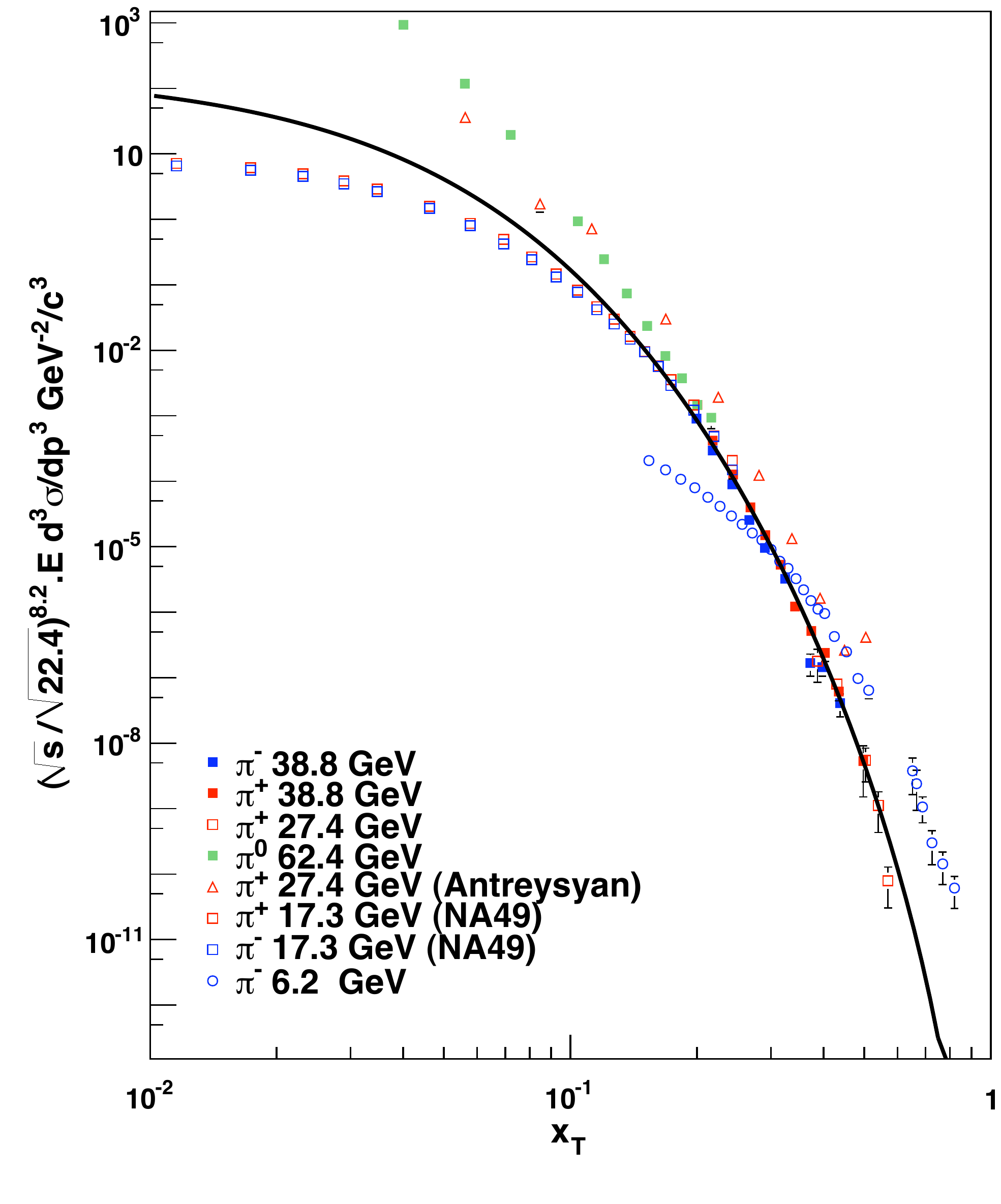} 
\end{center}
\caption{Scaled invariant cross sections for several energies. The full drawn curve is the best fit to the 22.4  GeV data.}
\label{Fig:pion_xt}
\end{minipage}
\end{figure}

\subsection{RHIC 62.4 GeV}\label{SubSec:pp62}

\noindent After the first RHIC A+A run at 62.4 GeV (Run 4) a compilation of p+p
data for that energy was made by d'Enterria  \cite{d'Enterria:2004rp} and used for the first $\raa$ results. The different measurements from the ISR show significant differences from the mean cross sections  (vs \pT) so a dedicated p+p run at 62.4 GeV  at RHIC
was warranted and carried out in 2006.
The ISR data agree reasonably well with data from PHENIX~\cite{Adare:2008qb} and unpublished BRAHMS data on \piplus\ and \piminus. It noted that PHENIX changed the energy scale
of the data from the CCOR experiment~\cite{Angelis:1978ec}.

\subsection{Energy dependence of high-\pT cross sections}\label{SubSec:ppHighpT}

\noindent In Fig.~\ref{Fig:pion_pt} we show some selected data for pion
production from AGS to ISR energies plotted vs. \pT. The typical decrease of the hard scattering cross section as the energy is lowered is observed. Ref.~\cite{Beier:1978ux} presents an overall function that describes
the available data as of 1978.
It gives a good overall description of data from AGS to ISR energies. 
We refer to the reference article for the functional form, which is
quite complicated. Since its publication additional data has been reported and for the
specific formulae to be used as a reliable interpolation between energies  \sqrts$=5-30$ GeV it will
have to be revisited and refit. However, it is  expected to be well suited for a first order estimate of cross section in this energy regime.

~

\noindent In Fig.~\ref{Fig:pion_xt} we present the data from Fig.~\ref{Fig:pion_pt} in terms of the variable $x_T$ as has been customary.  STAR and PHENIX demonstrated in ~\cite{Adams:2006nd,Adare:2008qb} that the 62 and 200 GeV data falls on a common curve at large $x_T$ when the cross sections are scaled by \sqrts$^{n}$ with n=6.5 $\pm$ 08 and 6.38 respectively. At the lower ISR and SPS energies an increased n factor is needed to describe the data well.  In Fig.~\ref{Fig:pion_xt} the cross sections have been scaled by $\sqrt{22.4}^{n}$ with n=8.2 as indicated by the CCOR data~\cite{Angelis:1978ec}. The very low energy data do not follow this trend, but correspond to an even higher factor of n indicative of non-perturbative effects dominating at lower energies. A value of n$\approx 11$ would allow scaling at high values of $x_t  >  0.4$ from the two energies. We conclude that for $x_t > 0.15$ good scaling with energy is observed; thus the scaling function from Ref.\cite{Arleo:2008zd} (solid curve in Fig.~\ref{Fig:pion_xt}) is a usable description. Of course it cannot be used at lower $x_T$.

~

\noindent In addition the NA61 (SHINE) experiment is approved to run at the CERN SPS~\cite{Gazdzicki:2008kk}. In 2009 they collected 
between $2-6 \times 10^6$ \pp events at \sqrts = 6.4, 7.8, 9.0, 12.7 and 17.8 GeV. These new data should supplement those currently available, providing high statistics data at very similar collision energies to those planned for the RHIC BES.

~

\noindent In conclusion existing \pp data and parameterizations for \sqrts $>17$ GeV gives an adequate, albeit not perfect, description. At lower energies the higher \pT cross sections are not well known and subject to large interpolation errors. This may not be a big concern since at RHIC with the low luminosities at the low energies of the first BES this momentum range will not be accessible.

\section{The STAR detector}

\begin{figure}[htb]
\begin{center}
\includegraphics[width=0.5\linewidth]{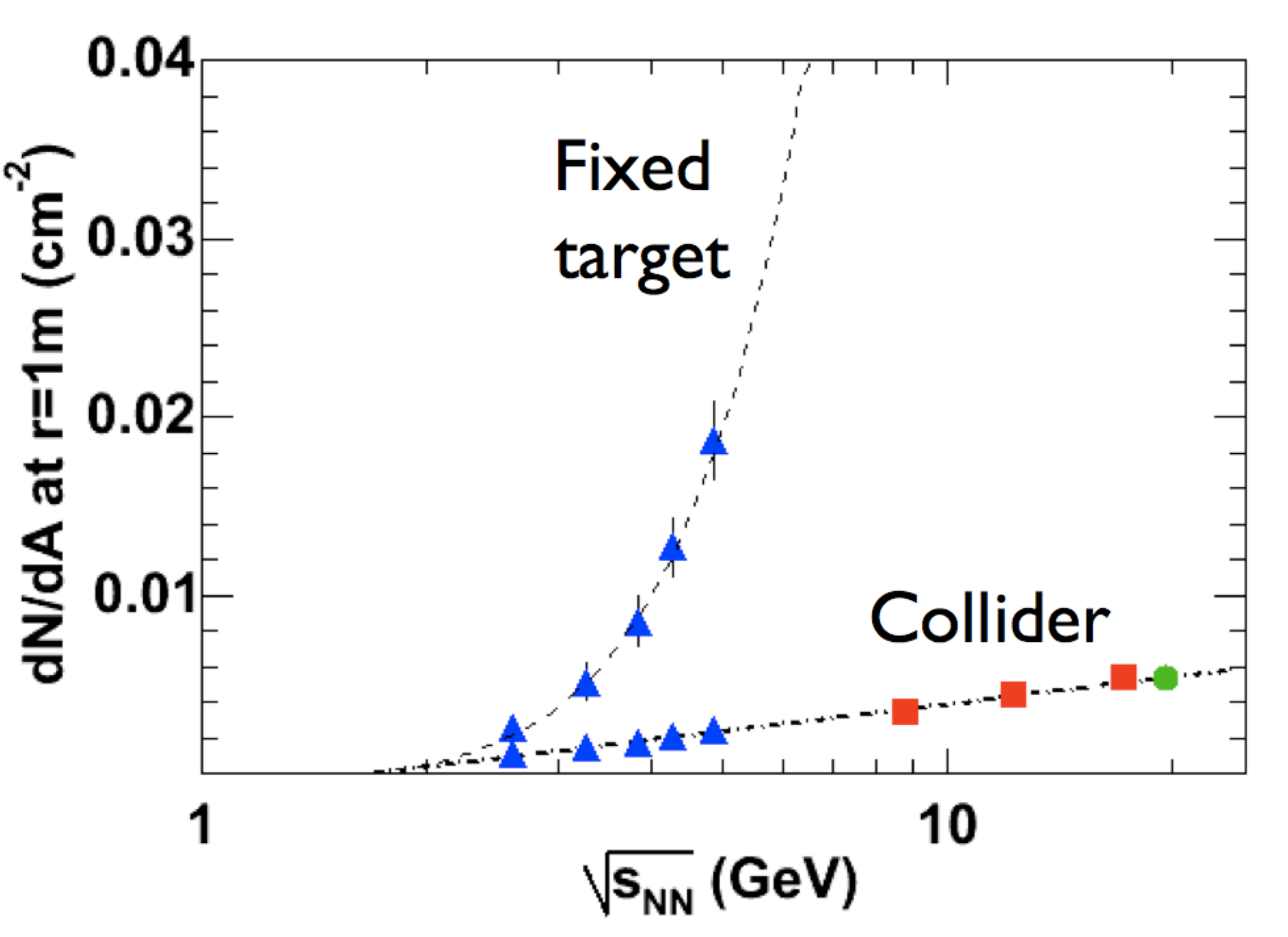}
\caption{The mid-rapidity track density at 1 m radius  in a collider setup compared to that of a fixed target as a function of \sqrts. Figure from \cite{Gunter:Accept}.}
\label{Fig:DetectOccup}
\end{center}
\end{figure}

\begin{figure}[htbp]
\begin{center}
\includegraphics[width=0.7\linewidth]{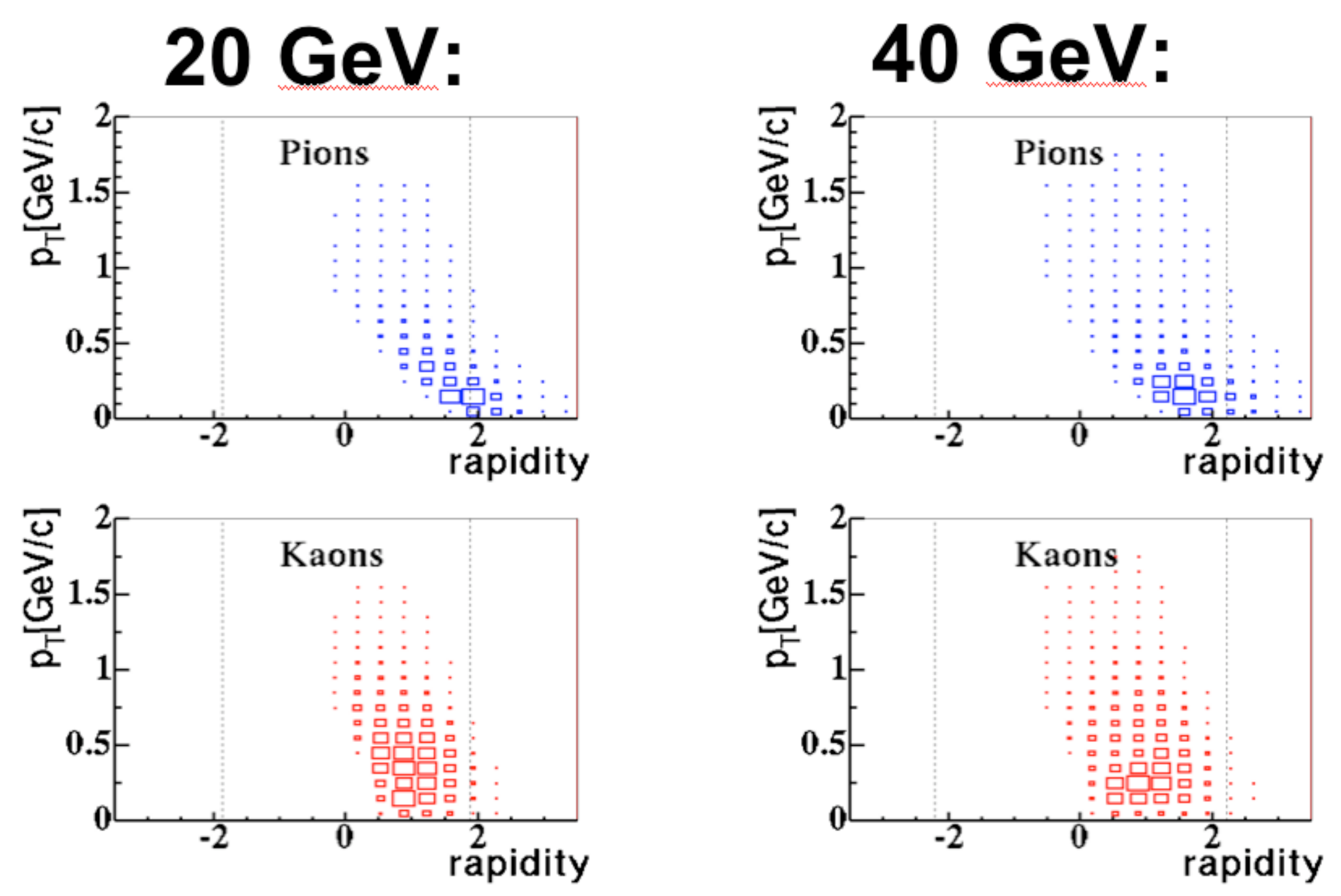}
\caption{The acceptance for $\pi$ and K used in the fluctuation analyses as a function of  \pT and rapidity for NA49 at SPS fixed target collision energies of 20 and 40 GeV. Figure from \cite{Gunter:Accept}. The acceptance is greater for the inclusive spectra analyses.  }
\label{Fig:AcceptFixed}
\end{center}
\end{figure}

\begin{figure}[htb]
\begin{minipage}{0.46\linewidth}
\begin{center}
\includegraphics[width=0.9\linewidth]{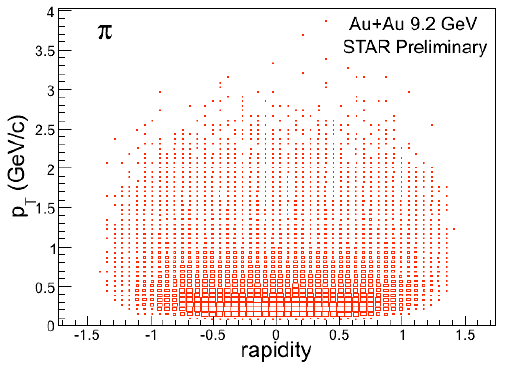}
\caption{ The acceptance for $\pi$ as  a function of  \pT and rapidity for  STAR for \sqrts = 9.2 GeV. }
\label{Fig:AcceptCollidPi}
\end{center}
\end{minipage}
\hspace{1cm}
\begin{minipage}{0.46\linewidth}
\begin{center}
\includegraphics[width=0.9\linewidth]{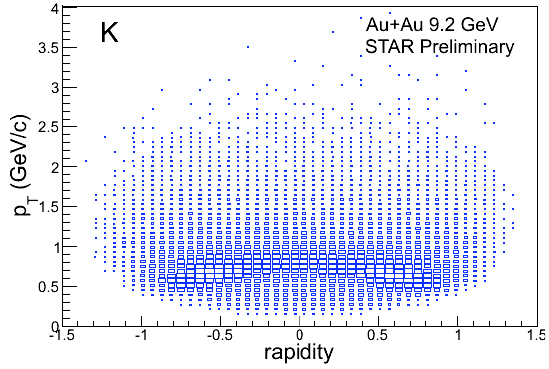}
\end{center}
\caption{ The acceptance for K as a function of  \pT and rapidity for STAR for  \sqrts = 9.2 GeV. }
\label{Fig:AcceptCollidKaon}
\end{minipage}
\end{figure}

\noindent The STAR collider geometry immediately provides advantages in comparison to the SPS fixed target experiments. These advantages are two fold~\cite{Gunter:Accept}. The first is that the detector occupancy  at mid-rapidity increases much faster as a function of \sqrts in a fixed target experiment than in a collider environment, as shown in Fig.~\ref{Fig:DetectOccup}. The second is that the particle acceptances  are dependent on the beam energy in fixed target experiments, e.g. Fig.~\ref{Fig:AcceptFixed}, whereas they are constant at colliders and similar, at mid-rapidity, for all particle types, Fig.~\ref{Fig:AcceptCollidPi} and Fig.~\ref{Fig:AcceptCollidKaon}. The reduction in track density means that for a given collision energy a detector in a  collider environment has less problems with charge sharing of hits, and track merging resulting in a better reconstruction efficiency and cleaner particle identification. The constant acceptance also means that there is better control of the systematics of the measurements and that a number of the uncertainties  cancel when comparisons at different \sqrts are made. This is especially important for the energy scan as the identification of the CP is  likely to be made by observing how various measures alter as a function of \sqrts.

\subsection{STAR's Subsystems}\label{SubSec:STARSubsystems}

\begin{figure}[htb]
\begin{center}
\includegraphics[width=\linewidth]{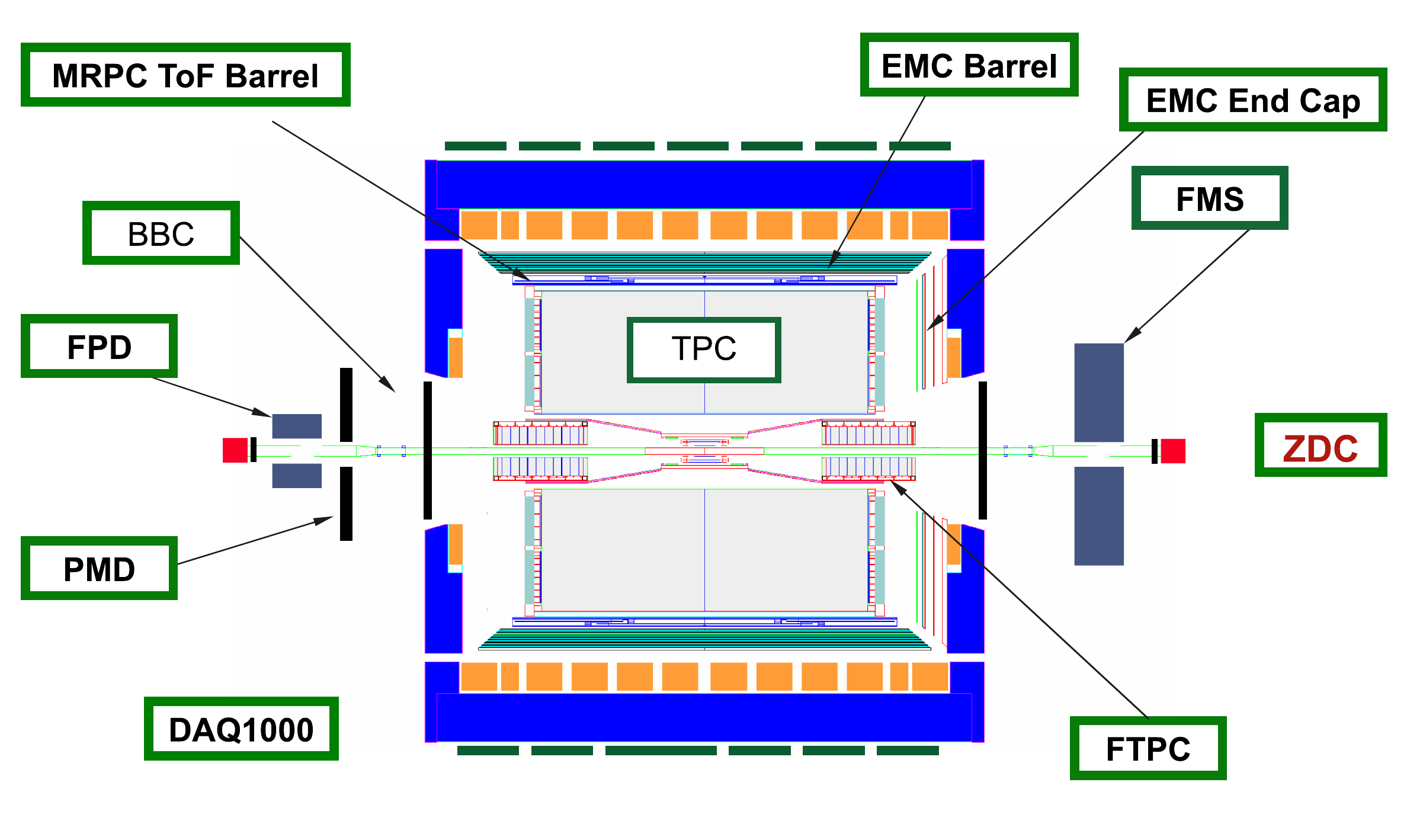}
\caption{Schematic of the STAR detector showing all the subsystems, including those currently being proposed and undergoing research and development studies.}
\label{Fig:STAR}
\end{center}
\end{figure}

\noindent This section briefly describes the various subsystems of STAR relevant to the BES, Fig.~\ref{Fig:STAR}. Of particular interest to the CP search is the installation of the full barrel ToF which was  completed during the 2009 summer shutdown. For a full description of STAR see \cite{Ackermann:2002ad}
and references therein.

~

 \subsubsection{Trigger Detectors}\label{SubSubSec:STARTrig}
 
\noindent We propose to use the Beam-Beam Counters (BBCs) for triggering during the low energy runs.   The collision rates at these low \sqrts are only a few Hz so all events can be recorded and no selective trigger is required. Therefore it is proposed to use only a minimum bias trigger, i.e. that of a  coincidence in the two BBCs.

~

\begin{figure}[htb]
\begin{center}
\includegraphics[width=0.3\linewidth]{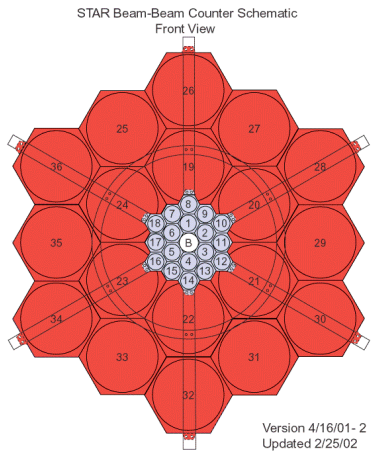}
\caption{A schematic diagram of one of the BBCs. }
\label{Fig:BBC}
\end{center}
\end{figure}

\noindent The BBCs, shown schematically in Fig.~\ref{Fig:BBC}, are  positioned at $\pm$ 3.5 m from the center of the TPC along the beam direction. The small inner tiles, shown in blue in Fig.~\ref{Fig:BBC},  cover  3.8 $<$ $|\eta|$ $<$ 5.2 and  can all be inscribed within a circle of 9.64 cm. The outer (red) tiles cover 2.2 $<|\eta|<$ 3.8.  The RHIC beam  passes through the center of the BBC (labeled B  in Fig.~\ref{Fig:BBC}. These annular scintillator detectors are sensitive to charged tracks down to the single minimum ionizing particle (MIP).  Even at the lowest collision energies the number of produced particles in heavy-ion collisions is larger than in \pp collisions at \sqrts= 200 GeV where they  perform well.  Table~\ref{Table:BBCs} shows estimations of the number of particles hitting the inner and outer sections of the BBCs as a function of the interaction's centrality for Au+Au collisions at \sqrts = 5 and 8.75 GeV.

\begin{table}[htb]
\caption{Estimation of number of charged particles in the BBCs in Au+Au collisions at \sqrts = 5 and 8.75 GeV for various impact parameters.} 
\begin{center}
\begin{tabular}{|c|c|c|c|c|}
\hline
Impact Parameter  &  \sqrts = 5 GeV & \sqrts = 5 GeV & \sqrts = 8.75 GeV & \sqrts= 8.75 GeV \\
(fm)& BBC Inner  & BBC Outer  & BBC Inner & BBC Outer  \\
\hline
0$<$b$<$3 	&	5	&	27	&	12	          & 54 \\
3$<$b$<$6 	&	11	&	30	&	21		& 57 \\
6$<$b$<$9	& 	22	&	35	&	39		& 40 \\
b$>$9	         &	44	&	30	&	66		& 8 \\
\hline
\end{tabular}
\end{center}
\label{Table:BBCs}
\end{table}

~

\noindent The BBC's will also be used to calculate the first order event plane, although with reduced resolution to that calculated from the FTPCs. More details are discussed below.

\subsubsection{Time Projection Chamber}\label{SubSubSec:STARTPC}

\noindent The main tracking detector in STAR remains the Time Projection Chamber (TPC). The active  dimensions of the TPC radially are from 50 cm to 200 cm and it is 400 cm in length. This gives the TPC full azimuthal coverage and a uniform acceptance within $|\eta|$ $<$ 1. It is placed within a uniform solenoidal magnetic field which allows for the determination of the momenta of charged tracks. When the magnet is set to its maximum value of 0.5 T  the transverse momentum resolution is $\Delta p_{T}/p_{T} \sim 0.012 p_{T}$ (\pT in GeV/c). 

 \subsubsection{Time of Flight}\label{SubSubSec:STARToF}

\noindent The Multi-Gap Resistive Plate Chamber Time of Flight (MRPC ToF)~\cite{Geurts:2004fn,Llope:2005yw} will completely surround the outer radius of the TPC, $|\eta|$$<$0.9, 0$<$$\phi$$<$2$\pi$. It consists of 23K channels from 120 modules. The first MRPC tray was installed in Run 3 and it was full installed for Run 10.  The ToF has an intrinsic stop time resolution of 80ps. After  folding in STAR's start signal a total resolution of at least 120 ps will be  achieved. This 120 ps total resolution is for the lowest \sqrts values and will decrease significantly for higher energies. This gives the ToF an excellent mass resolution, see for example Fig.~\ref{Fig:ToFPID}.
 
 \subsubsection{Forward Time Projection Chamber}\label{SubSubSec:STARFTPC}
 
\noindent The Forward Time Projection Chambers (FTPCs)  are two radial drift TPCs which cover the region 2.5$<|\eta|<$4.0. They have a  2-track resolution of 1Ð2 mm and a momentum resolution of 12-15$\%$. Unfortunately due to the small number of hit points, there are only 10  recorded per track,  in the FTPCs no PID is possible. It is planned to use the FTPCs as the main detectors used in the determination of the event plane, see later. 
 
 \subsubsection{Photon Multiplicity Detector}\label{SubSubSec:STARPMD}
 
\noindent The Photon Multiplicity Detector (PMD)~\cite{Aggarwal:2002bm} measures the inclusive number of photons (dominantly from $\pi^{0}$ decays) produced in the pseudorapidity regions of  -2.3 to -3.8, with full azimuthal coverage.  It also provides the spatial distribution of the photons in ($\eta,\phi$) phase space. 
 
 ~
 
 \noindent  The PMD therefore provides a unique opportunity of addressing  physics  at a higher baryon chemical potential ($\mu_{B}$)  compared to midrapidity measurements at a fixed beam energy. It also allows us to continue the studies of various scalings with N$_{part}$, y Ð y$_{beam}$, started by the now decommissioned  PHOBOS and BRAHMS experiments. In conjunction with the FTPCS (both of which are needed), it will provide a unique opportunity to look for a Disoriented Chiral Condensate and hence chiral phase transition. Measurements of v$_{1}$ and v$_{2}$ at forward rapidity for photons and neutral pions can be made.

\subsection{Particle Identification}\label{SubSubSec:STARPID}

\noindent With completion of the full barrel ToF STAR will have excellent mid-rapidity PID capabilities. Good quality track-by-track particle identification is necessary for our proposed fluctuation measurements.

~

\begin{figure}[htb]
\begin{minipage}{0.46\linewidth}
\begin{center}
\includegraphics[width=\linewidth]{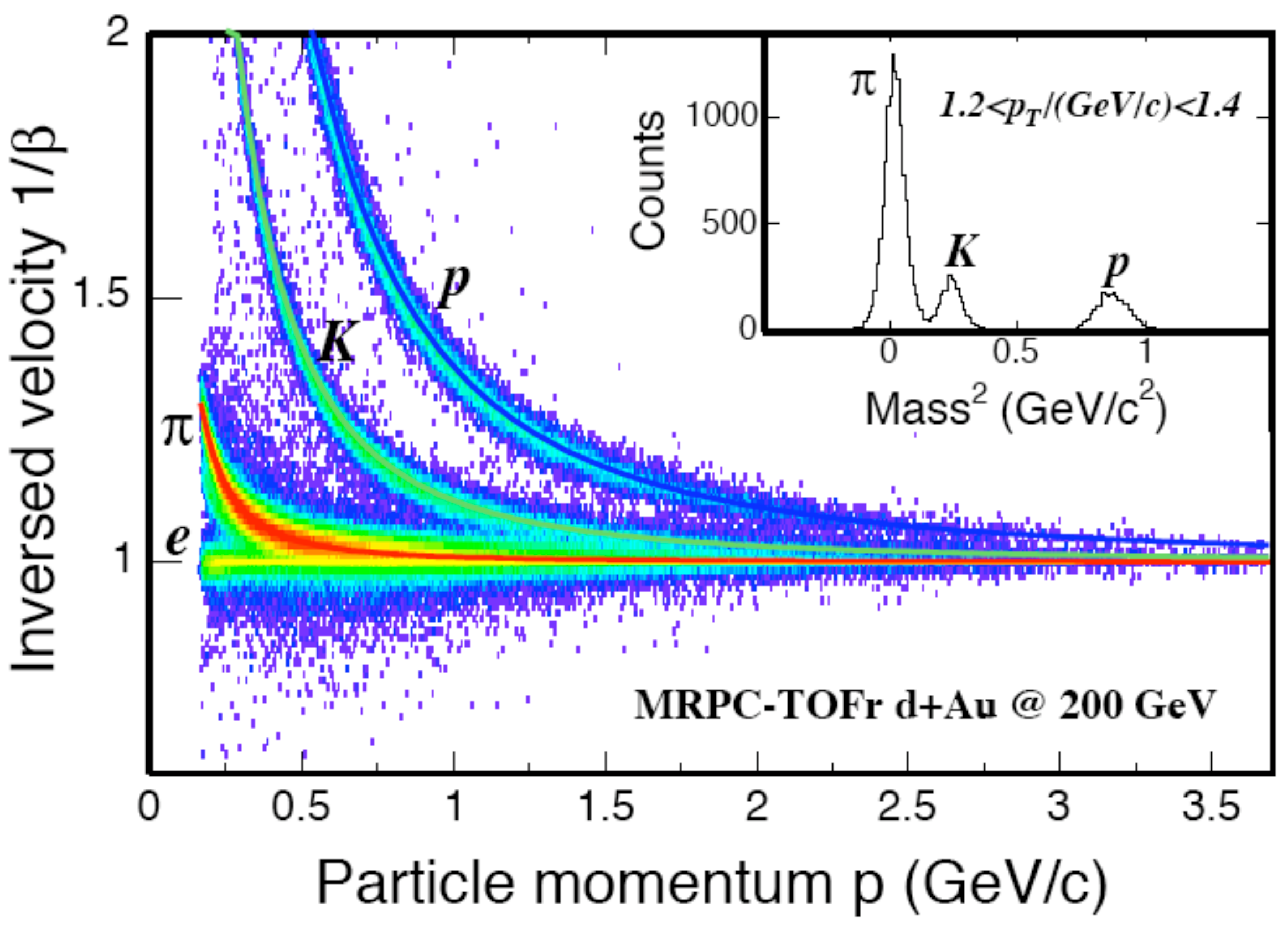}
\caption{Particle identification using the ToF in d-Au collisions at \sqrts= 200GeV. $\pi$, K, p, and electron are clearly identifiable. The inset shows the projection of the ToF mass$^{2}$ measurements in the 1.2 $< $\pT $<$ 1.4 GeV/c range.}
\label{Fig:ToFPID}
\end{center}
\end{minipage}
\hspace{1cm}
\begin{minipage}{0.46\linewidth}
\begin{center}
\includegraphics[width=\linewidth]{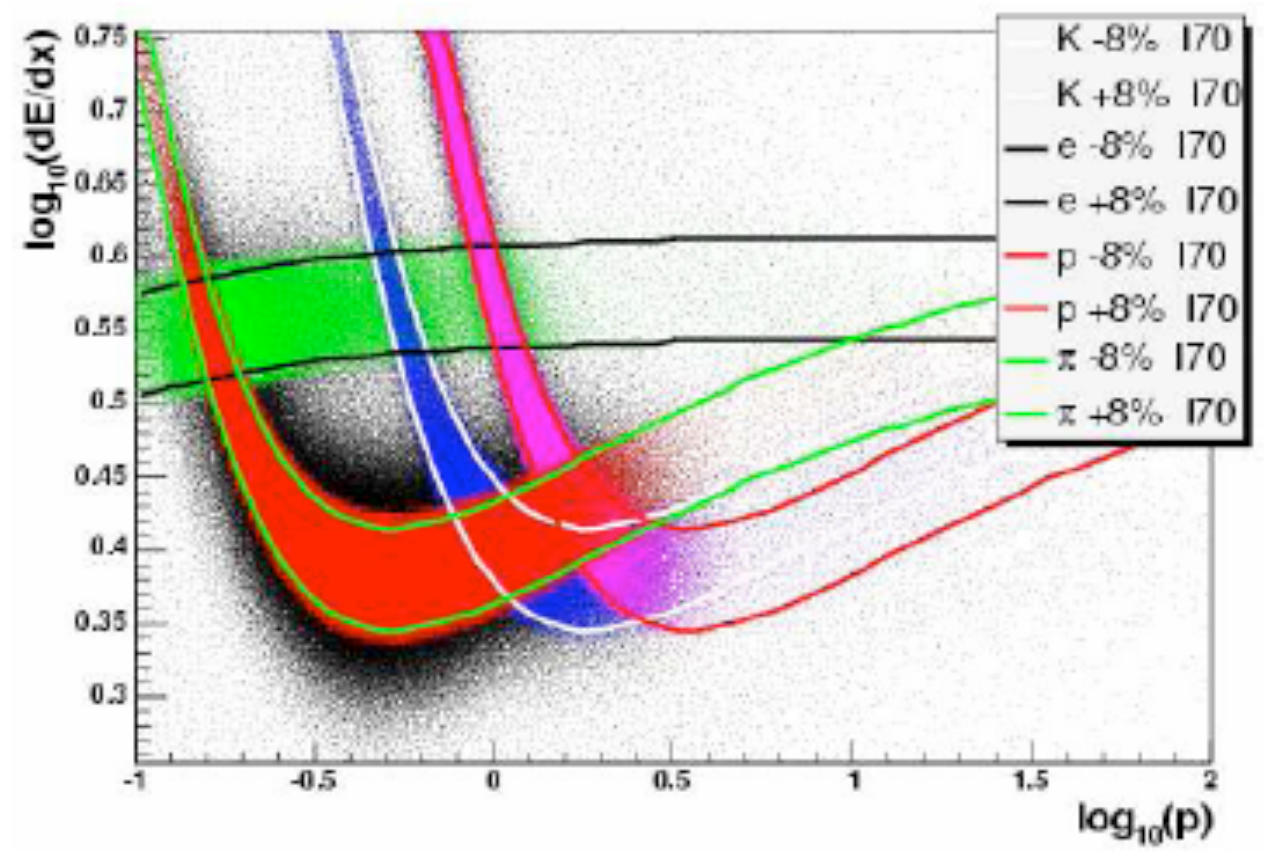}
\caption{dE/dx measurements using the TPC.  The curves represent  the $\pm$8$\%$ deviation from dE/dx calculations using the Bichsel parameterization for each particle.}
\label{Fig:TPCPID}
\end{center}
\end{minipage}
\end{figure}

\noindent  Charged particle identification will be performed via a combination of measured ionization in the TPC (dE/dx) and its ToF. The addition of the ToF allows for the identification of particles where their dE/dx  as a function of momentum merge. Using these two techniques STAR will be able to resolve K and $\pi$ up to momenta of 1.6 GeV/c and protons up to 3 GeV/c. Fig.~\ref{Fig:ToFPID} shows the measured 1/$\beta$  of the ToF in d-Au collisions at \sqrts= 200 GeV and Fig.~\ref{Fig:TPCPID} shows the measured dE/dx of tracks in the TPC,  the $\pi$, K, p and electron bands are clear. The colored curves indicated the $\pm$ 8$\%$ deviation from dE/dx calculations using the Bichsel parameterization~\cite{Bichsel:2006cs}. Using ionization measurements in the relativistic region of the TPC and the data from the ToF also allows the statistical identification of protons and $\pi$ up to 12 GeV/c~\cite{Shao:2005iu}. 

~

\noindent In addition STAR can use secondary vertexing to identify K$^{0}_{s}$, $\Lambda$, $\overline{\Lambda}$, $\Xi^{-}$, $\overline{\Xi^{-}}$, $\Omega^{-}$, and $\overline{\Omega^{-}}$ over the whole \pT range, and, via invariant mass mixed event techniques, a  variety of resonances such as the $\phi$ and $K^{*}$  and $\rho$ ( e.g. ~\cite{Wang:2009vma} and references therein).

\subsection{Event Plane Resolution}\label{SubSec:STAREventPlane}

\begin{figure}[htb]
\begin{minipage}{0.46\linewidth}
\begin{center}
\includegraphics[width=\linewidth]{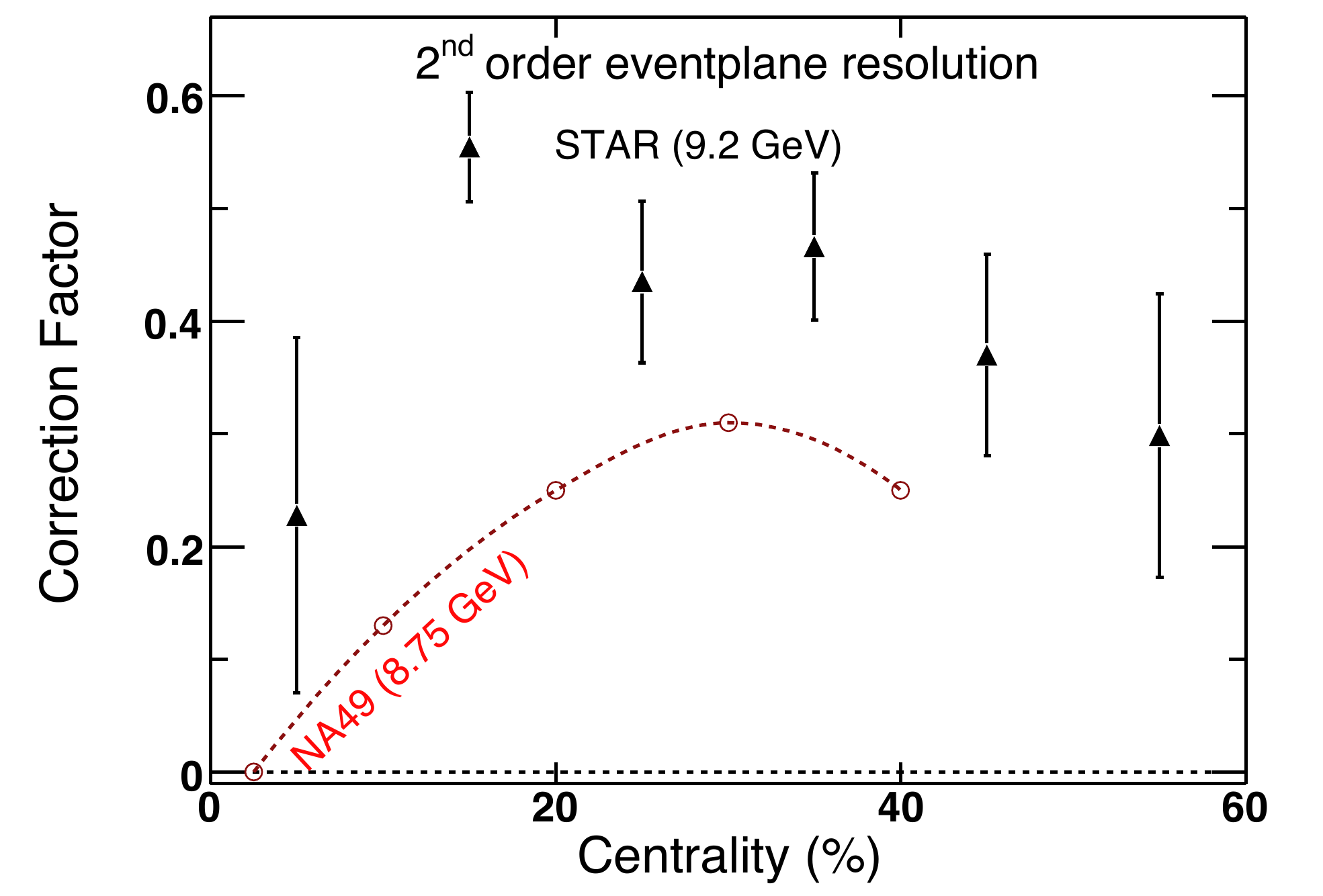}
\caption{Second order event plane resolutions for Au+Au events at \sqrts=9.2 GeV compared to that of NA49 at \sqrts=8.75 GeV~\cite{Alt:2003ab}.}
\label{Fig:EventPlane}
\end{center}
\end{minipage}
\hspace{1cm}
\begin{minipage}{0.46\linewidth}
\begin{center}
\includegraphics[width=\linewidth]{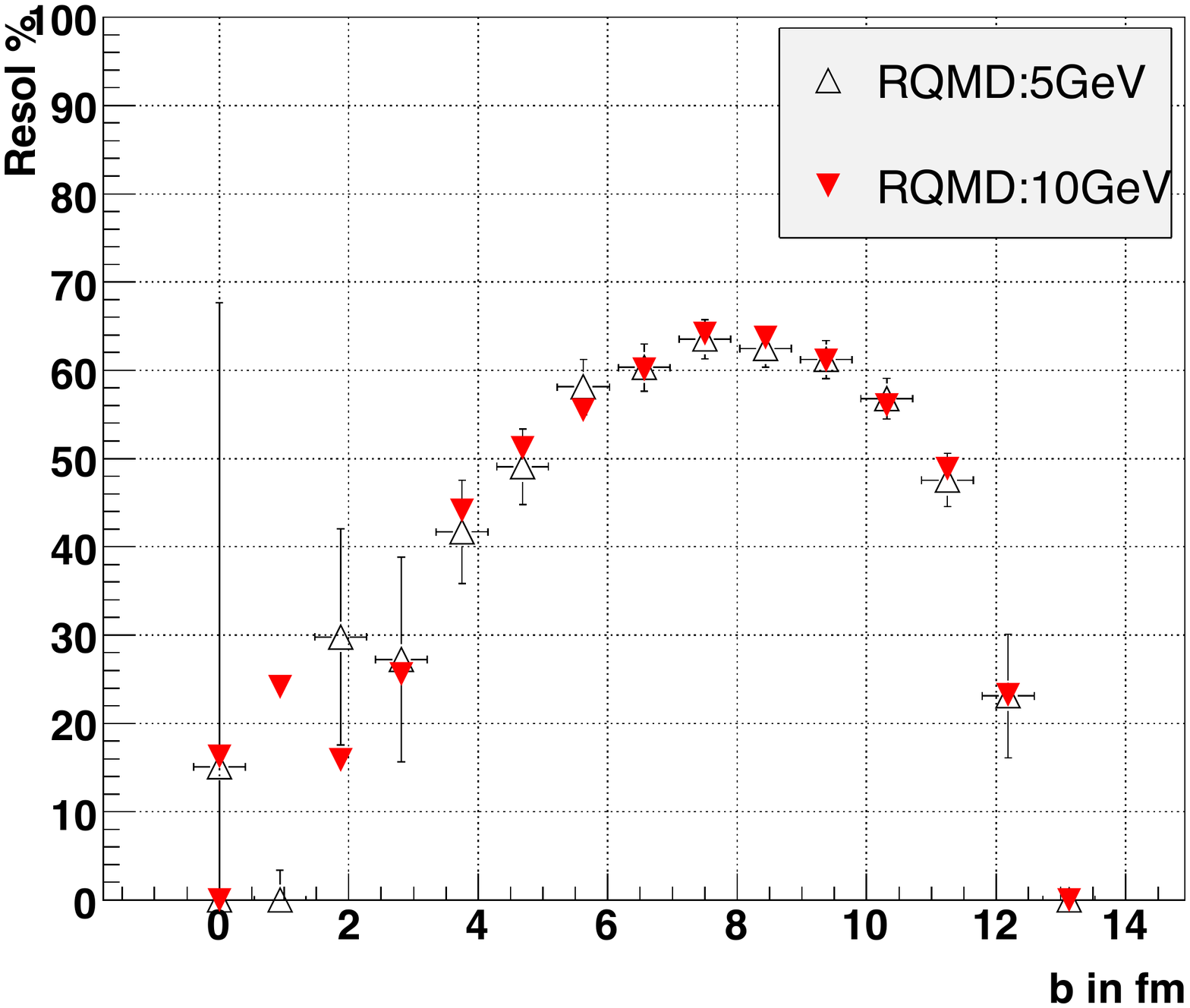}
\caption{Estimated first order event plane resolutions as a function of impact parameter from RQMD simulations of  Au+Au events at \sqrts= 5 and 10 GeV using the BBCs.  Similar results are obtained for all \sqrts. See text for details.}
\label{Fig:BBCEventPlane}
\end{center}
\end{minipage}
\end{figure}

\noindent STAR has excellent event plane resolution even at low \sqrts. Fig.~\ref{Fig:EventPlane} shows  the measured second order event plane resolution using the TPC as a function of centrality for Au+Au events at \sqrts= 9.2 GeV. Also shown for comparison is the event plane resolution for NA49 at \sqrts=8.75 GeV~\cite{Alt:2003ab}. It can be seen that  STAR's larger acceptance results in a much superior resolution.
 While the v$_{2}$ event plane can be calculated using the TPC, for v$_{1}$ measurements an independent determination of the event plane is necessary, this can be done by utilizing the FTPCs  or with slightly less resolution the BBCs. Calculating the event plane in this away also minimizes several auto-correlation effects.

~

\noindent The first order event plane resolution of the BBCs is shown in Fig.~\ref{Fig:BBCEventPlane} as a function of impact parameter for RQMD simulations of  Au+Au events at \sqrts= 5 and 10 GeV. For this simulation the BBC was fully instrumented. For the 2008 test run at \sqrts = 9.2 GeV only the inner tiles were present so the resolution was degraded, (for details of the BBC please see section~\ref{SubSubSec:STARTrig}). The resolution  for this dataset was estimated to be $\sim$ 25$\%$ which is in good agreement to the resolution obtained from  RQMD when the BBCs were simulated without the outer tiles. This confirms that the simulations are a good representation of the data. The resolution is similar for all collision energies discussed in this proposal.

\subsection{Acceptance, Efficiency, and Particle Production Rates.}\label{SubSubSec:STARRates}

\noindent In order to make realistic estimates of the number of events needed to perform the analyses described in the previous sections it is important to be able to estimate the acceptances and tracking efficiencies of the various particles as well as their expected yields. 

~

\begin{figure}[htb]
\begin{center}
\includegraphics[width=0.5\linewidth, angle=90]{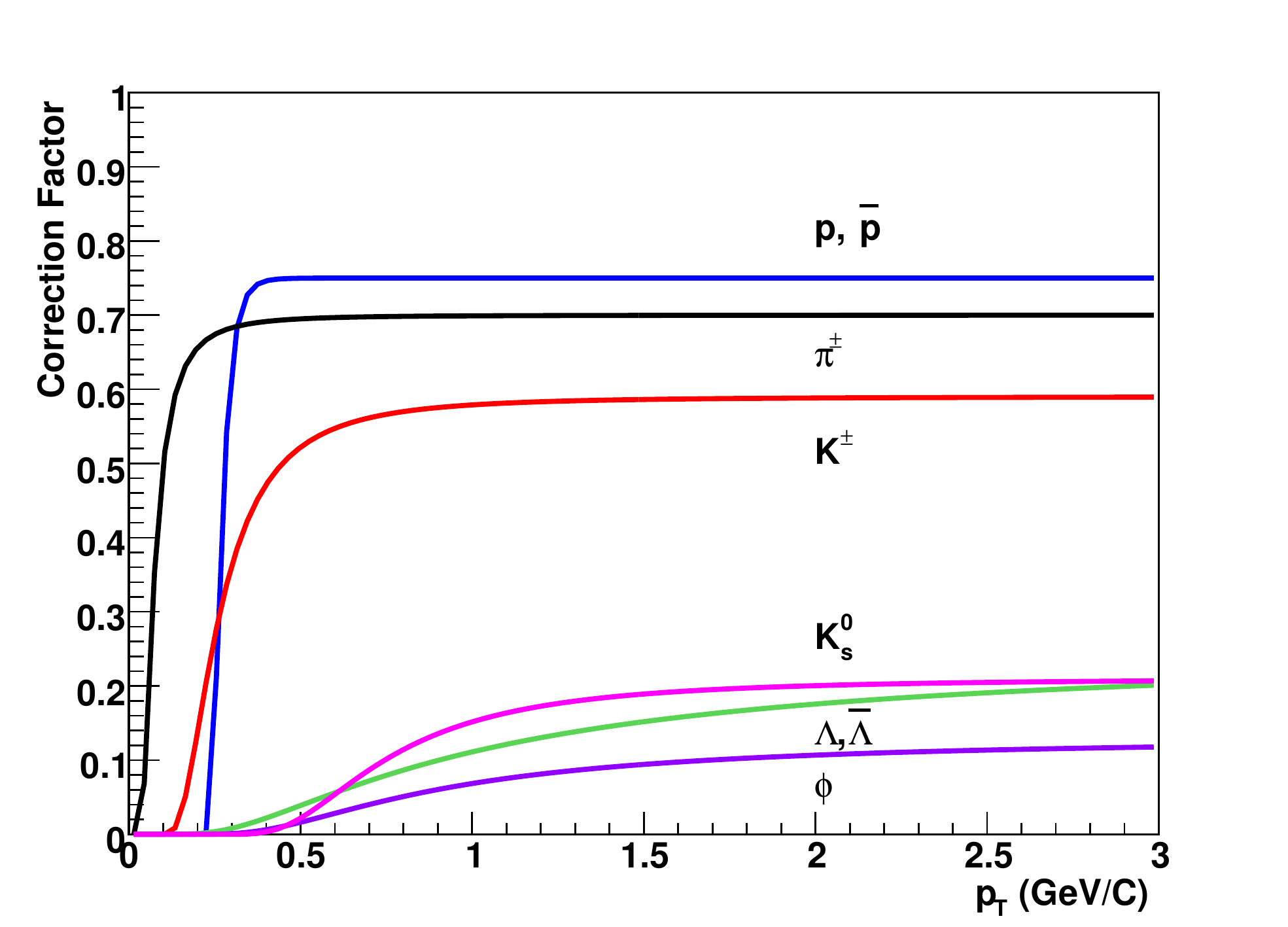}
\caption{The mid-rapidity correction factors as a function of \pT assumed for the BES. }
\label{Fig:EfficSTAR}
\end{center}
\end{figure}

\noindent The correction factors (acceptance*efficiency) used for $\pi$, K and protons are estimated from those  calculated for the \sqrts= 19.6 GeV test run.  For the $\phi$, $K^{0}_{s}$, $\Lambda$, and $\bar{\Lambda}$ results have not yet been obtained for the \sqrts= 19.6 GeV data.  Therefore the $\phi$ correction factors obtained for the most peripheral (70-80$\%$) Au+Au at \sqrts= 200 GeV data were used, and for the $K^{0}_{s}$, $\Lambda$, and $\bar{\Lambda}$ the Cu+Cu at \sqrts= 200 GeV values were used. For all particles that decay the branching ratio to  their charged daughters are included in the correction factors. All curves were fit to the parameterization

\begin{eqnarray*}
Eff = a \times exp[(-b/p_{T})^c]
\end{eqnarray*}

\noindent where a, b and c are free parameters. While there is a correction factor dependence on occupancy it is small over these energy ranges and therefore neglected in these approximations. The resulting correction factor curves  are shown in Fig~\ref{Fig:EfficSTAR}.  At the lowest \sqrts proposed the efficiencies may be slightly lower than those assumed here due to the expected slightly worse timing of the ToF, see section~\ref{SubSubSec:STARToF}.

~

\noindent  The estimated acceptances and efficiencies can be folded with the expected \pT spectra to determine the raw reconstructed particle yields as a function of \sqrts and centrality. To  estimate the rates and $\langle p_{T} \rangle$ at mid-rapidity for several \sqrts energies  and N$_{part}$ a statistical model can be used together with the following formulae, from Ref.~\cite{Cleymans:2005xv}. 

~

\noindent For a given \sqrts:

\begin{eqnarray}
\mu_{B} = \frac{1.308}{1+0.273 \sqrt{s_{NN}}}
\end{eqnarray}
\begin{eqnarray}
T =   0.166 - 0.39  \mu_{B}^{2} - 0.053 \mu_{B}^{4}     
\end{eqnarray}

\begin{figure}[htb]
\begin{center}
\includegraphics[width=0.5\linewidth]{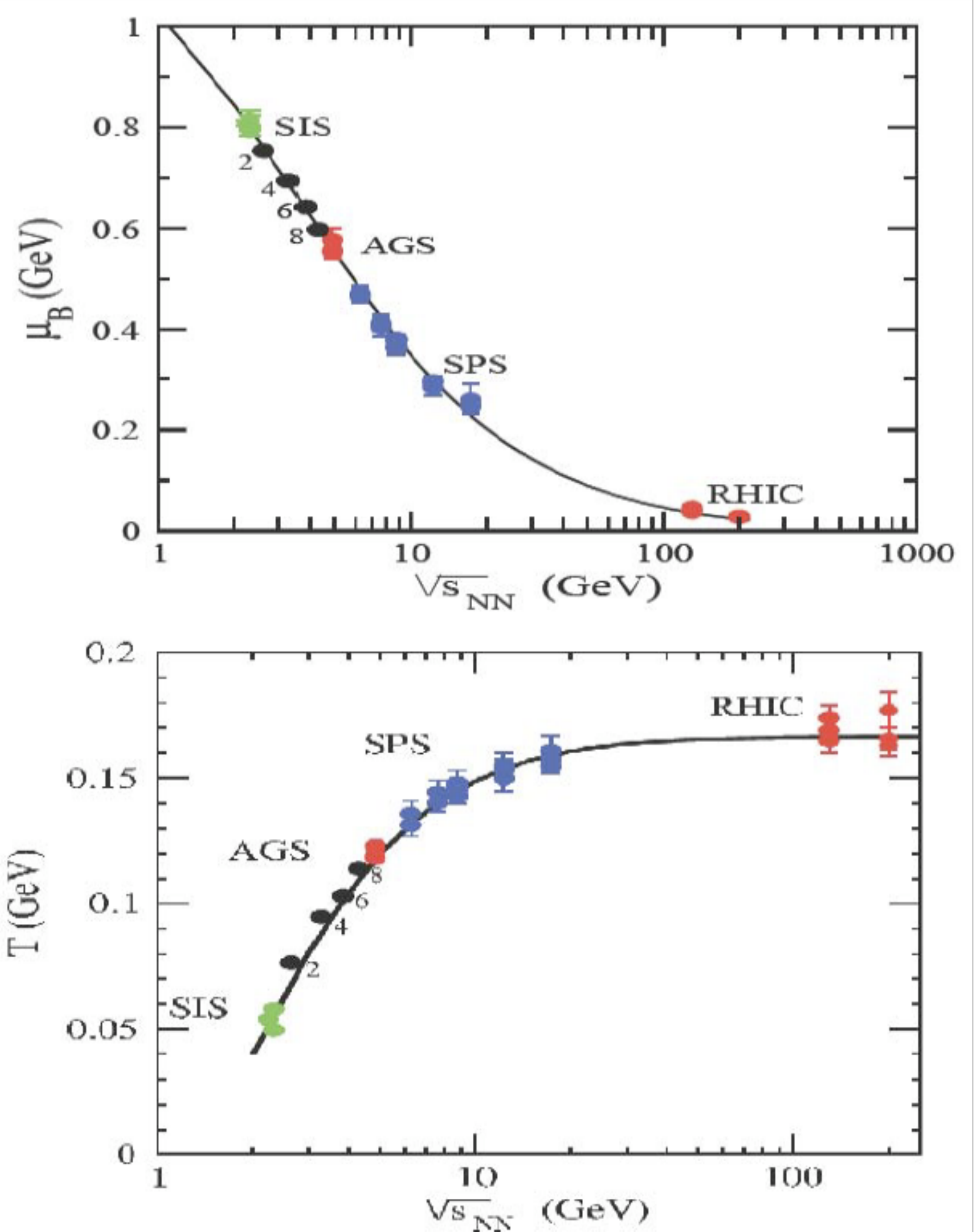}
\caption{The calculated chemical freeze-out temperature, T, and chemical potential, $\mu_{B}$,  as functions of \sqrts.  The curves are the phenomenological parameterizations  used in our estimates of the particles yields. Figures from \cite{Cleymans:2005xv}.}
\label{Fig:TMuB}
\end{center}
\end{figure}

\noindent This gives a very good phenomenological parameterization of the current data as shown in Fig.~\ref{Fig:TMuB}~\cite{Cleymans:2005xv}. Having calculated T and $\mu_{B}$ the statistical model can be  used to calculate each particle/N$_{ch}$  ratio.  N$_{ch}$ is then calculated via a phenomenological parameterization of Glauber model calculations of the number of binary collisions, N$_{bin}$, and the number of participants, N$_{part}$ as a function of \sqrts. 

\begin{eqnarray}
N_{bin} = (0.314 + 8.7e^{-4} \sqrt{s_{NN}} - 1.8e^{-6}  \sqrt{s_{NN}}^{2})  N_{part}^{4/3}
\end{eqnarray}

~

\begin{eqnarray}
N_{ch} = (0.5933 \xspace ln(\sqrt{s_{NN}}) - 0.4153) \times ((1-0.11) N_{part}/2+ 0.11 N_{bin})
\end{eqnarray}

\noindent  Given the particle/N$_{ch}$ ratio and N$_{ch}$ one can then estimate the mid-rapidity particle yields. A Blast-Wave parameterization is used to calculated the $\langle p_{T} \rangle$ where T$_{kin}$, the kinetic freeze-out temperature can be assumed to be independent of \sqrts and set to  0.1 GeV/c while the mean transverse velocity can be  estimated from:

\begin{eqnarray}
\langle \beta \rangle = (0.388 +  0.186\ln( 0.7928 N_{part} + 0.0129 N_{part}^{2}- 
			       3.31600e^{-5}  N_{part}^{3} + 
			       4.01681e^{-8}  N_{part}^{4})/N_{part} ) \nonumber \\			       
			       (0.05727  \ln(\sqrt{s_{NN}}) + 0.2933)/0.592     
\end{eqnarray}

\section{Summary and the Run Plan}

\noindent  For the first BES we propose to run a wide range of energies, from \sqrts= 5-39 GeV, to truly scan over the available collision energies. We have selected lower \sqrts values that give the greatest discovery potential for the CP as well as higher energies to cover the current gap between the SPS and RHIC. This will allow us to explore the questions about "turn on/off" effects of  signatures of partonic media such as the high and low  \pT ridges and constituent quark scaling of the elliptic flow. Table~\ref{Table:EventsNeeded} summarizes the measurements that drive the number of events we request at each collision energy.

\begin{table}[htb]
\caption{Estimate of number of events needed at each collision energy. The most statistics hungry measurements are listed. } 
\begin{center}
\begin{tabular}{|c|c|c|c|c|c|c|}
\hline
\sqrts / Number of events & 5 GeV& 7.7 GeV  &11.5 GeV  &17.3 GeV & 27 GeV & 39 GeV \\
\hline

  v$_{2}$ up to $\sim$ 1.5 GeV/c  &  0.3 M & 0.2 M & 0.1 M & 0.1 & 0.1 M &  0.1 M \\ \hline
 v$_{1}$ &  0.5 M & 0.5 M & 0.5 M & 0.5 & 0.5 M &  0.5 M \\ \hline
Azimuthally sensitive HBT &  4 M & 4 M & 3.5 M & 3.5 & 3 M &  3 M \\ \hline
PID fluctuations &  1 M & 1 M & 1 M & 1 M & 1 M &  1 M \\ \hline
Net proton kurtosis &  4 M & 4 M & 4 M & 4 M& 4 M &  4 M \\ \hline
Differential corr. and fluc. vs centrality &  4 M & 4 M & 4 M & 4 M& 4 M &  4 M \\  \hline
 N$_{q}$ scaling $\pi$, K, p, $\Lambda$ to $m_{T}-m_{0}$/N$_{q} \sim 2 GeV/c$ &  N/A & 6 M (marginal)& 5 M & 4.5 M & 4.5 M &  4.5 M\\  \hline
N$_{q}$ scaling $\phi, \Omega$  to $p_{T}$/N$_{q} \sim 2 GeV/c$ &  N/A & 56 M & 25 M & 18 M & 13 M &  12 M\\  \hline
R$_{cp}$   to $p_{T} \sim$ 4.5 GeV/c (17.3), 5.5 GeV/c (27), 6 GeV/c (39) &  N/A & N/A & N/A & 15 M & 33 M &  24 M\\  \hline
Un-triggered ridge &  N/A & 27 M & 13 M & 8 M & 6 M &  6 M \\  \hline
Local Parity Violation &  N/A & 4 M & 4 M & 4 M & 4 M &  44 M \\
\hline
\end{tabular}
\end{center}
\label{Table:EventsNeeded}
\end{table}

~

\noindent Rate estimates are based on a number of ingredients. Most important are the data from the Au+Au beam test at \sqrts= 9.2 GeV taken during Run8. In particular, these data allow a determination of the actual rate of events on which STAR can trigger and that lie within a range of vertex positions for which STAR can perform the desired physics analyses. Thus, rates can be estimated as the number of good physics events with less uncertainty of the usable fraction of the total luminosity. The test run was very short with the emphasis on demonstrating capability rather than tuning for the highest possible event rate. The BNL Collider Accelerator Division (CAD) staff have indicated a high degree of confidence that these rates can be increased by a factor of about 6 via improvements in injection efficiency and increasing the number of bunches in the machine. Additional tuning is likely to provide further incremental rate improvements. Another option being explored is continuous injection to fill the bunches and extend the beam lifetime. This is expected to increase the integrated rate by a factor of about 2. Very significant rate increases may be possible using electron cooling. Until the details of the beam loss mechanisms at these low energies are determined, the exact gain that could be achieved is unknown but factors of about an order of magnitude are considered very possible. To be very conservative, none of these further possible enhancements are included in the estimates presented below.

~

\noindent Again based on consultation with CAD~\cite{Satogata:2007aa}, it is estimated that the rates will scale up and down from the 9.2 GeV values by $\gamma^3$ up to the injection energy of and by roughly $\gamma^2$ above that. To cross-check these estimates, there are actual data rates taken by STAR during the early injection-energy run and also measurements of the injection energy luminosity under current operating conditions.
The  values typically quoted are total energy, so the $\gamma$ of each beam is found by dividing by 2 and then by the average nucleon mass for a gold nucleus. These factors are independent of energy and therefore, the rate below injection energy simply scales with the cube of the beam energy. Using the measured values from the test run increased by the conservative values listed above, the rates are the following: 
Below injection energy: 6.5$\times$10$^{-3}\times\left( \sqrt{s_{_{NN}}}\right ) ^3$ events/sec
At injection energy: 48 events/sec
Above injection energy: 12.6$\times$10$^{-2}\times\left( \sqrt{s_{_{NN}}}\right ) ^2$~events/sec
Note that, as mentioned above, these are estimates of the actual rate of events for which STAR can perform all of the desired physics analyses and are therefore not estimates of the total luminosity available from the accelerator. These estimates imply the following times required for 1 million events assuming 10 hours per day of beam.
5 GeV   0.8 evts/sec   35 days
10 GeV   6.4 evts/sec   4.3 days
20 GeV   50 evts/sec   0.6 days
40 GeV   200 evts/sec   0.14 days

~

\noindent A suggested run plan  is shown in Table~\ref{Table:RunPlan}. The length of the run at each energy is determined by the minimum required number of events to perform the detailed measurements discussed in the previous sections. The lower beam energies are specifically chosen to map out region around the "horn" in the K/$\pi$ ratio observed by the SPS experiments~\cite{Gazdzicki:2004ef}. All the selected energies allow collisions at both STAR and PHENIX; the lower \sqrts values are also mapped as closely as possible to those already provided at the SPS.

~

\begin{table}[htb]
\caption{A run plan for STAR assuming a 10 hour day.} 
\label{Table:RunPlan}
\begin{center}
\begin{tabular}{|c|c|c|c|c|}
\hline
$\sqrt{s_{NN}}$ (GeV) & $\mu_{B}$ (MeV) & Rate (Hz) & Event Count & Run Time  (days)\\
\hline
5.0 	&	550	&	0.8	&    Beam development	& 7 \\
7.7	& 	410	&	3	&	4M	& 36 \\
11.5	&	300	&	10	&	4M	 & 15\\
17.3	&	229	&	33	&	15M	  & 13 \\
27	&	151	&	92	&	33M	& 10 \\
39	&	112	&	190	&	24M	& 4\\
\hline
\end{tabular}
\end{center}
\end{table}

\noindent After  analysis of the first run period is mature we propose that a second scan is performed focussed more specifically on a few collisions energies. These energies and physics topics will be chosen to explore in more depth the most interesting regions found via this first scan. We propose two runs since the second run, more data at fewer energies, can take advantage of further luminosity upgrades proposed by the RHIC accelerator division.

~

 \noindent In summary there is great discovery potential for a  Beam Energy Scan at RHIC. The most exciting result would be the identification of a Critical Point. Equally interesting would be the proof that at lower \sqrts energies the transition to a QGP occurs via a first order phase transition. Either of these results would enable us to make precise entries on the Phase Diagram of QCD matter. Currently the most promising predictions for identifying these phenomena are critical point fluctuations of either \pT or conserved quantum numbers (baryon or strangeness), pair correlations or non-monatonic deviations in the energy dependence of flow characteristics. A scan will also provide results that would enable a more precise determination of the region where sQGP medium effects, such as constituent quark scaling of  elliptic flow, (dis)appear.

  ~

\noindent  Finally we have demonstrated that not only is STAR committed to performing a Beam Energy Scan but that the detector system is ideally suited for such an explorative endeavor.  With the run plans detailed above we can make significant improvements not only to the existing SPS data but also plan to extend the investigations into regimes not yet probed by any experiments, and/or to include unexplored physics observations.

\section*{ Acknowledgements}

\noindent We especially thank the RHIC CA-D for all their help getting the RHIC BES program underway. They enthusiastically picked up the challenge to run RHIC at low energies, in particular at sub-injection energies. The Run-10 BES could not have been success without their hard work. We also thank the RHIC Operations Group and RCF at BNL, the NERSC Center at LBNL and the Open Science Grid consortium for providing resources and support. This work was supported in part by the Offices of NP and HEP within the U.S. DOE Office of Science, the U.S. NSF, the Sloan Foundation, the DFG cluster of excellence `Origin and Structure of the Universe'of Germany, CNRS/IN2P3, STFC and EPSRC of the United Kingdom, FAPESP CNPq of Brazil, Ministry of Ed. and Sci. of the Russian Federation, NNSFC, CAS, MoST, and MoE of China, GA and MSMT of the Czech Republic, FOM and NWO of the Netherlands, DAE, DST, and CSIR of India, Polish Ministry of Sci. and Higher Ed., Korea Research Foundation, Ministry of Sci., Ed. and Sports of the Rep. Of Croatia, Russian Ministry of Sci. and Tech, and RosAtom of Russia.

\begin{footnotesize}
\bibliographystyle{} 
{\raggedright
\bibliography{INT2008-STAR-Latest}
}
\end{footnotesize}

\end{document}